# Individual and Contextual Variables of Cyber Security Behaviour

**An empirical analysis of national culture, industry, organisation, and individual variables of (in)secure human behaviour**

Submitted as part of the requirements for the award of the MSc in Information Security of the University of London

**Student:** Marten de Bruin (student number 190230551)
**Supervisor:** Konstantinos Mersinas

**Date of submission**: April 30, 2022

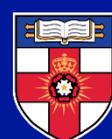

UNIVERSITY OF LONDON

# Table of contents

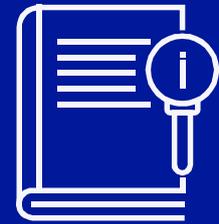

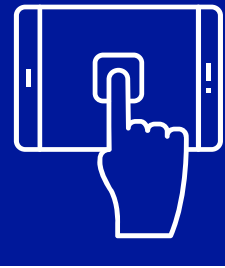

# Background to the research

Scan the QR code below to obtain additional information on the background to this research.

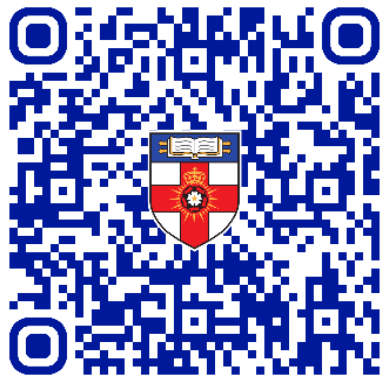

# Preface

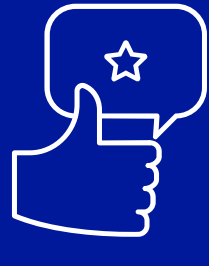

The dissertation you are about to read is the result of a long process of researching, analysing, discussing and writing. It was quite the journey and along the way there were quite a few people that supported me and enabled me to write this report. Therefore, I would like to start off by saying thanks to those that supported me.

First off, I want to start off by thanking my wife, as would any wise man, for her patience and support while I was busy working on this very interesting and enjoyable, but very time-consuming, project. Second, I would like to thank my employer for enabling me to pursue the MSc Information Security study which I am about to finish. Third, I would like to give a shout out to my supervisor, Konstantinos, for his support throughout this demanding process and his valuable review comments. Fourth, I would like to thank KnowBe4 for supporting my research, and Jacopo in particular.

Finally, I would like to close off this preface with a quote that closely connects with the origin and reason for this research:

> "*Amateurs hack systems,*
> *professionals hack people*"
>
> ~ **Bruce Schneier**

# Executive summary / abstract

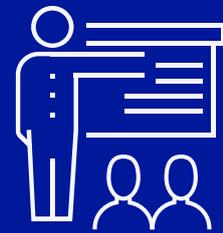

Data breaches are increasing and humans play an important role in reducing the likelihood and impact of data breaches. We identify limitations in our literature review which lead to the presumption that there is a focus in literature on technical aspects of information security and factors influencing the secure behaviour of individuals require additional research. These factors are both at the individual level and at the contextual level in which the individual is situated. At the contextual level we identify that national culture, industry type and organisational security culture play an important role in influencing individuals' secure behaviour. At the individual level we identify demographics (age, gender and level or urbanisation) and security-specific factors (security awareness, security knowledge and prior experience with security incidents) as important factors to influence individuals' secure behaviour.

This research has important implications for both research and practise. For research our findings fill a gap in literature and provide opportunities for further research. For organisations we provide valuable insights in what groups may be most likely to display insecure behaviour and what groups may require additional attention. Based on our insights, organisations may tailor their security training and awareness efforts (e.g. by building employee profiles), adapt their communication (e.g. of information security policies) based on national culture and/or gain an understanding of, and influence, their organisation's security culture in order to improve secure behaviour of employees. Additionally, organisations could consider if they may better leverage security incidents that occur to improve secure behaviour of employees. Finally, industry type is found to influence secure behaviour. Organisations may want to consider whether any factors at industry level (e.g. level of regulation or IT dependency) are negatively affecting secure behaviour within their organisation.

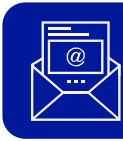

**Questions? Click here to email the author**

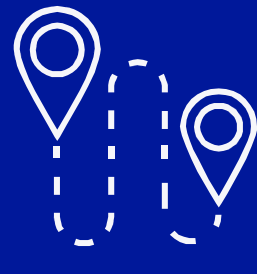

# Overview and navigation

For ease of navigation this document contains three elements that facilitate easy navigation throughout the document:

- Bold blue text indicates a cross reference that is clickable and, when clicked, navigates to the respective section. Note, however, that non-bold light blue text is not a cross reference, but indicates a reference to a source (included in the reference list in chapter 11).

- Sections within chapters (e.g. 1.1, 1.2 etc.) are indicated with blue bars which contain section jump arrows ( ⇧ and ⇩ ). When clicked, these arrows navigate to the previous or next section. For example, an arrow up in section 1.2 will navigate to section 1.1 whilst an arrow down will navigate to section 1.3.

- On the bottom of each page there is navigation pane that indicates, through a dark blue marking, the chapter the reader is reading. Additionally, the icons in this navigation pane can be clicked to navigate to the corresponding chapter.

Now that we've cleared this up, you can start reading in chapter 1 or use the navigation wheel below to quickly navigate to a chapter of choice.

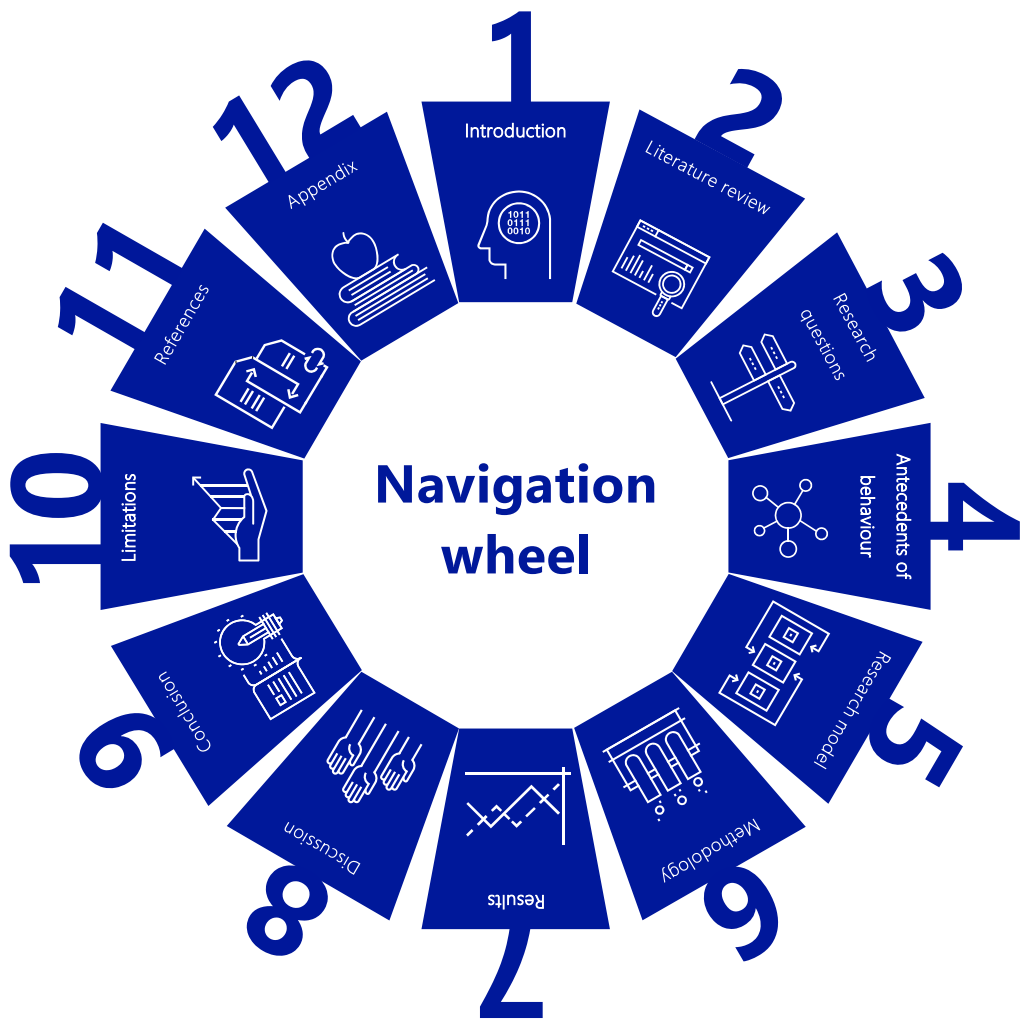

Navigation wheel

1 Introduction
2 Literature review
3 Research questions
4 Antecedents of behaviour
5 Research model
6 Methodology
7 Results
8 Discussion
9 Conclusion
10 Limitations
11 References
12 Appendix

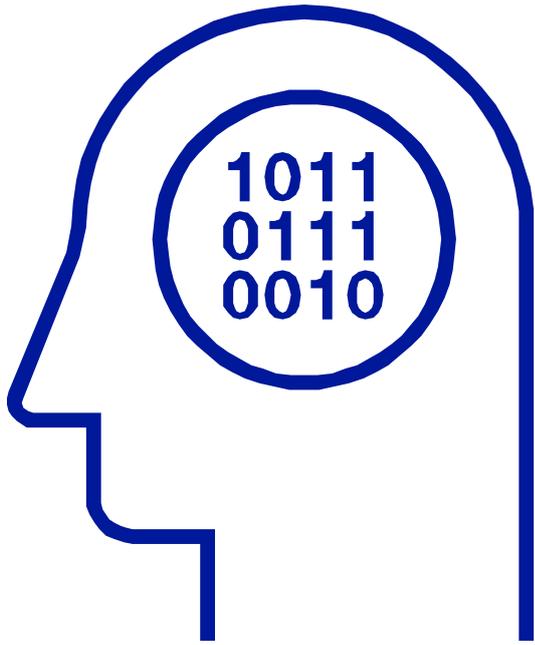

**1**

# Introduction

# 1   Introduction

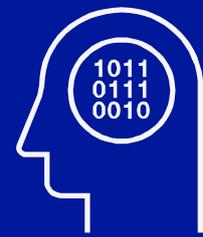

## 1.1   Cyber attacks and data breaches: trends and definitions

Cyber attacks have increased over the past years (Nallainathan, 2021). Cyber security incidents are reported on a daily basis (Abbiati et al., 2021). This trend is not surprising as people's reliance on systems, and their interconnection via the internet, is growing (Arachchilage and Love, 2014). The December 2021 Apache Log4j open source logging library vulnerability has once again shown the scale of impact that cyber attacks (and vulnerabilities related thereto) can have on organisations worldwide (Cooney, 2021). The European Union Agency for Cybersecurity (ENISA), which reports on the cyber threat developments, reported an increase of various cyber threats which resulted in a new record in data breaches reported in 2018 (ENISA, 2019). This increasing trend of cyber attacks and data breaches is confirmed by researchers (Khando et al., 2021 and Abbiati et al., 2021).

Before continuing it's important to establish a common understanding of what is meant when we talk about "cyber attacks" and "data breaches", and what the relation is between the two. For cyber attacks we use the definition proposed by the U.S. National Institute of Standards and Technology (NIST, 2012) since NIST is a prominent institute in the development of cyber security standards and guidance. NIST, 2012 defines cyber attacks as:

-   *"An attack, via cyberspace, targeting an enterprise's use of cyberspace for the purpose of disrupting, disabling, destroying, or maliciously controlling a computing environment/infrastructure; or destroying the integrity of the data or stealing controlled information."*

For data breaches we use the definition proposed by the International Organization for Standardization (ISO, 2015) since ISO, similar to NIST, is a prominent organisation which has provided a variety of commonly accepted standards (e.g. ISO27001). ISO, 2015 defines a data breach as:

-   *"Compromise of security that leads to the accidental or unlawful destruction, loss, alteration, unauthorized disclosure of, or access to protected data transmitted, stored or otherwise processed."*

Although closely related, a distinction between cyber attacks and data breaches (sometimes also referred to as (cyber) security breaches) is important. As noted by ENISA data breaches should be considered not as a cyber attack or cyber threat in itself, but a possible consequence of one of a variety of successful cyber attacks (ENISA, 2019). Therefore it could be said that data breaches are a possible outcome of different successful cyber attacks. However, other outcomes of a cyber attack, not covered by the definition of a data breach, are possible. An example is unavailability of data (when no data is destructed, lost, altered, unauthorised

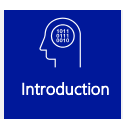 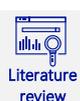 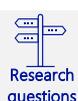 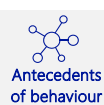 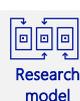 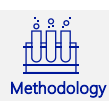 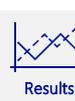 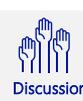 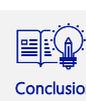 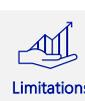 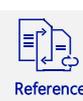 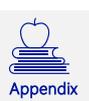

Introduction | Literature review | Research questions | Antecedents of behaviour | Research model | Methodology | Results | Discussion | Conclusion | Limitations | References | Appendix

disclosed or accessed), for example, resulting from a Denial of Services (DoS) attack. Hence, cyber attacks can have multiple outcomes one of which is a data breach.

## 1.2  The importance of data breaches in information security

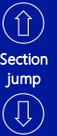

A commonly used concept in the field of information security is the "CIA triad" (Lundgren and Möller, 2019) which states that the primary information security goals are comprised of the protection of confidentiality (C), integrity (I) and availability (A) of information. Lundgren and Möller, 2019 provide the following definitions for each of these three:

- Confidentiality: *"property that information is not made available or disclosed to unauthorized individuals, entities, or processes"*;
- Integrity: *"property of accuracy and completeness"*;
- Availability: *"property of being accessible and usable upon demand by an authorized entity"*.

In addition to the traditional CIA triad, additional security services have been introduced over the years (Imran et al., 2021). Examples of such security services and their corresponding definitions are the following (Imran et al., 2021):

- Authorisation: "*ensure and verify that the user have the required control permissions or privilege to perform the operation or certain action*";
- Access control: "*a security mechanism to handle and grant access rights to only authorized entities*";
- Authenticity: "*deals with personal information or identification ... includes validating the incoming request against certain identifying credentials*";
- Non-repudiation: "*evidence to prove certain actions in order to ensure that it can't be repudiated later*";
- Inter-operability: "*ability of several systems to connect, exchange and share information with one another, without restrictions*".

When comparing the definitions of the information security services with the definition of data breaches it becomes apparent that data breaches can have a significant effect on information security by affecting one or more of these information security services. For example: destruction, loss and/or alteration of information (systems) may breach integrity, availability, non-repudiation and/or inter-operability. Similarly, unauthorized disclosure of, or access to, information may breach confidentiality, authorisation, access control and/or authenticity. The aforementioned illustrates that Information security (services) and data breaches are closely related and explains why data breaches are often mentioned in information security research (Chen et al., 2022, Abbiati et al., 2021, Ponemon Institute, 2021, Kam et al., 2020, Failla, 2020, Li et al., 2019, Bavel et al., 2019, Chua et al., 2018 and Menard et al., 2017).

The definition of data breaches we presented includes "Compromise of security that leads to the accidental or unlawful…". This reveals that there are two primary sources for data breaches: accidental and unlawful (malicious / deliberate). According to Ingham, 2018 the vast majority (88%) of data breaches (in the UK) are caused by human error (accidental) rather than cyber attacks (unlawful). Additionally, researches on cyber attacks by ENISA have revealed that in the majority of types of cyber attacks it is human vulnerabilities that are exploited (ENISA, 2017 and ENISA, 2019). Examples of such cyber attacks are social engineering and phishing (ENISA, 2017).

Regardless of whether a data breach is accidental or deliberate it is apparent that humans play a significant role in causing (and preventing) data breaches (White et al., 2017). This is confirmed by ENISA's estimate that "*about 77% of the companies' data breaches are due to exploitation of human weaknesses*" (ENISA, 2019). Various researches confirm the importance of the human aspect in data breaches and security incidents. Haeussinger and Kranz, 2013b estimate that 50-70% of information security incidents result from employee's intentional or accidental behaviour. McCormac et al., 2018 cite a research estimating that 95% of data breaches are accidental and the result of human error. Humaidi and Balakrishnan, 2015 cite a research stating a percentage of 80% of security incidents in organisations are the result from organisation employees. Although numbers vary, it is clear that humans can pose a significant vulnerability to information security and are a significant root cause to the occurrence of many data breaches (Khando et al., 2021, Humaidi and Balakrishnan, 2015, McCormac et al., 2018).

The identification of the human aspect as a primary vulnerability to information security has not gone unnoticed by cyber attackers. ENISA noted that in 2018 attack tactics have shifted increasingly towards malware type attacks which primarily use email and impersonation as a way of infecting information systems (ENISA, 2019). In doing so "*threats which abuse weaknesses in the people part are increasing*" (ENISA, 2019). Additionally, cyber attacks on humans are increasingly refined. Some examples are selective phishing and targeted attacks tailored to, for example, sectors or business functions (ENISA, 2019). A 2021 report by Ponemon Institute reveals that a focus on the human vulnerabilities in information security seems to pay off (Ponemon Institute, 2021). Amongst the top 5 costliest types of data breaches are mostly data breaches that are primarily resulting from insecure human behaviour such as business email compromise, phishing and social engineering (Ponemon Institute, 2021).

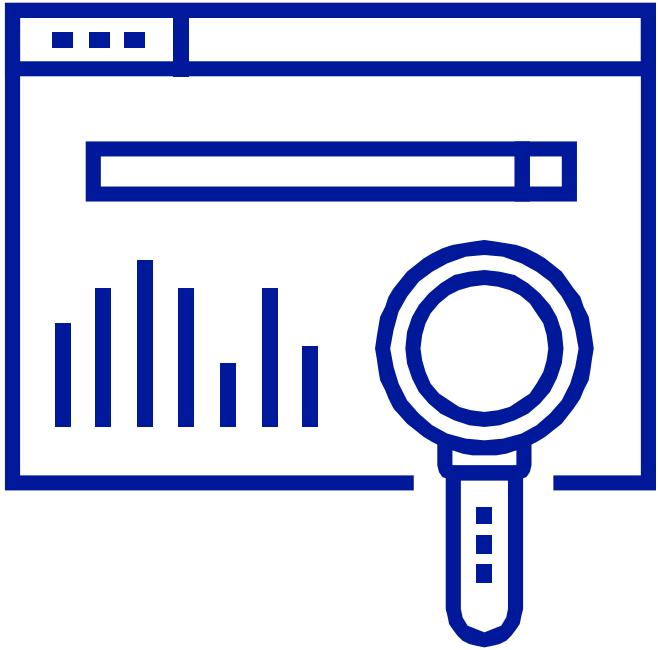

# Literature review

# 2   Literature review

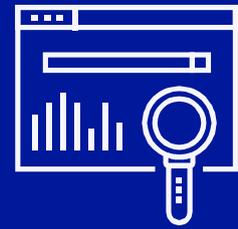

Much research is performed in the field of information security. Research can be broadly be categorised into technical research (e.g. strength of an encryption algorithm), organisational research (e.g. security awareness) or a combination of both. We start off our literature review by focusing on organisational research within the information security field. Based on this literature review we conclude that focus within the information security field is focused on technical aspects rather than human aspects of information security (**limitation 1**). Subsequently we focus on research on secure behaviour within organisational research and conclude that within organisational research limited research is performed on secure behaviour (**limitation 2**). Furthermore, we explore the performed research on secure behaviour in more detail by focusing on the way in which secure behaviour was measured. We conclude that, possibly inaccurate, indicators of secure behaviour are often researched rather than actual secure behaviour itself (**limitation 3** and **limitation 4**). Finally, we explore the antecedents to secure behaviour (factors influencing secure behaviour) that are incorporated in research and conclude that abstract antecedents are researched (**limitation 5**) and/or contextual antecedents are insufficiently taken into account (**limitation 6**). Based on our literature review of information security we argue that information security research suffers from six primary limitations with respect to the human factor in information security, and secure behaviour in particular. We will discuss each of these in the following sections.

## 2.1   Limitation 1: Focus on technical aspects rather than human aspects of information security

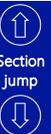

According to McCormac et al., 2018 it is commonly known and documented that data breaches "*cannot be prevented through the implementation of solely technical solutions*", although technical solutions can help reduce the risk and impact of breaches. For example the use of encryption can reduce the impact of a data breach (ENISA, 2019). Therefore, it is apparent that the human aspect has an important role in preventing data breaches and reinforcing the information security of organisations. But if this is the case, why does research focus so much on security through technical measures?

A review of literature on information security provides evidence that the focus on technical measures would benefit from a shift towards research focused on the human aspect of information security (Ng et al., 2009, Waly et al., 2012, Khando et al., 2021). Ng et al., 2009 state that "*technology solutions alone are not sufficient*" and "*security behavior of employees play an important role, and this calls for more research studying the factors that influence individual's decision to practice computer security*". Waly et al., 2012 noted that "*more attention needs to be paid to human, organisational and training factors if the problem of security breaches is to be managed effectively*". According to an extensive information security literature review by Khando et al., 2021

"*merely focusing on the technical aspects of information security is not enough as information security is multidisciplinary in nature and the human aspect plays a major part in it*". This focus on technical aspects is found not to be limited to just research, but also practice. Marks and Rezgui, 2009 state that "*most IS security managers pay more attention to technical aspects and solutions…, and tend to overlook socio-organizational issues such as the hazards caused by end users' lack of IS awareness*". This begs the question: what research has been performed on the human aspect of information security and why is it in need of expansion?

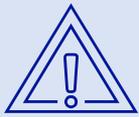 Considering the abovementioned we arrive at the following limitation:

- **Limitation 1**: Information security research should focus more on human aspects of information security rather than technical aspects.

## 2.2  Limitation 2: Limited research on secure behaviour

Besides the need for behavioural research the aforementioned shows that there are multiple ways of identifying the human aspect in information security. For example: human weaknesses or weaknesses in the people part (ENISA, 2019), security behaviour (Ng et al., 2009), human factor (Waly et al., 2012 and McCormac et al., 2018), human elements and human aspect (Khando et al., 2021). Essentially, all these descriptions refer to certain security behaviour being displayed either securely or insecurely. As such, within this research we adopt the term "secure behaviour" which is described by Menard et al., 2017 as:

- "*Behaviour that supports in reducing information security risks; for example refraining from sharing passwords, using weak passwords, clicking on unfamiliar links, and/or downloading e-mail attachments without proper scrutiny*"

As mentioned in limitation 1, there appears to be a lack of focus on the human aspect in information security research. When exploring the research performed on the human aspect of information security, it becomes apparent that within this research area the topic of secure behaviour would benefit from more attention. This lack of research on secure behaviour has increasingly been noted by scholars (Gratian et al., 2018, Bauer et al., 2017, Hickmann Klein and Mezzomo Luciano, 2016, Hovav and D'Arcy, 2012, Ng et al., 2009 and Herath and Rao, 2009). Research on secure behaviour and how it is influenced can be greatly beneficial as this could be used to maximise the "*effectiveness of SA training programs, while minimizing the costs of such efforts. Given the fact that overall information security spending continues to increase*" (Hanus et al., 2018).

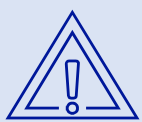 Considering the abovementioned we arrive at the following limitation:

- **Limitation 2**: Within information security research limited research is performed into secure behaviour of individuals.

Despite the need for more focus in information security research on the human aspect of information security, there are some related areas that have received considerable attention by scholars. One of these areas is information security awareness. Examples of research on information security awareness include McCormac et al., 2018, McCormac et al., 2017, Haeussinger and Kranz, 2013a, Kruger et al., 2011, Alfawaz, 2011 and Da Veiga and Eloff, 2010. Information security awareness is defined as (McCormac et al., 2018):

- *"The extent to which employees understand the significance of information security policies, rules and guidelines in their organisation and the extent to which their behaviour is congruent with these policies, rules and guidelines"*

The definition proposes that security awareness is linked to secure behaviour leading to the inherent assumption that security awareness influences secure behaviour. Indeed there is empirical evidence that security awareness is linked to secure behaviour in the sense that an increase in information security awareness tends to have a positive association with secure behaviour (Chua et al., 2018 and Choi et al., 2008). However, research on security awareness (effort) and its impact on behaviour yield several conclusions that suggest that security awareness is not a one-to-one predictor of secure behaviour:

- Khando et al., 2021 find various evidence in their literature review that supports the notion that information security awareness efforts are failing to change employees' secure behaviour for various reasons;

- Hanus et al., 2018 note that there could be factors that prevent individuals from translating security awareness into more secure behaviour;

- Karjalainen et al., 2013 reveal in their research that "*employees in different countries prefer different means for learning IS security behaviors. These culture-dependent reasons explain employees' IS security behavior and mean that different learning paradigms*";

- Talib et al., 2010 acknowledge that security awareness activities do not always translate to much safe behaviour.

Other researchers (Abawajy, 2014, Annetta, 2010 and Cone et al., 2007) arrive at the same conclusion, that is, that security awareness is not always a reliable predictor of secure behaviour. Gratian et al., 2018 does note, however, that security awareness efforts may be more effective in translating to secure behaviour when they are tailored to account for individual differences between users. Hence, more tailored and effective security awareness efforts may be more successful in creating security awareness and making the user translate this security awareness into secure behaviour.

Considering the abovementioned we arrive at the following limitation:

- <u>Limitation 3</u>: Within information security research information security awareness is used as an indicator of secure behaviour of individuals whilst this may not be an accurate indicator.

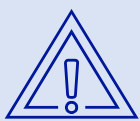

Besides literature that has focused on security awareness as an indicator of secure behaviour there has also been research in which secure behaviour was the object of interest. Although this may suggest that limitation 2 is invalid, this is not the case; an extensive literature review revealed that the vast majority of research focuses on the intention to behave securely rather than actual behaviour. Some examples include Hanus et al., 2018, Hovav and D'Arcy, 2012, Vance et al., 2020, Ameen et al., 2020, Crespo-Pérez, 2021, Kim and Han, 2019, Chua et al., 2018, Waly et al., 2012, Bulgurcu et al., 2010, Al-Omari et al., 2012, Haeussinger and Kranz, 2013b and Menard et al., 2018. Below we have included a summary of researches and the way that secure behaviour was measured in the research.

| Type behaviour variable | Measurement secure behaviour | Description | Articles |
|---|---|---|---|
| Behavioural intention | Information security policy compliance intention | A user's intention to comply with information security policies as measured by, for instance, statements or a rating in a survey. | - Bauer et al., 2017<br>- Hanus et al., 2018<br>- Hovav and D'Arcy, 2012<br>- Da Veiga, 2015<br>- Vance et al., 2020<br>- Ameen et al., 2020<br>- Crespo-Pérez, 2021<br>- Chua et al., 2018<br>- Humaidi and Balakrishnan, 2015<br>- Waly et al., 2012<br>- Bulgurcu et al., 2010<br>- Yazdanmehr and Wang, 2016<br>- Al-Omari et al., 2012<br>- Haeussinger and Kranz, 2013b<br>- Chen et al., 2022<br>- Safa et al., 2015<br>- Karjalainen et al., 2013<br>- Kam et al., 2015<br>- Menard et al., 2018<br>- Kim and Han, 2019<br>- Hu et al., 2012<br>- Zhang and Borden, 2020 |
| | Security measures intended | A user's intention to implement / perform certain security measures as measured by, for instance, a degree of likelihood that a certain security measure would be implemented indicated in a survey. | - Aurigemma and Mattson, 2018<br>- Vedadi et al., 2021<br>- Dinev and Hu, 2007<br>- Kumar et al., 2008 |
| | Information system misuse intention | A user's intention to misuse information systems measured by a user's intention to gain unauthorised access to and/or perform unauthorised modifications of computerised data as indicated in a survey. | - D'Arcy and Hovav, 2009 |

| Type behaviour variable | Measurement secure behaviour | Description | Articles |
|---|---|---|---|
| Actual secure behaviour | Observed security behaviour | Behaviour observed through monitoring of the user; e.g. a user's response to a phishing mail (clicking the e-mail, opening attachments etc.). | - Bavel et al., 2019<br>- Hanus et al., 2018<br>- Hickmann Klein and Mezzomo Luciano, 2016<br>- Arachchilage and Love, 2014 |
| | Security measures taken | Security measures taken by the user as measured, for example, through a survey of users. | - Tam et al., 2022<br>- Nallainathan, 2021<br>- Breitinger et al., 2020<br>- Gratian et al., 2018<br>- White et al., 2017<br>- Ng et al., 2009<br>- Talib et al., 2010<br>- Marks and Rezgui, 2009 |

The focus on behavioural intention rather than actual behaviour is confirmed in the literature review by Cram et al., 2019 and Chen et al., 2022. Although some research such as Waly et al., 2012 argue that intention is a predictor of behaviour, many scholars argue that research should focus on actual behaviour rather than behavioural intentions:

- Khando et al., 2021 argue that actual secure behaviour of an employee is influenced by other factors such as personality and culture;

- Vedadi et al., 2021 emphasise the necessity for examining users' actual behavior instead of behavioral intentions;

- Yazdanmehr and Wang, 2016 also note their approach to measure behaviour by measuring self-reported (intended) compliance behaviour as a limitation and suggest future studies to measure behaviour differently;

- Crossler et al., 2013 state that "*studies suggest that in an information security context, it is preferable to measure actual behaviors rather than intentions … measuring intentions rather than behaviors for the dependent variable is especially troubling because intentions do not always lead to behaviors. For example, a person has to fail to perform a protective behavior only once in order for a threat to manifest itself. When it comes to behaviors people should be employing to protect their computers from security threats, it is the behavior that matters and not the intention to perform the behavior*";

- Haeussinger and Kranz, 2013b acknowledge the limitations of using intention to measure behavior and recommend future research to measure actual behaviour;

- Al-Omari et al., 2012 mention, in accordance with Crossler et al., 2013, that "*a major limitation to measuring "intention" is that it is self-reported, and so some employees might not express their true intention for different reasons*";

- Bulgurcu et al., 2010 note their own approach to measure behavioural intention as a limitation and note that some respondents, having reported their behavioural intention themselves, may have "*concealed their true intentions because they perceived noncompliance as socially undesirable*";

- Talib et al., 2010 argue that future research is required on factors that influence an employee's actual secure behaviour in practice.

Another limitation of behavioural intentions is that these are often measured at an abstract level. For example by using generic statements such as "I intent to comply with the organisation's information security policy". Below are some examples of statements used in researches on behavioural intention for secure behaviour.

| Articles | Statements used to measure behavioural intentions |
|---|---|
| - Hanus et al., 2018<br>- Haeussinger and Kranz, 2013b<br>- Bulgurcu et al., 2010<br>- Wall et al., 2013 | 1. I intend to comply with the requirements of the information security policy of my organization in the future.<br>2. I intend to protect information and technology resources according to the requirements of the information security policy of my organization in the future.<br>3. I intend to carry out my responsibilities prescribed in the information security policy of my organization when I use information and technology in the future. |
| - Ameen et al., 2020<br>- Herath and Rao, 2009 | 1. I intend to follow the smartphone security policies and practices for using smartphones at work.<br>2. I intend to use the smartphone security technologies for using smartphones at work.<br>3. I intend to use common sense on good smartphone security practices for using smartphones at work. |
| - Zhang and Borden, 2020 | 1. I would follow the advice in the corporate message to reduce my risk of private information theft.<br>2. I would pay closer attention to similar instructions to prevent information theft.<br>3. I would take the steps necessary to protect myself to avoid private data theft. |
| - Chua et al., 2018 | 1. I intend to comply with the requirement of the rules and regulations for personal data protection of my company in future.<br>2. I intend to protect information and technology resources according to the requirement of the rules and regulations for personal data protection of my company.<br>3. I intend to carry out my responsibilities prescribed in the rules and regulations for personal data protection of my company.<br>4. I will comply with the rules and regulations for protecting personal data of my company to protect organization's Information Systems. |
| - Al-Omari et al., 2012 | 1. I intend to comply with the requirements of the ISP of my organization.<br>2. I intend to protect information resources according to the requirements of the ISP of my organization.<br>3. I intend to protect technology resources according to the requirements of the ISP of my organization.<br>4. I intend to carry out my responsibilities prescribed in the ISP of my organization when I use technology resources. |

Measuring behavioural intentions through such generic statements creates potential research problems. For example:

- Does the respondent sufficiently understand what is being meant by the statements?
- Even if the respondent understands the statements: is the respondent sufficiently aware of the requirements imposed by the information security policy and implications of non-compliance? This may be even more problematic as an information security policy may be comprised of / supported by a wide array of operational policies such as password policies and logical access control policies.

- Even if the respondent is aware of the various aspects to the information security policy: are the responsibilities that stem from these policies sufficiently clear?

As such, questions could be raised about the reliability of such statements for measuring intentions, let alone actual behaviour. Some scholars suggest the use of scenario's to make intentions more specific since these can provide more detailed explanations in regard to various information security policies such as password policies, remote access policies etc. (Hovav and D'Arcy, 2012 and Bulgurcu et al., 2010).

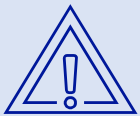
Considering the abovementioned we arrive at the following limitation:
- **Limitation 4**: Within information security research behavioural intention is used as an indicator of secure behaviour of individuals whilst this may not be an accurate indicator. Additionally, behavioural intention is often measured in an abstract manner casting doubt over research results.

## 2.5 Limitation 5: Abstract constructs as antecedents to secure behaviour

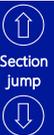

Theories in research on secure behaviour

The fifth limitation we argue that is present in information security literature is related to the constructs that are used as antecedents to secure behaviour (factors influencing secure behaviour). Much research on secure behaviour (or intentions related thereto) incorporates abstract constructs which are often at an individual level and hard to measure. These constructs are often based on, or derived from, prominent theories in other fields of research, such as psychology and sociology, and have found their way into information security research. Some examples of prominent theories in information security research (stemming from other fields of research) with a focus on human aspects are (Hanus et al., 2018 and Chen et al., 2022):

- Theory of Planned Behaviour (TPB): states that people's behaviour is formed based on three aspects: 1) someone's attitude and beliefs about certain behaviour, 2) someone's normative principles about other people's expectations and 3) someone's belief of control over what is done (Tam et al., 2022 and Bulgurcu et al., 2010);
- Theory of Reasoned Action (TRA): states that an intention can be seen as a direct determinant of behaviour and that it is influenced by someone's attitude towards performing a certain behaviour and subjective norms (which could be described as "*social pressures to perform certain behaviour*") (Alfawaz, 2011);
- Technology Acceptance Model (TAM): states that attitude towards technology is determined by "*perceived usefulness and perceived ease of use*" (Kumar et al., 2008);
- Protection Motivation Theory (PMT): states that "*people protect themselves based on two factors: threat appraisal and coping appraisal*" (Maddux and Rogers, 1983). Threat appraisal, on the one hand, is related to the assessment of the severity of a situation and assesses the seriousness of a situation. Coping appraisal, on the other hand, refers to how someone is responding to the assessed situation. Threat appraisal consists of the perceived severity of a threatening event and the perceived probability of the occurrence. Coping appraisal consists of someone's expectation that carrying out a certain action will successfully remove the threat (perceived response efficacy) and someone's belief in his/her ability to successfully execute the recommended actions (perceived self-efficacy) (Maddux and Rogers, 1983);

- <u>Deterrence Theory</u>: states that deterring measures can explain employees' compliance (e.g. sanctions and monitoring) (Karjalainen et al., 2013);
- <u>Rational Choice Theory</u>: states that an individual determines how to act based on a trade-off between the costs and benefits of the behavioural options (Bulgurcu et al., 2010).

<u>Demographics: easy to measure individual-level antecedents to secure behaviour</u>

Ng et al., 2009 argue that "*organisational security awareness programs and activities may not be attaining their desired effectiveness. There is a need to re-look at the design and implementation of security awareness campaigns so that users are effectively educated on threat information and skills to mitigate security threats, thereby improving the security climate of the organization.*" They propose that organisations use a customised approach to targeting security awareness messages at individuals and attempting to change their security behaviour. Gratian et al., 2018 agree by stating that "*security messaging, educational campaigns, and training might also be more effective if they appeal to individual differences in users*". Similarly, Talib et al., 2010 suggest organisations should use psychological profiling of employees in order to maximise the translation of security awareness to behaviour and the research by Karjalainen et al., 2013 reveals that "*employees in different countries prefer different means for learning IS security behaviors*".

However, when looking at the aforementioned theories it becomes apparent that within these theories there is a large focus on abstract constructs such as attitude, norms and beliefs, and threat assessment. All of which are hard to measure due to the abstract level and complexity of the construct. This makes it hard for research to be implemented in practice. For example: how will security awareness efforts (e.g. trainings) be tailored to specific users? Based on the abovementioned this would require organisations to have insights about their employees' attitude, norms and beliefs, and/or threat assessment. These are all difficult to measure, especially if these are to be measured for each individual separately. Although, Safa et al., 2015 propose to take into account individual attitudes and personalities when trying to influence secure behaviour we question the viability of such an approach.

We argue that measuring individual demographics (e.g. age and gender) would be a good starting point as these are acknowledged by literature to require attention and are easy to collect. For example, Chua et al., 2018 note that "*an enormous amount of research work has been carried out to understand the factors affecting* *employees' intention and decision to comply with information security policies. Nevertheless, there is a lack of study exploring the impact of demographics specifically on ISP compliance*". Additionally they argue that "*numerous investigations have also confirmed the influence of demographic differences on compliance behavior*" (Chua et al., 2018). This is emphasised by Gratian et al., 2018 who note that "*further evaluation of demographics is necessary*".

<u>Applying constructs which are specific to secure behaviour</u>

Another problem with using general concepts such as attitude, norms and beliefs is that these are not specific to information security, and more specifically: secure behaviour. This means that, although constructs such as attitude, norms and beliefs may influence secure behaviour, they may also influence other variables (e.g. security culture) which in turn influence secure behaviour. Measuring a clear relationship between such general constructs and secure behaviour may therefore be particularly challenging. As such, we argue that, in addition to easy to measure demographic variables (as mentioned above), using constructs which are more specific to secure behaviour might be beneficial for research on secure behaviour. Examples of such constructs are security awareness and security knowledge. Although there is much research on secure behaviour that already incorporates constructs like these, much of this research suffers from limitation 3 (using security awareness as an indicator of behaviour) (Haeussinger and Kranz, 2013a, Kruger et al., 2011, ENISA, 2017, McCormac et al., 2017, McCormac et al., 2018, Da Veiga and Eloff, 2010 and Alfawaz, 2011) or limitation 4 (measuring behavioural intention rather than actual behaviour) (Hanus et al., 2018, Hovav and D'Arcy, 2012 Chua et al., 2018, Humaidi and Balakrishnan, 2015, Bulgurcu et al., 2010, Yazdanmehr and Wang, 2016, Al-Omari et al., 2012, Haeussinger and Kranz, 2013b and Chen et al., 2022).

Contrary to demographics, data on constructs such as security awareness and security knowledge may not be as readily available. However, since these constructs are more specific to secure behaviour it's easier to interpret and link the results to secure behaviour. For example: a level of security awareness or security knowledge can more easily be tested against / linked to secure behaviour, but the same cannot be said for abstract constructs such as attitude, ethics or norms and values.

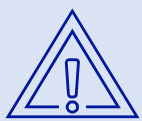 Considering the abovementioned we arrive at the following limitation:

- <u>Limitation 5</u>: Within information security behavioural research abstract antecedents are measured making translation to practice difficult.

## 2.6 Limitation 6: Limited focus on contextual antecedents to secure behaviour

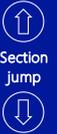

In the previous limitation we focused on individual-level antecedents to secure behaviour. However, there are many contextual variables that may have an effect on an individual's behaviour. For example, Dhillon et al., 2021 conclude that context influencing behaviour is complex and is comprised of a wide array of variables. However, measuring these complex variables leaves many challenges for measurement and interpretation of results. Although individual characteristics matter, individuals cannot be considered only in isolation of their environment, but it's important to account for the context in which individuals operate (Wiley et al., 2020, Vedadi et al., 2021 and Ameen et al., 2020).

Supported by Cram et al., 2019, we note that much research focusing on secure behaviour focuses on individual-level variables such as normative beliefs, attitude, self-efficacy and personal ethics. However behaviour is argued to be greatly influenced by contextual factors such as culture (Wiley et al., 2020, Miao et al., 2020, Menard et al., 2018 and Crossler et al., 2013). As such, research would benefit from research on the impact of such contextual factors on secure behaviour.

Of course, contextual factors would preferably be easy to measure through, for instance, demographics, but contrary to individual-level variables this is less of an issue due to scalability. For example, culture is measured for a larger group of individuals and as such conclusions are applicable to multiple individuals.

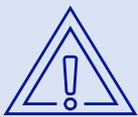
Considering the abovementioned we arrive at the following limitation:
- <u>Limitation 6</u>: Within information security behavioural research contextual antecedents, such as culture, appear to be overlooked.

## 2.7 Conclusions of the literature review

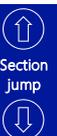

Based on the abovementioned limitations we argue that research would benefit from:
- An increased focus on human aspects instead of technical aspects (limitation 1);
- An increased focus on secure behaviour (limitation 2);
- Measuring actual behaviour rather than using indicative variables such as security awareness (limitation 3) and secure behavioural intentions (limitation 4);
- Incorporating antecedents at an individual level: both general demographic antecedents and antecedents specific to secure behaviour (limitation 5);
- Taking into account antecedents regarding the context in which a group of individuals manifests their behaviour, such as culture (limitation 6).

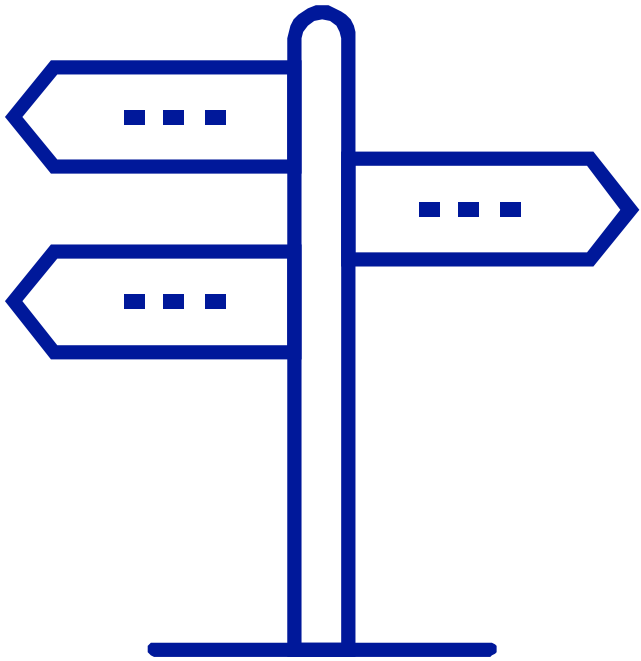

# Research questions

# 3   Research questions

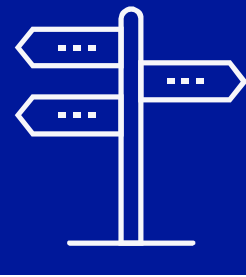

Based on the literature review we arrive at the following research questions:

**1** Which are the practically easy-to-measure antecedents at an individual level (demographics) that influence actual secure behaviour?

**2** Which are the secure behaviour specific antecedents that influence actual secure behaviour at an individual level?

**3** What are antecedents at the contextual level of the individual (e.g. culture) that influence secure behaviour?

**4** How do the antecedents at individual and contextual level influence secure behaviour?

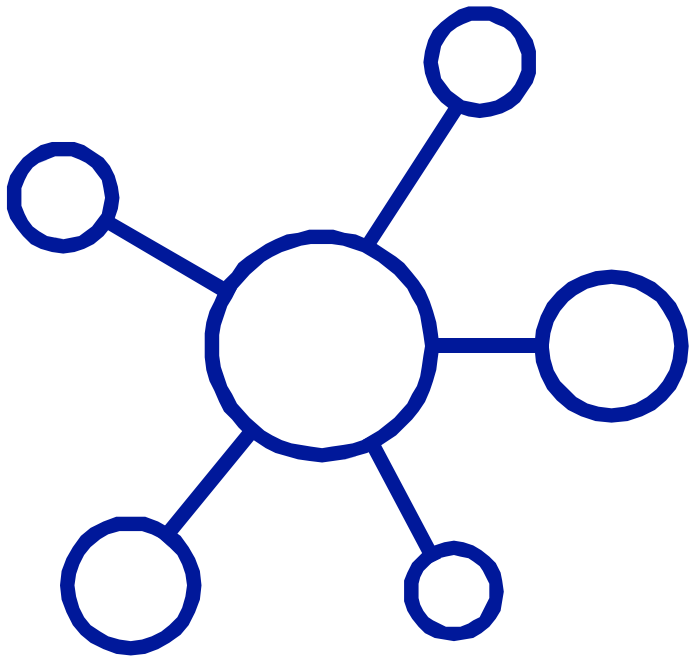

**4**

# Antecedents of secure behaviour

# 4 Antecedents of secure behaviour

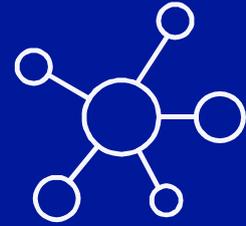

We discussed that secure behaviour is influenced by both individual-level variables (antecedents) and context-level variables of the context that the individual is operating in. But if this is the case: which variables should we consider? We start off by examining the context that individuals find themselves in and then we will continue with individual-level variables.

## 4.1 Context-level antecedents to secure behaviour

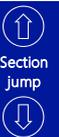

With respect to context-level antecedents to secure behaviour we argue that such antecedents are situated at least at the following three levels:

- National level
- Industry level
- Organisation level

We examine each of these and the respective antecedents in more detail in the following sections. This is in accordance with Burton-Jones and Gallivan, 2007 which state that "…*researchers tend to study information system usage at three levels: individual, group, and organization…*" and state that "… *some studies mention system usage at higher levels (e.g., at the industry or national level)*".

## 4.2 National level

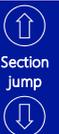

### 4.2.1 National culture

The country in which an individual operates could have great impact on the way an individual behaves (Bavel et al., 2019). A primary reason for this is that the national culture of the country from which the individual stems and/or resides in could have a great influence on secure behaviour. Indeed, much research has noted that the influence of culture on information security (Khando et al., 2021, Wiley et al., 2020 and Haeussinger and Kranz, 2013a), and secure behaviour in particular (Chen et al., 2022, Tam et al., 2022, Vance et al., 2020, Miao et al., 2020, Jaeger, 2018, Menard et al., 2018, ENISA, 2017, Yazdanmehr and Wang, 2016, Da Veiga, 2015, Crossler et al., 2013, Haeussinger and Kranz, 2013b and Karjalainen et al., 2013) should be further examined. Other scholars have argued that variances in information security awareness and behaviour may be explained by differences in country and culture (Bavel et al., 2019, Chua et al., 2018, Kruger et al., 2011). As an example,

ENISA, 2017 illustrates this by stating *"for example, people from the United States were more willing to share information online than people from India and the UAE"*.

## 4.2.2  Hofstede's cultural dimensions

When discussing national culture the work of Hofstede (Hofstede, 1980, Hofstede, 1983 and Hofstede, 1991) cannot be ignored as it dominates cultural research in general and this is no different for the research on culture in the field of information security (Tung and Verbeke, 2010 and Myers and Tan, 2002). Examples of information security research applying Hofstede's dimensions are Vance et al., 2020, Ameen et al., 2020, Aurigemma and Mattson, 2018, Menard et al., 2018, Kam et al., 2015, Flores et al., 2014, Karjaleinen et al., 2013, Hovav and D'Arcy, 2012 and Alfawaz, 2011. However, most research on the impact of culture on secure behaviour is criticised for being performed in a limited set of (developed / Western) countries (Khando et al., 2021, Ameen et al., 2020, Jaeger, 2018, Crossler et al., 2013 and Haeussinger and Kranz, 2013b). A review of literature on the role of culture on information security reveals that performed research on this topic has some limitations or characteristics that call for additional research in the field of national culture and its impact on secure behaviour in particular. Some examples of this are:

- Vance et al., 2020 apply Hofstede's cultural variables as moderating variable. Additionally, their research focuses on the behavioural intention rather than actual behaviour (limitation 4);
- Ameen et al., 2020 apply a part of Hofstede's cultural variables (power distance, uncertainty avoidance and collectivism). Additionally, their research focuses on the behavioural intention rather than actual behaviour (limitation 4) and limit their focus on smartphone security;
- Li et al., 2019 apply a part of Hofstede's cultural variables (long-term orientation). Additionally, their research focuses on the behavioural intention rather than actual behaviour (limitation 4);
- Connolly et al., 2019 apply a part of Hofstede's cultural variables (uncertainty avoidance, power distance and collectivism) and limit their research to Ireland and the United States of America. Additionally, this research does not specifically focus on measuring secure behaviour, although some elements of this are covered;
- Aurigemma and Mattson, 2018 apply a part of Hofstede's cultural variables (uncertainty avoidance) and state that *"future research can investigate …. other cultural dimensions"*. Additionally, their research focuses on the behavioural intention rather than actual behaviour (limitation 4);
- Menard et al., 2018 apply a part of Hofstede's cultural variables (collectivism) and they state that the further influence of this variable might also be further explored. Additionally, their research focuses on the behavioural intention rather than actual behaviour (limitation 4);
- Kam et al. , 2015 limit their research to two countries: the United States of America and South Korea. Although these countries are selected based on their cultural distance in terms of Hofstede, the impact of culture on behaviour is not explicitly tested. Additionally, their research focuses on the behavioural intention rather than actual behaviour (limitation 4);

- Flores et al., 2014 apply Hofstede's cultural variables as moderating effect, but focus on security knowledge sharing and do not research secure behaviour (limitation 2);

- Karjaleinen et al., 2013 limit their research to four countries (Finland, Switzerland, the UAE and China) and use a limited sample size (80 samples). Additionally, their research is qualitative and as such they do not test statistical associations;

- Hovav and D'Arcy, 2012 limit their research to two countries (United States of America and South Korea) which makes generalisation of the research outcomes difficult. Additionally, they focus on information system misuse intention rather than actual secure behaviour (limitation 4);

- Alfawaz, 2011 performs a single case study to support the research. Additionally, the research is limited to only Saudi Arabia and does not research secure behaviour (limitation 2).

Based on this, we find that within information security behavioural research Hofstede's cultural dimensions have not been considered as independent variables influencing secure behaviour (Vance et al., 2020, Flores et al., 2014), are only partly considered in research (Ameen et al., 2020, Li et al., 2019, Connolly et al., 2019, Aurigemma and Mattson, 2018 and Menard et al., 2018), are considered only for a limited set of countries (Kam et al., 2015, Karjaleinen et al., 2013, Hovav and D'Arcy, 2012 and Alfawaz, 2011) and/or secure behaviour was measured based on behavioural intention rather than actual behaviour (Vance et al., 2020, Ameen et al., 2020, Li et al., 2019, Aurigemma and Mattson, 2018, Menard et al., 2018, Kam et al., 2015 and Hovav and D'Arcy, 2012). As such, we conclude that when researching actual secure behaviour incorporating Hofstede's cultural dimensions (for a sufficiently large set of countries) would be beneficial for information security research.

The work of Hofstede has received much criticism over the years (Karjaleinen et al., 2013 and Walsham, 2002). One of the main criticisms is that the dimensions of Hofstede are simplistic and are not applicable to specific context issues like information security behaviour and compliance with information security policies (Karjaleinen et al., 2013 and Walsham, 2002). Other criticism is related to the way in which the cultural dimensions were measured; cultural dimensions were measured through a survey and amongst employees of a single company (Alfawaz, 2011). Most of this criticism is addressed (Alfawaz, 2011). For example, the Hofstede cultural dimensions model has been expanded over the years to make it more comprehensive (Hofstede, 1980, Hofstede, 1991 and Hofstede et al., 2010). Additionally, we note that Hofstede is still the most frequently applied way of measuring national culture (Tung and Verbeke, 2010 and Failla, 2020). As such we will apply Hofstede's cultural dimensions in this research. The initial work of Hofstede (Hofstede, 1980) contained four dimensions, but was later expanded with a fifth dimension of long-term orientation (Hofstede, 1991) and a sixth dimension of indulgence (Hofstede et al., 2010). Below we describe each of these variables and examine how these may affect secure behaviour.

## 4.2.2.1 Power distance (PDI)

Power distance was first introduced in Hofstede, 1980 and *"expresses the degree to which the less powerful members of a society accept and expect that power is distributed unequally. The fundamental issue here is how a society handles inequalities among people"* (Hofstede Insights, 2022). Power distance is one of the four initial cultural dimensions proposed by Hofstede and is therefore amongst the most often used cultural dimensions in cultural research in information security. Within the examined research we noted that power distance is mentioned as both a (potentially) moderating variable and independent variable.

Moderating variable

Vance et al., 2020 argue that high power distance moderates negatively between sanctions and a user's intention to violate information security policy (insecure behaviour). In other words they argue that sanctions have an increasing positive impact on secure behaviour the higher the power distance becomes. One argument here is that one of the aspects of cultures with a higher power distance is the accepted use of sanctions. This is because punishments to enforce a leader's wishes are acceptable behaviour (Vance et al., 2020). Additionally Vance et al., 2020 cite a research which finds that *"individuals from cultures with a high level of power distance were more likely to accept abusive supervision, which includes the use of formal sanctions"*.

Hovav and D'Arcy, 2012 applied power distance as a moderator between age and information system misuse intention (insecure behaviour). They argue that in countries with a high power distance age will have a positive association with IS misuse intention (negative association with secure behaviour). The argument here is that in a culture with a higher power distance age is often translated to seniority and senior people receive lighter punishments given their social status.

Independent variable

Ameen et al., 2020 researched behavioural intention towards smartphone security and noted that high power distance has a positive effect on secure behaviour. The reasoning is somewhat different from Vance et al., 2020 in the sense that they argue that employees *"may be more influenced by the views and opinions of their managers as they espouse high power distance cultural values"*. The argument that power distance positively impacts secure behaviour is supported by other scholars (Crespo-Pérez, 2021, Miao et al., 2020 and Connolly, et al., 2019). Alfawaz, 2011 agrees that power distance may positively impact secure behaviour since employees *"will take more from their managers"*. However, he also notes that high power distance could also negatively impact secure behaviour since this could also mean *"less involvement of employees in decision-making which has been linked to information system failure"* and reduced accountability since management is considered responsible for decisions and actions.

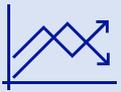

> Hypothesis
>
> Considering the majority of the examined research predicts a positive association between power distance and secure behaviour we adopt the following hypothesis:
>
> - **Hypothesis 1**: Power distance will have a positive association with secure behaviour.

## 4.2.2.2 Individualism vs. collectivism (INV)

Individualism vs. collectivism was first introduced in Hofstede, 1980. Individualism is defined as "*a preference for a loosely-knit social framework in which individuals are expected to take care of only themselves and their immediate families*" (Hofstede Insights, 2022). Collectivism on the other hand represents "*a preference for a tightly-knit framework in society in which individuals can expect their relatives or members of a particular ingroup to look after them in exchange for unquestioning loyalty*" (Hofstede Insights, 2022). Individualism vs. collectivism is one of the four initial cultural dimensions proposed by Hofstede and is therefore amongst the most often used cultural dimensions in cultural research in information security. Within the examined research we note that individualism vs. collectivism is mentioned as both a (potentially) moderating variable and independent variable.

Moderating variable

Vance et al., 2020 argue that individualism vs. collectivism moderates negatively between sanctions and intention to violate the information security policy. They state that the higher a culture scores on collectivism the lower the intention to violate the information security policy (the more secure the behaviour is). Their argument is that in collectivistic culture shame, sanctions and moral beliefs have a strong impact on secure behaviour. Additionally, they argue that to individuals in collectivistic cultures the "*norms of the collective body have greater salience, and, these individuals have been found to be more compliant with these norms*" (Vance et al., 2020).

Independent variable

Ameen et al., 2020 researched behavioural intention towards smartphone security and noted that collectivism has a positive effect on secure behaviour. They argue that "*individuals with an espoused collectivist culture tend to work together and have a higher level of awareness of security issues related to their use of technology. Individuals with an espoused collectivist culture tend to define themselves in terms of their relationships and social groups and avoid behaviours that cause social disruption*" (Ameen et al., 2020). Some other scholars (Crespo-Pérez, 2021 and Miao et al., 2020) agree that collectivistic cultures are more likely to display secure behaviour / refrain from counterproductive behaviour.

However, Menard et al., 2018 argue that collectivistic cultures are more likely to not protect information (display insecure behaviour). Their reasoning is that collectivistic cultures are characterised by sharing and this also applies to sharing of information. Additionally, they note that personal accountability is lowered in collectivistic cultures leading to a reduced incentive to behave securely (Menard et al., 2018). They go on to state that "*collectivists have been repeatedly shown to give up control, subordinate goals, and personal ambitions for what they believe the larger group wants*". However, this raises the question: if the group agrees on the importance of secure behaviour wouldn't this mean that collectivistic cultures can indeed have a positive influence on secure behaviour? Connolly, et al., 2019 agrees with the argument that collectivistic cultures are more likely to display insecure behaviour (negatively impact secure behaviour) since the tolerance for mistakes is higher in such countries due to lower personal accountability.

Alfawaz, 2011 agrees that collectivism may have a twofold influence on secure behaviour. In accordance with Crespo-Pérez, 2021, Ameen et al., 2020 and Miao et al., 2020 he states that in collectivistic cultures individuals are more likely to follow information security policies (display secure behaviour). On the other hand he agrees with Menard et al., 2018 stating that indeed in collectivistic cultures information is more easily shared leading to less secure behaviour. As such the influence of the individualism vs. collectivism dimension is twofold.

Hovav and D'Arcy, 2012 argue that the influence of individualism vs. collectivism on secure behaviour is similar to a parabola graph in the sense that individuals in a highly individualistic culture have a higher perceived severity of sanction (greater concern over severity of punishment as it's carried by an individual, but threat certainty is lower due to personal accountability) and individuals in a highly collectivistic culture have a higher perceived certainty of sanctions (severity is considered lower since it is spread over the group, but threat of embarrassment is much higher since it is carried by the group).

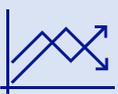

Hypothesis

Considering the aforementioned discussion there seem to be two sides to the individualism vs. collectivism dimension. As such we arrive at the following hypotheses:

- **Hypothesis 2a**: Individualism will have a negative association with secure behaviour.
- **Hypothesis 2b**: Individualism will have a positive association with secure behaviour.

### 4.2.2.3 Uncertainty avoidance (UAI)

Uncertainty avoidance was first introduced in Hofstede, 1980 and "*expresses the degree to which the members of a society feel uncomfortable with uncertainty and ambiguity. The fundamental issue here is how a society deals with the fact that the future can never be known: should we try to control the future or just let it happen?*" (Hofstede Insights, 2022). Uncertainty avoidance is one of the four initial cultural dimensions proposed by Hofstede and is therefore amongst the most often used cultural dimensions in cultural research in information security. Within the examined research we noted that uncertainty avoidance is mentioned as both a (potentially) moderating variable and independent variable.

<u>Moderating variable</u>

Vance et al., 2020 argue that uncertainty avoidance moderates negatively between sanctions and intention to violate the information security policy. They state that the higher a culture scores on uncertainty avoidance the lower the intention to violate the information security policy (the more secure the behaviour is). Their argument is that in cultures with a high uncertainty avoidance individuals will be "*more responsive to the threat of formal and informal sanctions, as well as to shame*" leading to more secure behaviour (Vance et al., 2020).

<u>Independent variable</u>

Similar to Vance et al., 2020 various scholars (Ameen et al., 2020, Aurigemma and Mattson, 2018, Crespo-Pérez, 2021, Miao et al., 2020, Connolly, et al., 2019 and Alfawaz, 2011) agree that a culture characterised by a higher uncertainty avoidance is expected to display securer behaviour. For example, Aurigemma and Mattson, 2018 argue that the desire to avoid uncertainty is translated into an increased protection motivation (Protection Motivation Theory (PMT)). Alfawaz, 2011 argues that individuals in culture that score high on uncertainty avoidance are more likely to apply secure settings and listen to expert when it comes making decisions on how to behave.

<u>Hypothesis</u>

Considering the aforementioned there seems to be a consensus regarding the impact of uncertainty avoidance on secure behaviour. As such we arrive at the following hypothesis:

- **Hypothesis 3**: Uncertainty avoidance will have a positive association with secure behaviour.

#### 4.2.2.4 Masculinity vs. Femininity (MAS)

Masculinity vs. femininity was first introduced in Hofstede, 1980. Masculine cultures are characterised by "*a preference in society for achievement, heroism, assertiveness, and material rewards for success. Society at large is more competitive*" (Hofstede Insights, 2022). Feminine cultures, on the other hand, are characterised by "*a preference for cooperation, modesty, caring for the weak and quality of life. Society at large is more consensus-oriented*" (Hofstede Insights, 2022). Although it is one of the four initial cultural dimensions proposed by Hofstede, masculinity vs. femininity appears to have been limitedly applied in cultural research in information security.

Although some research and arguments have been made on the impact of the masculinity vs. femininity dimension on secure behaviour there seems to be no consensus. Crespo-Pérez, 2021, on the one hand, argues that the more masculine a culture the more likely it is to display secure behaviour (comply with an information security policy) since such cultures are more focused on results / success rather than feelings. Miao et al., 2020, on the other hand, argue that the more feminine a culture the more likely it is to display secure behaviour (refrain from counterproductive / harmful behaviour). Ameen et al., 2020 argues, similar to the argument made

Introduction | Literature review | Research questions | Antecedents of behaviour | Research model | Methodology | Results | Discussion | Conclusion | Limitations | References | Appendix

by Hovav and D'Arcy, 2012 on individualism vs. collectivism, that the influence of masculinity vs. femininity is twofold. They argue that in masculine cultures "*individuals are more likely to avoid failure in their job if any security attack occurs due to their actions*" and in feminine cultures "*employees follow security procedures as they care for their organisations and other employees since they are more people-oriented*" (Ameen et al., 2020).

Hypothesis

Considering the aforementioned discussion there seem to be two sides to the masculinity vs. femininity dimension. As such we arrive at the following hypotheses:

- **Hypothesis 4a**: Masculinity will have a positive association with secure behaviour.
- **Hypothesis 4b**: Masculinity will have a negative association with secure behaviour.

## 4.2.2.5 Long-term orientation vs. short-term orientation (LTO)

Long-term orientation vs. short-term orientation is a dimension that was later added in Hofstede, 1991 to the initial four dimensions presented in Hofstede, 1980. Long-term oriented cultures "*encourage thrift and efforts in modern education as a way to prepare for the future*" (Hofstede Insights, 2022). Long-term oriented cultures live with a focus on the future. Short-term oriented cultures, on the other hand, are more focused on the past or present and "*prefer to maintain time-honoured traditions and norms while viewing societal change with suspicion*" (Hofstede Insights, 2022). Within the examined research we noted that long-term orientation vs. short-term orientation is mentioned as both a (potentially) moderating variable and independent variable.

Moderating variable

Hovav and D'Arcy, 2012 argue that in cultures that score higher on the long-term orientation dimension age is positively correlated with information system misuse intention (age is negatively correlated with secure behaviour). They argue that in long-term oriented cultures seniority is for an important part determined by age, and senior people tend to have higher privileges / their behaviour will be tolerated more which results in them being less inclined to follow information security rules. As such long-term oriented cultures in itself do not necessarily invoke less secure behaviour, but the combination with a higher age might.

Independent variable

Many other scholars (Crespo-Pérez, 2021, Miao et al., 2020, Kim and Han, 2019 and Li et al., 2019) argue that cultures that score higher on long-term orientation correlate positively with secure behaviour / refraining from counterproductive work. The main argument is that employees with a high degree of long-term orientation "*have high consideration of future consequences are more likely to engage in prosocial behaviour*" (Li et al., 2019). Additionally, it is argued that "*costs of compliance are experienced in the present*" and "*the costs of noncompliance are experienced in the future, after it is detected*" (Kim and Han, 2019). As such, for short-term oriented individuals the costs of compliance in the present are more likely to outweigh the costs of noncompliance in the future and they are less likely to follow information security policies (display secure behaviour) due to these present costs of compliance.

**Hypothesis**

Considering the aforementioned there seems to be a consensus regarding the impact of long-term orientation vs. short-term orientation on secure behaviour. As such we arrive at the following hypothesis:

- **Hypothesis 5**: Long-term orientation will have a positive association with secure behaviour.

### 4.2.2.6 Indulgence vs. restraint (IND)

Indulgence vs. restraint is, as of time of writing, the latest addition to Hofstede's cultural dimensions and was introduced in Hofstede et al., 2010. Cultures that are more closely related to indulgence allow "*relatively free gratification of basic and natural human drives related to enjoying life and having fun*" (Hofstede Insights, 2022). Restraint cultures, on the other hand, represents societies which suppress "*gratification of needs and regulates it by means of strict social norms*" (Hofstede Insights, 2022). Cultures in which control over desires and impulses is weak are called indulgent whilst culture in which this control is strong are called restraint. Due to the relative recent addition of this cultural dimension (at the time of writing) there appears to be limited research on the impact of culture on secure behaviour which incorporates this variable.

A research by Miao et al., 2020 argues that cultures that score high on restraint correlate positively with secure behaviour. The reason for this is that indulgent cultures permit more freedom for enjoyment and having fun which leads to a relatively weak control of desires whilst restraining cultures are more suppressing of human needs and regulate via social norms which leads to a strong control of desires and impulses (Miao et al., 2020). This is said to also translate into secure behaviour (refraining from counterproductive / harmful behaviour), because individuals do not prioritise their own needs over those of others (e.g. those of information security requirements).

**Hypothesis**

Considering the aforementioned we arrive at the following hypothesis:

- **Hypothesis 6**: Indulgence will have a negative association with secure behaviour.

### 4.2.3 Meyer's cultural dimensions

Earlier we noted the criticism to Hofstede's approach of measuring culture. An example of this criticism is Hofstede's simplistic way of measuring culture (Walsham, 2002). Another potential limitation of Hofstede's research is the age. The majority of Hofstede's approach to measuring culture was developed in the early 1980s (Hofstede, 1980). Culture is said to gradually change over time (Da Veiga and Eloff, 2010). Additionally, Tarabar, 2019 find that factors such as economic development have impact on, and change, national culture in terms of Hofstede's dimensions. This raises the question as to whether Hofstede's (initial) dimensions are still as valid as they were back when they were measured.

Meyer, 2014 proposes an alternative approach to measuring culture. She proposes seven dimensions to measure national culture. Her research on culture, as published in her book "The Culture Map" (Meyer, 2014), is well received and has been cited and applied in research (Shandler et al., 2021, Dewi and Adiarsi, 2020, Masai and Zanni-Merk, 2016 and Tung and Verbeke, 2010). However, since her work at the time of writing is very limitedly applied in research there are insufficient grounds to apply Meyer's work as a primary method for measuring the impact of national culture on secure behaviour. Regardless, due to the criticism on Hofstede's work we will include Meyer's cultural dimensions in our research model and analysis for exploratory purposes. As mentioned in the introduction of this chapter, culture is argued to potentially have an impact on secure behaviour (Chen et al., 2022, Tam et al., 2022, Vance et al., 2020, Miao et al., 2020, Jaeger, 2018, Menard et al., 2018, ENISA, 2017, Yazdanmehr and Wang, 2016, Da Veiga, 2015, Crossler et al., 2013, Karjalainen et al., 2013). As such we expect Meyer's cultural dimensions to have an association with secure behaviour. In the next sections we describe the dimensions that are at the core of Meyer's culture research (Meyer, 2014) and discuss the possible impact that these cultural dimensions may have on secure behaviour. Due to the very limited application of Meyer's variables in research we will not discuss the way in which each of the dimensions is considered in research (e.g. independent, moderating, control), but (similar to Hofstede's variables) consider all variables to be potential independent variables to secure behaviour.

### 4.2.3.1 Communicating

Meyer's first dimension is "Communicating": a scale ranging from low-context cultures to high-context cultures. "*In low-context cultures good communication is precise, simple, explicit, and clear. Messages are understood at face value. Repetition is appreciated for purposes of clarification, as is putting messages in writing. In high-context cultures, communication is sophisticated, nuanced, and layered.*" (Meyer, 2014). Communication is found to be an important aspect in increasing information security policy compliance (making behaviour more secure) (Barlow et al., 2018). According to Rantao and Njenga, 2020 "*an ambiguous task may cause conflicting interpretation because people lack the necessary information to process such a task*". Thus, communication plays an important role in making people understand what information is relevant to perform a certain task (e.g. what is required to display secure behaviour).

However, a balance must be struck between appropriate communication and a communication overload. On the one hand, a high number of information security policies is said to create more uncertainty and negatively impact compliance (Rantao and Njenga, 2020). On the other hand, according to Rantao and Njenga, 2020, clear messages can be preferred for certain tasks; e.g. certain information security policy areas (e.g. password policies for selecting secure passwords). According to Goo et al., 2014, communication with employees is an important interaction that contributes to creating an information security climate that positively affects employees' compliance with information security policies (display more secure behaviour). Similarly, Puhakainen and Siponen, 2010 note that communication, and way of communicating, can have a positive impact on employee behaviour in terms of information security. The opposite is said to be equally true; a lack of communication is said to contribute to development of bad habits and behaviour (Waly et al., 2012).

As described above, cultures with low-context communication generally communicate more precise, simple, explicit, and clear. Additionally, in low-context communication cultures repetition of communication is appreciated for clarification. As a result of the clarity and repetition of communication in such cultures, we expect communication in such cultures to be less ambiguous. Resulting from this, individuals are expected to have a better understanding of what is expected of them and are more likely to possess over the necessary information required to behave securely.

Hypothesis

Considering the aforementioned we arrive at the following hypothesis:

- **Hypothesis 7**: Low-context communication cultures (low score on Communicating) will have more secure behaviour than high-context communication cultures (high score on Communicating). As such a negative association is expected between communicating and secure behaviour.

### 4.2.3.2 Evaluating

Meyer's second dimension is "Evaluating": a scale ranging from direct negative feedback to indirect negative feedback. "*This scale measures a preference for frank versus diplomatic negative feedback. Evaluating is often confused with Communicating*" (Meyer, 2014). Feedback is said to be an important mechanism that supports security awareness (Bauer et al., 2017) and information security policy compliance / secure behaviour (Jaeger, 2018). According to coping theory, a positive feedback loop for information security behaviour incorporates "*two cognitive processes: threat appraisal (primary) and coping appraisal (secondary)*" (Mejias, 2012). As such, feedback can be useful for improving both threat identification and the way that these threats are dealt with. Similarly, according to Waly et al., 2012, a lack of feedback is said to lead to the development of bad habits (insecure behaviour).

Feedback can be beneficial in multiple ways. Feedback can be useful towards individuals to set clear expectations and provide individuals with the necessary information to behave securely. As a simple example: an application could show a pop-up with password guidance if an individual has selected a password with poor strength. However, feedback can also be valuable from individuals towards the information security policy maker. According to Waly et al., 2012, "*user feedback on policies and procedures is essential to improve their effectiveness*". Feedback can also have a deterring effect; for example, Straub and Welke, 1998 argue that feedback through "*on-going dissemination of security actions taken and policies deployed*" can deter potential information security policy abusers by making them aware of the consequences of such abuse. Therefore, it is clear that feedback can have an important role in increasing secure behaviour.

When considering the manner in which feedback is provided: according to Meyer's evaluating scale feedback can provided both frankly and diplomatically. On the one hand (similar to communicating), frank feedback may set clearer expectations than diplomatic feedback and may therefore make behaviour more secure. On the other hand, frank feedback may reduce an individual's self-confidence with respect to information security as a result of which behaviour is less secure. We discuss the impact of self-confidence with respect to information security in section 4.5.3.2. As a result of this we predict that evaluating may be both positively (diplomatic feedback leading to more secure behaviour) and negatively (frank feedback leading to more secure behaviour) associated with secure behaviour.

<u>Hypothesis</u>

Considering the aforementioned we arrive at the following hypotheses:

- **Hypothesis 8a**: Cultures in which feedback is provided frankly (low score on Evaluating) will have more secure behaviour than cultures in which feedback is provided diplomatically (high score on Evaluating). As such a negative association is expected between evaluating and secure behaviour.
- **Hypothesis 8b**: Cultures in which feedback is provided diplomatically (high score on Evaluating) will have more secure behaviour than cultures in which feedback is provided frankly (low score on Evaluating). As such a positive association is expected between evaluating and secure behaviour.

### 4.2.3.3 Leading

Meyer's third dimension is "Leading": a scale ranging from egalitarian to hierarchical. "*This scale measures the degree of respect and deference shown to authority figures, placing countries on a spectrum from egalitarian to hierarchical. The Leading scale is based partly on the concept of power distance*" (Meyer, 2014). This dimension is similar to Hofstede's power distance dimension (Hofstede, 1980). As such, for our discussion of this variable we refer to section 4.2.2.1 and we arrive at a hypothesis that is similar to the one with respect to Hofstede's power distance dimension.

<u>Hypothesis</u>

Considering the aforementioned we arrive at the following hypothesis:

- **Hypothesis 9**: Hierarchical cultures (high score on Leading) will have more secure behaviour than egalitarian cultures (low score on Leading). As such a positive association is expected between leading and secure behaviour.

### 4.2.3.4 Deciding

Meyer's fourth dimension is "Deciding": a scale ranging from consensual to top-down. "*This scale measures the degree to which a culture is consensus-minded.*" (Meyer, 2014). Cultures which are not consensus-minded are said to be top-down; decisions are made at a higher level and flow "down" (e.g. towards those that need to execute the decisions made). This dimension is closely linked to Hofstede's power distance dimension (Hofstede, 1980) and Meyer's leading dimension in the sense that in cultures with a high power distance decision-making is more often made top-down. As such, for our discussion of this variable we refer to section 4.2.2.1 and we arrive at a hypothesis that is similar to the one with respect to Hofstede's power distance dimension and Meyer's leading dimension.

Hypothesis

Considering the aforementioned we arrive at the following hypothesis:

- **Hypothesis 10**: Top-down decision-making cultures (high score on Deciding) will have more secure behaviour than consensus-minded cultures (low score on Deciding). As such a positive association is expected between deciding and secure behaviour.

### 4.2.3.5 Trusting

Meyer's fifth dimension is "Trusting": a scale ranging from task-based to relationship-based. "*In task-based cultures, trust is built cognitively through work. If we collaborate well, prove ourselves reliable, and respect each other's contributions, we come to trust each other. In a relationship-based society, trust is a result of weaving a strong affective connection. If we spend time laughing and relaxing together, get to know each other at a personal level, and feel a mutual liking, and then come to trust each other.*" (Meyer, 2014). Trust is a mechanism that is closely related to secure behaviour. The relation between trust and secure behaviour can be viewed as a multilateral relation. What we mean by this is that trust can be present from management towards employees, from employees towards management, and from one employee towards another employee or others.

Employee towards management: Karjalainen et al., 2013 argue that the way in which employees' security behaviour is influenced may affect trust. For example, inappropriately influencing security behaviour of employees by management may negatively affect employee trust in management. The opposite is equally true: Jalali et al., 2020 state that "*trust in the management reduces the risk that employees perceive security policies as a sign of management distrust in them and their abilities*". As such, employee trust towards management increases the likelihood that secure behaviour is displayed.

Management towards employee: Li et al., 2019 argue that people care whether they are viewed by others (e.g. management) as trustworthy or not and that such trust may in turn elicit securer intentions and behaviour. As such, when employees feel trusted by management they are more likely to comply to information security policies (display securer behaviour).

<u>Employee towards employees and others</u>: Bryce and Fraser, 2014 argue that trust in others is related to risk evaluation of an individual and may influence secure behaviour. Similarly, Jalali et al., 2020 argue that trust "*influences how individuals assess cost-benefit considerations and make decisions, and ultimately their behavior*" and Hosseini Seno and Mazaheri, 2018 argue that "*trust has a significant impact on information sharing behavior*". The central notion here is that when an employee's trust in others is higher, he is more likely to evaluate the risk of information sharing as low and is more likely to share information. This could lead to undesirable situations and is often exploited by, for example, social engineering where trust is created which leads to a higher likelihood of information sharing (e.g. password sharing). This suggests that trust may provoke insecure behaviour. However, there are also potential benefits of trust between employees. Safa and Von Solms, 2016 find support for their hypothesis that trust among employees increases the extent to which employees share information security knowledge with others. Similarly, Dang-Pham et al., 2017 find support for their hypothesis that employees that are trusted "*tend to influence others security behaviours*". In this sense, employee trust in others may be beneficial for secure behaviour as long as the individual that is trusted is trustworthy (and does not abuse trust put in him).

When considering Meyer's trusting dimension we note that trust can be built through tasks performed (low trusting score) or through relationship building (high trusting score). We argue that in cultures which tend towards task-based trust (low trusting score) trust is being built by tasks performed and therefore trust increases when secure behaviour is displayed. Similarly, behaving insecurely could negatively impact the trust that is put in an individual in task-based trust cultures. In accordance with Li et al., 2019, the assumption here is that individuals want to be trusted. In relationship-based cultures, trust is not as dependent on the tasks performed (and therefore behaviour displayed is not as important), but rather on the relationships that an individual has with others. As a result of this, in a relationship-based culture displaying secure behaviour is not as important for an individual for gaining trust and therefore the displayed behaviour is less secure.

<u>Hypothesis</u>

Considering the aforementioned we arrive at the following hypothesis:

- **Hypothesis 11**: Task-based trust cultures (low score on Trusting) will have more secure behaviour than relationship-based cultures (high score on Trusting). As such a negative association is expected between trusting and secure behaviour.

### 4.2.3.6 Disagreeing

Meyer's sixth dimension is "Disagreeing": a scale ranging from confrontational to avoids confrontation. "*This scale measures tolerance for open disagreement, and views on whether it is likely to improve or destroy collegial relationships.*" (Meyer, 2014). This dimension is closely linked to Meyer's evaluating dimension in the sense that in cultures with frank feedback it is more likely that there is a higher tolerance for open disagreement whilst in cultures with diplomatic feedback this is more likely to be detrimental for relationships. As such, for our discussion of this variable we refer to section 4.2.3.2 and we arrive at a hypothesis that is similar to the one with respect to Meyer's evaluating dimension.

Hypothesis

Considering the aforementioned we arrive at the following hypotheses:

- **Hypothesis 12a**: Cultures with a higher tolerance for open disagreement (low score on Disagreeing) will have more secure behaviour than cultures in which open disagreement is less tolerated (high score on Disagreeing). As such a negative association is expected between disagreeing and secure behaviour.
- **Hypothesis 12b**: Cultures with a lower tolerance for open disagreement (high score on Disagreeing) will have more secure behaviour than cultures in which open disagreement is more tolerated (low score on Disagreeing). As such a positive association is expected between disagreeing and secure behaviour.

## 4.2.3.7 Scheduling

Meyer's seventh dimension is "Scheduling": a scale ranging from linear-time to flexible-time. "*This scale assesses how much value is placed on operating in a structured, linear fashion versus being flexible and reactive.*" (Meyer, 2014). Our literature review revealed that there appears to be little research on time perception and its impact on secure behaviour. This is supported by Chowdhury et al., 2019 who state "*despite concerns about the impact of time pressure on human cybersecurity (HCS) behaviour, research on this matter is scant and there is no literature review available that may inform researchers and practitioners about the current body of knowledge*". As such, it is difficult to base our hypothesis on literature. Nevertheless, we argue that in cultures that consider time flexible the goals to be achieved (e.g. certain information security goals) and corresponding deadlines are less strictly defined. Whereas in cultures that consider time linear goals and deadlines are more strictly defined resulting in better achievement of these goals. This achievement of information security goals (e.g. implementation of multi-factor authentication or raising information security awareness) could ultimately lead to securer behaviour.

Hypothesis

Considering the aforementioned we arrive at the following hypothesis:

- **Hypothesis 13**: Cultures with linear-time scheduling (low score on Scheduling) will have more secure behaviour than cultures with flexible-time scheduling (high score on Scheduling). As such a negative association is expected between scheduling and secure behaviour.

### 4.2.3.8 Persuading

Meyer's eighth dimension is "Persuading": a scale ranging from principles-first reasoning to applications-first reasoning. "*The ways in which you persuade others and the kinds of arguments you find convincing are deeply rooted in your culture's philosophical, religious, and educational assumptions and attitudes. ... people from southern European and Germanic cultures tend to find deductive arguments (... principles-first arguments) most persuasive, whereas American and British managers are more likely to be influenced by inductive logic (... applications-first logic)*" (Meyer, 2014). Persuasion can have an important influence on secure behaviour displayed by, for example, employees from both a management and an attacker's perspective. From management's perspective, persuasion can be used to positively influence an employee's intention to comply with information security policies (Johnston et al., 2019). From an attacker's perspective, persuasion can be used to negatively influence an employee's secure behaviour by abusing it to increase digital scam (e.g. phishing) success rate (Ferreira and Teles, 2019). In the latter case: the higher an employee's susceptibility to persuasion the more likely he is to be successfully scammed / display insecure behaviour (Modic et al., 2018). This emphasises the importance of persuasion with respect to secure behaviour.

When considering Meyer's persuasion scale (application-first reasoning versus principles-first reasoning) we argue that in cultures that prefer application before principles the information security practices are better as there is more focus on the "how" rather than the "why", as is the case with principles-first reasoning. In other words, application-first cultures could be argued to be more pragmatic (e.g. focus on appropriate information security measures) rather than principle-focused (e.g. focus on appropriate information security policies). This pragmatic approach leads to a higher extent of implementation of security measures (securer behaviour).

Hypothesis

Considering the aforementioned we arrive at the following hypothesis:

- **Hypothesis 14**: Cultures with application-first reasoning (high score on Persuading) will have more secure behaviour than cultures with principles-first reasoning (high score on Persuading). As such a positive association is expected between persuading and secure behaviour.

## 4.3   Industry level

Section Jump

### 4.3.1   **Industry type**

The research by Abbiati et al., 2021 indicates that the industry type (sector) of an organisation can greatly influence an organisation's likelihood of being subject to cyber attacks. There are various reasons that can explain why this is the case. Some examples include:

Click to navigate

Introduction   Literature review   Research questions   Antecedents of behaviour   Research model   Methodology   Results   Discussion   Conclusion   Limitations   References   Appendix

- Some industries may have data which is more appealing to cyber attackers (e.g. healthcare or financial services industry) than others (e.g. non-profit);
- Some industries (e.g. industrial production) may have systems (e.g. SCADA systems that are used for controlling manufacturing processes) which may be more interesting to interrupt than others (e.g. public sector).

When looking back at the CIA triad (Lundgren and Möller, 2019) and additional security services (Imran et al., 2021) we introduced earlier, it becomes apparent that different industries may have different prioritisations of information security requirements. For example: for the healthcare industry availability (continuity) and confidentiality (e.g. through authorisation, authenticity and/or access control) of information may be the primary requirement (e.g. for a hospital monitoring patients) whilst in the public sector integrity of data (e.g. through authorisation and/or authenticity) and non-repudiation (e.g. for a civilian filing his taxes) may be the primary requirement (e.g. financial information used for taxation purposes).

An industry's likelihood of being subject to a cyber attack could significantly impact its occurrence of data breaches. After all, cyber attacks are one of the origins of data breaches, as we discussed earlier. Some scholars (Khando et al., 2021, Chua et al. 2018 and Straub and Welke, 1998) argue that the industry an individual works in can greatly impact security awareness and in doing so impact secure behaviour. An explanation of this could be that some industries acknowledge that they face an increased risk of being subject to a cyber attack and therefore respond by focusing more efforts on increasing security awareness. This is supported by Straub and Welke, 1998 who note that "*managerial concern about the organization's security*" is partly determined by the inherent risk of the industry. This could explain why secure behaviour appears to vary between industries (Chua et al., 2018 and Waly et al., 2012) and why industry as an antecedent to secure behaviour is acknowledged by scholars such as Haeussinger and Kranz, 2013b.

Bauer et al., 2017 researched information security policy compliance (secure behaviour) within the banking industry. They argue that "*users' ISP compliance still needs to be explored empirically in more depth in banks and other industries*" (Bauer et al., 2017). As an argument they add that due to the banking industry's heavily regulated nature the observed secure behaviour may differ from other industries. Hanus et al., 2018 researched users' secure behaviour and argue that their results may not be generalised since the research data is limited that that of a government-owned organisation and results may differ for organisations which are not.

Although McCormac et al, 2018 find no support that industry (employment sector) influences information security awareness and behaviour, various scholars (Kam et al., 2020, Chua et al., 2018, Flores et al., 2014 and Choi et al., 2008) have noted the need to further research the impact of industry on information security decisions and secure behaviour. Kam et al., 2015 argue that differences between industries may exist due to different industry cultures. An example of this could be highly regulated industries such as banking in which more strict procedures could be expected to result in a more formal culture. Some scholars (Kam et al., 2015 and Leidner and Kayworth, 2006) argue that cultures at various levels; national, industry and organisation may exist and interact. Burton-Jones and Gallivan, 2007 also note the potential impact of industry on the way in which information systems are used, but go on to state that few detailed studies have been performed on this.

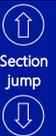

Hypothesis

As described, research on the effect of industry type on secure behaviour is lacking, but various scholars encourage researching the effect of industry type (Kam et al., 2020, Chua et al., 2018, Flores et al., 2014 and Choi et al., 2008). Since it is unclear which characteristics of industries (e.g. degree of regulation in the industry or industry culture dimensions) may influence secure behaviour we limit our research at industry level to the industry type, and arrive at the following hypothesis:

- **Hypothesis 15**: Industry type has an association with secure behaviour.

## 4.4   Organisation level

### 4.4.1   Organisational security culture

At national level we argued that national culture may have impact on the extent to which secure behaviour is displayed. However, Flores et al., 2014 argue that differences in culture may exist between firms in the same country and that research on culture may need to take this into consideration when researching information handling. This is supported by Khando et al., 2021 who argue that secure behaviour varies per type of industry (sector). Kam et al., 2015 argue that cultures exist at various levels; national, industry and organisation, and that these may be interacting. This presumption that culture exists at various levels is also supported by other scholars (Wiley et al., 2020, Vance et al., 2020, Kappos and Rivard, 2008 and Leidner and Kayworth, 2006). This raises the question: at which levels should we consider culture besides national? At industry level (Kam et al., 2015), organisation level (Tang et al., 2016, Kam et al., 2015, Flores et al., 2014 and Alfawaz, 2011) or even individual level (Flores et al., 2014)?

Culture at (least at) national and organisation level appears to be frequently mentioned levels in information security research (Vance et al., 2020, Tang et al., 2016, Alfawaz, 2011, Kappos and Rivard, 2008, Leidner and Kayworth, 2006). The need to include organisational culture in future information security research is supported by various scholars (Khando et al., 2021, Vance et al., 2020, McCormac et al., 2018, McCormac et al., 2017, Yazdanmehr and Wang, 2016, Tang et al., 2016, Kam et al., 2015, Flores et al., 2014, Hu et al., 2012, Alfawaz, 2011, Da Veiga and Eloff, 2010, Kappos and Rivard, 2008 and Leidner and Kayworth, 2006). Various researchers go one step further by describing a culture specific to information security and argue that such a "*security culture*" can be changed through, for instance, training and increasing of awareness (Wiley et al., 2020, ENISA, 2017, Tang et al., 2016, Da Veiga, 2015, Waly et al., 2012 and Da Veiga and Eloff, 2010). This security culture is in turn influenced by an organisation's culture (Tang et al., 2016).

The research by Da Veiga and Eloff, 2010 shows the complexity of variables that may influence an information security culture. Due to the lack of research on the impact of organisational security culture on information security, and secure behaviour in particular, it is difficult to make accurate predictions regarding the impact of specific dimensions of organisational culture on secure behaviour. As noted by Wiley et al., 2020 "*a publicly available, comprehensive, validated and reliable security culture instrument*" is not available. In addition, there appear to be opposing views as to whether:

- a security culture influences secure behaviour (Waly et al., 2012);
- secure behaviour influences a security culture (Da Veiga and Eloff, 2010);
- the influence between a security culture and secure behaviour is both ways (Kam et al., 2015), or;
- a security culture acts as a moderating factor between, for instance, security awareness and secure behaviour (Wiley et al., 2020).

Therefore, we will not use a publicly available and validated instrument for assessing organisational security culture.

Scholars have also noted that the interplay between national culture, organisational (security) culture and information security requires additional research (Wiley et al., 2020, Vance et al., 2020 and Kam et al., 2015). However, culture in isolation (be it national, organisational or different) and its impact on secure behaviour is complex, let alone the interaction between various cultural levels in addition to this. Alfawaz, 2011 argues that "*in some cases national culture may override organisational culture or vice versa; cannot be easily controlled for one or another*". Adding to this that other factors (e.g. work experience (Volkova and Chiker, 2020)) may in turn influence perception of organisational (security) culture, it becomes apparent that considering the interplay between cultures and antecedents that influence culture would benefit from separate research. Therefore, in addition to the earlier noted limitations, we will not consider possible interplays that may exist between national culture and organisational (security) culture, or other factors such as perception of organisational (security) culture.

Like mentioned before, a publicly available and validated instrument for assessing organisational security culture is not available. The European Union Agency for Cybersecurity (ENISA) published a research on measuring security cultures in organisations in 2017 (ENISA, 2017). In this research ENISA proposes several approaches to measuring an organisation's security culture. One of these approaches is through the use of the Security Culture Toolkit provided by CLTRe, a company of the KnowBe4 corporation. KnowBe4 is a global leader in assessing and measuring security cultures within organisations with years of experience in the security culture industry. KnowBe (CLTRe) provides a framework containing seven dimensions to measure a security culture (Laycock et al., 2019). The seven dimensions are the following:

- Attitudes
- Behaviours
- Cognition
- Communication
- Compliance
- Norms
- Responsibilities

We will use the KnowBe4 (CLTRe) model for our research to measure the impact of an organisation's security culture on secure behaviour. In the following sections we will describe each of the dimensions and the corresponding hypothesis on the impact the dimensions are expected to have on secure behaviour.

### 4.4.1.1 Attitudes

Attitudes (in terms of information security) is defined as "*The feelings and beliefs that employees have toward the security protocols and issues.*" (Laycock et al., 2019). Attitudes are said to be expressions of relationships (that can be either positive or negative) between an individual and an attitude object (Jhangiani et al., 2014). An example of an attitude object could be the changing of passwords. A positive attitude could be one in which an individual has a positive attitude towards changing passwords by considering it an important mechanism to keep his information secure. A negative attitude could be one on which an individual has a negative attitude towards changing passwords by considering it annoying and cumbersome to remember many different passwords.

Attitudes can be developed in various ways, but are mostly developed through experience, either direct (first hand experience) or indirect (second or third hand), with the "attitude object" (De Houwer et al., 2001). Although experience can be either direct or indirect, a direct experience is argued to contribute stronger to the development of the attitude an individual has towards the attitude object (Ashenden, 2018). Another influencing factor towards the development of an attitude can be the awareness of risks (Bryce and Fraser, 2014). The latter can be particularly important in information security since information security is essentially about prevention of (information security) risks. Therefore, it is no surprise that so much research is performed on creating security awareness (refer to the literature review). As explained above, an attitude requires the individual to have developed an attitude through, for instance, (in)direct experience with the attitude object. This means that individuals' attitudes must be "activated" (Augoustinos et al., 2006). In other words: if limited to no factors have developed an attitude of an individual (e.g. no experience related to the attitude object or awareness of its risks), there may be no (significant) attitude (neither positive nor negative) towards the attitude object. This is of particular interest to information security since individuals are said to often have no "activated" attitude towards information security due to limited experience with information security and limited awareness of information security risks (Laycock et al., 2019).

Nevertheless, behavioural security research argues that attitudes are an important predictor of (in)secure behaviour, but at the same time are influenced by a multitude of factors (Siponen et al., 2014 and Tsohou et al., 2008). The relationship between attitudes and behaviour may not be a simple linear relationship; attitudes can influence behaviour, but behaviour can also influence attitudes. Let us consider two examples:

- <u>Attitudes influence behaviour</u>: An individual who smokes may develop a negative attitude towards smoking, because he argues that smoking is bad for your health. This attitude could lead to the individual stopping smoking. In this case attitudes clearly influence behaviour.

- <u>Behaviour influences attitudes</u>: An individual who smokes may also change his attitudes as a result of his behaviour. For example, an individual loves to smoke and therefore doesn't want change his behaviour and stop smoking. He argues that his grandpa also smoked and had a long life and therefore smoking is probably not very bad for his health. Essentially, the individual is justifying his behaviour and in doing so changes his attitude towards smoking toward an increasingly positive attitude. This phenomenon is also called cognitive dissonance.

Whether attitudes influence behaviour or the other way around, both seem to be related. This would lead us to assume that whichever way we consider the relationship between attitudes and behaviour, attitudes can be a predictor of behaviour. When applying this to secure behaviour, an individual:

- Either has a positive / negative attitude towards information security and secure behaviour, and changes his behaviour accordingly (behaves more / less secure);
- Or behaves secure / insecure and changes his attitudes accordingly (has a positive / negative attitude towards information security and secure behaviour).

Within this research, when we consider attitudes as the extent to which attitudes that are favourable to secure behaviour are part of the organisation's culture. For instance, the extent to which attitudes that are favourable to secure behaviour are part of the organisation's culture could be plotted on a scale of 0 (no (activated) attitudes or attitudes that are unfavourable to secure behaviour) to 100 (attitudes that are favourable to secure behaviour).

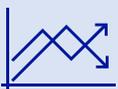

<u>Hypothesis</u>

Considering the aforementioned we arrive at the following hypothesis:

- **Hypothesis 16**: A positive attitude towards information security within an organisation's security culture (high score on attitudes) has a positive association with secure behaviour.

### 4.4.1.2 Behaviours

Behaviours is defined as the "*actions and activities of employees that have direct or indirect impact on the security of the organization*" (Laycock et al., 2019). When applied to information security and handling of information this refers to the extent to which an individual displays (in)secure behaviour. Since secure behaviour is the object of interest of this research (the dependent variable) we have not added this variable of organisational security culture to our research model as an independent variable and have not assessed the impact of the "behaviours" dimension on secure behaviour. Therefore, we have not drafted an hypothesis and have not described the relationship that this dimension / variable may have on secure behaviour.

### 4.4.1.3 Cognition

Cognition is defined as "*the employees' understanding, knowledge and awareness of security issues and activities*" (Laycock et al., 2019). This definition reveals that cognition is comprised of three factors, one of which is "knowledge". Knowledge is in important and dominant aspect within research on information security awareness (Farooq et al., 2015 and Kajtazi and Bulgurcu, 2013), but at the same time empirical research shows that there is no (significant) relation between knowledge and secure behaviour in terms of information security (Kaur and Mustafa, 2013). This suggests that knowledge on information security alone is not sufficient to change behaviour towards increasingly more secure behaviour. Knowledge Management Theory defines knowledge as "*the contextual and high-value form of information and experience that positively affect decisions and actions*" (Wang, 2010), but the aforementioned discussion reveals that knowledge in itself may not be sufficient to affect decisions and actions.

Cognition is a broader concept and is comprised of 1) knowledge of facts, processes and concepts, 2) the ability to apply this knowledge and 3) the ability to reason using knowledge when making decisions and taking actions (Farooq et al., 2015). Essentially, knowledge suggests that "knowing something" changes decisions and actions whilst cognition is broader and suggests that it is a combination of "knowing something", "understanding how to apply the knowledge" and "being able to reason with the available knowledge" that changes decisions and actions. This distinction suggests that, at organisational level, focusing solely on knowledge may be an incomplete approach to assessing the impact of knowledge on behaviour, and cognition is a more complete and reliable concept when predicting (in)secure behaviour. The aforementioned suggests that the higher the overall level of cognition of information security, the more secure behaviour is expected to be. This is confirmed by Laycock et al., 2019 who state that "*higher levels of cognition help employees understand critical factors in improving security culture, such as how important their behavior is in sustaining or endangering the security of the organization*". Within this research, when we consider cognition we consider the extent to which cognition with respect to information security is part of the organisation's culture. For instance, the extent to which cognition with respect to information security is part of the organisation's culture could be plotted on a scale of 0 (no cognition with respect to information security) to 100 (full cognition with respect to information security).

Hypothesis

Considering the aforementioned we arrive at the following hypothesis:

- **Hypothesis 17**: Cognition (of information security) within an organisation's security culture has a positive association with secure behaviour.

#### 4.4.1.4 Communication

Communication is defined as "*a mechanism for securing or compromising information through the management of people and technology*" (Backhouse and Dhillon, 1996). Communication is an essential part in information security as it supports discussions regarding information security (events), promotes a sense of belonging and supports in dealing with, and reporting about, security issues and incidents (Laycock et al., 2019). The importance of good communication is emphasised by research on data breaches that indicates that "*it takes in average 197 days for organizations to detect a breach and a further 69 days to resolve the situation and restore service*" (Ponemon Institute, 2018). Communication plays an important role in both prevention of data breaches (for instance through communication regarding information security tasks and responsibilities) and responding to data breaches (for instance through a crisis communication team) (Laycock et al., 2019 and Arhin and Wiredu, 2018). Additionally, communication means are said to be necessary for establishing and creating awareness regarding desired practices (Arhin and Wiredu, 2018).

Communication regarding desired practices helps employees in decision-making and identifying courses of action to deal with information security related issues (Laycock et al., 2019). Additionally, communication can be used to influence individuals' attitudes (and therefore their behaviour as argued earlier) towards, amongst others, information security (Laycock et al., 2019). Laycock et al., 2019 even go as far as stating that "*a well-informed person will always have better attitude than a less informed person*". Not only content, but also the time at which communication is provided can influence the way that it influences behaviour. For example: communicating tips for choosing safe passwords is likely to be more effective when the user is about to change his password than when his password change is not due for another 90 days. In this research we focus primarily on level / extent of communication rather than timing of communication. Despite the importance of communication in security culture empirical research on the role of communication in a security culture is rare (Laycock et al., 2019). Within this research, when we consider communication we consider the extent to which communication regarding information security is part of the organisation's culture. For instance, the extent to which communication regarding information security is part of the organisation's culture could be plotted on a scale of 0 (low level of communication) to 100 (high level of communication).

> Hypothesis
>
> Considering the aforementioned we arrive at the following hypothesis:
>
> - **Hypothesis 18**: Communication within an organisation's security culture has a positive association with secure behaviour.

## 4.4.1.5 Compliance

Compliance is defined as "*the knowledge of written security policies and the extent that employees follow them*" (Laycock et al., 2019). According to Al Kabani et al., 2017 adoption of compliance with respect to information security involves:

- *"Implementation of effective and balanced information security measures and mechanisms.*
- *Compliance with legal and security requirements and expectations of organizations.*
- *Maintaining both employees' and stakeholders' confidence and trust in the security."*

An organisation's security culture plays an important role in supporting employees to be compliant with information security policies (Laycock et al., 2019). However, supporting each of the aforementioned aspects of compliance through a security (compliant) culture is complex issue resulting in a challenge to improve employees' compliance with information security policies (Laycock et al., 2019 and Al Kalbani et al., 2014).

There are various approaches towards increasing compliance with (information security) policies. Some prominent examples are:

- The Theory of Planned Behaviour which focuses on concepts such as attitude, beliefs and norms to change behaviour (Bulgurcu et al., 2010);
- Focusing on sanctions and fear as mechanisms to change behaviour (Kankanhalli et al., 2003);
- Focusing on intrinsic motivations rather than extrinsic motivations (Son, 2011).

Whichever the approach to changing behaviour, it is clear that having documented policies and procedures alone is insufficient to change behaviour and sufficiently reduce the likelihood and impact of data breaches (Safa et al., 2016). Compliance with information security policies is closely related to secure behaviour since the extent of compliance with information security policies can (depending on the quality of the information security policies) be a strong indicator of secure behaviour. This could explain why (intention to) compliance is so often the object of interest within research (refer to the literature review). Within this research, when we consider compliance we consider the extent to which compliance to information security policies is part of the organisation's culture. For instance, the extent to which compliance to information security policies is part of the organisation's culture could be plotted on a scale of 0 (low level of compliance) to 100 (high level of compliance).

> Hypothesis
>
> Considering the aforementioned we arrive at the following hypothesis:
>
> - **Hypothesis 19**: Compliance within an organisation's security culture has a positive association with secure behaviour.

### 4.4.1.6 Norms

Norms is defined as "*the knowledge and adherence of unwritten rules of conduct in the organization, i.e. how security-related behaviors are perceived by employees as normal and accepted or unusual and unaccepted*" (Laycock et al., 2019). Norms are considered one of the most important factors that influences behaviour that individuals display (Hechter and Opp, 2001). Although norms are argued to be powerful in terms of the influence they have on behaviour, they are said to be difficult to influence due to the fact that norms are a "*relatively stable set of unwritten rules regarding what is good, right and important*" (Bicchieri, 2016).

There are two types of norms (Laycock et al., 2019):
- Social norms: these can be described as "*a set of (unwritten) rules that are based on common beliefs about how people act in a particular situation*" (Yazdanmehr and Wang, 2016). Such norms are based on social interactions and adherence to these norms, or lack thereof, is guided by social rewards and sanctions rather than law. Examples of social rewards are praise and increase in reputation, and examples of social sanctions are ignorance or mocking of the individual (Laycock et al., 2019).
- Personal norms: these can be described as a set of (unwritten) rules that are based on an individual's beliefs and personal values. In that sense personal norms are self-imposed and, similarly, rewarding and sanctioning of (lack of) adherence to these personal norms is self-imposed (Laycock et al., 2019).

This distinction between these two types of norms is important since individuals that follows social norms may do this in order to prevent the social sanctions that incompliance may cause and not necessarily because the individual agrees that the social norms are the best way of doing things (Laycock et al., 2019). Similarly, personal norms may be followed by an individual, because he believes this is the best way of doing things and because this is in accordance with his own personal values, and not because others think this is the right way of doing something (Laycock et al., 2019).

Personal values can be changed through external influences such as social norms, awareness of consequences and assigning of responsibilities (Yazdanmehr and Wang, 2016). However, changing social norms may be a more efficient approach for organisations since these could be incorporated in the organisation's culture (Laycock et al., 2019). Norms can have both a positive and negative effect on (secure) behaviour. For example: an organisation could have social norms that indicate that password and account sharing is accepted practice whilst in another organisation this could be considered forbidden (Laycock et al., 2019). Within this research, when we consider norms we consider the extent to which norms that are favourable to secure behaviour are part of the organisation's culture. For instance, the extent to which norms that are favourable to secure behaviour are part of the organisation's culture could be plotted on a scale of 0 (no norms or norms that are unfavourable to secure behaviour) to 100 (norms that are favourable to secure behaviour).

Hypothesis

Considering the aforementioned we arrive at the following hypothesis:

- **Hypothesis 20**: Norms within an organisation's security culture that are favourable to secure behaviour (high score on norms) have a positive association with secure behaviour.

### 4.4.1.7 Responsibilities

Responsibilities is defined as "*how employees perceive their role as a critical factor in sustaining or endangering the security of the organization*" (Laycock et al., 2019). Responsibilities can be described as employee practices and performances such as "*monitoring and control, reward and deterrence and acceptance of responsibility*" (Al-Hogail, 2017). Secure behaviour in terms of information security is one of those practices, and employees should know and be aware of their responsibility to behave securely (Al-Hogail, 2015). Organisations are argued to be unable to protect their assets if employees do not have a clear understanding of their role and responsibilities to fulfil, and are not appropriately trained to perform the tasks that are expected from them (displaying secure behaviour) (Furnell and Thomson, 2009).

Therefore, employee awareness regarding information security roles and responsibilities is an important aspect of an organisation's security culture (Laycock et al., 2019). Additionally, employees should understand the importance of their tasks and responsibilities and how these contribute to information security and the organisation as a whole (Laycock et al., 2019). According to Laycock et al., 2019 employees with insufficient understanding or their tasks and responsibilities are "*less likely to follow the necessary steps and procedures to make the organization safe*". Within this research, when we consider responsibilities we consider the extent to which responsibilities with respect to information security are part of the organisation's security culture and therewith are well understood by an organisation's employees. For instance, the extent to which responsibilities with respect to information security are part of the organisation's security culture could be plotted on a scale of 0 (employees have no to little understanding of their responsibilities with respect to information security) to 100 (employees have a good understanding of their responsibilities with respect to information security).

Hypothesis

Considering the aforementioned we arrive at the following hypothesis:

- **Hypothesis 21**: Responsibilities within an organisation's security culture that are well understood (high score on responsibilities) have a positive association with secure behaviour.

### 4.5.1    Introduction to individual-level antecedents to secure behaviour

In the previous sections we focused on context-level antecedents to secure behaviour at various levels: national level, industry level and organisation level. As discussed in the literature review, the extent to which an individual displays secure behaviour is a result of both contextual factors and factors specific to the individual. In the end, the choice to behave (in)securely is made by individuals and therefore individual-level antecedents cannot be overlooked. Earlier we argued that research would benefit from both insights into the effect of demographic variables and variables which are specific to information security. This presumption is supported by Chua et al., 2018 who argue that more research on demographics is required and future research should aggregate both demographic and non-demographic influential factors. Additionally, Yazdanmehr and Wang, 2016 and Jaeger, 2018 emphasise the opportunity for future research to further explore factors such as personal characteristics and demographics. In the next sections we discuss both demographic antecedents and information security specific antecedents with a focus on secure behaviour.

### 4.5.2    Demographic antecedents

#### 4.5.2.1  Age

Age is a core demographic which is argued to require more research with respect to information security in general and secure behaviour in specific (Ameen et al., 2020, Chua et al., 2018, Jaeger, 2018, McCormac et al., 2017 and D'Arcy and Hovav, 2009). Despite the need for additional research many scholars have noted the impact that age may have on security awareness and secure behaviour (Khando et al., 2021, Breitinger et al., 2020, Vance et al., 2020, Bavel et al., 2019 and McCormac et al., 2017). White et al., 2017 argue that youth behave different than adults since "*youth are still in the learning phase of life while adults are in the productive phase of life. The motivations for computer usage is different*". Hovav and D'Arcy, 2012 argue that age can have different effects based on the cultural context; for example they argue that age has a different effect in the United States than it does in South Korea. Bavel et al., 2019 agree in the sense that age may not be a straightforward "*linear relationship*" and argue that "*the relationship between age and cybersecurity is highly complex. For example, older adults are more vulnerable than younger adults to certain types of phishing attack, but less vulnerable to others*". This may explain why some scholars find support for the impact of age on secure behaviour (Chen et al., 2022, Vance et al., 2020, Hovav and D'Arcy, 2012, D'Arcy and Hovav, 2009) whilst other do not (Vedadi et al., 2021, Li et al., 2019, Bavel et al., 2019, Yazdanmehr and Wang, 2016, Haeussinger and Kranz, 2013b).

Further analysis of research on secure behaviour reveals that the majority of research has considered age as a control variable (Chen et al., 2022, Vedadi et al., 2021, Vance et al., 2020, Bavel et al., 2019, Li et al., 2019, Yazdanmehr and Wang, 2016, Haeussinger and Kranz, 2013b and D'Arcy and Hovav, 2009) rather than an independent variable (McCormac et al., 2017 and Hovav and D'Arcy, 2012) within their research. This is confirmed by Hovav and D'Arcy, 2012 who note that "*contrary to the bulk of deterrence studies that included age … as control variables, we included them as variables of theoretical significance*". Interestingly enough only few of the researchers find statistical support (significance) for age as a control variable on secure behaviour (Chen et al., 2022, Vance et al., 2020 and D'Arcy and Hovav, 2009) while most do not (Vedadi et al., 2021, Bavel et al., 2019, Li et al., 2019, Yazdanmehr and Wang, 2016, Haeussinger and Kranz, 2013b). McCormac et al., 2017 and Hovav and D'Arcy, 2012, who both consider age to be an independent variable to secure behaviour, find support for this hypothesis in their research results. Considering the results from earlier research applying age as a control variable (limited support found), and the limited application of age as an independent variable in earlier research, we argue that the results of applying age as an independent variable to secure behaviour will be more beneficial to research.

However, when looking at the potential impact age may have on secure behaviour what type of effect can we expect? Nallainathan, 2021 determined that security awareness and implementation of preventive security mechanisms are found to be much lower for individuals aged over 30. In contrast to this, other scholars find that age has a positive association with security awareness (McCormac et al., 2018 and McCormac et al., 2017), more secure behaviour (Gratian et al., 2018, McCormac et al., 2018 and McCormac et al., 2017) and less misuse of information systems (D'Arcy and Hovav, 2009). Considering research results most scholars seem to agree that generally speaking age has a positive association with secure behaviour (Chen et al., 2022, Chua et al., 2018, McCormac et al., 2018, Gratian et al., 2018, McCormac et al., 2017 and D'Arcy and Hovav, 2009). Chen et al., 2022 note, however, that the impact may vary per type of behaviour. In their research they find that voluntary compliance behaviour (behaviour out of desire) increased with age, but instrumental compliance behaviour (behaviour out of fear) decreased. Despite the argument of Bavel et al., 2019 that the relation between age and secure behaviour may not be "*straightforward*" linear, the research results of McCormac et al., 2017 contradict this presumption.

Hypothesis

Considering the aforementioned we arrive at the following hypothesis:

- **Hypothesis 22**: Age has a positive association with secure behaviour.

## 4.5.2.2 Gender

Gender, similar to age, is another core demographic which is argued to be of importance and require more research with respect to its impact on secure behaviour and the use of information systems (Ameen et al., 2020, McCormac et al., 2018, Jaeger, 2018, McCormac et al., 2017 and D'Arcy and Hovav, 2009). Research results are divided when it comes to the effect which gender may have on secure behaviour. On the one hand many scholars argue that no support is found for gender's impact on secure behaviour (Khando et al., 2021, Vedadi et al., 2021, Vance et al., 2020, Bavel et al., 2019 and Chua et al., 2018). On the other hand many other scholars argue and/or find support for gender's impact on secure behaviour (Chen et al., 2022, Gratian et al., 2018, McCormac et al., 2017, Haeussinger and Kranz, 2013b, Hovav and D'Arcy, 2012 and D'Arcy and Hovav, 2009). Some scholars (Ameen et al., 2020 and Hovav and D'Arcy, 2012) argue that the effect of gender may be different depending on the cultural context. For example, in some cultures females suffer from "*economic, social and legal rights ... and they are less informed of the security issues*" (Ameen et al., 2020). Although there may be merit to this presumption, we argue that another reason could be the manner in which gender is considered in research.

Similar to age, the majority of research on secure behaviour appears to consider gender a control variable (Hovav and D'Arcy, 2012). However, when examining the result of researches in which gender is considered a control variable (Chen et al., 2022, Vedadi et al., 2021, Vance et al., 2020, Li et al., 2019, Bavel et al., 2019, Yazdanmehr and Wang, 2016, Haeussinger and Kranz, 2013b and D'Arcy and Hovav, 2009) it becomes apparent that only few find (partial) support (Chen et al., 2022 and Haeussinger and Kranz, 2013b) and most other do not find support (Vedadi et al., 2021, Vance et al., 2020, Li et al., 2019, Bavel et al., 2019, Yazdanmehr and Wang, 2016 and D'Arcy and Hovav, 2009). Another way of considering gender could be by applying it as a moderating factor; for example Ameen et al., 2020 find partial support for a moderating effect of gender between self-efficacy and culture on secure behaviour. However, the majority of research, not applying gender as a control variable to secure behaviour, consider gender to be an independent variable to secure behaviour (Chua et al., 2018, McCormac et al., 2018, Gratian et al., 2018, McCormac et al., 2017 and Hovav and D'Arcy, 2012). Interestingly, all of the aforementioned researches find support for this presumption. Considering the results from earlier research, we argue that the results of applying gender as an independent variable to secure behaviour will be more beneficial to research.

When we further examine the effect of gender on secure behaviour contrasting arguments are made that the on the effect of gender on secure behaviour. Gratian et al., 2018 argue that females may display poorer security awareness, security practices and less secure behaviour than their male counterparts. Gratian et al., 2018 even go further to state that "*women may be a demographic group in need of additional cyber security training and guidance*". McCormac et al., 2017 and ENISA, 2017 posit a more nuanced view by arguing that the effect of gender may be twofold. McCormac et al., 2017 find that females obtained higher information security awareness and behaviour scores, but were found to be more "*susceptible to phishing emails than males*". ENISA, 2017 states that "*men tend to be more confident in their security behaviour and privacy attitude online than women, although women generally perceive vulnerability more and are more likely to behave securely*". The argument that females behave more securely than their male counterparts appears to be supported by Introduction | Literature review | Research questions | Antecedents of behaviour | Research model | Methodology | Results | Discussion | Conclusion | Limitations | References | Appendix

many researchers; females were argued and found to have higher information security awareness (McCormac et al., 2018, McCormac et al., 2017), display more secure behaviour (Chen et al., 2022, Haeussinger and Kranz, 2013b) and engage less in information system misuse (Hovav and D'Arcy, 2012 and D'Arcy and Hovav, 2009).

<u>Hypothesis</u>

Considering the aforementioned we arrive at the following hypothesis:

- **Hypothesis 23**: Females are more likely to display secure behaviour than males.

### 4.5.2.1 Level of urbanisation

Another potential demographic that may be of interest to research on secure behaviour is the level of urbanisation. This demographic appears to have received little consideration in research at time of writing. Kruger et al., 2011 note that the area in which an individual grows up can have impact on security awareness and secure behaviour. Nallainathan, 2021 notes that "*periphery areas (i.e. rural areas) are prone and susceptible to cyber security threats*" and state that there is "*significance and importance of vulnerability awareness assessment especially among rural area residents*". Interestingly Nallainathan, 2021 finds that news regarding cyber security has a high penetration in rural areas, but applicability of security principles appears to be very low. Nallainathan, 2021 argues that this may be the case since "*newspapers explain the news but don't explain how to prevent it*". Nallainathan, 2021 finds that a significant percentage of population living is rural areas is victim to cyber attacks. A possible reason for this could be that individuals living in rural areas are less likely to have technology-intensive jobs (e.g. farmers) in comparison to individuals living in urbanised areas where such jobs may be prevalent (e.g. banking). Therefore, it could be expected that individuals in urbanised areas are more likely to display secure behaviour than individuals living in rural areas.

<u>Hypothesis</u>

Considering the aforementioned we arrive at the following hypothesis:

- **Hypothesis 24**: Individuals living in rural areas are less likely to display secure behaviour than individuals living in urbanised areas.

### 4.5.3 Security behaviour specific antecedents

Earlier we argued that research would benefit from applying antecedents specific to information security. In the following section we discuss the variables which we will incorporate in our research model and analysis.

### 4.5.3.1 Security education and training

Security education and training is a key component in promoting secure behaviour within organisations, but also at other levels. For instance, at national level the United States' Cybersecurity & Infrastructure Security Agency (CISA) has a Cybersecurity Awareness Programme with the primary goal to increase the "*understanding* *of cyber threats and empowering the American public to be safer and more secure online"* (CISA, 2022). Similar efforts can also be seen in other countries. When we discuss security education and training in this section we refer to the security education and training provided to the individual by the organisation where the individual is working.

When discussing security education and training (sometimes also referred to as Security Education, Training and Awareness (SETA) programmes) many scholars argue that additional research should be performed on its impact on security awareness and secure behaviour (Khando et al., 2021, McCormac et al., 2017 and Haeussinger and Kranz, 2013a). This is confirmed by White et al., 2017 who state that "*many believe education training will improve protective behavior and lower security breaches and incidents… however, the literature mainly lacks any evidence showing training and education lowers security breaches/incidents"*. While many scholars argue that security education and training play an important role in increasing employee security awareness and promoting secure behaviour (Khando et al., 2021, Jaeger, 2018, White et al., 2017, ENISA, 2017, Haeussinger and Kranz, 2013a, Waly et al., 2012, Al-Omari et al., 2012, Hu et al., 2012, Da Veiga and Eloff, 2010 and Marks and Rezgui, 2009) it is noted in the literature review of Khando et al., 2021 that many scholars at the same time argue that "*ISA campaigns and training are failing to change employees' behaviour for various reasons"*. Similarly Breitinger et al., 2020 state that "*while education is important, our results show that even advanced users (higher security familiarity) follow weak practices"*. This emphasises some of the limitations we identified in the literature review: 1) that security awareness may not always be a good indicator of behaviour and 2) that practical insights into what influences secure behaviour can be beneficial to tailoring security awareness efforts to make them more effective. Nevertheless, other scholars still find support for the hypothesis that security education and training efforts promote security awareness and secure behaviour (White et al., 2017, Haeussinger and Kranz, 2013a and Waly et al., 2012). Therefore, paraphrasing Hu et al., 2012, we pose that "*properly designed training programs improve employee awareness about the possible consequences of noncompliance towards established information security policies, resulting in an increased level of compliance"* (more secure behaviour).

When analysing the way security education and training is considered in research it becomes apparent that most scholars seem to consider security education and training to be an independent variable with respect to security awareness and secure behaviour (Jaeger, 2018, Haeussinger and Kranz, 2013a, Waly et al., 2012 and Al-Omari et al., 2012). White et al., 2017 argue that security education and training may have a moderating effect on independent variables influencing secure behaviour, but considering other research we will consider security education and training as an independent variable to secure behaviour.

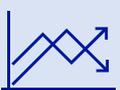

Hypothesis

Considering the aforementioned we arrive at the following hypothesis:

- **Hypothesis 25**: Security education and training has a positive association with secure behaviour

**Click to navigate**

### 4.5.3.2 Security consciousness and confidence

Security consciousness and confidence, often referred to in information security literature as self-efficacy, refers to an individual's "*self-confidence about the ability to perform a behavior*" (Cram et al., 2019) or an individual's "*perceived ability to carry out the needed response in order to cope with the risk*" (White et al., 2017). Essentially, the idea is that an individual needs self-confidence to perform a desired behaviour in order to be able to actually perform such behaviour. This is said to also apply to information security and therefore self-efficacy is an important concept that is frequently applied in research (Cram et al., 2019). This is not entirely surprising since the concept is closely related to the concept of attitude, an important aspect of many of the frequently applied theories in information security behavioural research (refer to limitation 5 in our literature review).

Despite its extensive coverage in literature we will include this variable in our research model, because 1) much research suffers from one or more of the limitations we described in the literature review and 2) due to its dominance in information security literature it cannot be overlooked. Some examples of research incorporating (security) consciousness and confidence (self-efficacy) in relation to information security include Vedadi et al., 2021, Ameen et al., 2020, Menard et al., 2018, Jaeger, 2018, Hanus et al., 2018, White et al., 2017, Safa et al., 2015, Arachchilage and Love, 2014, Waly et al., 2012, Al-Omari et al., 2012, Bulgurcu et al., 2010, D'Arcy and Hovav, 2009, Galvez, 2009, Ng et al., 2009 and Dinev and Hu, 2007. While some scholars apply the concept of self-efficacy in the general sense of an individual's self-consciousness and confidence (Hanus et al., 2018, White et al., 2017 and Arachchilage and Love, 2014), others apply it specifically to information security through the concept of computer self-efficacy (Jaeger, 2018, Galvez, 2009 and D'Arcy and Hovav, 2009). In this study we will consider self-efficacy specific to information security through the concept of security consciousness and confidence.

D'Arcy and Hovav, 2009 argue that an increase in computer self-efficacy may negatively impact secure behaviour by stating "*individuals with higher computer self-efficacy have lower perceptions of threats pertaining to IS misuse*" and "*... it can be expected that higher computer self-efficacy users will be less deterred by security countermeasures*". Despite they find support for this in their research most other research suggests that self-efficacy actually leads to more secure behaviour (Vedadi et al., 2021, Ameen et al., 2020, Menard et al., 2018, Jaeger, 2018, Hanus et al., 2018, White et al., 2017, Safa et al., 2015, Arachchilage and Love, 2014, Waly et al., 2012, Al-Omari et al., 2012, Bulgurcu et al., 2010, Galvez, 2009, Ng et al., 2009 and Dinev and Hu, 2007). As such, we expect self-efficacy to have a positive impact on secure behaviour.

Another area in which literature appears consistent is the way self-efficacy is applied in research with respect to secure behaviour. Although D'Arcy and Hovav, 2009 consider self-efficacy a moderating variable to secure behaviour, most scholars consider it to be an (indirect) independent variable with respect to secure behaviour (Vedadi et al., 2021, Ameen et al., 2020, Menard et al., 2018, Hanus et al., 2018, White et al., 2017, Safa et al., 2015, Arachchilage and Love, 2014, Al-Omari et al., 2012, Bulgurcu et al., 2010 and Ng et al., 2009).

### 4.5.3.3 Security awareness

Security awareness can be defined as "*the extent to which an employee understands the importance and implications of InfoSec policies, rules and guidelines, and, the extent to which they behave in accordance with these policies, rules and guidelines*" (McCormac et al., 2017). Similar to security consciousness and confidence, the concept of security awareness is frequently applied in information security behavioural research (Tam et al., 2022, Breitinger et al., 2020, Chua et al., 2018, Hanus et al., 2018, Yazdanmehr and Wang, 2016, Humaidi and Balakrishnan, 2015, Haeussinger and Kranz, 2013b, Al-Omari et al., 2012, Hovav and D'Arcy, 2012, Bulgurcu et al., 2010, Kumar et al., 2008 and Dinev and Hu, 2007). Nevertheless, research suggests that further insights can be gained by researching the impact of security awareness on secure behaviour (Jaeger, 2018 and Haeussinger and Kranz, 2013) and the impact on prevention of data breaches (Safa et al., 2015).

There are multiple ways in which security awareness can be interpreted and incorporated into research. Security awareness can be considered in terms of the way an individual perceives a threat (e.g. severity and certainty of sanctions (Mejias, 2012 and Hovav and D'Arcy, 2012)), the extent to which an individual is familiar with policies (Chua et al., 2018) or the extent to which an individual is familiar with (technical) measures and responsibilities related thereto (Breitinger et al., 2020). We argue that all of these three are important elements that may influence an individual's security behaviour. More specifically, 1) familiarity with security policies enables an individual to know *what* to do, 2) familiarity with (technical) measures enables an individual to know *how* to do what is needed, and 3) perceiving / identifying a threat is important to know *when* do what is needed. However, we assess the second aspect (familiarity with (technical) measures) through the separate concept of security knowledge in the next section. As such, we will apply security awareness as a concept that refers to the extent to which an individual is familiar with security policies and is able to identify / perceive a threat (first and third aspect).

Regardless of how security awareness is interpreted and incorporated in research, literature suggests that, despite different interpretations and incorporations of security awareness, there seems to be consensus in literature that security awareness has a(n indirect) positive effect on secure behaviour (Khando et al., 2021, Breitinger et al., 2020, Hanus et al., 2018, Jaeger, 2018, Chua et al., 2018, ENISA, 2017, Yazdanmehr and Wang, 2016, Humaidi and Balakrishnan, 2015, Arachchilage and Love, 2014, Haeussinger and Kranz, 2013b, Mejias, 2012, Hovav and D'Arcy, 2012, Al-Omari et al., 2012, Alfawaz, 2011, Bulgurcu et al., 2010, Galvez, 2009, Marks and Rezgui, 2009, Kumar et al., 2008 and Dinev and Hu, 2007). As such, we expect security awareness to have a positive impact on secure behaviour.

Similar to research' supposed consensus regarding the impact of security awareness on secure behaviour we identify a clear pattern in the way security awareness is incorporated in literature. Tam et al., 2022 and Yazdanmehr and Wang, 2016 consider security awareness to be a moderating variable towards secure behaviour. However, most scholars consider security awareness to be an independent variable towards secure behaviour (Breitinger et al., 2020, Chua et al., 2018, Hanus et al., 2018, Humaidi and Balakrishnan, 2015, Haeussinger and Kranz, 2013b, Al-Omari et al., 2012, Hovav and D'Arcy, 2012, Bulgurcu et al., 2010, Kumar et al., 2008 and Dinev and Hu, 2007).

Hypothesis

Considering the aforementioned we arrive at the following hypothesis:

- **Hypothesis 27**: Security awareness has a positive association with secure behaviour.

### 4.5.3.4 Security knowledge

Security awareness could have a large impact on the extent of individuals' secure behaviour. However, being aware of the necessity of security alone may not be sufficient to display secure behaviour. This is confirmed by the literature review of Khando et al., 2021 who note that many scholars argue that "*ISA campaigns and training are failing to change employees' behaviour for various reasons*". Similarly Nallainathan, 2021, who considers both awareness and behaviour regarding information security, argues "*newspapers explain the news but don't explain how to prevent it*" resulting sometimes in a high awareness level, but low level of secure behaviour. Even if the individual is aware of the importance of security (e.g. with respect to passwords), does he know how to reduce the security risk (e.g. implement multi-factor authentication)?

An important factor to consider here is the IT knowledge (Khando et al., 2021) / security knowledge (Kruger et al., 2011) / security familiarity (Ng et al., 2009) / experience with computers (Dinev and Hu, 2007) of an individual. For sake of simplicity we will refer to this as "security knowledge". In accordance with Khando et al., 2021 we define security knowledge as "*general knowledge of the basic IT applications used in daily business, such as computers, email systems, and the internet.*". Indeed the importance of security knowledge is stated by various scholars (Khando et al., 2021, Breitinger et al., 2020 and Kruger et al., 2011). Other scholars have stated the need to further research the impact of security knowledge on behaviour (Jaeger, 2018, Ng et al., 2009 and Dinev and Hu, 2007).

D'Arcy and Hovav, 2009 argue that "*computer savvy users*" are more likely to display information system misuse (less secure behaviour), because security training and awareness (SETA) efforts are less deterring for such individuals. Despite this, the majority of research seems to argue that security knowledge has a positive effect on secure behaviour (Khando et al., 2021, Breitinger et al., 2020, Jaeger, 2018, Arachchilage and Love, 2014, Haeussinger and Kranz, 2013a, Al-Omari et al., 2012, Mejias, 2012 and Dinev and Hu, 2007). As such, we expect security knowledge to have a positive impact on secure behaviour.

The manner in which security knowledge is incorporated with respect to its (indirect) impact on secure behaviour appears to range from a minority applying security knowledge as control variable (Li et al., 2019, Yazdanmehr and Wang, 2016 and Ng et al., 2009) or moderating variable (White et al., 2017) to a majority applying it as an independent variable (Khando et al., 2021, Breitinger et al., 2020, Jaeger, 2018, Arachchilage and Love, 2014, Haeussinger and Kranz, 2013a, Al-Omari et al., 2012, Mejias, 2012 and Dinev and Hu, 2007). Therefore, we will consider security knowledge to be an independent variable to secure behaviour.

Hypothesis

Considering the aforementioned we arrive at the following hypothesis:

- **Hypothesis 28**: Security knowledge has a positive association with secure behaviour.

#### 4.5.3.5 Prior experience with security incidents

From a neurocognitive perspective various scholars such as Kress and Aue, 2017, Sharot and Dolan, 2011 and Weinstein, 1980 argue that humans "*tend to overestimate the likelihood of future positive events and underestimate the likelihood of future negative events*". This is called the "optimism bias" or sometimes also referred to as the "illusion of Invulnerability". This bias could be one of the factors explaining why security awareness does not always translate to more secure behaviour (Khando et al., 2021, Hanus et al., 2018, Abawajy, 2014, Talib et al., 2010, Annetta, 2010 and Cone et al., 2007): individuals may be aware of a security risk, but do not always consider it likely for this risk to manifest and therefore do not always adapt their behaviour accordingly. This raises the question: if an individual has experienced a security incident (either first, second or third hand), does this change his perception of security risks and ultimately improve his behaviour (display more secure behaviour)? Various scholars have noted the potential impact of prior experience with security incidents and the requirement for further research into this (Hickmann Klein and Mezzomo Luciano, 2016, Haeussinger and Kranz, 2013b and Dinev and Hu, 2007).

Various scholars have argued that prior experience with security incidents increases security awareness and secure behaviour (Khando et al., 2021, Hanus et al., 2018, Jaeger, 2018, Haeussinger and Kranz, 2013a, Haeussinger and Kranz, 2013b and Straub and Welke, 1998). Interestingly, alternative views such as White et al., 2017 argue that the relation could just as well be the reverse; they argue that secure behaviour leads to an increase in perceived security incidents (and therefore a highest likelihood of experience with security incidents), because individuals are more capable of identifying security incidents. However, research by Introduction | Literature review | Research questions | Antecedents of behaviour | Research model | Methodology | Results | Discussion | Conclusion | Limitations | References | Appendix

Mersinas et al., 2015 found no effect of prior experience with security incidents on the risk attitude of professionals. Considering the earlier described optimism bias and the arguments made within the research that is performed in the field of experience with security incidents in relation to security awareness and behaviour, we expect experience with security incidents to have a positive impact on secure behaviour. Since experience with security incidents appears to have been primarily considered as an independent variable to security awareness and secure behaviour (Jaeger, 2018, Hanus et al., 2018, Haeussinger and Kranz, 2013a and Haeussinger and Kranz, 2013b) we will consider this variable as such in this study.

Hypothesis

Considering the aforementioned we arrive at the following hypothesis:

- **Hypothesis 29**: Prior experience with security incidents has a positive association with secure behaviour.

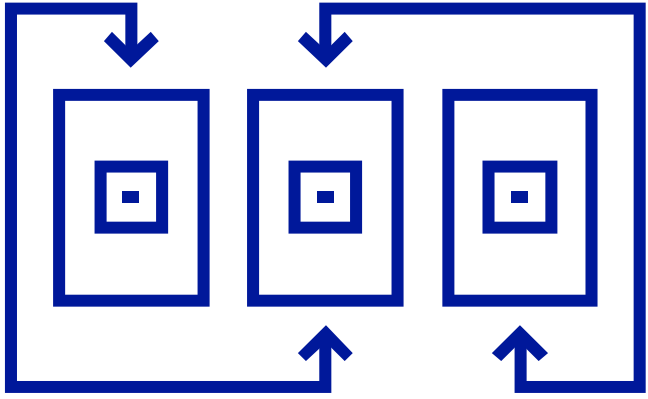

# Research model

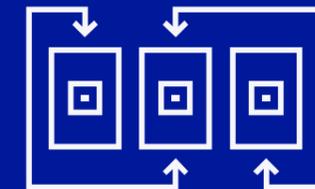

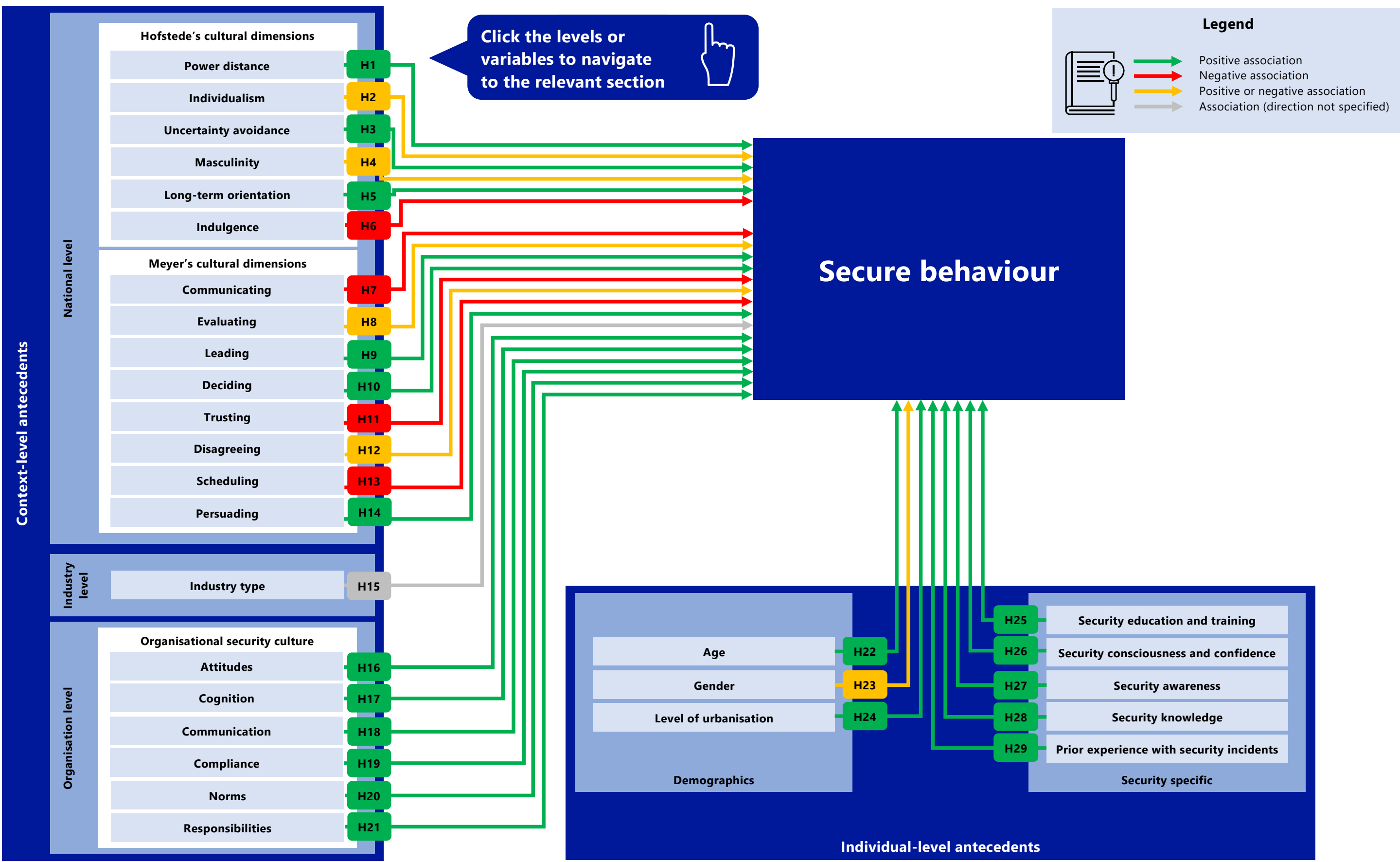

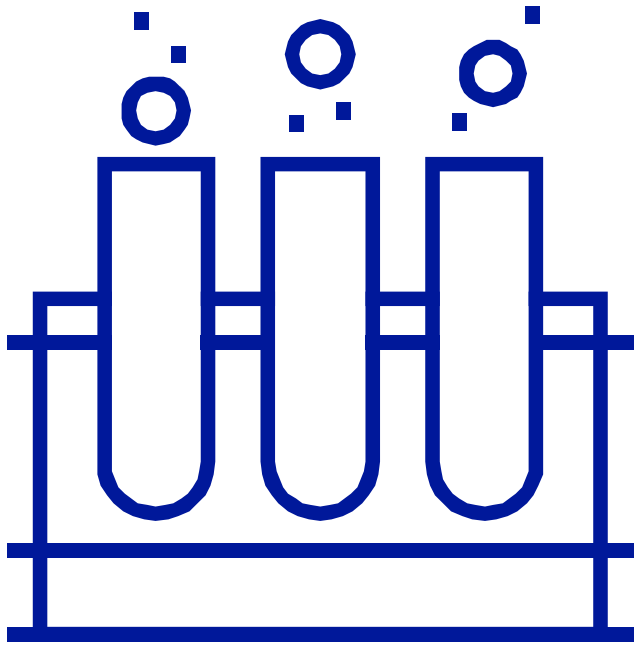

# 6

# Methodology

# 6 Methodology

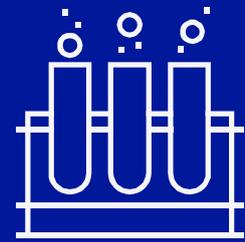

## 6.1 Data sources and interpretation

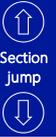

Section Jump

### 6.1.1 Background of the data sources

Within this research we use two separate datasets to test the variables listed in the research model. Since both datasets are distinct, they cannot be combined. In the following section we briefly discuss the datasets, their origin and the variables which they cover.

#### 6.1.1.1 Dataset 1: European Commission Barometer 92.2 (2019)

The first dataset is provided by the European Commission as part of the European Union (EU) Barometer 2019. The EU yearly performs various researches throughout Europe. These researches, sometimes called barometers, are reports based on elaborate surveys amongst European inhabitants. For this research we use the data collected for the 2019 European Barometer 92.2 (dataset name ZA7580) which includes a section regarding Europeans' attitudes towards cyber security. The dataset and questionnaires of European barometers are provided by Gesis; the Leibniz Institute for the Social Sciences which supports the EU in performing these researches (European Commission, 2021). The dataset contains 27.607 records and contains data which we will use for the following variables:

- Nationality of the interviewee (used for determining the cultural dimensions);
- Age of the interviewee;
- Gender of the interviewee;
- Level of urbanisation where the interviewee lives;
- Prior experience with security incidents;
- Data on actual (in)secure behaviour through security measures taken.

Further along this chapter a table is included which contains all variables, the questions used for measuring these variables, the possible answers, and any data processing performed (including interpretation of the data).

#### 6.1.1.2 Dataset 2: KnowBe4 dataset (2022)

The second dataset is provided by KnowBe4. KnowBe4 is a listed multinational organisation and is one of the largest organisations in the field of information security training and monitoring. We were provided with data on organisations (including organisational culture and industry type) and individuals (e.g. security knowledge, security awareness and security behaviour). This dataset is particularly valuable since it provides insights into individuals' behaviours through monitoring of their response to simulated phishing emails, as we will discuss later on. We limit our focus to users who have received 5 or more simulated phishing e-mails over the period

Introduction  Literature review  Research questions  Antecedents of behaviour  Research model  **Methodology**  Results  Discussion  Conclusion  Limitations  References  Appendix of 2019 until and including 2021. The dataset contains 9.468 records and contains data which we will use for the following variables:

- Industry type the organisation is operating in;
- Organisational cultural variables;
- Security education and training of individuals;
- Security consciousness and confidence of individuals;
- Security awareness of individuals;
- Security knowledge of individuals;
- Data on actual (in)secure behaviour through monitoring of phishing email response.

Further along this chapter a table is included which contains all variables, the questions used for measuring these variables, the possible answers, and any data processing performed (including interpretation of the data).

### 6.1.1.3  Important notes on the data sources

Measuring secure behaviour

With respect to the datasets there are a few important aspects that require attention. First, as mentioned earlier, the two datasets are distinct and can therefore not be combined. As a result of this both datasets include data which we use to measure the dependent variable; secure behaviour. The way in which we measure secure behaviour varies slightly between both dataset. However, earlier we discussed the importance of measuring actual behaviour rather than intentions:

- <u>EU dataset</u>: we measure secure behaviour through the security measures taken, as reported by the interviewee.
- <u>KnowBe4 dataset</u>: we measure secure behaviour through monitoring of individuals' behaviour upon receiving simulated phishing emails in accordance with the proposed approach by Jaeger, 2018.

Although both ways of measuring secure behaviour are different, we argue that they still meet the objective of measuring actual behaviour rather than intentions or other indications of secure behaviour such as awareness.

Direct vs. indirect prior experience with security incidents

Second, earlier we discussed that individuals can have prior experience with security incidents either first hand (experienced themselves) or second / third hand (experienced by other around them). Within the hypothesis we did not distinguish between either of these since the expected association is the same. However, within the data we have separate data on first hand prior experience with security incidents and second / third hand. As such, we will test these separately in our data analysis.

**Click to navigate**

## 6.1.2 Overview of data sources and interpretation

| Source | Variable | Question(s) used to measure the variable | Possible answers | Data interpretation |
|---|---|---|---|---|
| **Context-level** | | | | |
| National level | | | | |
| National culture | | | | |
| Hofstede's cultural dimensions | | | | |
| EU | Power distance | What is your nationality? | 1-28 correspond with countries in the EU, 29 corresponds with another country not listed (outside of the EU), 30 corresponds with don't know. | Answers from 1 to 28 are translated to the respective EU country as listed in the questionnaire. Answers 29 and 30 are excluded from the dataset since they do not reveal a country. |
| EU | Individualism | | | |
| EU | Uncertainty avoidance | | | |
| EU | Masculinity | | | |
| EU | Long-term orientation | | | |
| EU | Indulgence | | | The resulting countries are enriched with the cultural dimensions by Hofstede (when available) as published on Hofstede Insights, 2022. |
| Meyer's cultural dimensions | | | | |
| EU | Communicating | What is your nationality? | 1-28 correspond with countries in the EU, 29 corresponds with another country not listed (outside of the EU), 30 corresponds with don't know. | Answers from 1 to 28 are translated to the respective EU country as listed in the questionnaire. Answers 29 and 30 are excluded from the dataset since they do not reveal a country. |
| EU | Evaluating | | | |
| EU | Leading | | | |
| EU | Deciding | | | |
| EU | Trusting | | | |
| EU | Disagreeing | | | |
| EU | Scheduling | | | The resulting countries are enriched with the cultural dimensions by Meyer (when available) as published on Erin |
| EU | Persuading | | | |

Introduction 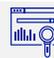 Literature review 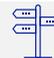 Research questions 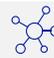 Antecedents of behaviour 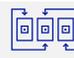 Research model 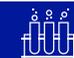 Methodology 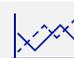 Results 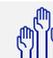 Discussion 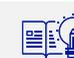 Conclusion 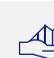 Limitations 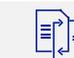 References 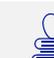 Appendix

| Source | Variable | Question(s) used to measure the variable | Possible answers | Data interpretation |
|---|---|---|---|---|
| | | | | Meyer, 2022. Meyer's dimensions are by default not expressed in numbers, but visually indicated on a scale. We plotted the visualisation on a scale of 1-10 and included the resulting numbers in the dataset. |
| **Industry level** | | | | |
| KnowBe4 | Industry type | Measured based on the industry type that is entered when the organisation is registered in the KnowBe4 database. | - Business Services<br>- Construction<br>- Consulting<br>- Consumer Services<br>- Education<br>- Financial Services<br>- Government<br>- Healthcare & Pharmaceuticals<br>- Hospitality<br>- Insurance<br>- Legal<br>- Manufacturing<br>- Non-profit<br>- Retail & Wholesale<br>- Technology<br>- Transportation<br>- Other | Industries are interpreted as they are provided. No further data processing is performed. |
| **Organisation level** | | | | |
| **Organisational security culture** | | | | |
| KnowBe4 | Attitudes | Organisational security culture is measured combined for all organisational security culture dimensions through a set of 28 questions in the Security Culture Survey (SCS). Due to intellectual property rights we have not been provided insights into all questions and the precise | Answer are provided based on a Likert scale and result in a score on a scale of 1-100 per organisational security culture dimension. | A perfect score of 100 indicates a strong positive attitude towards security, a score of 50 indicates an indifferent attitude towards security, and a score of 1 indicates |

| Source | Variable | Question(s) used to measure the variable | Possible answers | Data interpretation |
|--------|----------|------------------------------------------|------------------|---------------------|
| | | measurements that are used to measure the scores in the SCS. | | a strong negative attitude towards security. |
| KnowBe4 | Cognition | | | A perfect score of 100 indicates that security awareness and knowledge are strongly embedded in the organisation's security culture and a score of 1 indicates that security awareness and knowledge are poorly embedded in the organisation's security culture. |
| KnowBe4 | Communication | | | A perfect score of 100 indicates strong and well established communication channels (e.g. for incident reporting) and a score of 1 indicates very weak and poorly established communication channels (e.g. for incident reporting). |
| KnowBe4 | Compliance | | | A perfect score of 100 indicates a strong (intention to) compliance to information security policies and a score of 1 indicates a very weak (intention to) compliance to information security policies. |
| KnowBe4 | Norms | | | A perfect score of 100 indicates norms in the security culture that strongly support secure behaviour and a score of 1 indicates norms in the security culture do not support secure behaviour. |
| KnowBe4 | Responsibilities | | | A perfect score of 100 indicates very good understanding of individuals with respect to their |

| Source | Variable | Question(s) used to measure the variable | Possible answers | Data interpretation |
|--------|----------|------------------------------------------|------------------|---------------------|
| | | | | security responsibilities and how these responsibilities contribute to the security of the organisation as a whole. A score of 1 indicates very poor understanding of individuals with respect to their security responsibilities and how these responsibilities contribute to the security of the organisation as a whole. |
| **Individual-level** | | | | |
| **Demographics** | | | | |
| EU | Age | How old are you? | Any number, 99 = refuse to answer (no age indicated). | All numbers except for 99 (no age indicated) were included in the dataset. |
| EU | Gender | Gender | 1 (Man), 2 (Woman) | We included 1 as "Male" and 2 as "Female" in our dataset. |
| EU | Level of urbanisation | Would you say you live in a…? | Rural area or village (1), small or medium-sized town (2), large town/city (3), Don't know (4) | Entries with an answer other than 1, 2 or 3 (e.g. don't know ("DK"; 4)) were excluded from the dataset. |
| **Security specific** | | | | |
| KnowBe4 | Security education and training | Number of trainings performed by the user before the first phishing e-mail was sent out. | Any number. | We only considered trainings that were provided <u>before</u> the first phishing e-mail was sent out, because this provides all trainings with equal opportunity to impact the secure behaviour score.<br><br>Training provided <u>during</u> the period of the five phishing mails (we use to measure secure behaviour) can only partly affect secure behaviour score (e.g. when |

| Source | Variable | Question(s) used to measure the variable | Possible answers | Data interpretation |
|--------|----------|------------------------------------------|------------------|---------------------|
| | | | | a training is provided after phishing e-mail 3 it can only affect the user's behaviour for phishing email 4 and 5). Since the secure behaviour score is based on an average of the five phishing e-mails, trainings provided during the five phishing e-mails period will therefore have lesser impact on the secure behaviour score and may distort our results.<br><br>Trainings provided <u>after</u> the phishing e-mails (we use to measure secure behaviour) will not impact the secure behaviour score. As such, the aforementioned two categories of trainings were excluded. |
| KnowBe4 | Security consciousness and confidence | Security consciousness and confidence is measured by KnowBe4 through the Security Awareness Proficiency Assessment (SAPA) which consists of 28 questions that test, amongst others, the security consciousness and confidence of users.<br><br>Due to intellectual property rights we have not been provided insights into all questions and the precise measurements that are used to measure the scores in the SAPA. | Multiple choice answers that represent a score. For example: 0 points may be given for an answer indicating low security consciousness and confidence, 1 point for an answer indicating medium security consciousness and confidence, and 2 points an answer indicating high security consciousness and confidence. | Based on the answers provided by a user, a security consciousness and confidence score is calculated ranging from 0 to 100%. |
| KnowBe4 | Security awareness | Security awareness is measured by KnowBe4 through the Security Awareness Proficiency Assessment (SAPA) which | Multiple choice answers that represent a score. For | Based on the answers provided by a user, a security awareness |

| Source | Variable | Question(s) used to measure the variable | Possible answers | Data interpretation |
|--------|----------|------------------------------------------|------------------|---------------------|
| | | consists of 28 questions that test, amongst others, the security awareness and the security knowledge (proficiency) of users. Examples of questions are:<br>- I receive a friend request from someone I don't know; what should I do?<br>- Which of the following the most secure practice when creating passwords?<br><br>The SAPA measures security awareness and knowledge for the following seven domains:<br>- Passwords & Authentication:<br>- Email Security:<br>- Internet Use:<br>- Social Media:<br>- Mobile Devices:<br>- Incident Reporting<br>- Security Awareness<br><br>Due to intellectual property rights we have not been provided insights into all questions and the precise measurements that are used to measure the scores in the SAPA. | example: 0 points may be given for an incorrect answer, 1 point for a partly correct answer, and 2 points for the best answer. | proficiency score is calculated for each of the seven domains and an overall SAPA score ranging from 0 to 100%.<br><br>Since the SAPA data represents both security awareness and security knowledge we will use these scores to test our hypothesis for both security awareness and security knowledge. Considering that we have similar hypothesis for both security awareness and security knowledge (both are expected to positively influence secure behaviour) combining security awareness and security knowledge is not expected to impair our data interpretation. We will, however, include this approach as a limitation in the limitations and future research section. |
| KnowBe4 | Security knowledge | Refer to "Security awareness"; we test both Security Awareness and Security Knowledge through the same data. | Refer to "Security awareness"; we test both Security Awareness and Security Knowledge through the same data. | Refer to "Security awareness"; we test both Security Awareness and Security Knowledge through the same data. |
| EU | Prior experience with security incidents – direct experience | In the last three years, how often have you personally experienced or been a victim of each of the following situations?<br>1. Receiving fraudulent emails or phone calls asking for your personal details (including | Per situation: Once (1), two or three times (2), More than three times (3), Never (4), Don't know (5). | Refuse to answer answers (9) were excluded from the dataset.<br><br>Situations 2, 6 and 7 and their respective scores are excluded |

Click to navigate

| Source | Variable | Question(s) used to measure the variable | Possible answers | Data interpretation |
|---|---|---|---|---|
| | | access to your computer, logins, banking or payment information) <br> 2. Online fraud where goods purchased are not delivered, are counterfeit or are not as advertised <br> 3. Cyber attacks which prevent you from accessing online services like banking or public services <br> 4. Discovering malicious software (viruses, etc.) on your device <br> 5. Identity theft (somebody stealing your personal data and impersonating you) <br> 6. Accidentally encountering child sexual abuse material online <br> 7. Accidentally encountering material which promotes racial hatred or religious extremism <br> 8. Your social network or email account being hacked <br> 9. Being a victim of bank card or online banking fraud <br> 10. Being asked for payment in return for getting back control of your device | | from the data analysis since these are not closely related to our definition of cyber attacks and data breaches as introduced in the introduction section. <br><br> The scores of all included situations are added up to compute an overall "prior experience with security incidents score". For example: a once (1) on situation 1, a two or three times (2) on situation 2, and a more than three times (3) on situation 3 would yield a total score of 6 (1+2+3). The maximum number of points that can be obtained is the sum of all in scope situations (7) which are not 9 (refuse to answer situations are excluded) multiplied by 3 (maximum points that can be gained per situation); which yields a total of maximum 21 (when no answers are refused). <br><br> Never (4) and Don't know (5) answers were interpreted as 0 points per situation where this applies (no direct prior experience with this type of security incident). |
| EU | Prior experience with security incidents – indirect experience | In the last three years, has anybody in your family, amongst your friends or acquaintances experienced or been a victim of any of these situations? Please tell me all that apply (multiple answers possible). | - Receiving fraudulent emails or phone calls asking for their personal details (including access to | Refuse to answer answers (9) were excluded from the dataset. |

Click to navigate

Introduction  Literature review  Research questions  Antecedents of behaviour  Research model  Methodology  Results  Discussion  Conclusion  Limitations  References  Appendix

| Source | Variable | Question(s) used to measure the variable | Possible answers | Data interpretation |
|--------|----------|------------------------------------------|------------------|---------------------|
| | | | their computer, logins, banking or payment information) (1)<br>- Online fraud where goods purchased are not delivered, are counterfeit or are not as advertised (2)<br>- Cyber attacks which prevent them from accessing online services like banking or public services (3)<br>- Discovering malicious software (viruses, etc.) on their device (4)<br>- Identity theft (somebody stealing their personal data and impersonating them) (5)<br>- Accidentally encountering child sexual abuse material online (6)<br>- Accidentally encountering material which promotes racial hatred or religious extremism (7)<br>- Their social network or email account being hacked (8)<br>- Being a victim of bank card or online banking fraud (9) | Answers 2, 6 and 7 and their respective scores are excluded from the data analysis since these are not closely related to our definition of cyber attacks and data breaches as introduced in the introduction section.<br><br>Each indicated answer is interpreted as 1 point. For example: if 3 answers apply, a total score of 3 would be interpreted. The maximum number of points that can be obtained is the sum of all in scope situations (10) which are not 9 (refuse to answer situations are excluded) multiplied by 1 (maximum points that can be gained per situation); which yields a total of maximum 10 (when no answers are refused).<br><br>No, nothing (12) and Don't know (13) answers were interpreted as 0 points (no indirect prior experience with security incidents). |

Click to navigate

| Source | Variable | Question(s) used to measure the variable | Possible answers | Data interpretation |
|--------|----------|------------------------------------------|------------------|---------------------|
| | | | - Being asked for payment in return for getting back control of their device (10) <br> - Other cybercrimes or any other illegal online behaviour (cyberattack, online harassment or bullying) (11) <br> - No, nothing (12) <br> - DK (Don't Know) (13) | |
| **Secure behaviour** | | | | |
| Secure behaviour | | | | |
| EU | Secure behaviour | Has concern about security issues made you change the way you use the Internet in any of the following ways (multiple answers possible)? | - You are less likely to buy goods or services online (1) <br> - You are less likely to bank online (2) <br> - You are less likely to enter personal information on websites (3) <br> - You have changed your security settings (e.g. on your browser, online social network, search engine) (4) <br> - You only visit websites you know and trust (5) <br> - You use different passwords for different sites (6) <br> - You do not open emails from people you don't know (7) <br> - You only use your own computer (8) | Refuse to answer answers (9 on all answers) were excluded from the dataset. Don't know answers (1 on situation 18) were excluded from the dataset. <br><br> Each indicated answer is interpreted as 1 point. For example: if 3 answers apply, a total score of 3 would be interpreted. The maximum number of points is 16 (number of answers indicating security measures taken) unless there is a refuse to answer (9 on all answers) or the interviewee didn't know the answers (1 on situation 18) in which case 0 points could be gained. |

| Source | Variable | Question(s) used to measure the variable | Possible answers | Data interpretation |
|---|---|---|---|---|
| | | | - You have installed anti-virus software (9)<br>- You have cancelled an online purchase because of suspicions about the seller or website (10)<br>- You regularly change your passwords (11)<br>- You use more complex passwords than in the past (12)<br>- You use a password manager (13)<br>- You use biometric features (e.g. facial recognition, fingerprint) (14)<br>- You do not connect to the Internet through unsecured hotspots (15)<br>- Other (16)<br>- None / You are not concerned about online security (17)<br>- DK (Don't Know) (18) | None (17) answers were interpreted as 0 points (no secure behaviour observed). |
| KnowBe4 | Secure behaviour | 1. Was the phishing email opened? | Yes (0), No (0) | As mentioned before, we have limited our focus to all individuals that have received at least 5 simulated phishing e-mails over the period of 2019 up until and including 2021. We based the secure behaviour score on the average score of the first five phishing e-mails, because as |
| | | 2. Was the phishing email link clicked? | Yes (0), No (1) | |
| | | 3. Was the phishing email attachment opened? | Yes (0), No (0) | |
| | | 4. Was the exploited phishing email exploited by enabling the macro's in the attachment? | Yes (0), No (0) | |
| | | 5. Was data entered on phishing email landing page? | Yes (0), No (1) | |
| | | 6. Was a reply sent on the phishing email? | Yes (0), No (1) | |
| | | 7. Was the phishing email reported? | Yes (2), No (0) | |

| Source | Variable | Question(s) used to measure the variable | Possible answers | Data interpretation |
|---|---|---|---|---|
| | | | | more phishing e-mails are sent out the secure behaviour of the individual is expected to increase.

A simulated phishing email is sent to multiple recipients. The data collected concerns behaviour of the receiver as he handles the phishing e-mail. The answer to questions 1 was not included in the calculation since a phishing email often is opened before the recipient can make a risk assessment and determine the course of action. The answers to question 3 and 4 were not included in the calculation since a large part of the simulated phishing e-mails did not include an attachment.

The answer to question 7 is weighted double (2 points can be obtained instead of 1) since this action clearly demonstrates that the user recognised the email as phishing e-mail. These 2 points are only assigned if the user has not clicked the phishing e-mail link (question 2), has not entered data on the phishing e-mail landing page (question 5) and has not replied to the phishing e-mail (question 6). The reason for this is |

| Source | Variable | Question(s) used to measure the variable | Possible answers | Data interpretation |
|--------|----------|------------------------------------------|------------------|---------------------|
| | | | | that when a user clicks on the phishing e-mail link, enters data on the phishing e-mail landing page and/or replies to the phishing e-mail he or she will be notified that the e-mail is a simulated phishing e-mail. In total 5 points can be obtained.<br><br>The answers to the questions all yield a score that depends on the behaviour of the user that was observed. Two illustrative examples:<br><br>**Example 1:**<br>A user who receives a phishing e-mail clicks the phishing e-mail link (0 out of 1 points for question 2), but doesn't enter data on the e-mail landing page (1 out of 1 points for question 5) and doesn't reply to the phishing email (1 out of 1 points for question 6). After noticing that the e-mail link is from a phishing e-mail he reports the phishing e-mail, but he receives no points for this (0 out of 2 points for question 7). In total 2 points were obtained by the user (0 + 1 + 1 + 0). This results in a secure behaviour score of 2 out of 5 points, which is 40%. |

Introduction | Literature review | Research questions | Antecedents of behaviour | Research model | Methodology | Results | Discussion | Conclusion | Limitations | References | Appendix

| Source | Variable | Question(s) used to measure the variable | Possible answers | Data interpretation |
|--------|----------|------------------------------------------|------------------|---------------------|
|        |          |                                          |                  | Example 2: A user who receives a phishing e-mail doesn't click the phishing e-mail link (1 out of 1 points for question 2), doesn't enter data on the e-mail landing page (1 out of 1 points for question 5) and doesn't reply to the phishing email (1 out of 1 points for question 6). The user reports the phishing e-mail (2 out of 2 points for question 7). In total 5 points were obtained by the user (1 + 1 + 1 + 2). This results in a secure behaviour score of 5 out of 5 points, which is 100%. |

# 6.2 Data analysis

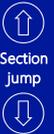

Section
jump

## 6.2.1 Data types

When performing our analyses we make use of multiple types of data which require different types of analyses. Within our data we use the following data types:

- Non-metric data:
    o <u>Nominal data</u>: this is non-metric data that cannot be ordered (for example from low to high). An example of this type of data is the Gender variable.
    o <u>Ordinal data</u>: this is non-metric data that can be ordered (for example from low to high). An example of this type of data is level of urbanisation. This could be ordered from low level of urbanisation (e.g. small village) to high level of urbanisation (e.g. large city).
- Metric data:
    o <u>Interval data (scale)</u>: this is metric data on which computations can be performed. For example averages and means can be computed on this data; this permits comparison of different data values. However, this type of data does not have a "absolute zero point" which means that the zero point is chosen arbitrarily and, theoretically, scores below zero are possible.
    o <u>Ratio data (scale)</u>: this is metric data on which computations can be performed. For example averages and means can be computed on this data; this permits comparison of different data values. This type of data has an "absolute zero point" which means that at this point nothing of the variable exists and scores below zero are not possible.

In the next sections we describe the data types and data analysis types used. In the next section we briefly describe all applied statistical analyses.

## 6.2.2 Data analysis approach

**Background to the data analysis approach**

Within quantitative information security research a multiple (linear) regression, in which multiple variables are combined into one analysis, is often applied (Hickmann Klein and Mezzomo Luciano, 2016, White et al., 2017, Wiley et al., 2020, Arachchilage and Love, 2014 and McCormac et al., 2017). Within our data analysis we also apply a multiple regression on scale variables, but in addition we perform single variable analysis. This is done because: 1) not all variables are metric (scale) variables and can therefore not be included in the multiple regression as independent variables, 2) multicollinearity was detected in the analysis of the sample (for example: refer to 6.2.3.2) as a result of which all variables cannot be analysed in one multiple regression, 3) different analyses may offer additional insights (e.g. the linear regression on age vs. a comparison of different

Click to navigate

age groups through a one-way ANOVA) and 4) although not common, this approach is also applied by some other researchers such as Lucas et al., 2021. We note that when analysing a variable through both single and multiple variable testing (e.g. simple and multiple regression), the multiple variable results are considered more important due to the higher level of comprehension of the statistical model / analysis. We elaborate further on the types of single variable and multiple variable testing that we performed for each variable (data) type.

## Single variable testing

The dependent variable (secure behaviour) is a metric variable (interval (scale)). The independent variables are nominal, ordinal, interval (scale) or ratio (scale). We test all independent variables individually through:

- <u>Independent samples t-test</u>: this is a statistical test to compare two groups of nominal or ordinal data with respect to a dependent variable, and is therefore suitable for analysing nominal / ordinal data variables with two values. An example of this is gender which can be either "Male" or "Female".

- <u>One-way ANOVA</u>: this is a statistical test which is appropriate when comparing more than two groups of nominal or ordinal data (within a single variable) with respect to a dependent variable, and is therefore suitable for analysing nominal / ordinal data variables with more than two values. An example of this is prior experience with security incidents which can be "Little experience", "Some experience", "Medium experience" and "Large experience".

- <u>Simple regression</u>: this is a statistical test to analyse the impact that one metric (interval or ratio (scale)) independent variable has on a metric dependent variable. An example of this is age which can be analysed for its influence on secure behaviour.

## Multiple variable testing

As described, we start off our data analysis with an analysis of each individual variable. However, within our data there are multiple variables which are closely related. For example, (some of) the Hofstede cultural dimensions may have significant predictive accuracy of the dependent variable (secure behaviour). Since these dimensions are closely related to one central concept (national culture) the dimensions combined may provide valuable insights into the predictive accuracy of national culture with respect to secure behaviour. Therefore, we combine such variables in a single analysis to assess whether this model can more accurately predict secure behaviour than the variables in isolation. For such an analysis we use a <u>multiple regression</u> in which multiple independent metric variables are analysed against one single dependent metric variable. There are multiple approaches in which such a multiple regression can be performed, but we will use the stepwise approach to multiple regression data analysis in which independent variables are included and excluded in the analysis model to determine the best model fit. An exception is multiple regressions on models with only two independent variables (e.g. Prior experience with security incidents); in such cases we use the default enter approach (in which variables are entered into the model one by one) due to the limited number of variables.

**Confidence interval**

For all analyses we apply a (two tailed) confidence interval of 95% (α = 0.05 / p > 0.05). If a variable yields a significance of below 0.05 we consider the variable to be *significant* in relation to the dependent variable (secure behaviour). If a variable yields a significance above 0.05, but below 0.10, we consider the variable to be *marginally significant* in relation to the dependent variable (secure behaviour) since it is not below the 95% (α = 0.05) confidence interval, but still below the alternative 90% confidence interval (α = 0.10) interval.

## 6.2.3  Appropriateness of the samples and data analyses

### 6.2.3.1  General

When using data for research purposes it is important to validate that the data to be used is appropriate. As such, we assess the appropriateness of the sample size and the distribution of the sample. Additionally, we validate the assumptions (where applicable) for the statistical tests that we intend to perform. In the following paragraphs we elaborate on each of these aspects and thresholds that we apply when performing our assumption testing. In the next sections we elaborate on each of the two datasets.

**Sample testing**

Sample size

There are varying arguments regarding the minimum required sample size for statistical analysis; for example: varying numbers are proposed ranging from 100-1000 (Mundfrom, 2005). Considering the sample sizes of our two datasets (EU: 27.607 records, KnowBe4: 9.468 records) we consider sample sizes sufficient.

Distribution of the sample

When analysing a sample it is important that the data entries are not skewed towards a single target group. For example, the sample should not excessively represent one gender or age group. To this purpose we perform an analysis of the distribution within the samples in the next sections.

**Assumption testing**

We perform assumption testing to test the appropriateness of our analyses. We perform various tests to assess whether the assumptions underlaying our analyses are fulfilled. Assumptions differ per type of statistical test. Amongst others, we apply statistical tests to make this assessment. In this section we indicate the thresholds we use  when we apply statistical tests in our assumption testing:

Levene's test of equal variances

We apply the standard confidence interval of .05 which means that values below .05 are considered statistically significant (reject the null hypothesis that equal variances can be assumed).

Independence of observations

When performing data analysis (e.g. when running a multiple regression) there is the risk that observations (e.g. items in our dataset) are related. When this may be the case we need to test for independence of observations. When applicable, we apply a Durbin-Watson test to test for so-called "first order autocorrelations" which indicate that, for example adjacent observations, may not be independent. Where we need to analyse whether the independence of residuals assumption is met we use the Durbin-Watson statistic and consider scores between the threshold of 1.5 and 2.5 as acceptable (Draper and Smith, 1998).

Unusual points that may affect data analysis outcomes

When performing data analysis (e.g. multiple regressions) we need to consider unusual points that may distort our data analysis findings. For regressions we may need to test whether the following are present:

- **Significant outliers**: these are observations that do not follow the usual pattern. We use a threshold of 3 times the standard deviation (SD) for the studentised deleted residual (SDR, a statistic calculated on the dataset) (Laerd Statistics, 2022a) to identify such outliers. When outliers are identified this may indicate possible distortion of data analysis results, but we use the leverage points and influential points to further assess the impact on our results.

- **Leverage points**: these are points that may cause a large change in the outcomes of the regression model. We assess whether there are items with a leverage value (LEV, a statistic calculated on the dataset) above 0.2 (Huber, 1981) to identify such leverage points.

- **Influential points**: these are outlier points that may have a significant impact on the regression slope and therefore excessively change our analysis results. We assess whether there are items with a Cook's Distance value (COOK, a statistic calculated on the dataset) of above 1 (Cook and Weisberg, 1982) to identify such influential points.

For independent samples T-test and One-way ANOVA we have to use a different testing method that doesn't involve thresholds.

Multicollinearity

When conducting statistical analyses with multiple (interval or scale) independent variables, as is the case with multiple regressions, there is the risk of multicollinearity. Multicollinearity is when correlations are present between independent variables and can distort statistical outcomes since independent variables are intended to function independently of each other. In case independent variables are found to correlate with other independent variables they can either be excluded from the model or be combined in a common factor through factor analysis to improve predictive accuracy of statistical tests (e.g. regressions). We choose to go for the former option by excluding such independent variables from the model. When performing a test for multicollinearity we examine the variance inflation factor (VIF) and, when multicollinearity is suspected, proceed to the condition index and values in the collinearity diagnostics. For multicollinearity testing we apply the following commonly applied thresholds (Kim, 2019):

- A <u>variance inflation factor</u> (VIF) of above 10 indicates potential issues with multicollinearity (Kim, 2019);
- A <u>condition index</u> between 15 and 30 indicates a potential multicollinearity issue and a condition index of above 30 indicates strong issues with multicollinearity (Kim, 2019);
- A <u>value in the collinearity diagnostics</u> of above .8 indicates a multicollinearity issue (Kim, 2019).

## 6.2.3.2 EU dataset

### Distribution of the sample

We performed an analysis of the distribution within the sample (EU dataset). The results of the analysis are included below. Based on this analysis we conclude that the sample is not skewed or biased towards one or more target groups. As such we consider the distribution of the sample appropriate for our research.

| Variable | Value | Count | Percentage |
|---|---|---|---|
| Gender | Male | 12413 | 45% |
| | Female | 15194 | 55% |
| Age | 15-20 years old | 1581 | 6% |
| | 21-29 years old | 2385 | 9% |
| | 30-39 years old | 3850 | 14% |
| | 40-49 years old | 4164 | 15% |
| | 50-59 years old | 4680 | 17% |
| | 60-69 years old | 5455 | 20% |
| | 70+ years old | 5492 | 20% |
| Level of urbanisation | Unknown | 10 | 0% |
| | Rural area or village | 9476 | 34% |
| | Small or medium-sized town | 10350 | 37% |
| | Large town or city | 7771 | 28% |
| Country | Austria | 1011 | 4% |
| | Belgium | 1086 | 4% |
| | Bulgaria | 1030 | 4% |
| | Croatia | 1016 | 4% |
| | Cyprus | 506 | 2% |
| | Czech Republic | 1003 | 4% |
| | Denmark | 1013 | 4% |
| | Estonia | 1027 | 4% |
| | Finland | 1012 | 4% |
| | France | 1018 | 4% |
| | Germany | 1506 | 5% |
| | Greece | 1012 | 4% |
| | Hungary | 1027 | 4% |
| | Ireland | 1015 | 4% |
| | Italy | 1023 | 4% |
| | Latvia | 1007 | 4% |
| | Lithuania | 1000 | 4% |
| | Luxembourg | 509 | 2% |
| | Malta | 511 | 2% |
| | Netherlands | 1017 | 4% |
| | Poland | 1027 | 4% |

| Variable | Value | Count | Percentage |
|---|---|---|---|
| | Portugal | 1007 | 4% |
| | Romania | 1089 | 4% |
| | Slovakia | 1013 | 4% |
| | Slovenia | 1030 | 4% |
| | Spain | 1005 | 4% |
| | Sweden | 1047 | 4% |
| | United Kingdom | 1040 | 4% |

## Testing of assumptions

### Independent samples T-test

For the EU dataset we intend to execute an independent samples T-test on Gender. We perform assumption testing for this analysis (Laerd Statistics, 2022c) and the outcomes are described below.

| Independent samples T-test Gender | |
|---|---|
| Assumption | Assumption evaluation |
| #1: Dependent variable is measured on a continuous scale | Yes, the dependent variable is secure behaviour score which is measured on a continuous scale. Therefore, assumption is met. |
| #2: One independent variable | Yes, Gender is the independent variable. Therefore, assumption is met. |
| #3: Independence of observations | The EU dataset is based on an elaborate survey throughout European member states. We have no reason to assume that observations in our dataset are related. As such we perform no further testing on the independence of observations and consider this assumption to be met for our data analysis. |
| #4: Unusual points and significant outliers | Per inspection of the box plot (refer to 12.1.1.1) we determined that there are some outliers in the data. We inspected these data entries and conclude that the data entries contain no unrealistic values or values that may otherwise lead to suspect that these must be incorrect. Additionally, the number of outliers on the total data set is limited as a result of which the impact is expected to be limited. Therefore, we consider this assumption to be met for our data analysis. |
| #5: Normal distribution of the dependent variable | Per inspection of the test of normality statistics (refer to 12.1.1.1) we determined that the sample is not normally distributed (significance of the Kolmogorov-Smirnov test is below the .05 threshold). However, this doesn't have to be an issue if the distribution is similar between the compared groups. Per inspection of the distribution histogram and the Normal Q-Q plots for both groups (refer to 12.1.1.1) we determined that the distribution of the dependent variable is similar between both groups (males and females). Therefore, we consider this assumption to be met for our data analysis. |
| #6: Homogeneity of variances is assumed | Per inspection of the statistics from Levene's test for equality of variances (refer to 12.1.1.1) we determined that the significance level is below the .05 threshold which means that the null hypothesis (homogeneity of variances) is rejected. This means that we cannot assume equal variances. Therefore, we will interpret the "equal variances not assumed" in the statistical output. |

### One-way ANOVA

For the EU dataset we intend to execute the following one-way ANOVA analyses:

- One-way ANOVA on level of urbanisation;
- One-way ANOVA on age groups;

- One-way ANOVA on direct prior experience categories;
- One-way ANOVA on indirect prior experience categories.

We perform assumption testing for each of the aforementioned (Laerd Statistics, 2022b) and the outcomes are described below.

| One-way ANOVA Level of urbanisation | |
|---|---|
| Assumption | Assumption evaluation |
| #1: Dependent variable is measured on a continuous scale | Yes, the dependent variable is secure behaviour score which is measured on a continuous scale. Therefore, assumption is met. |
| #2: Two or more categorical independent variables | Yes, the analysis consists of three levels of urbanisation which can be ordered. |
| #3: Independence of observations | The EU dataset is based on an elaborate survey throughout European member states. We have no reason to assume that observations in our dataset are related. As such we perform no further testing on the independence of observations and consider this assumption to be met for our data analysis. |
| #4: Unusual points and significant outliers | Per inspection of the box plot (refer to 12.1.1.2) we determined that there are some outliers in the data. We inspected these data entries and conclude that the data entries contain no unrealistic values or values that may otherwise lead to suspect that these must be incorrect. Additionally, the number of outliers on the total data set is limited and similar between groups as a result of which the impact is expected to be limited. Therefore, we consider this assumption to be met for our data analysis. |
| #5: Normal distribution of the dependent variable | Per inspection of the distribution histogram (refer to 12.1.1.2) we determined that the sample is not normally distributed. However, this doesn't have to be an issue if the distribution is similar between the compared groups. Per inspection of the distribution histogram and the Normal Q-Q plots for all three groups (refer to 12.1.1.2) we determined that the distribution of the dependent variable is similar between all groups. Therefore, we consider this assumption to be met for our data analysis. |
| #6: Homogeneity of variances is assumed | Per inspection of the statistics from Levene's test for equality of variances (refer to 12.1.1.2) we determined that the significance level is between .048 and .079 (depending on way of measuring). The value is around the .05 threshold, but not convincingly below the .05 level. Therefore, we will assume equal variances in the statistical output. |
| One-way ANOVA Age groups | |
| Assumption | Assumption evaluation |
| #1: Dependent variable is measured on a continuous scale | Yes, the dependent variable is secure behaviour score which is measured on a continuous scale. Therefore, assumption is met. |
| #2: Two or more categorical independent variables | Yes, the analysis consists of seven age groups which can be ordered. |
| #3: Independence of observations | The EU dataset is based on an elaborate survey throughout European member states. We have no reason to assume that observations in our dataset are related. As such we perform no further testing on this assumption to be met for our data analysis. |
| #4: Unusual points and significant outliers | Per inspection of the box plot (refer to 12.1.1.3) we determined that there are some outliers in the data. We inspected these data entries and conclude that the data entries contain no unrealistic values or values that may otherwise lead to suspect that these must be incorrect. Additionally, the number of outliers on the |

| | total data set is limited and similar between groups as a result of which the impact is expected to be limited. Therefore, we consider this assumption to be met for our data analysis. |
|---|---|
| #5: Normal distribution of the dependent variable | Per inspection of the distribution histogram (refer to 12.1.1.3) we determined that the sample is not normally distributed. However, this doesn't have to be an issue if the distribution is similar between the compared groups. Per inspection of the distribution histogram and the Normal Q-Q plots for all seven groups (refer to 12.1.1.3) we determined that the distribution of the dependent variable is similar between all groups. Therefore, we consider this assumption to be met for our data analysis. |
| #6: Homogeneity of variances is assumed | We perform this analysis to gain additional insights into means per group, but not as a primary method to statistically analyse differences between groups (e.g. F-statistic). We use a regression to analyse the impact of age on secure behaviour. As such, we will not perform testing for homogeneity of variances. |

| One-way ANOVA Direct prior experience with security incidents | |
|---|---|
| **Assumption** | **Assumption evaluation** |
| #1: Dependent variable is measured on a continuous scale | Yes, the dependent variable is secure behaviour score which is measured on a continuous scale. Therefore, assumption is met. |
| #2: Two or more categorical independent variables | Yes, the analysis consists of four levels of direct prior experience which can be ordered. |
| #3: Independence of observations | The EU dataset is based on an elaborate survey throughout European member states. We have no reason to assume that observations in our dataset are related. As such we perform no further testing on the independence of observations and consider this assumption to be met for our data analysis. |
| #4: Unusual points and significant outliers | Per inspection of the box plot (refer to 12.1.1.5) we determined that there are some outliers in the data. We inspected these data entries and conclude that the data entries contain no unrealistic values or values that may otherwise lead to suspect that these must be incorrect. Additionally, the number of outliers on the total data set is limited and similar between groups as a result of which the impact is expected to be limited. Therefore, we consider this assumption to be met for our data analysis. |
| #5: Normal distribution of the dependent variable | Per inspection of the distribution histogram (refer to 12.1.1.5) we determined that the sample is not normally distributed. However, this doesn't have to be an issue if the distribution is similar between the compared groups. Per inspection of the distribution histogram and the Normal Q-Q plots for all four groups (refer to 12.1.1.5) we determined that the distribution of the dependent variable is similar between all groups (disregarding frequencies). Therefore, we consider this assumption to be met for our data analysis. |
| #6: Homogeneity of variances is assumed | We perform this analysis to gain additional insights into means per group, but not as a primary method to statistically analyse differences between groups (e.g. F-statistic). We use a regression to analyse the impact of direct prior experience with security incidents on secure behaviour. As such, we will not perform testing for homogeneity of variances. |

| One-way ANOVA Indirect prior experience with security incidents | |
|---|---|
| **Assumption** | **Assumption evaluation** |
| #1: Dependent variable is measured on a continuous scale | Yes, the dependent variable is secure behaviour score which is measured on a continuous scale. Therefore, assumption is met. |
| #2: Two or more categorical independent variables | Yes, the analysis consists of four levels of indirect prior experience which can be ordered. |

Introduction | Literature review | Research questions | Antecedents of behaviour | Research model | Methodology | Results | Discussion | Conclusion | Limitations | References | Appendix

| #3: Independence of observations | The EU dataset is based on an elaborate survey throughout European member states. We have no reason to assume that observations in our dataset are related. As such we perform no further testing on the independence of observations and consider this assumption to be met for our data analysis. |
|---|---|
| #4: Unusual points and significant outliers | Per inspection of the box plot (refer to 12.1.1.6) we determined that there are some outliers in the data. We inspected these data entries and conclude that the data entries contain no unrealistic values or values that may otherwise lead to suspect that these must be incorrect. Additionally, the number of outliers on the total data set is limited and similar between groups as a result of which the impact is expected to be limited. Therefore, we consider this assumption to be met for our data analysis. |
| #5: Normal distribution of the dependent variable | Per inspection of the distribution histogram (refer to 12.1.1.6) we determined that the sample is not normally distributed. However, this doesn't have to be an issue if the distribution is similar between the compared groups. Per inspection of the distribution histogram and the Normal Q-Q plots for all four groups (refer to 12.1.1.6) we determined that the distribution of the dependent variable is similar between all groups (disregarding frequencies). Therefore, we consider this assumption to be met for our data analysis. |
| #6: Homogeneity of variances is assumed | We perform this analysis to gain additional insights into means per group, but not as a primary method to statistically analyse differences between groups (e.g. F-statistic). We use a regression to analyse the impact of indirect prior experience with security incidents on secure behaviour. As such, we will not perform testing for homogeneity of variances. |

Simple and multiple regressions

For the EU dataset we intend to execute the following regressions:

- Various simple linear (single variable) regressions;
- Multiple regression of the Hofstede cultural dimensions;
- Multiple regression of Meyer's cultural dimensions;
- Multiple regression of direct and indirect prior experience with security incidents;
- Multiple regression of all metric variables in the EU dataset (Hofstede cultural dimensions, Meyer cultural dimensions, age and (in)direct prior experience with security incidents).

We perform assumption testing for each of the aforementioned (Laerd Statistics, 2022a) and the outcomes are described below. An exception to this is the various simple linear regressions, since these are executed on the same data as the multiple regressions. Therefore, the assumption testing is performed with the multiple regressions.

| Multiple regression of Hofstede cultural dimensions | |
|---|---|
| Assumption | Assumption evaluation |
| #1: Dependent variable is measured on a continuous scale | Yes, the dependent variable is secure behaviour score which is measured on a continuous scale. Therefore, assumption is met. |
| #2: Two or more independent variables | Yes, Hofstede's 6 cultural dimensions. Therefore, assumption is met. |
| #3: Independence of observations | The EU dataset is based on an elaborate survey throughout European member states. We have no reason to assume that observations in our dataset are |

| | related. As such we perform no further testing on the independence of observations and consider this assumption to be met for our data analyses. |
|---|---|
| #4: Linearity of data | Per inspection of the partial regression plots and the overall regression plot (refer to 12.1.2.1) we determined that it is reasonable to assume that the data is linear. |
| #5: Homoscedasticity of data | Per inspection of the partial regression plots and the overall regression plot (refer to 12.1.2.1) we determined that it is reasonable to assume that the homoscedasticity assumption is met for the data since no heteroscedasticity was observed in the inspected plots. |
| #6: Multicollinearity of data | Our analysis for multicollinearity on Hofstede's cultural dimensions (refer to 12.1.2.1) reveals that none of Hofstede's cultural dimensions has a VIF of above 10. Therefore, we conclude that there is no (significant) risk of multicollinearity and we consider this assumption met. |
| #7: Unusual points and significant outliers | Per inspection of studentised deleted residuals (SDR_1 in the dataset) we determined that that 145 out of 21321 records exceed the SDR threshold of 3 times the standard deviation. Per inspection of the leverage point values (LEV_1 in the dataset) we determined that none of the items had a leverage point value above the 0.20 threshold (highest value was .001). Finally, per inspection of the Cook Distance Values (COO_1 in the dataset)) we determined that none of the items had a Cook Distance Value above the threshold of 1 (highest value was .00293). As such we conclude there were no unusual points and significant outliers that may significantly affect our data analysis results. |
| #8: Residual errors are normally distributed | Per inspection of the distribution histogram we determined that the data is normally distributed with a mean of close to 0 (-0.000000000000000231) and a standard deviation of 1. Additionally, per inspection of the normal P-P Plot we determined that the residuals are normally distributed. Refer to 12.1.2.1. Therefore, we consider this assumption to be met. |

| Multiple regression of Meyer cultural dimensions | |
|---|---|
| Assumption | Assumption evaluation |
| #1: Dependent variable is measured on a continuous scale | Yes, the dependent variable is secure behaviour score which is measured on a continuous scale. |
| #2: Two or more independent variables | Yes, Meyer's 8 cultural dimensions |
| #3: Independence of observations | The EU dataset is based on an elaborate survey throughout European member states. We have no reason to assume that observations in our dataset are related. As such we perform no further testing on the independence of observations. |
| #4: Linearity of data | Per inspection of the partial regression plots and the overall regression plot (refer to 12.1.2.2) we determined that it is reasonable to assume that the data is linear. |
| #5: Homoscedasticity of data | Per inspection of the partial regression plots and the overall regression plot (refer to 12.1.2.2) we determined that it is reasonable to assume that the homoscedasticity assumption is met for the data since no heteroscedasticity was observed in the inspected plots. |
| #6: Multicollinearity of data | Our analysis for multicollinearity on Meyer's cultural dimensions (refer to "INITIAL MODEL" in 12.1.2.2) reveals that multiple variables have a VIF of above 10. We inspected the condition index and identified various variables that have a condition index of above 15. Based on the values in the collinearity diagnostics we identified one value of .79 for variable 9 ("Persuading"). Since this value is close to the .8 threshold we excluded this variable and repeated the analysis for multicollinearity. Based on this we additionally excluded variable 6 ("Trusting") and re-ran the analysis for multicollinearity (refer to "IMPROVED MODEL" in 12.1.2.2). No variables remained with a VIF of above 10 and therefore we |
| | conclude that this improved model of Meyer's cultural dimensions is appropriate for performing a multiple regression analysis. Therefore, we conclude that there is no remaining (significant) risk of multicollinearity and we consider this assumption met. |
| #7: Unusual points and significant outliers | Per inspection of studentised deleted residuals (SDR_2 in the dataset) we determined that that 110 out of 19988 records exceed the SDR threshold of 3 times the standard deviation. Per inspection of the leverage point values (LEV_2 in the dataset) we determined that none of the items had a leverage point value above the 0.20 threshold (highest value was .00085). Finally, per inspection of the Cook Distance Values (COO_2 in the dataset)) we determined that none of the items had a Cook Distance Value above the threshold of 1 (highest value was .00194). As such we conclude there were no unusual points and significant outliers that may significantly affect our data analysis results. |
| #8: Residual errors are normally distributed | Per inspection of the distribution histogram we determined that the data is normally distributed with a mean of close to 0 (-0.00000000000000248) and a standard deviation of 1. Additionally, per inspection of the normal P-P Plot we determined that the residuals are normally distributed. Refer to 12.1.2.2. Therefore, we consider this assumption to be met. |

**Multiple regression of Prior experience with security incidents**

| Assumption | Assumption evaluation |
|---|---|
| #1: Dependent variable is measured on a continuous scale | Yes, the dependent variable is secure behaviour score which is measured on a continuous scale. |
| #2: Two or more independent variables | Yes, the 2 variables direct and indirect experience with security incidents |
| #3: Independence of observations | The EU dataset is based on an elaborate survey throughout European member states. We have no reason to assume that observations in our dataset are related. As such we perform no further testing on the independence of observations. |
| #4: Linearity of data | Per inspection of the partial regression plots and the overall regression plot (refer to 12.1.2.3) we determined that it is reasonable to assume that the data is linear. |
| #5: Homoscedasticity of data | Per inspection of the partial regression plots and the overall regression plot (refer to 12.1.2.3) we determined that it is reasonable to assume that the homoscedasticity assumption is met for the data since no heteroscedasticity was observed in the inspected plots. |
| #6: Multicollinearity of data | Our analysis for multicollinearity on direct and indirect prior experience with security incidents (refer to 12.1.2.3) reveals that neither the variable "Direct experience score" nor "Indirect experience score" have a VIF of above 10. Therefore, we conclude that there is no (significant) risk of multicollinearity. |
| #7: Unusual points and significant outliers | Per inspection of studentised deleted residuals (SDR_3 in the dataset) we determined that that 142 out of 21662 records exceed the SDR threshold of 3 times the standard deviation. Per inspection of the leverage point values (LEV_3 in the dataset) we determined that none of the items had a leverage point value above the 0.20 threshold (highest value was .00236). Finally, per inspection of the Cook Distance Values (COO_3 in the dataset) we determined that none of the items had a Cook Distance Value above the threshold of 1 (highest value was .00595). As such we conclude there were no unusual points and significant outliers that may significantly affect our data analysis results. |
| #8: Residual errors are normally distributed | Per inspection of the distribution histogram we determined that the data is normally distributed with a mean of close to 0 (-0.000000000000000137) and a standard deviation of 1. Additionally, per inspection of the normal P-P Plot we determined that the residuals are normally distributed. Refer to 12.1.2.3. Therefore, we consider this assumption to be met. |

| Multiple regression of All metric variables | |
|---|---|
| **Assumption** | **Assumption evaluation** |
| #1: Dependent variable is measured on a continuous scale | Yes, the dependent variable is secure behaviour score which is measured on a continuous scale. |
| #2: Two or more independent variables | Yes, 17 independent variables from the research model |
| #3: Independence of observations | The EU dataset is based on an elaborate survey throughout European member states. We have no reason to assume that observations in our dataset are related. As such we perform no further testing on the independence of observations. |
| #4: Linearity of data | Not applicable, refer to #6. |
| #5: Homoscedasticity of data | Not applicable, refer to #6. |
| #6: Multicollinearity of data | Our analysis for multicollinearity on all metric variables in the EU dataset (refer to 12.1.2.4) reveals that the majority of variables (12 out of 17) has a VIF of above 10. Additionally, condition index values were significant ranging up to 145,5. Therefore we conclude that this model is not appropriate for performing a multiple regression analysis and variables are best analysed separately (through separate simple regressions and/or separate multiple regressions). Therefore, we will not perform a single multiple regression including all metric variables in the EU dataset. |
| #7: Unusual points and significant outliers | Not applicable, refer to #6. |
| #8: Residual errors are normally distributed | Not applicable, refer to #6. |

### 6.2.3.3 KnowBe4 dataset

#### Distribution of the sample

We performed an analysis of the distribution within the sample (KnowBe4 dataset). The results of the analysis are included below. Based on this analysis we conclude that the sample is somewhat skewed towards one or more target groups. At the country level, the United Kingdom is overrepresented with a 71% share of the total number of entries (users) in the sample. At the industry level, the industries "Financial Services" and "Transportation" each have a share of 33% of the total number of entries (users) in the sample. As such we acknowledge the sample distribution of the KnowBe4 dataset as a limitation to our research. However, considering the size and source of this dataset we feel that this dataset is still of value to our analysis.

| Variable | Value | Count | Percentage |
|---|---|---|---|
| Users per country | Belgium | 148 | 2% |
| | France | 25 | 0% |
| | Germany | 21 | 0% |
| | Gibraltar | 35 | 0% |
| | Greece | 18 | 0% |
| | Ireland | 83 | 1% |

| Variable | Value | Count | Percentage |
|---|---|---|---|
| | Jersey | 105 | 1% |
| | Latvia | 84 | 1% |
| | Lithuania | 67 | 1% |
| | Luxembourg | 101 | 1% |
| | Malta | 62 | 1% |
| | Netherlands | 715 | 8% |
| | Norway | 66 | 1% |
| | Portugal | 377 | 4% |
| | Slovakia | 504 | 5% |
| | Spain | 222 | 2% |
| | Sweden | 78 | 1% |
| | Switzerland | 11 | 0% |
| | United Kingdom | 6746 | 71% |
| Companies per country | Belgium | 3 | 3% |
| | France | 1 | 1% |
| | Germany | 1 | 1% |
| | Gibraltar | 1 | 1% |
| | Greece | 1 | 1% |
| | Ireland | 4 | 4% |
| | Jersey | 1 | 1% |
| | Latvia | 1 | 1% |
| | Lithuania | 1 | 1% |
| | Luxembourg | 1 | 1% |
| | Malta | 2 | 2% |
| | Netherlands | 11 | 11% |
| | Norway | 1 | 1% |
| | Portugal | 1 | 1% |
| | Slovakia | 1 | 1% |
| | Spain | 2 | 2% |
| | Sweden | 1 | 1% |
| | Switzerland | 1 | 1% |
| | United Kingdom | 65 | 65% |
| Users per industry type | Business Services | 174 | 2% |
| | Construction | 80 | 1% |
| | Consulting | 2 | 0% |
| | Consumer Services | 136 | 1% |
| | Education | 92 | 1% |
| | Financial Services | 3082 | 33% |
| | Government | 6 | 0% |
| | Healthcare & Pharmaceuticals | 96 | 1% |
| | Hospitality | 592 | 6% |
| | Insurance | 52 | 1% |
| | Legal | 14 | 0% |

| Variable | Value | Count | Percentage |
|---|---|---|---|
| | Manufacturing | 969 | 10% |
| | Non-profit | 86 | 1% |
| | Retail & Wholesale | 229 | 2% |
| | Technology | 3131 | 33% |
| | Transportation | 688 | 7% |
| | Other | 39 | 0% |
| Companies per industry type | Business Services | 6 | 6% |
| | Construction | 1 | 1% |
| | Consulting | 1 | 1% |
| | Consumer Services | 2 | 2% |
| | Education | 1 | 1% |
| | Financial Services | 20 | 20% |
| | Government | 2 | 2% |
| | Healthcare & Pharmaceuticals | 1 | 1% |
| | Hospitality | 1 | 1% |
| | Insurance | 5 | 5% |
| | Legal | 1 | 1% |
| | Manufacturing | 10 | 10% |
| | Non-profit | 2 | 2% |
| | Retail & Wholesale | 5 | 5% |
| | Technology | 36 | 36% |
| | Transportation | 5 | 5% |
| | Other | 1 | 1% |

## Testing of assumptions

### One-way ANOVA

For the KnowBe4 dataset we intend to execute the following one-way ANOVA analyses:

- One-way ANOVA on industry type.

We perform assumption testing the aforementioned (Laerd Statistics, 2022b) and the outcomes are described below.

| One-way ANOVA Industry type | |
|---|---|
| Assumption | Assumption evaluation |
| #1: Dependent variable is measured on a continuous scale | Yes, the dependent variable is secure behaviour score which is measured on a continuous scale. Therefore, assumption is met. |
| #2: Two or more categorical independent variables | Yes, the analysis consists of 17 industry types (ordinal data). |
| #3: Independence of observations | The KnowBe4 dataset is based on an elaborate data collection over a large number of organisations. Through individual organisation ID's data is linked to organisations. Therefore, observations for individual organisations are independent of one another (e.g. the same organisation is not represented |
| | multiple times in the data). As such we perform no further testing on the independence of observations and consider this assumption to be met for our data analysis. |
| #4: Unusual points and significant outliers | Per inspection of the box plot (refer to 12.2.1.1) we determined that there are some outliers in the data. We inspected these data entries and conclude that the average over all industries is not significantly affected since the total number of data entries is 9468. However, at the industry level there are some industries with low record counts. As such, this may distort industry averages and we will include this as a limitation in the discussion section. Nevertheless, we don't consider this a limitation prohibiting us from comparing means through a one-way ANOVA. Therefore, we consider this assumption, apart from the limitation we noted, to be met for our data analysis. |
| #5: Normal distribution of the dependent variable | Per inspection of the distribution histogram (refer to 12.2.1.1) we determined that the sample is approximately normally distributed. There is, however, a small distortion on the left side of the distribution. Per inspection of the Normal Q-Q plot (refer to 12.2.1.1) we determined, similarly, that the distribution of the dependent variable is somewhat distorted, but distribution appears to be marginally sufficient to perform our analysis. Therefore, we consider this assumption to be met for our data analysis. |
| #6: Homogeneity of variances is assumed | Per inspection of the statistics from Levene's test for equality of variances (refer to 12.2.1.1) we determined that the significance level is below the .05 level. This is likely to be due to the fact that some industries have a low record count and significant outliers (refer to assumption above). Therefore, we don't consider this assumption to be met for our analysis. As a result, we will not be consider the coefficients in the one-way ANOVA (e.g. significance level) to be particularly valuable, but we will primarily use the one-way ANOVA to compare means between the different industries. |

<u>Simple and multiple regressions</u>

For the KnowBe4 dataset we intend to execute the following regressions:

- Simple linear (single variable) regression of security consciousness and confidence;
- Simple linear (single variable) regression of security education and training;
- Various simple linear (single variable) regressions with respect to dimensions in organisational security culture and security awareness and knowledge;
- Multiple regression of Organisation security culture dimensions;
- Multiple regression of Security awareness and security knowledge dimensions.

We perform assumption testing for each of the aforementioned (Laerd Statistics, 2022a) and the outcomes are described below. An exception to this is the assumption testing for the simple linear regressions with respect to dimensions in organisational security culture and security awareness and knowledge since assumption testing for these simple regressions is performed as part of the multiple regression assumption testing.

| Simple regression of Security consciousness and confidence | |
|---|---|
| Assumption | Assumption evaluation |
| #1: Dependent variable is measured on a continuous scale | Yes, the dependent variable is secure behaviour score which is measured on a continuous scale. Therefore, assumption is met. |
| #2: One independent variable | Yes, one variable that indicates a user's security consciousness and confidence. Therefore, assumption is met. |

| #3: Independence of observations | The KnowBe4 dataset is based on an elaborate data collection over a large number of users. Through individual user ID's data is linked to individuals. Therefore, observations for individual users are independent of one another (e.g. the same user is not represented multiple times in the data). As such we perform no further testing on the independence of observations and consider this assumption to be met for our data analysis. |
|---|---|
| #4: Linearity of data | Per inspection of the regression scatter plot (refer to 12.2.1.17) we determined that it is reasonable to assume that the data is linear. |
| #5: Homoscedasticity of data | Per inspection of the regression scatter plot (refer to 12.2.1.17) we determined that it is reasonable to assume that the homoscedasticity assumption is met for the data since no significant heteroscedasticity was observed in the inspected plot. We did observe, however, a limited set of significant outliers, but due to the record count (2094) this is not expected to significantly affect the outcomes (also refer to assumption #6 below). |
| #6: Unusual points and significant outliers | Per inspection of studentised deleted residuals we determined that that 73 out of 2094 records exceed the SDR threshold of 3 times the standard deviation. Per inspection of the leverage point values we determined that none of the items had a leverage point value above the 0.20 threshold. Finally, per inspection of the Cook Distance Values we determined that none of the items had a Cook Distance Value above the threshold of 1. As such we conclude there were no unusual points and significant outliers that may significantly affect our data analysis results. |
| #7: Residual errors are normally distributed | Per inspection of the distribution histogram we determined that the data is approximately normally distributed with a mean of close to 0 and the majority of the sample within a standard deviation of 1. Per inspection of the normal Q-Q Plot we did, however, determine that in the residuals towards the high end of the plot there are deviations from the plot trend line. Refer to 12.2.1.17. We consider this assumption to be largely met, but we will include the finding with respect to the distribution of standardised residuals in the normal Q-Q plot as a limitation in our discussion section. |

| Simple regression of Security education and training | |
|---|---|
| **Assumption** | **Assumption evaluation** |
| #1: Dependent variable is measured on a continuous scale | Yes, the dependent variable is secure behaviour score which is measured on a continuous scale. Therefore, assumption is met. |
| #2: One independent variable | Yes, one variable that indicates a user's security education and training. Therefore, assumption is met. |
| #3: Independence of observations | The KnowBe4 dataset is based on an elaborate data collection over a large number of users. Through individual user ID's data is linked to individuals. Therefore, observations for individual users are independent of one another (e.g. the same user is not represented multiple times in the data). As such we perform no further testing on the independence of observations and consider this assumption to be met for our data analysis. |
| #4: Linearity of data | Per inspection of the regression scatter plot (refer to 12.2.1.19) we determined that it is reasonable to assume that the data is linear. |
| #5: Homoscedasticity of data | Per inspection of the regression scatter plot (refer to 12.2.1.19) we determined that it is reasonable to assume that the homoscedasticity assumption is met for the data since no heteroscedasticity was observed in the inspected plot. |
| #6: Unusual points and significant outliers | Per inspection of studentised deleted residuals we determined that that 145 out of 9208 records exceed the SDR threshold of 3 times the standard deviation. Per inspection of the leverage point values we determined that none of the items had a leverage point value above the 0.20 threshold. Finally, per inspection of the Cook Distance Values we determined that none of the items had a Cook Distance Value above the threshold of 1. As such we conclude there were no |
| | unusual points and significant outliers that may significantly affect our data analysis results. |
| #7: Residual errors are normally distributed | Per inspection of the distribution histogram we determined that the data is approximately normally distributed with a mean of close to 0 and the majority of the sample within a standard deviation of 1. Per inspection of the normal Q-Q Plot we did, however, determine that in the residuals towards the high end of the plot there are deviations from the plot trend line. Refer to 12.2.1.19. We consider this assumption to be largely met, but we will include the finding with respect to the distribution of standardised residuals in the normal Q-Q plot as a limitation in our discussion section. |

| Multiple regression of Organisation security culture dimensions | |
|---|---|
| **Assumption** | **Assumption evaluation** |
| #1: Dependent variable is measured on a continuous scale | Yes, the dependent variable is secure behaviour score which is measured on a continuous scale. Therefore, assumption is met. |
| #2: Two or more independent variables | Yes, 6 Organisation security culture dimensions. Therefore, assumption is met. |
| #3: Independence of observations | The KnowBe4 dataset is based on an elaborate data collection over a large number of users. Through individual user ID's data is linked to individuals. Therefore, observations for individual users are independent of one another (e.g. the same user is not represented multiple times in the data). As such we perform no further testing on the independence of observations and consider this assumption to be met for our data analysis. |
| #4: Linearity of data | Per inspection of the partial regression plots (refer to 12.2.2.1) we determined that it is reasonable to assume that the data is linear. |
| #5: Homoscedasticity of data | Per inspection of the partial regression plots (refer to 12.2.2.1) we determined that it is reasonable to assume that the homoscedasticity assumption is met for the data since no (significant) heteroscedasticity was observed in the inspected plots. |
| #6: Multicollinearity of data | Our analysis for multicollinearity on Security Culture's dimensions (refer to 12.2.2.1) reveals that none of Security Culture's dimensions has a VIF of above 10. Therefore, we conclude that there is no (significant) risk of multicollinearity and we consider this assumption met. |
| #7: Unusual points and significant outliers | Per inspection of studentised deleted residuals we determined that that 3 out of 9208 records exceed the SDR threshold of 3 times the standard deviation. Per inspection of the leverage point values we determined that none of the items had a leverage point value above the 0.20 threshold. Finally, per inspection of the Cook Distance Values we determined that none of the items had a Cook Distance Value above the threshold of 1. As such we conclude there were no unusual points and significant outliers that may significantly affect our data analysis results. |
| #8: Residual errors are normally distributed | Per inspection of the distribution histogram we determined that the data is approximately normally distributed with a mean of close to 0 and the majority of the sample within a standard deviation of 1. Per inspection of the normal Q-Q Plot we did, however, determine that in the residuals towards the high end of the plot there are deviations from the plot trend line (refer to 12.2.2.1). However, considering the histogram and the fact that the majority of studentised residuals is close to the Normal Q-Q plot trend line (frequency towards the high end is low compared to the total sample), we consider this assumption to be met. |

| Multiple regression of Security awareness and security knowledge dimensions | |
|---|---|
| **Assumption** | **Assumption evaluation** |
| #1: Dependent variable is measured on a continuous scale | Yes, the dependent variable is secure behaviour score which is measured on a continuous scale. Therefore, assumption is met. |

| #2: Two or more independent variables | Yes, 7 Security awareness and security knowledge dimensions. Therefore, assumption is met. |
|---|---|
| #3: Independence of observations | The KnowBe4 dataset is based on an elaborate data collection over a large number of users. Through individual user ID's data is linked to individuals. Therefore, observations for individual users are independent of one another (e.g. the same user is not represented multiple times in the data). As such we perform no further testing on the independence of observations and consider this assumption to be met for our data analysis. |
| #4: Linearity of data | Per inspection of the partial regression plots (refer to 12.2.2.3) we determined that it is reasonable to assume that the data is linear. |
| #5: Homoscedasticity of data | Per inspection of the partial regression plots (refer to 12.2.2.3) we determined that it is reasonable to assume that the homoscedasticity assumption is met for the data since no (significant) heteroscedasticity was observed in the inspected plots. |
| #6: Multicollinearity of data | Our analysis for multicollinearity on Security awareness and knowledge dimensions (refer to 12.2.2.3) reveals that none of Security awareness and knowledge dimensions has a VIF of above 10. Therefore, we conclude that there is no (significant) risk of multicollinearity and we consider this assumption met. |
| #7: Unusual points and significant outliers | Per inspection of studentised deleted residuals we determined that that 158 out of 9208 records exceed the SDR threshold of 3 times the standard deviation. Per inspection of the leverage point values we determined that none of the items had a leverage point value above the 0.20 threshold. Finally, per inspection of the Cook Distance Values we determined that none of the items had a Cook Distance Value above the threshold of 1. As such we conclude there were no unusual points and significant outliers that may significantly affect our data analysis results. |
| #8: Residual errors are normally distributed | Per inspection of the distribution histogram we determined that the data is approximately normally distributed with a mean of close to 0 and the majority of the sample within a standard deviation of 1. Per inspection of the normal Q-Q Plot we did, however, determine that in the residuals towards the high end of the plot there are deviations from the plot trend line (refer to 12.2.2.3). However, considering the histogram and the fact that the majority of studentised residuals is close to the Normal Q-Q plot trend line (frequency towards the high end is low compared to the total sample), we consider this assumption to be met. |

### 6.2.4 Data analysis and statistical tests

#### 6.2.4.1 EU dataset

Below an overview is included of all variables and data types in the EU dataset that are incorporated into our research. Additionally, this overview indicates which type of data analyses we will apply to the variable. For the variables "Age" and "Prior experience with security incidents" (both direct and indirect) we use the scale data to deduce ordinal data for additional analysis:

- **Age**: we want to analyse both the influence of age on secure behaviour, but also differences in secure behaviour between age groups. Therefore, we deduce age groups, largely in accordance with Chua et al., 2018, to enable comparison of different age groups. These age groups are 15-20, 21-29, 30-39, 40-49, 50-59, 60-69 and 70+ years old.

- **Prior experience with security incidents**: we want to we want to analyse both the influence of prior experience on secure behaviour, but also analyse differences in secure behaviour between groups with different levels of prior experience with security incidents. We use the following experience categories for both direct and indirect prior experience with security incidents:

    o <u>Little experience</u>: 0-25% of the prior experience scenario's in the question apply to the respondent;

    o <u>Some experience</u>: 26-50% of the prior experience scenario's in the question apply to the respondent;

    o <u>Medium experience</u>: 51-75% of the prior experience scenario's in the question apply to the respondent;

    o <u>Large experience</u>: 76-100% of the prior experience scenario's in the question apply to the respondent.

| Variable | Data type | Statistical test |
|---|---|---|
| **Context-level** | | |
| National level | | |
| National culture | | |
| Hofstede's cultural dimensions | | |
| Power distance | Interval (scale) | Simple regression analysis (to analyse the impact of the individual cultural dimension on secure behaviour) and (stepwise) multiple regression (to analyse the impact of the various cultural dimensions combined on secure behaviour). |
| Individualism | Interval (scale) | Simple regression analysis (to analyse the impact of the individual cultural dimension on secure behaviour) and (stepwise) multiple regression (to analyse the impact of the various cultural dimensions combined on secure behaviour). |
| Uncertainty avoidance | Interval (scale) | Simple regression analysis (to analyse the impact of the individual cultural dimension on secure behaviour) and (stepwise) multiple regression (to analyse the impact of the various cultural dimensions combined on secure behaviour). |
| Masculinity | Interval (scale) | Simple regression analysis (to analyse the impact of the individual cultural dimension on secure behaviour) and (stepwise) multiple regression (to analyse the impact of the various cultural dimensions combined on secure behaviour). |
| Long-term orientation | Interval (scale) | Simple regression analysis (to analyse the impact of the individual cultural dimension on secure behaviour) and |

| Variable | Data type | Statistical test |
|---|---|---|
| | | (stepwise) multiple regression (to analyse the impact of the various cultural dimensions combined on secure behaviour). |
| Indulgence | Interval (scale) | Simple regression analysis (to analyse the impact of the individual cultural dimension on secure behaviour) and (stepwise) multiple regression (to analyse the impact of the various cultural dimensions combined on secure behaviour). |
| Meyer's cultural dimensions | | |
| Communicating | Interval (scale) | Simple regression analysis (to analyse the impact of the individual cultural dimension on secure behaviour) and (stepwise) multiple regression (to analyse the impact of the various cultural dimensions combined on secure behaviour). |
| Evaluating | Interval (scale) | Simple regression analysis (to analyse the impact of the individual cultural dimension on secure behaviour) and (stepwise) multiple regression (to analyse the impact of the various cultural dimensions combined on secure behaviour). |
| Leading | Interval (scale) | Simple regression analysis (to analyse the impact of the individual cultural dimension on secure behaviour) and (stepwise) multiple regression (to analyse the impact of the various cultural dimensions combined on secure behaviour). |
| Deciding | Interval (scale) | Simple regression analysis (to analyse the impact of the individual cultural dimension on secure behaviour) and (stepwise) multiple regression (to analyse the impact of the various cultural dimensions combined on secure behaviour). |
| Trusting | Interval (scale) | Simple regression analysis (to analyse the impact of the individual cultural dimension on secure behaviour) and (stepwise) multiple regression (to analyse the impact of the various cultural dimensions combined on secure behaviour). |
| Disagreeing | Interval (scale) | Simple regression analysis (to analyse the impact of the individual cultural dimension on secure behaviour) and (stepwise) multiple regression (to analyse the impact of the various cultural dimensions combined on secure behaviour). |
| Scheduling | Interval (scale) | Simple regression analysis (to analyse the impact of the individual cultural dimension on secure behaviour) and (stepwise) multiple regression (to analyse the impact of the various cultural dimensions combined on secure behaviour). |

| Variable | Data type | Statistical test |
|---|---|---|
| Persuading | Interval (scale) | Simple regression analysis (to analyse the impact of the individual cultural dimension on secure behaviour) and (stepwise) multiple regression (to analyse the impact of the various cultural dimensions combined on secure behaviour). |
| **Individual-level** | | |
| Demographics | | |
| Age | Ratio (scale) | Simple regression (to analyse the impact of age on secure behaviour) |
| Age groups | Ordinal | One-way ANOVA (to analyse the difference in secure behaviour between various age groups. |
| Gender | Nominal | Independent samples T-test (to analyse the difference in secure behaviour between genders). |
| Level of urbanisation | Ordinal | One-way ANOVA (to analyse the difference in secure behaviour between various levels of urbanisation). |
| Prior experience – Direct | Interval (scale) | Simple regression (to analyse the impact of direct experience on secure behaviour) and (enter) multiple regression (to analyse the impact of direct and indirect experience combined on secure behaviour). |
| Prior experience – Direct per category | Ordinal | One-way ANOVA (to analyse the difference in secure behaviour between various direct experience categories). |
| Prior experience – Indirect | Interval (scale) | Simple regression (to analyse the impact of indirect experience on secure behaviour) and (enter) multiple regression (to analyse the impact of direct and indirect experience combined on secure behaviour). |
| Prior experience – Indirect per category | Ordinal | One-way ANOVA (to analyse the difference in secure behaviour between various indirect experience categories). |
| **Secure behaviour** | | |
| Secure behaviour | | |
| Secure behaviour | Interval (scale) | Not applicable, concerns the dependent variable. |

### 6.2.4.2  KnowBe4 dataset

Below an overview is included of all variables and data types in the KnowBe4 dataset that are incorporated into our research. Additionally, this overview indicates which type of data analyses we will apply to the variable.

| Variable | Data type | Statistical test |
|---|---|---|
| **Context-level** | | |
| **Industry type** | | |
| Industry type | Ordinal | One-way ANOVA (to analyse the difference in secure behaviour between various industries). |
| **Organisation level** | | |
| Organisation security culture | | |
| Attitudes | Interval (scale) | Simple regression analysis (to analyse the impact of the individual organisational security culture dimension on secure behaviour) and (stepwise) multiple regression (to analyse the impact of the various organisational security culture dimensions combined on secure behaviour). |
| Cognition | Interval (scale) | Simple regression analysis (to analyse the impact of the individual organisational security culture dimension on secure behaviour) and (stepwise) multiple regression (to analyse the impact of the various organisational security culture dimensions combined on secure behaviour). |
| Communication | Interval (scale) | Simple regression analysis (to analyse the impact of the individual organisational security culture dimension on secure behaviour) and (stepwise) multiple regression (to analyse the impact of the various organisational security culture dimensions combined on secure behaviour). |
| Compliance | Interval (scale) | Simple regression analysis (to analyse the impact of the individual organisational security culture dimension on secure behaviour) and (stepwise) multiple regression (to analyse the impact of the various organisational security culture dimensions combined on secure behaviour). |
| Norms | Interval (scale) | Simple regression analysis (to analyse the impact of the individual organisational security culture dimension on secure behaviour) and (stepwise) multiple regression (to analyse the impact of the various organisational security culture dimensions combined on secure behaviour). |
| Responsibilities | Interval (scale) | Simple regression analysis (to analyse the impact of the individual organisational security culture dimension on secure behaviour) and (stepwise) multiple regression (to analyse the |

| Variable | Data type | Statistical test |
|---|---|---|
| | | impact of the various organisational security culture dimensions combined on secure behaviour). |
| **Individual-level** | | |
| Security specific | | |
| Security education and training | Interval (scale) | Simple regression analysis (to analyse the impact of the an individual's security education and training on secure behaviour) |
| Security consciousness and confidence | Interval (scale) | Simple regression analysis (to analyse the impact of the an individual's security consciousness and confidence on secure behaviour) |
| Security awareness | Interval (scale) | Simple regression analysis (to analyse the impact of the individual security awareness and security knowledge dimensions on secure behaviour) and (stepwise) multiple regression (to analyse the impact of the various security awareness and security knowledge dimensions combined on secure behaviour). |
| Security knowledge | Interval (scale) | Refer to "Security awareness"; as indicated earlier, we test security awareness and security knowledge combined based on a combined set of data. |
| **Secure behaviour** | | |
| Secure behaviour | | |
| Secure behaviour | Interval (scale) | Not applicable, concerns the dependent variable. |

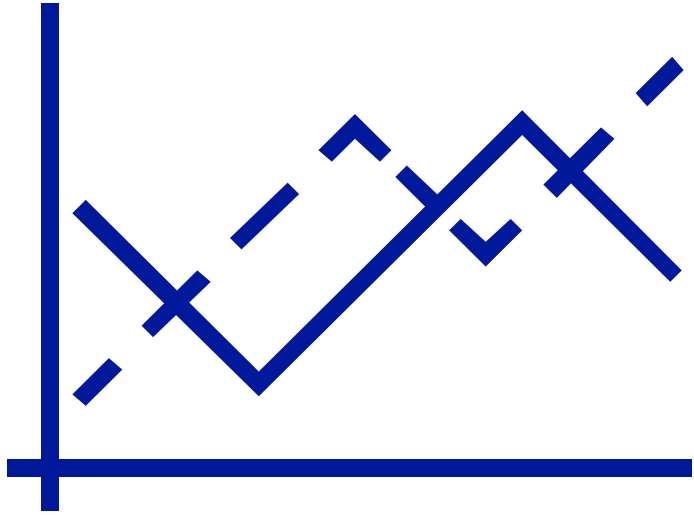

# 7

# Results

# 7 Results

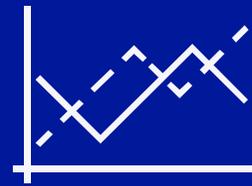

## 7.1 National level: National culture

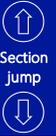
Section jump

### 7.1.1 Hofstede cultural dimensions

Single variable testing through simple regressions

We analysed Hofstede's cultural dimensions through a simple regression (refer to 12.1.1.9 (Power distance), 12.1.1.10 (Individualism), 12.1.1.11 (Uncertainty avoidance), 12.1.1.12 (Masculinity), 12.1.1.13 (Long-term orientation) and 12.1.1.14 (Indulgence)). These analyses reveal that:

- Power distance has a statistically significant (<.001) negative relation with secure behaviour. Therefore, we conclude that there is **no support for hypothesis 1**, but a statistically significant relation is found in the opposite direction than predicted.

- Individualism has a statistically significant (<.001) positive relation with secure behaviour. Therefore, we conclude that there is **no support for hypothesis 2a**, but there is **support for hypothesis 2b**.

- Uncertainty avoidance has a statistically significant (<.001) negative relation with secure behaviour. Therefore, we conclude that there is **no support for hypothesis 3**, but a statistically significant relation is found in the opposite direction than predicted.

- Masculinity has a statistically significant (<.001) negative relation with secure behaviour. Therefore, we conclude that there is **no support for hypothesis 4a** and **support for hypothesis 4b**.

- Long-term orientation has a statistically significant (.003) positive relation with secure behaviour. Therefore, we conclude that there is **support for hypothesis 5**.

- Indulgence has a statistically significant (<.001) positive relation with secure behaviour. Therefore, we conclude that there is **no support for hypothesis 6**, but a statistically significant relation is found in the opposite direction than predicted.

Overall model of Hofstede's cultural dimensions

We analysed Hofstede's various cultural dimensions through a multiple regression to take into account possible interrelations between various Hofstede cultural dimensions and assess the overall model fit (refer to 12.1.2.5). We tested the model based on our earlier assumption tests and our multicollinearity test in particular (refer to section 6.2.3.2). As a result of this we included all of Hofstede's cultural dimensions. The stepwise multiple regression yields four different statistical models; we selected the one with the highest model fit (an R Square of .092) since this is the model that can explain the most variance with respect to the dependent variable (secure behaviour). Based on this model (model 4) we conclude that Hofstede dimensions Individualism and Uncertainty avoidance were excluded from the model due to lack of explanatory power when it comes to variance of the dependent variable. The four remaining variables were all in accordance with the result we found through our separate simple regressions. As such we conclude that for the various dimensions, as far as included in the multiple regression model, our conclusions regarding the respective hypotheses do not change.

### 7.1.2  Meyer cultural dimensions

**Single variable testing through simple regressions**

We analysed Meyer's cultural dimensions through a simple regression (refer to Communicating (12.1.1.15), Evaluating (12.1.1.16), Leading (12.1.1.17), Deciding (12.1.1.18), Trusting (12.1.1.19), Disagreeing (12.1.1.20), Scheduling (12.1.1.21) and Persuading (12.1.1.22)). These analyses reveal that:

- Communicating has a statistically significant (<.001) negative relation with secure behaviour. Therefore, we conclude that there is **support for hypothesis 7**.
- Evaluating has a statistically significant (<.001) negative relation with secure behaviour. Therefore, we conclude that there is **support for hypothesis 8a** and **no support for hypothesis 8b**.
- Leading has a statistically significant (<.001) negative relation with secure behaviour. Therefore, we conclude that there is **no support for hypothesis 9**.
- Deciding has a statistically significant (.000) negative relation with secure behaviour. Therefore, we conclude that there is **no support for hypothesis 10**.
- Trusting has a statistically significant (<.001) negative relation with secure behaviour. Therefore, we conclude that there is **support for hypothesis 11**.
- Disagreeing has a statistically insignificant (.726) relation with secure behaviour. Therefore, we conclude that there is **no support for hypothesis 12 and 12b**.
- Scheduling has a statistically significant (<.001) negative relation with secure behaviour. Therefore, we conclude that there is **support for hypothesis 13**.
- Persuading has a statistically significant (<.001) positive relation with secure behaviour. Therefore, we conclude that there is **support for hypothesis 14**.

**Overall model of Meyer's cultural dimensions**

We analysed Meyer's various cultural dimensions through a multiple regression to take into account possible interrelations between various Meyer cultural dimensions and assess the overall model fit (refer to 12.1.2.6). We tested the model based on our earlier assumption tests and our multicollinearity test in particular (refer to section 6.2.3.2). As a result of this we included all of Meyer's cultural dimensions except for "Trusting" and "Persuading". The stepwise multiple regression yields six different statistical models; we selected the one with the highest model fit (an R Square of .097) since this is the model that can explain the most variance with respect to the dependent variable (secure behaviour). Based on this model (model 6) we conclude that none of the initially included Meyer dimensions were excluded from the model due to lack of explanatory power when it comes to variance of the dependent variable. As such, all six variables were included in the model with the highest model fit. For the dimensions "Leading", "Deciding" and "Scheduling" the statistical significance and direction of the relation were similar to the ones identified in the simple regressions performed for each of these dimensions (**no support for hypothesis 9 and 10**, but **support for hypothesis 13**). For the dimensions "Communicating" and "Evaluating" a significant relation was found, but in the opposite direction when compared to the simple regressions (**no support for hypothesis 7 and 8a**, but **support for hypothesis 8b**). For the dimension "Disagreeing" a statistically significant negative relation was identified whilst the simple regression showed no relation (**no support for hypothesis 12b**, but **support for hypothesis 12a**).

As mentioned, <u>Communicating</u> and <u>Evaluating</u> were found to have a negative relationship with secure behaviour in the simple regression (single variable test) and a positive relationship with secure behaviour in the multiple regression (multiple variable test). A possible explanation for this could be that there are variables that, when included in the multiple regression model, manipulate the relationship that "Communicating" and "Evaluating" have with secure behaviour. An example of this could be "suppression" in a regression analysis. We ran different analyses to determine whether there is a single variable that, when excluded from the multiple regression model, may explain the shift from a negative relation to a positive relation for the "Communicating" and "Evaluating" variables. We found no such variable. However, through exploratory analysis we identified, by means of an additional **multiple regression**, that when excluding the variables "Disagreeing" and "Scheduling" from the multiple regression model, all remaining variables (including "Communicating" and "Evaluating") yield the same results (a negative relationship) as they did in the simple regression (single variable test). As such, we conclude that different cultural dimensions of Meyer may be influencing each other when combined into a multiple regression model. We further explored this area through partial correlation analyses to identify if controlling for one or more of Meyer's dimensions (variables included in the multiple regression model) would explain why Communicating and Evaluating are switching in terms of their direction of the statistical relationship. Through exploratory analysis we identified that:

- <u>Communicating</u> changes from a statistically significant negative to a statistically significant positive relationship (.016 at p = .039) by controlling for the following of Meyer's cultural dimensions (refer to the **partial correlation analysis**): Evaluating, Leading, Deciding, Disagreeing and Scheduling.
- <u>Evaluating</u> changes from a statistically significant negative to a statistically significant positive relationship (.037 at p = .000) by controlling for the following of Meyer's cultural dimensions (refer to the **partial correlation analysis**): Disagreeing and Scheduling.

Although we have accounted for multicollinearity in our multiple regressions analysis, the findings from the additional multiple regression analysis and the partial correlation analyses warrant additional research on interrelationships between the dimensions of Meyer. Also refer to the **discussion** section where we interpret these results.

### 7.2.1  Industry type

**Industry type**

We analysed industry type through a one-way ANOVA (refer to 12.2.1.2). This analysis reveals that different industry types have a statistically significant (<.0000000000000002) difference in secure behaviour scores. We note however, a limitation in our assumption testing (refer to 6.2.3.3). As such, we focus primarily on the comparison of means between industries. Below we included a table of the secure behaviour means per industry type and the number of users within each of these industries in the sample. In accordance with our hypothesis, we find that industry type has an association with secure behaviour in the sense that secure behaviour varies statistically significant between industries. Therefore, we conclude that there is **support for hypothesis 15**.

| Industry type | N | Mean |
|---|---|---|
| Business Services | 174 | 16.00 |
| Construction | 80 | 15.21 |
| Consulting | 2 | 14.50 |
| Consumer Services | 136 | 15.67 |
| Education | 92 | 14.72 |
| Financial Services | 3082 | 14.95 |
| Government | 6 | 15.67 |
| Healthcare & Pharmaceuticals | 96 | 16.31 |
| Hospitality | 592 | 14.69 |
| Insurance | 52 | 14.40 |
| Legal | 14 | 16.14 |
| Manufacturing | 969 | 16.43 |
| Non-profit | 86 | 15.50 |
| Retail & Wholesale | 229 | 16.67 |
| Technology | 3131 | 16.25 |
| Transportation | 688 | 16.67 |
| Other | 39 | 14.93 |
| **Total** | **9468** | **15.73** |

### 7.3.1 Organisational security culture

**Single variable testing through simple regressions**

We analysed the organisational security culture dimensions through a simple regression (refer to 12.2.1.4 (Attitude), 12.2.1.5 (Cognition), 12.2.1.6 (Communication), 12.2.1.7 (Compliance), 12.2.1.8 (Norms) and 12.2.1.9 (Responsibilities). These analyses reveal that:

- <u>Attitude</u> has a statistically significant (<.001) negative relation with secure behaviour. Therefore, we conclude that there is **no support for hypothesis 16**, but a statistically significant relation is found in the opposite direction than predicted.

- <u>Cognition</u> has a statistically significant (<.001) negative relation with secure behaviour. Therefore, we conclude that there is **no support for hypothesis 17**, but a statistically significant relation is found in the opposite direction than predicted.

- <u>Communication</u> has a statistically significant (<.001) negative relation with secure behaviour. Therefore, we conclude that there is **support for hypothesis 18**.

- <u>Compliance</u> has a statistically significant (<.001) negative relation with secure behaviour. Therefore, we conclude that there is **no support for hypothesis 19**, but a statistically significant relation is found in the opposite direction than predicted.

- <u>Norms</u> has a statistically significant (<.001) negative relation with secure behaviour. Therefore, we conclude that there is **no support for hypothesis 20**, but a statistically significant relation is found in the opposite direction than predicted.

- <u>Responsibilities</u> has a statistically insignificant (.373) negative relation with secure behaviour. Therefore, we conclude that there is **no support for hypothesis 21**.

**Overall model of Organisational Security Culture dimensions**

We analysed the organisational security culture dimensions through a multiple regression to take into account possible interrelations between various dimensions and assess the overall model fit (refer to 12.2.2.2). We tested the model based on our earlier assumption tests and our multicollinearity test in particular (refer to section 6.2.3.3). As a result of this we included all organisational security culture dimensions. The stepwise multiple regression yields five different statistical models; we selected the one with the highest model fit (an R Square of .108) since this is the model that can explain the most variance with respect to the dependent variable (secure behaviour). This model (model 5) includes the following organisational security culture dimensions: Compliance, Communication, Cognition, Attitudes and Responsibilities. The dimension "Norms" was excluded due to model fit. For the dimensions "Cognition", "Communication" and "Compliance" the statistical significance and direction of the relation were similar to the ones identified in the simple regressions performed for each of these dimensions (**no support for hypothesis 17 and 19**, but **support for hypothesis 18**). For the dimension "Attitudes" a significant relation was found in accordance with the hypothesis, but in the opposite direction when compared to the simple regression (**support for hypothesis 16**). For the dimension "Responsibilities" a statistically significant positive relation was identified in accordance with the hypothesis whilst the simple regression showed no relation (**support for hypothesis 21**).

## 7.4 Individual level

### 7.4.1 Demographics

**Age**

We analysed <u>age</u> both through a one-way ANOVA of different age categories (refer to 12.1.1.3) and through a simple regression of the impact of age on secure behaviour (refer to 12.1.1.4). The one-way ANOVA of different age categories reveals a lower secure behaviour score (3,31 or 20,7%) for young adults (15-20 years old) compared to the overall mean (3,54 or 22,1%). The age group of 60-69 years old has the highest security behaviour score with a score of 3,64 or 22,8%, and the age group of 70+ years old also has a higher than average secure behaviour score with 3,59 or 22,4%. Below we included a table of the secure behaviour means per age group. These outcomes suggest that there is support for our hypothesis that secure behaviour increases with age. This presumption is supported by the simple regression on age in relation to secure behaviour. Through the simple regression we find that age has a statistically significant (significance = .007) relation with secure behaviour, although the strength of the relation is limited as evidenced by the B of .003. Therefore we conclude that there is **support for hypothesis 22** that age has a statistically significant and positive (albeit weak) relation with secure behaviour.

| Age category | N | Mean | Percentage |
|---|---|---|---|
| 15-20 years old | 1543 | 3,31 | 20,7% |
| 21-29 years old | 2333 | 3,54 | 22,1% |
| 30-39 years old | 3754 | 3,56 | 22,3% |
| 40-49 years old | 3942 | 3,52 | 22,0% |
| 50-59 years old | 4030 | 3,5 | 21,9% |
| 60-69 years old | 3717 | 3,64 | 22,8% |
| 70+ years old | 2343 | 3,59 | 22,4% |
| Total | 21662 | 3,54 | 22,1% |

**Gender**

We analysed <u>gender</u> through an independent samples T-test (refer to 12.1.1.1). This analysis reveals that males have a slightly higher average secure behaviour score (3,67 or 22,9%) compared to females (3,43 or 21,4%).

This finding contradicts our hypothesis which stated that females display more secure behaviour than males. Therefore, we conclude that there is **no support for hypothesis 23**. However, the different genders are found to have statistically significant (<.001) different mean secure behaviour scores.

### Level of urbanisation

We analysed level of urbanisation through a one-way ANOVA (refer to 12.1.1.2). This analysis reveals that different levels of urbanisation have a statistically significant (.007) difference in secure behaviour scores. Below we included a table of the secure behaviour means per level of urbanisation. In accordance with our hypothesis, rural areas and villages have the lowest secure behaviour score (3,45 or 21,6%). Small and medium-sized towns are found the have the highest secure behaviour score (2,59 or 22,4%) closely followed by large towns and cities (2,57 or 22,3%). According to the hypothesis, we would expect the secure behaviour score of large towns and cities to be the highest; this is not the case. However, since the difference between small / medium-sized towns and large towns / cities is limited (3,59 vs. 3,57), and since rural areas / villages clearly have a lower scores (3,45) than small / medium-sized towns and large towns / cities we conclude that there is **support for hypothesis 24**.

| Urbanisation | N | Mean | Percentage |
|---|---|---|---|
| Blanks | 8 | - | - |
| Rural area or village | 6896 | 3,45 | 21,6% |
| Small or medium-sized town | 8311 | 3,59 | 22,4% |
| Large town or city | 6447 | 3,57 | 22,3% |
| Total | 21662 | 3,54 | 22,1% |

## 7.4.2 Security specific

### Security education and training

We analysed security education and training through a simple regression (refer to 12.2.1.20). This analysis revealed no statistically significant (p = .202) relation between security education and training and secure behaviour. For exploratory purposes we performed a one-way ANOVA (refer to 12.2.1.3) between various security education and training groups. We divided the sample into three groups: 1) individuals with 1-2 security trainings, 2) individuals with 3-4 security training and 3) individuals with 5 or more security trainings. We noted a negligible difference in the means of these three groups. This seems to confirm the results of the simple regression in the sense that we find no (statistically significant) increase in secure behaviour (mean) with an increase in security education and training. Therefore we conclude that there is **no support for hypothesis 25**.

**Security consciousness and confidence**

We analysed security consciousness and confidence through a simple regression (refer to 12.2.1.18). This analysis revealed no statistically significant (p = .290) relation between security consciousness and confidence and secure behaviour. Therefore we conclude that there is **no support for hypothesis 26**.

**Security awareness and security knowledge**

Single variable testing through simple regressions

We analysed security awareness and security knowledge through the results from the Security Awareness Proficiency Assessment (SAPA) measured for different users. We measured the impact of security awareness and security awareness through simple linear regressions that analyse the individual impact of each of the seven security awareness and security knowledge domains on secure behaviour:

- Passwords & Authentication has a statistically significant (.000206) positive relation with secure behaviour (refer to the simple regression in 12.2.1.10);

- Email Security has a statistically significant (.0.000000000000579) positive relation with secure behaviour (refer to the simple regression in 12.2.1.11);

- Internet Use has a statistically significant (.0000000318) positive relation with secure behaviour (refer to the simple regression in 12.2.1.12);

- Social Media has a statistically significant (.0101) positive relation with secure behaviour (refer to the simple regression in 12.2.1.13);

- Mobile Devices has a statistically significant (.0000000000000002) positive relation with secure behaviour (refer to the simple regression in 12.2.1.14);

- Incident Reporting has a statistically significant (.00141) positive relation with secure behaviour (refer to the simple regression in 12.2.1.15);

- Security Awareness has a statistically significant (.0000000168) positive relation with secure behaviour (refer to the simple regression in 12.2.1.16).

Considering that our hypotheses (27 en 28) predicted a positive association between security awareness and knowledge and secure behaviour, and considering the statistically significant positive relations we identified, we conclude that there is **support for hypothesis 27** and **28**.

Overall model of Security awareness and security knowledge dimensions

Additionally, we analysed the impact of all seven security awareness and security knowledge domains combined through a multiple regression (refer to the multiple regression in 12.2.2.4). We tested the model based on our earlier assumption tests and our multicollinearity test in particular (refer to section 6.2.3.3). As a result of this we included all of the security awareness and knowledge dimensions as potential variables in the multiple regression model. The stepwise multiple regression yields five different statistical models; we selected the one with the highest model fit (an R Square of .017) since this is the model that can explain the most variance with respect to the dependent variable (secure behaviour). Based on this model (model 5) we conclude that the predictive power of the model is limited (low R Square), but 4 out of 5 variables in the model are statistically significant at the .05 level (all have a significance level of .000) and 1 out of 5 variables is marginally significant at the .10 level (significance level is .085). All 5 variables have positive relation with secure behaviour. As such, we conclude there is **support for hypothesis 27** and **28**.

### Prior experience with security incidents

We analysed prior experience with security incidents by dividing the variable into a score for direct prior experience with security incidents (first hand experience) and a score for indirect prior experience with security incidents (second or third hand experience). Additionally, we generated experience categories indicating whether the individual has little, some, medium or large (in)direct prior experience with security incidents. We analysed the resulting categories of direct and indirect prior experience with security incidents through one-way ANOVA analyses (refer to  and ). In addition, we analysed the direct and indirect prior experience scores individually through a simple regression (refer to  and ), and combined through a multiple regression (refer to ).

The one-way ANOVA analyses reveal that there is a statistically significant (<.001) difference in secure behaviour score mean between different categories of prior experience with security incidents, both direct and indirect. The table below shows the different means per category for both direct and indirect prior experience with security incidents. Interestingly, when it comes to direct prior experience with security incidents secure behaviour scores are lowest for individuals with medium (3,13 or 19,6%) to large (2,36 or 14,8%) direct prior experience with security incidents. Individuals with little direct prior experience have a slightly below average secure behaviour score (3,46 or 21,6%) and individuals with some direct prior experience have the highest secure behaviour score (4,82 or 30,1%). We see a different pattern when examining indirect prior experience with security incidents. Individuals with little indirect prior experience with security incidents are found to have the lowest secure behaviour score (3,31 or 20,7%). Individuals with large indirect prior experience are found to have significantly higher secure behaviour scores (4,57 or 28,6%), but lower than individuals with some indirect prior experience (4,93 or 30,8) and significantly lower than individuals with medium indirect prior experience (5,92 or 37,0%). Generally speaking, this suggests that as prior experience with security incidents, both direct and indirect, increases we see an initial increase in secure behaviour followed by a decrease in secure behaviour as prior experience with security incidents keeps increasing. As such, prior experience initially has a positive influence on secure behaviour, but as experience increases this positive influence reduces significantly.

| Direct prior experience | | | | Indirect prior experience | | | |
|---|---|---|---|---|---|---|---|
| Experience | N | Mean | Percentage | Experience | N | Mean | Percentage |
| Little experience | 19552 | 3,46 | 21,6% | Little experience | 18677 | 3,31 | 20,7% |
| Some experience | 1521 | 4,82 | 30,1% | Some experience | 2758 | 4,93 | 30,8% |
| Medium experience | 420 | 3,13 | 19,6% | Medium experience | 206 | 5,92 | 37,0% |
| Large experience | 169 | 2,36 | 14,8% | Large experience | 21 | 4,57 | 28,6% |
| Total | 21662 | 3,54 | 22,1% | Total | 21662 | 3,54 | 22,1% |

In addition to the one-way ANOVA analyses, as mentioned above, we performed two simple regressions on direct and indirect prior experience with security incidents, and one multiple regression in which both variables are incorporated into one statistical model (with a model fit of R Square .068). Both the simple regressions and the multiple regressions reveal a statistically significant (simple regressions: <.001 and multiple regression: .000) positive relation between direct and indirect prior experience with security incidents and secure behaviour. Therefore, we conclude that there is **support for hypothesis 29**. However, as discussed above, the relation between direct and indirect prior experience with security incidents and secure behaviour is found to be more complex than a simple linear relation.

## 7.5   Overview of outcomes

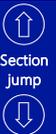

Below is an overview of the outcomes of our analyses. We have included, amongst other things, the significance level and the beta (B) that indicates the strength of the direction of the relation (when statistically significant). To put this into perspective, it is important to take into consideration that we measured secure behaviour in two ways. The mean secure behaviour score for **EU** variables is **3,54** and the mean secure behaviour score for **KnowBe4** variables is **15,73**. Earlier we indicated based on which data variables were measured (refer to 6.1.2). For clarity we have indicated this once more using colours in the column "Variable": yellow indicates variables measured based on EU data and green indicates KnowBe4.

| Level | Variable | Hypothesis | Relation predicted | Single variable test | Significance (p) and relationship (B) | Multiple variable test | Significance (p) and relationship (B) | Evaluation |
|---|---|---|---|---|---|---|---|---|
| National level: Hofstede | Power distance | 1 | + | - | p = <.001 B = -.025 | - | p = .000 B = -.009 | Significant, no support |
| | Individualism | 2a | - | + | p = <.001 B = .022 | Excluded (model fit) | Not applicable | Significant, no support |
| | | 2b | + | | | | | Significant, support |
| | Uncertainty avoidance | 3 | + | - | p = <.001 B = -.019 | Excluded (model fit) | Not applicable | Significant, no support |
| | Masculinity | 4a | + | - | p = <.001 B = -.014 | - | p = .000 B = -.011 | Significant, no support |
| | | 4b | - | | | | | Significant, support |

| Level | Variable | Hypothesis | Relation predicted | Single variable test | Significance (p) and relationship (B) | Multiple variable test | Significance (p) and relationship (B) | Evaluation |
|---|---|---|---|---|---|---|---|---|
| | Long-term orientation | 5 | + | + | p = .003 B = .003 | + | p = .000 B = .021 | Significant, support |
| | Indulgence | 6 | - | + | p = <.001 B = .031 | + | p = .000 B = .032 | Significant, no support |
| National level: Meyer | Communicating | 7 | - | - | p = <.001 B = -.429 | + | p = .039 B = .057 | Significant, partial support |
| | Evaluating | 8a | - | - | p = <.001 B = -.094 | + | p = .000 B = .166 | Significant, partial support |
| | | 8b | + | | | | | Significant, partial support |
| | Leading | 9 | + | - | p = <.001 B = -.282 | - | p = .000 B = -.173 | Significant, no support |
| | Deciding | 10 | + | - | p = .000 B = -.312 | - | p = .000 B = -.129 | Significant, no support |
| | Trusting | 11 | - | - | p = <.001 B = -.354 | Excluded (multicollinearity) | Not applicable | Significant, support |
| | Disagreeing | 12a | - | None | p = .726 B = Not applicable | - | p = .000 B = -.265 | Significant, partial support |
| | | 12b | + | | | | | Significant, no support |
| | Scheduling | 13 | - | - | p = <.001 B = -.296 | - | p = .000 B = -.194 | Significant, support |
| | Persuading | 14 | + | + | p = <.001 B = .196 | Excluded (multicollinearity) | Not applicable | Significant, support |
| Industry level | Industry type | 15 | ~ | ~ | p = <.000 B = Not applicable | Not applicable | Not applicable | Significant, support |
| Organisation level | Attitude | 16 | + | - | p = <.001 B = -.037 | + | p = .000 B = .139 | Significant, partial support |
| | Cognition | 17 | + | - | p = <.001 B = -.010 | - | p = .000 B = -.182 | Significant, no support |
| | Communication | 18 | + | + | p = <.001 B = .023 | + | p = .000 B = .145 | Significant, support |
| | Compliance | 19 | + | - | p = <.001 B = -.062 | - | p = .000 B = -.118 | Significant, no support |
| | Norms | 20 | + | - | p = <.001 B = -.027 | Excluded (model fit) | Not applicable | Significant, no support |
| | Responsibilities | 21 | + | None | p = .373 B = Not applicable | + | p = .000 B = .049 | Significant, partial support |
| Demographics | Age | 22 | + | + | p = .007 B = .003 | Not applicable | Not applicable | Significant, support |
| | Gender | 23 | Females behave more securely than males | Males behave more securely than females | p = <.001 B = Not applicable | Not applicable | Not applicable | Significant, no support |

| Level | Variable | Hypothesis | Relation predicted | Single variable test | Significance (p) and relationship (B) | Multiple variable test | Significance (p) and relationship (B) | Evaluation |
|---|---|---|---|---|---|---|---|---|
| | Level of urbanisation | 24 | + | Mixed results | p = .007 B = Not applicable | Not applicable | Not applicable | Significant, support |
| Security specific | Security education and training | 25 | + | None | p = .202 B = Not applicable | Not applicable | Not applicable | No support |
| | Security consciousness and confidence | 26 | + | None | p = .290 B = Not applicable | Not applicable | Not applicable | No support |
| | Security awareness | 27 | + | + | p = Various B = 0 to .007 | + | p = .000 B = .002 to .008 | Significant, support |
| | Security knowledge | 28 | + | | | | | |
| | Prior experience with security incidents - Direct | 29 | + | + | p = <.001 B = .138 | + | p = .000 B = .071 | Significant, support |
| | Prior experience with security incidents - Indirect | | | + | p = <.001 B = .595 | + | p = .000 B = .522 | |

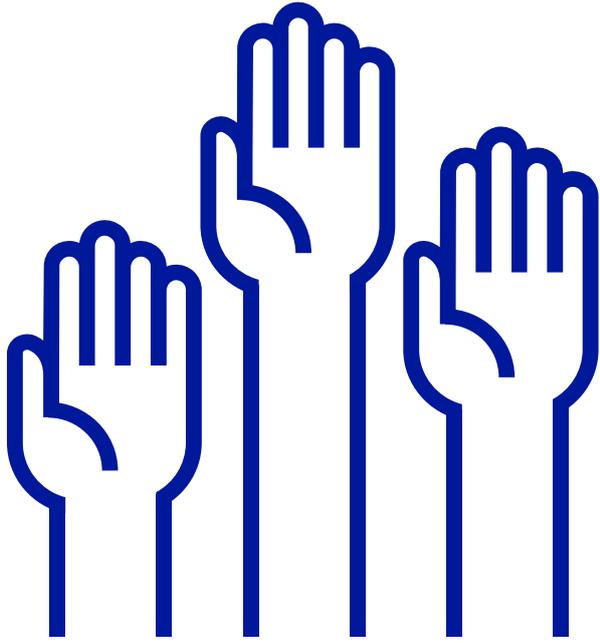

# Discussion

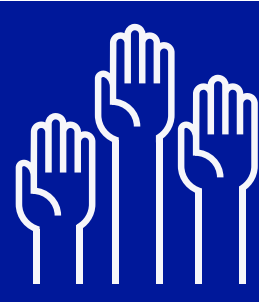

# 8 Discussion

In this research we identified the need for research on contextual-level factors (national culture, industry type and organisational culture) and individual-level factors (demographics and security aspects) that influence the behaviour of individuals in terms of handling information (in)securely. In the next sections we discuss the results of each of these factors based on the outcomes of our statistical analyses.

## 8.1 The influence of national culture on secure behaviour

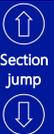

We measured and analysed the influence of national culture on secure behaviour in two ways: through the dimensions proposed by Hofstede and those proposed by Meyer.

**Hofstede's dimensions**

With respect to Hofstede's dimensions we noted, contrary to our hypothesis, that underline{power distance} was negatively associated with secure behaviour (simple and multiple regression). This is an interesting finding, especially since most scholars predict a positive relation between power distance and secure behaviour. A possible reason for this could be that in cultures with a higher power distance age is often translated to seniority and senior people receive lighter punishments given their social status which in turn results in less secure behaviour (Hovav and D'Arcy, 2012). Another explanation could be that power distance is influencing an employee's intention to behave more securely, but that intentions are not always translating into actual secure behaviour (Crossler et al., 2013). Additionally, in cultures where power distance is high, decision-making is more likely to be performed top-to-bottom. With respect to information security this could result in less organisation-wide support for, in example, information security policies.

With respect to individualism we argued that a relation was expected with secure behaviour, but that it could be either positive or negative. We found a positive relation (simple regression) which supports one of the two hypothesis on individualism and contradicts the argument of various scholars (Vance et al., 2020, Ameen et al., 2020, Crespo-Pérez, 2021 and Miao et al., 2020) stating that individuals in collectivistic cultures display more secure behaviour than individuals in individualistic cultures. This finding supports the arguments by Menard et al., 2018 and Connolly, et al., 2019 that individualistic cultures are less likely to share information to other individuals and have a higher accountability which results in more secure behaviour. Noteworthy is the fact that individualism was not statistically significant in the multiple regression and excluded in the multiple regression model due to a lack of contribution to the model fit. This outcome could support the argument by Hovav and D'Arcy, 2012 that individualism is not a simple linear relationship (as we measured it through simple and multiple linear regressions) with respect to secure behaviour, but is comparable with a parabola graph.

Uncertainty avoidance was found to have a negative relation with secure behaviour (**simple regression**). This finding contradicts our hypothesis and the arguments of various scholars (Vance et al., 2020, Ameen et al., 2020, Aurigemma and Mattson, 2018, Crespo-Pérez, 2021, Miao et al., 2020, Connolly, et al., 2019 and Alfawaz, 2011) that an increase in uncertainty avoidance is positively affecting secure behaviour. A possible explanation for this could be that, on the one hand, individuals in cultures with high uncertainty avoidance attempt to avoid a wide variety of uncertainties, one of which is related to information security. This leads to a diffused focus which results in less effective uncertainty avoidance with respect to information security. Individuals in cultures with low uncertainty avoidance, on the other hand, may only attempt to avoid those uncertainties which they deem undesirable. Such individuals, therefore, have a greater focus which makes them more effective in avoiding uncertainties with respect to information security (display more secure behaviour). Another (partial) explanation could be that there is an overall low level of awareness with respect to information security risks within our sample, as a result of which individuals from cultures with high uncertainty avoidance don't perceive information security as a (great) uncertainty that needs to be avoided.

With respect to masculinity we argued that this dimension could both have a positive and negative relation with secure behaviour (**simple** and **multiple regression**). Our results indicate that masculinity is negatively associated with secure behaviour which supports one of our hypotheses and the argument by Miao et al., 2020, but contradicts the argument by Crespo-Pérez, 2021. A possible explanation for this negative relation could be that in feminine cultures "*employees follow security procedures as they care for their organisations and other employees since they are more people-oriented*" (Ameen et al., 2020).

Long-term orientation was found to be positively associated with secure behaviour (**simple** and **multiple regression**) which confirms our hypothesis which is based on the argument that costs of compliance are experience in the present and costs for non-compliance are experienced in the future (Kim and Han, 2019), and that future focused cultures have a high consideration of future consequences and therefore display prosocial behaviour including securer behaviour (Li et al., 2019).

Indulgence was found to be positively associated with secure behaviour (**simple** and **multiple regression**) which contradicts our hypothesis. A possible explanation for this could be that restraining cultures are more suppressing of human needs and regulate via social norms which leads to a strong control of desires and impulses (Miao et al., 2020). If these social norms insufficiently cover desired information security behaviour, then an individual may refrain from displaying secure behaviour. In indulgent cultures, on the other hand, individuals may experience a need to handle information securely (e.g. to avoid being penalised) and therefore display more secure behaviour out of own desire as they are less dependent on social norms guiding their behaviour.

## Meyer's dimensions

Communicating was found to have both a negative (simple regression) and positive (multiple regression) relation with secure behaviour. The findings partly confirm our hypothesis. As described in the results, we identified through additional analysis, that this change in relationship direction may be explained by controlling for other Meyer cultural variables: Leading, Deciding, Disagreeing, Scheduling and Evaluating (in short: LDDSE). This suggests that, although we found no issues with multicollinearity, Communicating may be linked with the LDDSE variables. Considering that the relationship direction of the LDDSE dimensions in the multiple regression is mostly negative, we arrive at the following explanation. Communicating may be negatively associated with secure behaviour, but (most of) the LDDSE variables are also negatively associated with secure behaviour and at the same time the LDDSE variables may be negatively associated with Communicating as well (albeit below the multicollinearity threshold). Below we have included a schematic example for illustrative purposes. This example shows that when considering the relationship between Communicating and Secure behaviour in isolation we find a negative association (e.g. -5). However, when combining Communicating with the LDDSE variables into one multiple regression we find that an increase in the level of the LDDSE variables leads to a decrease in both Communicating (e.g. -10) and secure behaviour (e.g. -5). This leads to the conclusion that Communicating and secure behaviour may be positively associated since an increase of LDDSE leads to a decrease of 10 in Communicating and a decrease of 5 in secure behaviour. However, when controlling for the LDDSE variables we find a negative association between Communicating and secure behaviour. Based on this finding we identify the interrelationships between Meyer's cultural dimensions as an area for future research.

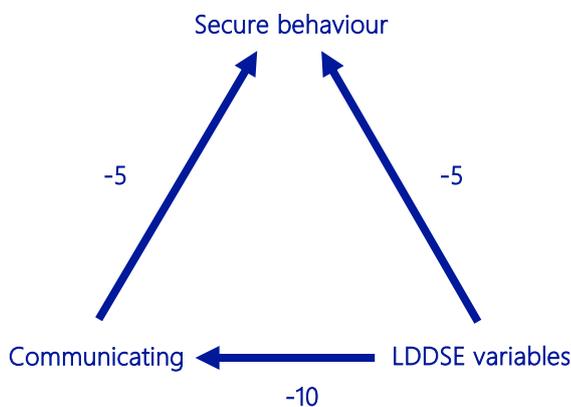

An argument in favour of the negative relation between Communicating and secure behaviour could be that in cultures with high context communication the information security expectations for individuals, and guidance on how to meet these expectations through secure behaviour, is less clear as communication is less precise, simple, explicit and clear (Meyer, 2014). An argument in favour of the positive relation could be that when information security expectations and guidance are insufficiently communicated, individuals in low-context communication cultures (low communicating score) may refrain from secure behaviour since they are not explicitly informed (how) to behave securely. Individuals from high-context communication cultures (high communicating score), on the other hand, may not always need explicit communication to understand what is expected from them and how they can achieve this. As a result of this, insufficient communication regarding expectations and guidance with respect to secure behaviour may be partially mitigated if a culture has a high context way of communicating.

Evaluating was found to have both a negative (simple regression) and positive (multiple regression) relation with secure behaviour. This partly confirms our hypotheses. As described in the results, we identified through additional analysis that this change in relationship direction may be explained by controlling for other Meyer cultural variables: Disagreeing and Scheduling. This suggests that, although we found no issues with multicollinearity, Evaluating may be linked with Disagreeing and Scheduling. Similar to what we discussed for Communicating, we argue that it may be that Evaluating, Disagreeing and Scheduling are all negatively associated with Secure behaviour, but Disagreeing and Scheduling are also negatively associated with Evaluating as well (refer to the illustrative example at Communicating). As mentioned before, this is an area for future research.

An argument in favour of the negative relation between Evaluating and secure behaviour could be that in cultures which tend towards indirect negative feedback (high evaluating score) the feedback recipient has a higher information security self-confidence (self-efficacy) than when negative feedback is provided directly (as criticism). Self-efficacy is argued by many scholars to have a positive relation with secure behaviour (Vedadi et al., 2021, Ameen et al., 2020, Menard et al., 2018, Jaeger, 2018, Hanus et al., 2018, White et al., 2017, Safa et al., 2015, Arachchilage and Love, 2014, Waly et al., 2012, Al-Omari et al., 2012, Bulgurcu et al., 2010, Galvez, 2009, Ng et al., 2009 and Dinev and Hu, 2007). An argument in favour of the positive relation could be that cultures which tend towards direct negative feedback (low evaluating score), similar to communicating, may provide clearer expectations of what improvements are needed with respect to information security, resulting in securer behaviour.

Leading and deciding were found to have a negative relation with secure behaviour (simple and multiple regression). This is in contrast to our hypotheses and means that the more hierarchical a culture is and the more decision-making is made top-to-bottom, the lower the secure behaviour displayed. This outcome is similar to Hofstede's power distance dimension where we found power distance to have a negative relation with secure behaviour. Therefore, for a possible explanation of the negative relation between leading and deciding and secure behaviour, we refer to our discussion above with respect to the Hofstede power distance dimension.

Trusting was found to have a negative relation with secure behaviour (simple regression). This means that relationship-based trust cultures display less secure behaviour than task-based trust cultures. This confirms our hypothesis that in cultures which tend towards task-based trust (low trusting score) trust is being built by tasks performed and therefore behaviour displayed. This means that behaving insecurely could negatively impact the trust that is put in an individual in task-based trust cultures. In relationship-based cultures, on the other hand, trust is not as dependent on the tasks performed and therefore behaviour displayed, but rather on the relationships that an individual has with others. As such, displaying secure behaviour is not as important for an individual for gaining trust and therefore the displayed behaviour is less secure.

Disagreeing was found to have a negative relation with secure behaviour (**multiple regression**). This confirms one of our hypothesis stating that cultures that don't avoid confrontation display more secure behaviour than those that avoid confrontation. The reasoning is that in cultures that don't avoid confrontation individuals are confronted when their information security behaviour is undesirable whereas in cultures that avoid confrontation this may be not (as much) the case.

Scheduling was found to have a negative relation with secure behaviour (**simple** and **multiple regression**). This confirms our hypothesis and means that cultures that consider time linear (linear-time scheduling) display more secure behaviour than cultures that consider time flexible (flexible-time scheduling). A possible explanation for this could be that in cultures that consider time flexible the goals to be achieved (e.g. certain information security goals) and corresponding deadlines are less strictly defined. Whereas in cultures that consider time linear goals and deadlines are more strictly defined resulting in better achievement of these goals. This achievement of information security goals (e.g. implementation of multi-factor authentication or raising information security awareness) could ultimately lead to securer behaviour.

Persuading was found to have a positive relation with secure behaviour (**simple regression**). This confirms our hypothesis and means that cultures that have an application-first reasoning display more secure behaviour than cultures that have a principles-first reasoning. A possible explanation for this could be that in cultures that prefer application before principles the information security practices are better as there is more focus on the "how" rather than the "why", as is the case with principles-first reasoning. In other words, application-first cultures could be argued to be more pragmatic (e.g. focus on appropriate information security measures) rather than principle-focused (e.g. focus on appropriate information security policies).

Finally, it is worth noting that Meyer's cultural variables appear to have a stronger predictive power than Hofstede's since Meyer's cultural variables have significantly higher Beta's than Hofstede's variables. Beta's for Hofstede's statistically significant variables in the multiple regression range from -.011 to .032 in relation to a mean secure behaviour score of the sample of 3,54. Beta's for Meyer's statistically significant variables in the multiple regression range from -.265 to .166 in relation to a mean secure behaviour score of the sample of 3,54. In other words: a change in Meyer's statistically significant cultural dimensions is expected to have a higher impact on secure behaviour scores than would be the case with Hofstede's statistically significant cultural dimensions. As such we recommend future research on culture and secure behaviour to consider other methods of measuring culture (such as Meyer's cultural variables) than Hofstede.

**Industry type**

Based on our analysis on <u>industry type</u> (<u>one-way ANOVA analysis</u>) we determined that, in accordance with our hypothesis, secure behaviour is different per industry type. The top 5 highest mean secure behaviour scores are obtained by the following industries: 1) Retail & Wholesale, 2) Transportation, 3) Manufacturing, 4) Healthcare & Pharmaceuticals and 5) Technology. There may be various reasons which these industries rank highest in mean secure behaviour score, for example:

- In a research by Jerbashian, 2021 the industries Retail & Wholesale ("Wholesale and Retail Trade"), Manufacturing ("Manufacturing") and Transport and Technology ("Transport, Storage, and Communication") rank highly on information technology (IT) dependence and require IT to perform their business. As such, continuity and security of their IT is of great importance since a lack of continuity and security of IT may impair business operations in such industries. This may translate into an increased focus on security awareness and secure behaviour to safeguard information security.

- The industries Healthcare & Pharmaceuticals and Technology could be characterised as industries that have sensitive data in terms of intellectual property (e.g. source code of software or pharmaceutical research information). Such intellectual property may put pressure (e.g. by internal management or stakeholders) on organisations to protect this information adequately. This may translate into an increased focus on security awareness and secure behaviour.

- The industry Healthcare & Pharmaceuticals could be characterised as an industry that has sensitive data in terms of personal data (e.g. medical records of patients). External pressures (e.g. by the regulator when enforcing privacy legislation such as the GDPR) may be put on organisations to protect this information adequately which in result may translate into an increased focus on security awareness and secure behaviour.

The top 5 lowest mean secure behaviour scores are obtained by the following industries: 1) Insurance, 2) Consulting, 3) Hospitality, 4) Education and 5) Other. Since it is unclear which type of organisations are in the "Other" industry, we have not explained the reason why Other may rank lower. According to Jerbashian, 2021 the Hospitality ("Hotels and Restaurants") ranks lower on IT dependence which may explain why less focus appears to be put on information security, and secure behaviour in particular. As for the other industries, Insurance, Consulting and Education could also be argued to be less reliant on IT since unavailability of IT would not (entirely) impair its primary business operations. However, integrity and confidentiality of (financial) data may be particularly important for Insurance and Consulting as a result of which it is surprising that these industries have the top 5 lowest secure behaviour scores. Future research may investigate which factors influence secure behaviour within industries. For example, IT dependence, level of regulation and type of information used may be considered. We do note, however, the limitation that some industries had significant outliers and low record counts which may, to a certain extent, distort our outcomes and findings.

**Click to navigate**

Introduction | Literature review | Research questions | Antecedents of behaviour | Research model | Methodology | Results | Discussion | Conclusion | Limitations | References | Appendix

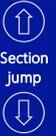

Organisational security culture

With respect to the organisational security culture dimensions we noted that attitude had both a negative (simple regression) and positive (multiple regression) relation with secure behaviour. This change in relationship direction may be explained by interrelationships between attitude and other dimensions in the organisation's security culture. Refer to Meyer's Communicating dimension for a further explanation on how such an interplay may function. The findings from the simple and multiple regression partly confirm our hypothesis that organisations with higher scores on attitudes (attitudes which are more positive toward information security) display more secure behaviour. However, the negative relation, as identified by the simple regression, raises the question why it may be that a higher score on attitudes (more positive attitude) may negatively influence secure behaviour and vice versa. As argued earlier, attitudes may influence behaviour, but behaviour may equally influence attitudes. As such, it could be that in organisations in which a high level of secure behaviour is displayed, there is a large focus on secure behaviour. This focus on security may lead to a culture in which individuals display secure behaviour not out of intrinsic motivation, but out of extrinsic motivation (e.g. fear for sanctions for non-compliance). As such, security may be considered to be a necessary burden by employees and there may be a culture in which there is a negative attitude towards security, but a higher level of secure behaviour.

Cognition was found to be negatively associated with secure behaviour (simple and multiple regression) which contradicts our hypothesis. A possible explanation for this could be the argument by D'Arcy and Hovav, 2009; they argue that "computer savvy users", who could be argued to have a higher level of information security cognition, are more likely to display information system misuse (less secure behaviour), because security training and awareness (SETA) efforts are less deterring for such individuals.

Communication was found to be positively associated with secure behaviour (simple and multiple regression). This confirms our hypothesis that communication may increase secure behaviour, for example through increased communication regarding information security policies and more effective communication regarding (potential) security incidents (reducing the impact should these occur).

Compliance was found to be negatively associated with secure behaviour (simple and multiple regression) which contradicts our hypothesis. This is a particularly interesting finding, because the indicated compliance level (within an organisation) is shown not to correspond with the actual displayed behaviour. This could be explained by the earlier arguments that we made that 1) intention to comply does not always translate into actual secure behaviour and 2) users within an organisation may indicate a desire to comply, but may not understand what this exactly entails and/or may not have the rights tools (e.g. security awareness and

**Click to navigate**

Introduction | Literature review | Research questions | Antecedents of behaviour | Research model | Methodology | Results | Discussion | Conclusion | Limitations | References | Appendix

knowledge) to behave more securely. In line with the aforementioned, users that have a good understanding of what information security compliance entails may indicate a lower desire to comply since they understand what implications (and burdens) this would have on them. Similarly, individuals with a limited understanding of what information security compliance entails may indicate a higher desire to comply since they do not fully understand what implications (and burdens) this would have on them. The result of this would be that individuals with a higher desire to comply actually display less secure behaviour.

Norms was found to be negatively associated with secure behaviour (**simple regression**) which contradicts our hypothesis. This is an interesting finding and suggests that organisations with positive norms towards information security may actually display less secure behaviour in practice. A possible explanation for this could be that organisations that have positive norms towards information security may feel that there is a good level of information security awareness and knowledge, because they have norms that are positive towards information security. However, in practice, these types of organisations focus less attention on increasing, for example, security awareness and knowledge as a result of which the actually displayed behaviour is less secure (e.g. because individuals are less aware of *how* and *when* to behave securely).

Responsibilities was found to be positively associated with secure behaviour (**multiple regression**). This confirms our hypothesis that employees that understand their tasks, roles and responsibilities with respect to information security are more likely perform the activities needed to keep the organisation and its information safe (display more secure behaviour).

## 8.4 The influence of demographics on secure behaviour

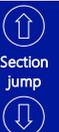

### Age

Age was found to have a positive relation with secure behaviour (**simple regression**). This means that as age increases displayed behaviour is more secure. This finding supports earlier findings made by other scholars (Chen et al., 2022, Chua et al., 2018, McCormac et al., 2018, Gratian et al., 2018, McCormac et al., 2017 and D'Arcy and Hovav, 2009). The **one-way ANOVA analysis** on age groups reveals that the highest variance in behaviour score is displayed by the youngest age group; 15-20 years old. The 15-20 years old age group was found to have a significantly lower secure behaviour score than the other age groups. This difference in secure behaviour for this age group could be explained by the argument made by White et al., 2017 that youth behave different than adults since "*youth are still in the learning phase of life while adults are in the productive phase of life. The motivations for computer usage is different*".

Interesting to note is that the age group of 60-69 years old has the highest security behaviour score with a score of 3,64 or 22,8%, and the age group of 70+ years old also has a higher than average secure behaviour score with 3,59 or 22,4%. This is somewhat surprising as elderly people could be expected to have less technology competences limiting them in displaying secure information security behaviour. A possible reason for this could be that cybercrime has increasingly focused on elderly people (Zulkipli et al., 2021) which has resulted in increased efforts (i.e. by governments and banks) to increase security awareness in these age groups and making their behaviour more secure.

### Gender

Based on our analysis on gender (independent samples T-test) we find that males have higher secure behaviour scores than females. This is contrary to our hypothesis and the arguments of various scholars that females have higher information security awareness (McCormac et al., 2018, McCormac et al., 2017), display more secure behaviour (Chen et al., 2022, Haeussinger and Kranz, 2013b) and engage less in information system misuse (Hovav and D'Arcy, 2012 and D'Arcy and Hovav, 2009). However, our finding is in accordance with the argument made by Gratian et al., 2018 that females may display poorer security awareness, security practices and less secure behaviour than their male counterparts. A possible explanation of our finding could be the way in which secure behaviour is measured in the EU dataset (as security measures taken). For example, McCormac et al., 2017 find that females obtained higher information security awareness and behaviour scores, but were found to be more "*susceptible to phishing emails than males*". This suggests that the way of measuring secure behaviour could have significant impact on whether males or females are found to obtain higher scores.

### Level of urbanisation

Based on our analysis on level of urbanisation (one-way ANOVA) we find that, in accordance with our hypothesis, rural areas and villages have the lowest secure behaviour score. However, there is limited difference in secure behaviour score between small or medium-sized towns and large towns or cities. This suggests that urbanised areas, regardless of whether these are small or large, display more secure behaviour than rural areas. As argued earlier, a possible reason for this could be that individuals living in rural areas are less likely to have technology-intensive jobs (e.g. farmers) in comparison to individuals living in urbanised areas where such jobs may be prevalent (e.g. banking). Due to this technology awareness (and awareness of risks related thereto) is lower in rural areas resulting in less secure behaviour.

## 8.5   The influence of security specific factors on secure behaviour

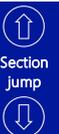

### Security education and training

Based on our analyses (both simple regression and one-way ANOVA) we found no (statistically) significant relation between extent of security education and training and secure behaviour. This is interesting since this suggests that security education and training, at least on the short term (since we only measured it over the period of 2019-2021), is not significantly changing user's behaviour. This finding is in accordance with Khando et al., 2021 who find that information security awareness efforts (e.g. security education and training) are failing to change employees' secure behaviour. An explanation for this may be the finding we made during our assumption testing: we noted a deviation in the normal Q-Q plot towards the end of the trend line that indicates that standardised residuals may not be entirely normally distributed (refer to section 6.2.3.3). We consider this a limitation of our analysis.

Additionally, we need to consider an important limitation when interpreting these results. We only included training and education before the first phishing e-mail and that took place in the period of 2019-2021. However, if users were subject to security training and education before 2019 then this may mean that the significant change (increase) in secure behaviour may have taken place before 2019. As a result of this, additional security education and training may be found not to have any significant effects in further improve secure behaviour.

Another factor to consider is that security education and training may be increasing security awareness, but this awareness is not translated into more secure behaviour for various reasons. This presumption is supported by various scholars (Khando et al., 2021, Hanus et al., 2018, Karjalainen et al., 2013, Talib et al., 2010, Abawajy, 2014, Annetta, 2010 and Cone et al., 2007). For a further discussion on this, we refer to section 2.3. Either way, this is an interesting finding and provides opportunities for future research on what factors affect the success of security education and training efforts in improving security awareness and, ultimately, secure behaviour.

### Security consciousness and confidence

Based on our analyses (simple regression) we found no (statistically) significant relation between security consciousness and confidence and secure behaviour. This suggests that security consciousness and confidence is not affecting secure behaviour which is in contradiction with many other scholars (Vedadi et al., 2021, Ameen et al., 2020, Menard et al., 2018, Jaeger, 2018, Hanus et al., 2018, White et al., 2017, Safa et al., 2015, Arachchilage and Love, 2014, Waly et al., 2012, Al-Omari et al., 2012, Bulgurcu et al., 2010, Galvez, 2009, Ng et al., 2009 and Dinev and Hu, 2007). An explanation for this may be the finding we made during our assumption testing: we noted a deviation in the normal Q-Q plot towards the end of the trend line that indicates that standardised residuals may not be entirely normally distributed (refer to section 6.2.3.3). We consider this a limitation of our analysis. Another explanation for this finding may be that we should consider security consciousness and confidence as a moderating variables with respect to secure behaviour rather than as an independent variable (D'Arcy and Hovav, 2009). More research may be required to explain this finding.

### Security awareness and security knowledge

Security awareness and knowledge was found to have a positive relation with secure behaviour (simple and multiple regression). This means that as security awareness and knowledge increases displayed behaviour is more secure. This finding supports earlier findings made by other scholars that security awareness (Khando et al., 2021, Breitinger et al., 2020, Hanus et al., 2018, Jaeger, 2018, Chua et al., 2018, ENISA, 2017, Yazdanmehr and Wang, 2016, Humaidi and Balakrishnan, 2015, Arachchilage and Love, 2014, Haeussinger and Kranz, 2013b, Mejias, 2012, Hovav and D'Arcy, 2012, Al-Omari et al., 2012, Alfawaz, 2011, Bulgurcu et al., 2010, Galvez, 2009, Marks and Rezgui, 2009, Kumar et al., 2008 and Dinev and Hu, 2007) and security knowledge (Khando et al., 2021, Breitinger et al., 2020, Jaeger, 2018, Arachchilage and Love, 2014, Haeussinger and Kranz, 2013a, Al-Omari et al., 2012, Mejias, 2012 and Dinev and Hu, 2007) has a positive relation with secure behaviour. This does, however, contradict the notion that security awareness is not translated into more secure behaviour (Khando et al., 2021, Hanus et al., 2018, Karjalainen et al., 2013, Talib et al., 2010, Abawajy, 2014, Annetta, 2010 and Cone et al., 2007). In accordance with our discussion of security education and training, future research could further explore factors that influence the successful translation of security education and training into security awareness and/or security awareness to secure behaviour.

### Prior experience with security incidents

<u>Prior experience with security incidents</u> was found to have a positive relation with secure behaviour (**simple** and **multiple regression**). This means that as prior experience with security incidents increases displayed behaviour is more secure. We noticed this same relation for both direct prior experience (first hand experience) and indirect prior experience (second or third hand experience). It is noteworthy, however, that based on the Beta indirect prior experience appears to have a much stronger influence (B= .595 (simple regression) and B= .522 (multiple regression)) on secure behaviour (mean secure behaviour score in the sample is 3,54) than direct prior experience (B= .138 (simple regression) and B= .071 (multiple regression)). This suggests that experience with security incidents in someone's network (indirect experience) may more strongly change an individual's secure behaviour than actually experiencing the security incident himself (direct experience). The relationship we find confirms our hypothesis and the predictions by various scholars that prior experience with security incidents results in an increase in security awareness which is translated into more secure behaviour (Khando et al., 2021, Hanus et al., 2018, Jaeger, 2018, Haeussinger and Kranz, 2013a, Haeussinger and Kranz, 2013b and Straub and Welke, 1998).

Further analysis (**one-way ANOVA**) of categorisations of prior experience with security incidents (categorisations: little experience, some experience, medium experience and large experience), however, reveals that the relationship between prior experience with security incidents and secure behaviour may not be a simple linear one. With respect to direct prior experience we note a strong increase in secure behaviour when comparing individuals with little experience (21,6%) to individuals with some experience (30,1%). However, this is followed by a strong decrease in secure behaviour for individuals with medium experience (19,6%) and is followed by a further decrease for individuals with large experience (14,8%). A similar pattern is viewed for indirect prior experience; secure behaviour is found to increase from little experience (20,7%) to some experience (30,8%) and further on to medium experience (37,0%), but is followed by a strong decrease for individuals with large experience (28,6%). As such, prior experience initially has a positive influence on secure behaviour, but as experience increases this positive influence reduces significantly. A possible explanation for this could be that as prior experience with security incidents increases the perceived level of risk and uncertainty is reduced as individuals are more familiar with security incidents and/or may consider the (personal) impact more limited than expected. In other words: as frequency of experienced security incidents increases, the security awareness effect decreases through a lower perceived level of risk and uncertainty, which in turn results in a decrease in secure behaviour.

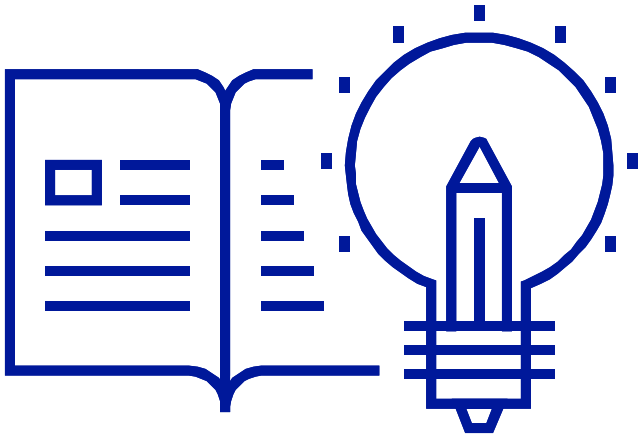

# Conclusion

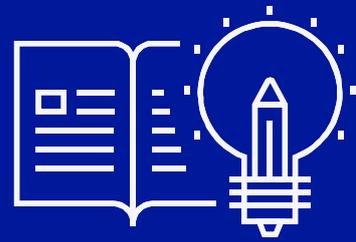

# 9   Conclusion

## 9.1   Reflecting on the research questions

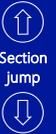
Section
jump

At the start of this research we identified the importance of the human factor in preventing data breaches and security incidents. We noted a research gap with respect to research on actual secure behaviour (rather than security awareness or secure behaviour intentions as indicators thereof). Additionally, we noted factors to consider that may influence secure behaviour at **contextual level** (national culture, industry type and organisation level) and **individual level** (demographics and security-specific aspects). Based on a literature review we arrived at a set of variables at both the **contextual level** and **individual level**. We conclude that both contextual level factors and individual level factors influence the extent to which secure behaviour is displayed by an individual.

### 9.1.1   Contextual level antecedents to secure behaviour

At the contextual level we identified national culture (and various dimensions that it contains), industry type and organisational culture as important antecedents to secure behaviour. We analysed **national culture** both through the cultural dimensions of Hofstede and the cultural dimensions of Meyer. Within our analysis of national culture by means of Hofstede's cultural dimensions we find that cultures with the following cultural characteristics display more secure behaviour: a low degree of power distance, high degree of individualism, low degree of uncertainty avoidance, high degree of femininity, high degree of long-term orientation and high degree of indulgence. With respect to Meyer's cultural dimensions we find that cultures with the following cultural characteristics display more secure behaviour: a low degree of hierarchy, consensus-based decision-making, task-based trust building, low confrontation avoidance, linear-time scheduling and application-first reasoning. We find mixed results for Meyer's communicating and evaluating dimensions as a result of which additional research is warranted. Additionally, we note that Meyer's cultural variables, when statistically significant, had much higher Beta's. In other words: a change in Meyer's statistically significant cultural dimensions is expected to have a higher impact on secure behaviour scores than would be the case with Hofstede's statistically significant cultural dimensions. As such we recommend future research on culture and secure behaviour to consider other methods of measuring culture than Hofstede.

With respect to **industry type** we find that there is a significant difference between industries when it comes to secure behaviour score. Industries such as Retail and Wholesale, Transportation and Manufacturing had much higher secure behaviour score than most other industries. This suggests that there may be factors at the industry level that affect secure behaviour. Some examples of such factors could be level of regulation and dependency on IT. Future research could focus on exploring factors at industry level that may affect secure behaviour.

We analysed the impact of an **organisational security culture** on the secure behaviour of individuals based on a set of dimensions. We find that organisational security culture has an important influence in predicting and/or influencing secure behaviour. As such, measuring and changing the organisation's security culture may be highly beneficial for organisations. Albeit that an organisation's security culture may be difficult to change. Additionally, the different security culture dimensions are found to have different (strengths and types of) relationships with secure behaviour. Furthermore, we did not take into account possible interrelations between various cultural layers (e.g. national level, organisation level), but research suggests that such interrelations may exist. This may as well be an area for future research.

### 9.1.2   Individual level antecedents to secure behaviour

At the individual level we identified demographic and security-specific factors as important antecedents to secure behaviour. With respect to **demographics** we find that secure behaviour increases with age, but that this increase is strongest at the lower end of the age spectrum. In other words: young adults are found to display significantly less secure behaviour than other age categories. We also find that gender has an impact on secure behaviour in the sense that males display more secure behaviour than females. However, the caveat here is that secure behaviour score per gender could differ strongly depending on the way in which secure behaviour is measured. With respect to level of urbanisation we find that individuals living in urbanised areas display significantly more secure behaviour than those that live in rural areas.

In addition to generic demographic factors we analysed **security-specific factors**. We find no evidence that security education and training increases secure behaviour. On the one hand, we argue that it may be the case that security education and training increases security awareness, but that this does not always translate to actual securer behaviour. On the other hand, we note that this finding may be the result of limitations in our dataset (refer to the discussion). Additionally, we find no evidence that security consciousness and confidence is affecting secure behaviour. This contradicts findings of many scholars. We do note, however, that this may be caused by the way in which we measured this variable (refer to the discussion). We find that security awareness and security knowledge are important factors that may positively influence secure behaviour. This supports our prediction and emphasises the importance thereof.

Interestingly, we find that prior experience with information security incidents initially has a strong positive effect on secure behaviour, but that secure behaviour decreases significantly for individuals with large prior experience with security incidents. This suggests that prior experience with security incidents initially has a positive influence, but after a certain point it has a negative influence on secure behaviour. Additionally, it is worth noting that we find that indirect prior experience with security incidents is much stronger related with secure behaviour than direct experience. This suggests that experience with security incidents in someone's network (indirect experience) may more strongly change an individual's secure behaviour than actually experiencing the security incident himself (direct experience).

**Click to navigate**

This research has various implications for both research and practise. Firstly, for research our findings fill a gap in research and provide opportunities for further research (refer to the next section).

Secondly, for organisations our findings provide valuable insights in what groups may be most likely to display insecure behaviour and what groups may require additional and/or tailored security training and awareness efforts. Based on our findings organisations may create employee profiles. Such employee profiles could, for example, be based on 1) national cultural values of the country the employee is working in, 2) age of the employee and/or 3) gender of the employee. Additionally, organisations may take into account factors such as the industry type the organisation is operating in (refer to below). Based on such employee profiles organisations could provide generic security trainings with additional security trainings tailored to these specific employee profiles (in accordance with Chua et al., 2018). Organisations may also need to adapt the way in which information security policies are communicated. For example, in low-context communication cultures information security policies may be required to be more explicit than in high-context communication cultures.

Thirdly, we find that industry type may affect secure behaviour. Although we have not researched specific factors at industry level that may affect secure behaviour, organisations may want to consider the impact that the industry they operate in has on the secure behaviour of their employees. Future research may provide useful insights into factors that organisations may consider.

Fourthly, another approach for organisations could be to gain an understanding of their organisation security culture and change those organisational security culture values that are most strongly negatively affecting secure behaviour. Some variables such as cognition may be more strongly related to secure behaviour than others such as responsibilities. For example, organisations may focus on cognition since this may be more easily changed (e.g. through learning and development) and have a more significant impact on secure behaviour than other organisational (security) culture variables such as norms.

Fifthly, security knowledge and security awareness are found to be important factors influencing secure behaviour. However, we find no evidence that security education and training is (significantly) improving secure behaviour. As such, organisations may want to reconsider their security education and training approach (also refer to the point above). We did find, however, that prior experience with security incidents may be greatly beneficial in improving secure behaviour. Organisations could consider if they may better leverage security incidents that occur to improve secure behaviour of employees. However, organisations should be aware that we find that prior experience with security incidents may not always be contributing positively to secure behaviour.

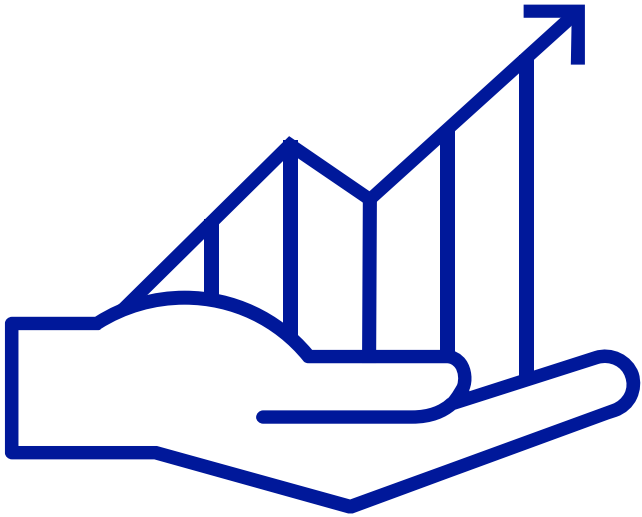

# 10

# Limitations and future research

# 10 Limitations and future research

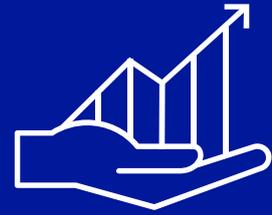

## 10.1 Limitations and future research

As with any research, our research has limitations. Firstly, our research is based on existing data that was not specifically collected for purposes of performing this research. As such, some variables may not be measured in an optimal way. For example, the secure behaviour score in the EU dataset was computed based on input on several questions with respect to security measures taken by the individual. Additionally, we noted for the KnowBe4 dataset that the data was skewed towards one country (at the country level) and two industries (at the industry level). Additionally, we noted some limitations in our assumption testing related to, primarily, the normal distribution (of residuals in) the sample. Future research could develop a set of questions / measurements tailored specifically to arrive at a secure behaviour score, and use this measurement to measure secure behaviour over a well distributed sample.

Secondly, our research was based on two separate datasets. This has limited our ability to tests for possible relations between variables that are in different datasets. Additionally, these two separate datasets has resulted in two different ways of measuring secure behaviour (security measures taken vs. observed phishing e-mail behaviour).

Thirdly, phishing e-mail response, as used in the KnowBe4 dataset to measure secure behaviour, could be argued to by one of many indicators of secure behaviour (Hickmann Klein and Mezzomo Luciano, 2016). For example, other indicators could be strength of passwords, online behaviour etc. Future research could develop and/or use a more comprehensive measurement of secure behaviour.

Fourthly, we measured organisational security culture through the CLTRe / KnowBe4 dimensions. However, these dimensions have not been (elaboratively) reviewed on their appropriateness and accuracy for measuring organisational security culture. Similarly, we measure security awareness and security knowledge combined through data in the Security Awareness Proficiency Assessment (SAPA). The SAPA is not an (elaboratively) reviewed research approach and it could be argued that security awareness and security knowledge are best measured separately. Future research could develop and/or apply a peer reviewed measurement for organisational security cultures and security awareness. Additionally, future research may measure security awareness and security knowledge separately.

Fifthly, possible interrelations may exist between the various levels that we identified. For example, national culture is expected to influence organisational (security) culture (Wiley et al., 2020, Vance et al., 2020 and Kam et al., 2015). Additionally, interrelationships may exist between variables within certain levels. For example, we found possible interrelationships between some of Meyer's cultural dimensions when combined into a single statistical model. We did not (fully) account for such interrelations. Future research could examine possible interrelations that may exist between various levels (e.g. cultural levels) and between various variables within these levels (e.g. cultural dimensions).

Sixthly, we found that secure behaviour varied statistically significantly between industries. We have not extensively investigated what may be the cause of this finding. Future research may investigate which factors influence secure behaviour within industries. For example, IT dependence, level of regulation and type of information used may be considered.

Seventhly, we have identified various levels and corresponding variables that influence secure behaviour. However, this overview is by no means exhaustive. Future research could focus on elaborating our model and incorporating variables suggested by other scholars.

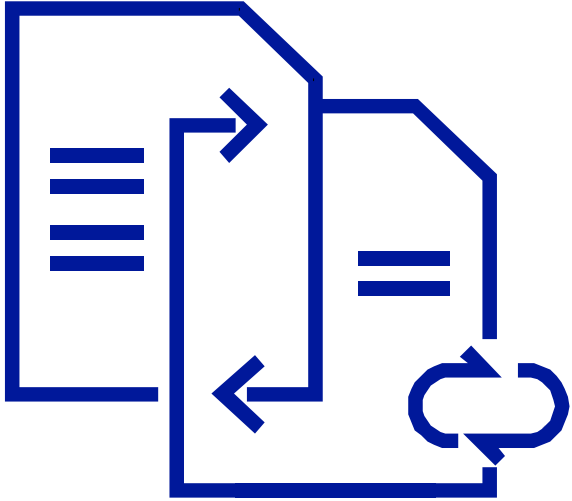

# 11

# References

# 11 References

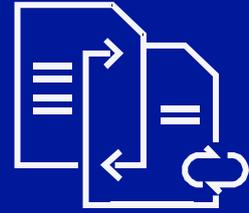

# A

# E

# F

# G

# H

# I

# J

# K

# L

# M

# S

# T

# V

# W

# Y

**Z**

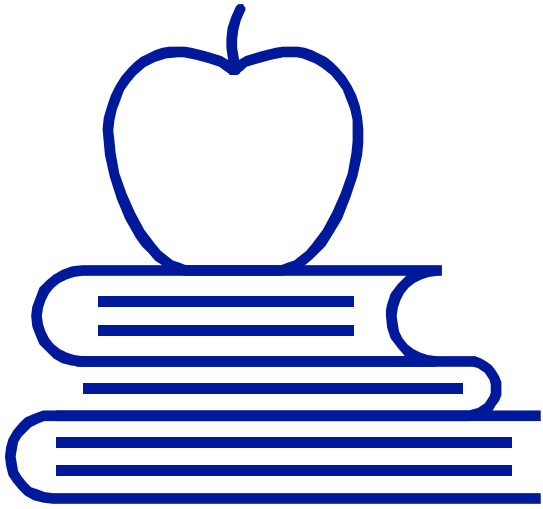

Appendix

# 12 Appendix

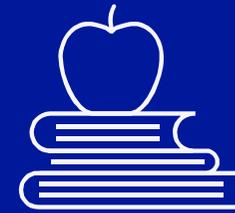

## 12.1  Statistical output data analyses EU Dataset

### 12.1.1  Single variable testing

#### 12.1.1.1  Independent samples T-test Gender

**Group Statistics**

|  | Gender | N | Mean | Std. Deviation | Std. Error Mean |
|---|---|---|---|---|---|
| Secure behaviour score | Male | 9875 | 3,67 | 2,629 | ,026 |
|  | Female | 11787 | 3,43 | 2,462 | ,023 |

**Independent Samples Test**

| | | Levene's Test for Equality of Variances | | t-test for Equality of Means | | | | | | | |
| | | F | Sig. | t | df | Significance One-Sided p | Two-Sided p | Mean Difference | Std. Error Difference | 95% Confidence Interval of the Difference Lower | Upper |
|---|---|---|---|---|---|---|---|---|---|---|---|
| Secure behaviour score | Equal variances assumed | 50,379 | <,001 | 7,126 | 21660 | <,001 | <,001 | ,247 | ,035 | ,179 | ,315 |

| | | | 7,085 | 20458,679 | <,001 | <,001 | ,247 | ,035 | ,179 | ,315 |
|---|---|---|---|---|---|---|---|---|---|---|
| Equal variances not assumed | | | 7,085 | 20458,679 | <,001 | <,001 | ,247 | ,035 | ,179 | ,315 |

**Independent Samples Effect Sizes**

| | | Standardizer[a] | Point Estimate | 95% Confidence Interval | |
|---|---|---|---|---|---|
| | | | | Lower | Upper |
| Secure behaviour score | Cohen's d | 2,539 | ,097 | ,070 | ,124 |
| | Hedges' correction | 2,539 | ,097 | ,070 | ,124 |
| | Glass's delta | 2,462 | ,100 | ,074 | ,127 |

a. The denominator used in estimating the effect sizes.

Cohen's d uses the pooled standard deviation.

Hedges' correction uses the pooled standard deviation, plus a correction factor.

Glass's delta uses the sample standard deviation of the control group.

Boxplot

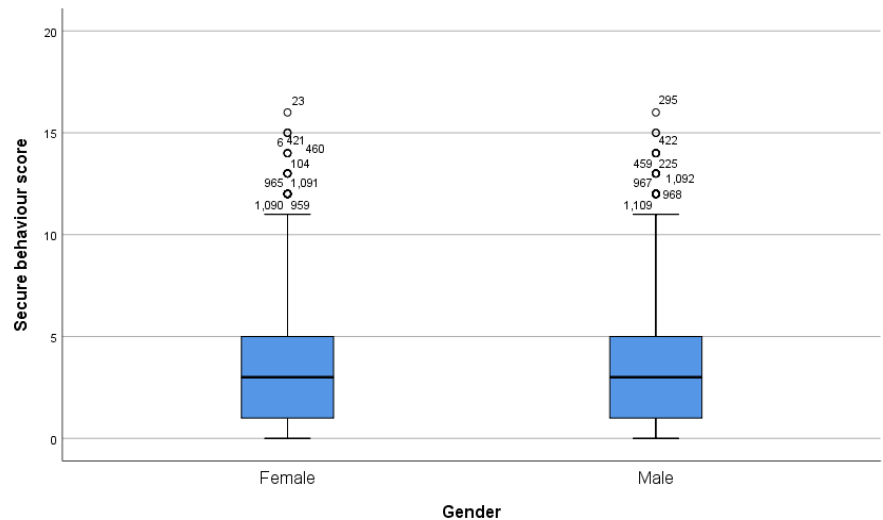

## Test of normality

### Tests of Normality

| | Gender | Kolmogorov-Smirnov[a] | | |
| | | Statistic | df | Sig. |
|---|---|---|---|---|
| Secure behaviour score | Female | .160 | 11787 | .000 |
| | Male | .156 | 9875 | .000 |

a. Lilliefors Significance Correction

## Distribution histograms

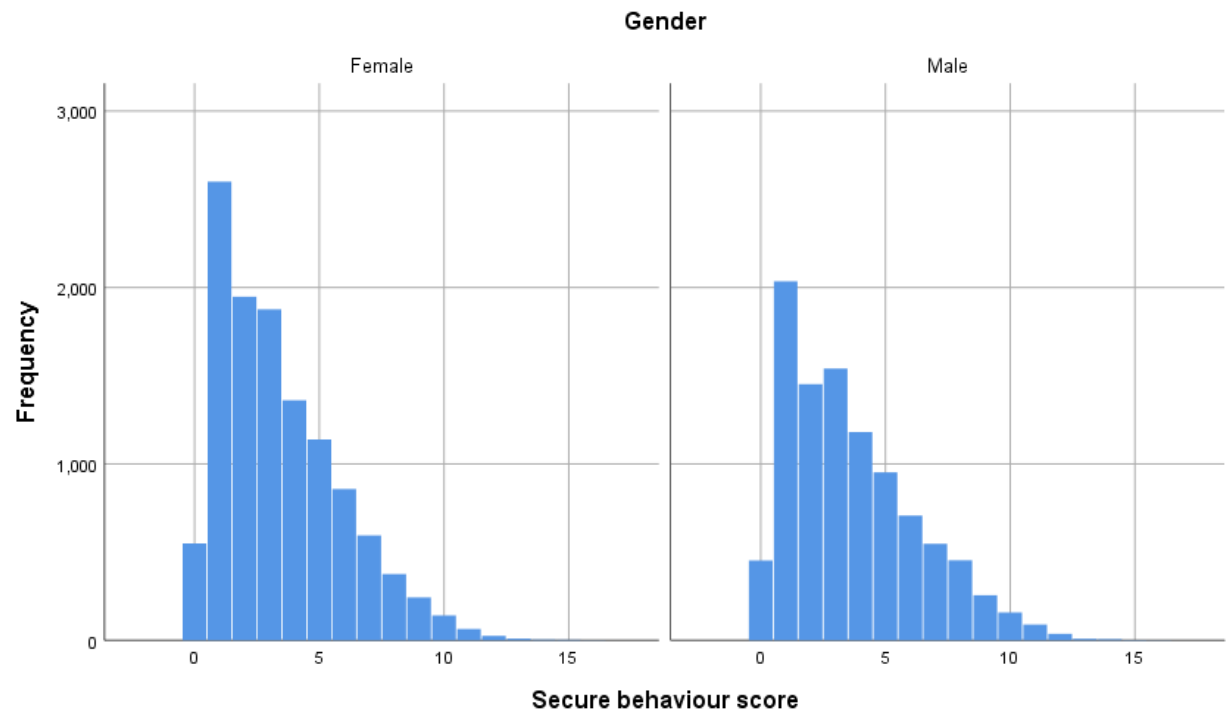

## Normal Q-Q Plots

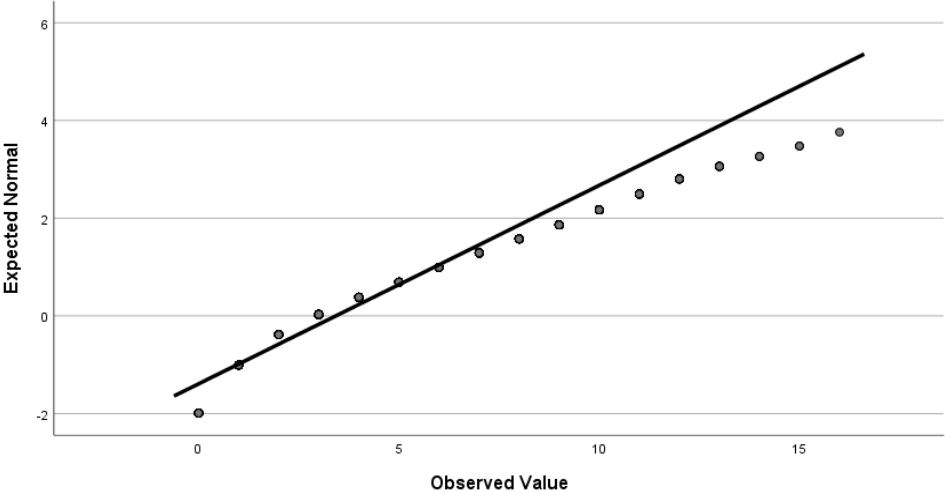

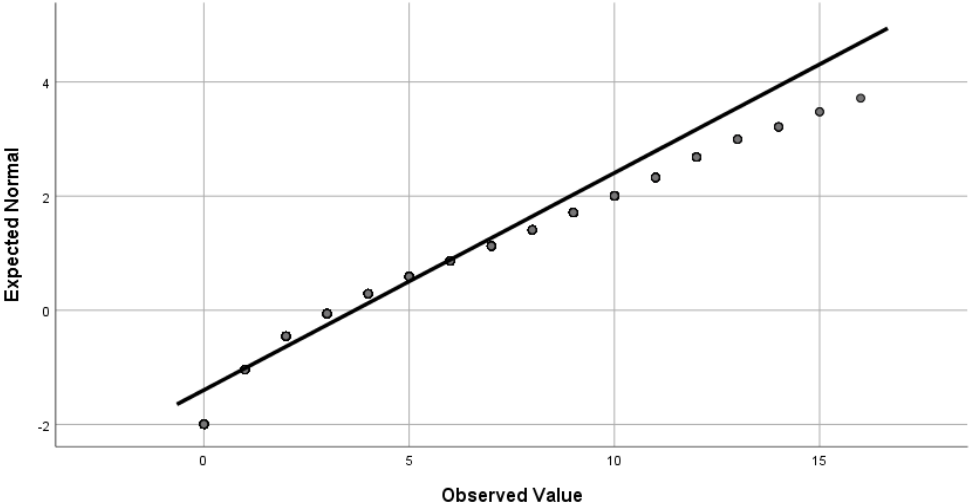

### 12.1.1.2 One-way ANOVA Level of urbanisation

**Descriptives**

Secure behaviour score

| | N | Mean | Std. Deviation | Std. Error | 95% Confidence Interval for Mean | | Minimum | Maximum |
|---|---|---|---|---|---|---|---|---|
| | | | | | Lower Bound | Upper Bound | | |
| 1 | 8 | 2,88 | 3,227 | 1,141 | ,18 | 5,57 | 0 | 10 |
| Large town or city | 6447 | 3,57 | 2,527 | ,031 | 3,51 | 3,63 | 0 | 16 |
| Rural area or village | 6896 | 3,45 | 2,514 | ,030 | 3,39 | 3,51 | 0 | 15 |
| Small or medium-sized town | 8311 | 3,59 | 2,575 | ,028 | 3,53 | 3,64 | 0 | 16 |
| Total | 21662 | 3,54 | 2,542 | ,017 | 3,51 | 3,57 | 0 | 16 |

**ANOVA**

Secure behaviour score

| | Sum of Squares | df | Mean Square | F | Sig. |
|---|---|---|---|---|---|
| Between Groups | 78,808 | 3 | 26,269 | 4,067 | ,007 |
| Within Groups | 139903,573 | 21658 | 6,460 | | |
| Total | 139982,382 | 21661 | | | |

**ANOVA Effect Sizes[a,b]**

| | | Point Estimate | 95% Confidence Interval | |
|---|---|---|---|---|
| | | | Lower | Upper |
| Secure behaviour score | Eta-squared | ,001 | ,000 | ,001 |
| | Epsilon-squared | ,000 | ,000 | ,001 |
| | Omega-squared Fixed-effect | ,000 | ,000 | ,001 |

| | | | | | Omega-squared Random-effect | ,000 | ,000 | ,000 |

| | | | |
|---|---|---|---|
| Omega-squared Random-effect | ,000 | ,000 | ,000 |

a. Eta-squared and Epsilon-squared are estimated based on the fixed-effect model.

b. Negative but less biased estimates are retained, not rounded to zero.

**Post Hoc Tests**

**Multiple Comparisons**

Dependent Variable: Secure behaviour score

Tukey HSD

| (I) Level of urbanisation | (J) Level of urbanisation | Mean Difference (I-J) | Std. Error | Sig. | 95% Confidence Interval | |
|---|---|---|---|---|---|---|
| | | | | | Lower Bound | Upper Bound |
| 1 | Large town or city | -,694 | ,899 | ,867 | -3,00 | 1,62 |
| | Rural area or village | -,579 | ,899 | ,918 | -2,89 | 1,73 |
| | Small or medium-sized town | -,712 | ,899 | ,858 | -3,02 | 1,60 |
| Large town or city | 1 | ,694 | ,899 | ,867 | -1,62 | 3,00 |
| | Rural area or village | ,116* | ,044 | ,043 | ,00 | ,23 |
| | Small or medium-sized town | -,018 | ,042 | ,975 | -,13 | ,09 |
| Rural area or village | 1 | ,579 | ,899 | ,918 | -1,73 | 2,89 |
| | Large town or city | -,116* | ,044 | ,043 | -,23 | ,00 |
| | Small or medium-sized town | -,133* | ,041 | ,007 | -,24 | -,03 |
| Small or medium-sized town | 1 | ,712 | ,899 | ,858 | -1,60 | 3,02 |
| | Large town or city | ,018 | ,042 | ,975 | -,09 | ,13 |
| | Rural area or village | ,133* | ,041 | ,007 | ,03 | ,24 |

*. The mean difference is significant at the 0.05 level.

**Homogeneous Subsets**

**Secure behaviour score**

Tukey HSD[a,b]

| Level of urbanisation | N | Subset for alpha = 0.05 |
|---|---|---|
| | | 1 |
| 1 | 8 | 2,88 |
| Rural area or village | 6896 | 3,45 |
| Large town or city | 6447 | 3,57 |
| Small or medium-sized town | 8311 | 3,59 |
| Sig. | | ,678 |

Means for groups in homogeneous subsets are displayed.

a. Uses Harmonic Mean Sample Size = 31,893.

b. The group sizes are unequal. The harmonic mean of the group sizes is used. Type I error levels are not guaranteed.

Levene's test of homogeneity

**Test of Homogeneity of Variances**

| | | Levene Statistic | df1 | df2 | Sig. |
|---|---|---|---|---|---|
| Secure behaviour score | Based on Mean | 2.632 | 3 | 21658 | .048 |
| | Based on Median | 2.263 | 3 | 21658 | .079 |
| | Based on Median and with adjusted df | 2.263 | 3 | 21624.308 | .079 |
| | Based on trimmed mean | 2.968 | 3 | 21658 | .031 |

## Boxplot

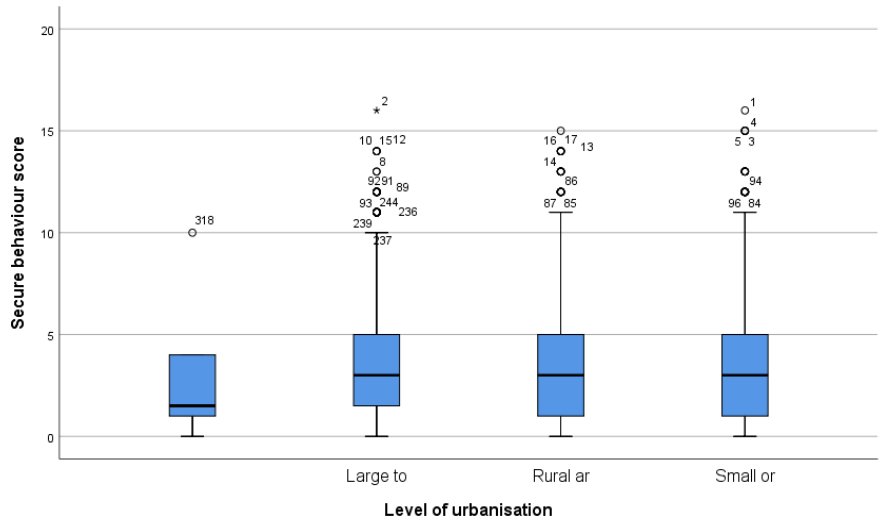

## Distribution histograms

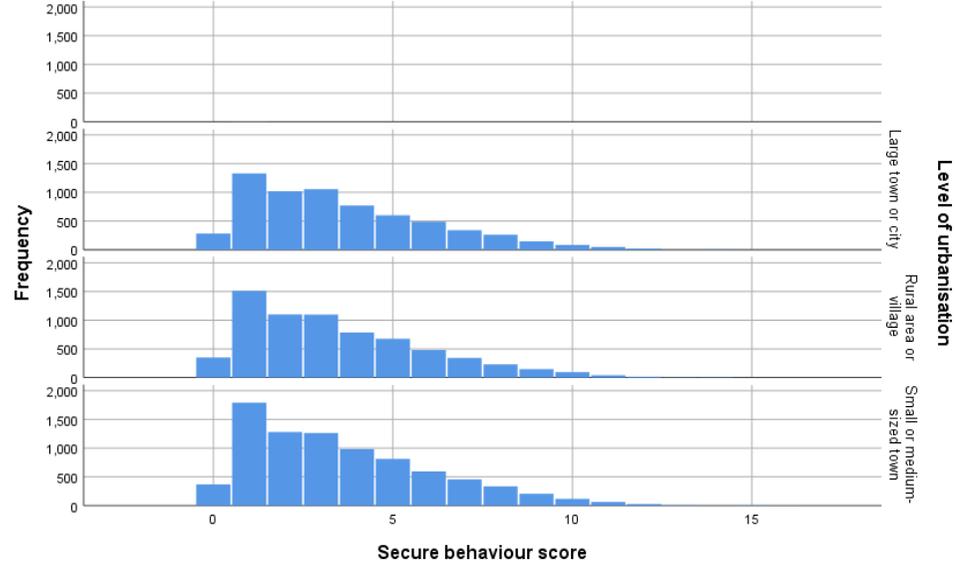

Click to navigate

Introduction  Literature review  Research questions  Antecedents of behaviour  Research model  Methodology  Results  Discussion  Conclusion  Limitations  References  Appendix

## Normal Q-Q Plots

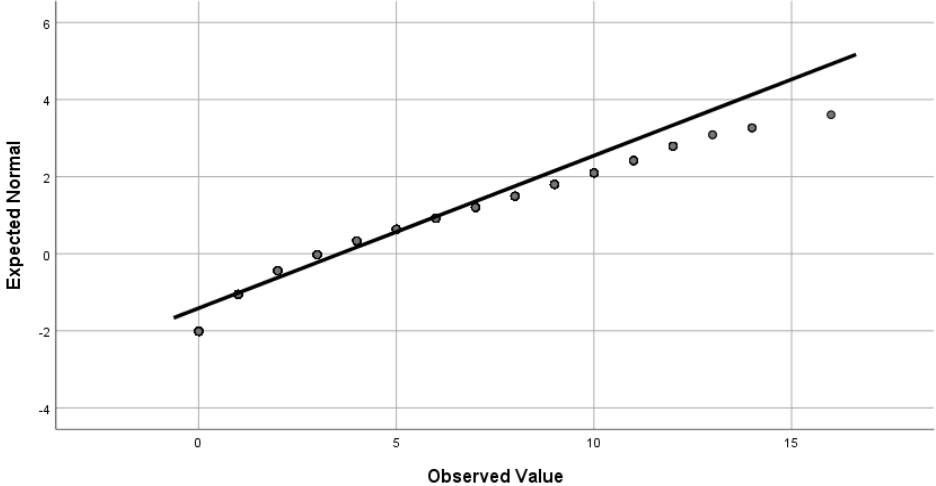

**Normal Q-Q Plot of Secure behaviour score**

for Levelofurbanisation= Large to

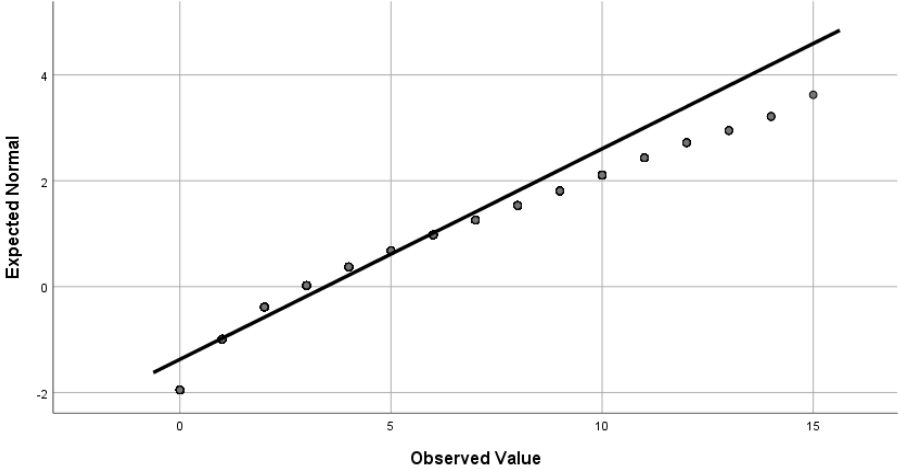

**Normal Q-Q Plot of Secure behaviour score**

for Levelofurbanisation= Rural ar

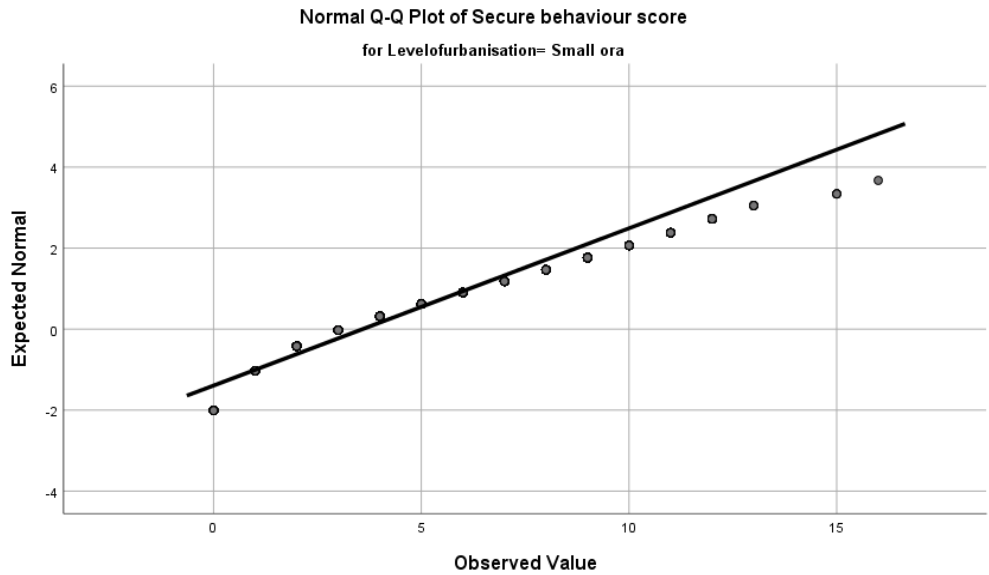

Normal Q-Q Plot of Secure behaviour score

for Levelofurbanisation= Small ora

### 12.1.1.3 One-way ANOVA Age groups

**Descriptives**

Secure behaviour score

| | N | Mean | Std. Deviation | Std. Error | 95% Confidence Interval for Mean | | Minimum | Maximum |
|---|---|---|---|---|---|---|---|---|
| | | | | | Lower Bound | Upper Bound | | |
| 15-20 years old | 1543 | 3,31 | 2,480 | ,063 | 3,19 | 3,44 | 0 | 15 |
| 21-29 years old | 2333 | 3,54 | 2,469 | ,051 | 3,44 | 3,64 | 0 | 16 |
| 30-39 years old | 3754 | 3,56 | 2,515 | ,041 | 3,48 | 3,64 | 0 | 16 |
| 40-49 years old | 3942 | 3,52 | 2,524 | ,040 | 3,44 | 3,60 | 0 | 15 |
| 50-59 years old | 4030 | 3,50 | 2,566 | ,040 | 3,42 | 3,58 | 0 | 14 |
| 60-69 years old | 3717 | 3,64 | 2,618 | ,043 | 3,56 | 3,73 | 0 | 14 |
| 70+ years old | 2343 | 3,59 | 2,558 | ,053 | 3,48 | 3,69 | 0 | 12 |
| Total | 21662 | 3,54 | 2,542 | ,017 | 3,51 | 3,57 | 0 | 16 |

**ANOVA**

Secure behaviour score

| | Sum of Squares | df | Mean Square | F | Sig. |
|---|---|---|---|---|---|
| Between Groups | 132,313 | 6 | 22,052 | 3,415 | ,002 |
| Within Groups | 139850,069 | 21655 | 6,458 | | |
| Total | 139982,382 | 21661 | | | |

**ANOVA Effect Sizes[a,b]**

| | | | 95% Confidence Interval | |
|---|---|---|---|---|
| | | Point Estimate | Lower | Upper |
| Secure behaviour score | Eta-squared | ,001 | ,000 | ,002 |
| | Epsilon-squared | ,001 | ,000 | ,001 |
| | Omega-squared Fixed-effect | ,001 | ,000 | ,001 |
| | Omega-squared Random-effect | ,000 | ,000 | ,000 |

a. Eta-squared and Epsilon-squared are estimated based on the fixed-effect model.

b. Negative but less biased estimates are retained, not rounded to zero.

**Post Hoc Tests**

<div align="center">

**Multiple Comparisons**

</div>

Dependent Variable:   Secure behaviour score

Tukey HSD

| (I) Age bracket | (J) Age bracket | Mean Difference (I-J) | Std. Error | Sig. | 95% Confidence Interval | |
|---|---|---|---|---|---|---|
| | | | | | Lower Bound | Upper Bound |
| 15-20 years old | 21-29 years old | -,222 | ,083 | ,107 | -,47 | ,02 |
| | 30-39 years old | -,247* | ,077 | ,022 | -,47 | -,02 |
| | 40-49 years old | -,206 | ,076 | ,098 | -,43 | ,02 |
| | 50-59 years old | -,186 | ,076 | ,179 | -,41 | ,04 |
| | 60-69 years old | -,328* | ,077 | <,001 | -,55 | -,10 |
| | 70+ years old | -,274* | ,083 | ,018 | -,52 | -,03 |
| 21-29 years old | 15-20 years old | ,222 | ,083 | ,107 | -,02 | ,47 |
| | 30-39 years old | -,025 | ,067 | 1,000 | -,22 | ,17 |
| | 40-49 years old | ,016 | ,066 | 1,000 | -,18 | ,21 |
| | 50-59 years old | ,036 | ,066 | ,998 | -,16 | ,23 |
| | 60-69 years old | -,106 | ,067 | ,700 | -,30 | ,09 |
| | 70+ years old | -,051 | ,074 | ,993 | -,27 | ,17 |
| 30-39 years old | 15-20 years old | ,247* | ,077 | ,022 | ,02 | ,47 |
| | 21-29 years old | ,025 | ,067 | 1,000 | -,17 | ,22 |
| | 40-49 years old | ,041 | ,058 | ,992 | -,13 | ,21 |
| | 50-59 years old | ,061 | ,058 | ,941 | -,11 | ,23 |
| | 60-69 years old | -,081 | ,059 | ,815 | -,25 | ,09 |
| | 70+ years old | -,027 | ,067 | 1,000 | -,22 | ,17 |
| 40-49 years old | 15-20 years old | ,206 | ,076 | ,098 | -,02 | ,43 |
| | 21-29 years old | -,016 | ,066 | 1,000 | -,21 | ,18 |
| | 30-39 years old | -,041 | ,058 | ,992 | -,21 | ,13 |
| | 50-59 years old | ,020 | ,057 | 1,000 | -,15 | ,19 |
| | 60-69 years old | -,122 | ,058 | ,356 | -,29 | ,05 |
| | 70+ years old | -,068 | ,066 | ,950 | -,26 | ,13 |
| 50-59 years old | 15-20 years old | ,186 | ,076 | ,179 | -,04 | ,41 |
| | 21-29 years old | -,036 | ,066 | ,998 | -,23 | ,16 |
| | 30-39 years old | -,061 | ,058 | ,941 | -,23 | ,11 |
| | 40-49 years old | -,020 | ,057 | 1,000 | -,19 | ,15 |
| | 60-69 years old | -,142 | ,058 | ,177 | -,31 | ,03 |
| | 70+ years old | -,088 | ,066 | ,839 | -,28 | ,11 |
| 60-69 years old | 15-20 years old | ,328[*] | ,077 | <,001 | ,10 | ,55 |
| | 21-29 years old | ,106 | ,067 | ,700 | -,09 | ,30 |
| | 30-39 years old | ,081 | ,059 | ,815 | -,09 | ,25 |
| | 40-49 years old | ,122 | ,058 | ,356 | -,05 | ,29 |
| | 50-59 years old | ,142 | ,058 | ,177 | -,03 | ,31 |
| | 70+ years old | ,054 | ,067 | ,984 | -,14 | ,25 |
| 70+ years old | 15-20 years old | ,274[*] | ,083 | ,018 | ,03 | ,52 |
| | 21-29 years old | ,051 | ,074 | ,993 | -,17 | ,27 |
| | 30-39 years old | ,027 | ,067 | 1,000 | -,17 | ,22 |
| | 40-49 years old | ,068 | ,066 | ,950 | -,13 | ,26 |
| | 50-59 years old | ,088 | ,066 | ,839 | -,11 | ,28 |
| | 60-69 years old | -,054 | ,067 | ,984 | -,25 | ,14 |

*. The mean difference is significant at the 0.05 level.

Introduction · Literature review · Research questions · Antecedents of behaviour · Research model · Methodology · Results · Discussion · Conclusion · Limitations · References · Appendix

**Homogeneous Subsets**

**Secure behaviour score**

Tukey HSD[a,b]

| Age bracket | N | Subset for alpha = 0.05 | |
|---|---|---|---|
| | | 1 | 2 |
| 15-20 years old | 1543 | 3,31 | |
| 50-59 years old | 4030 | 3,50 | 3,50 |
| 40-49 years old | 3942 | | 3,52 |
| 21-29 years old | 2333 | | 3,54 |
| 30-39 years old | 3754 | | 3,56 |
| 70+ years old | 2343 | | 3,59 |
| 60-69 years old | 3717 | | 3,64 |
| Sig. | | ,094 | ,371 |

Means for groups in homogeneous subsets are displayed.

a. Uses Harmonic Mean Sample Size = 2755,083.

b. The group sizes are unequal. The harmonic mean of the group sizes is used. Type I error levels are not guaranteed.

Click to navigate

## Boxplot

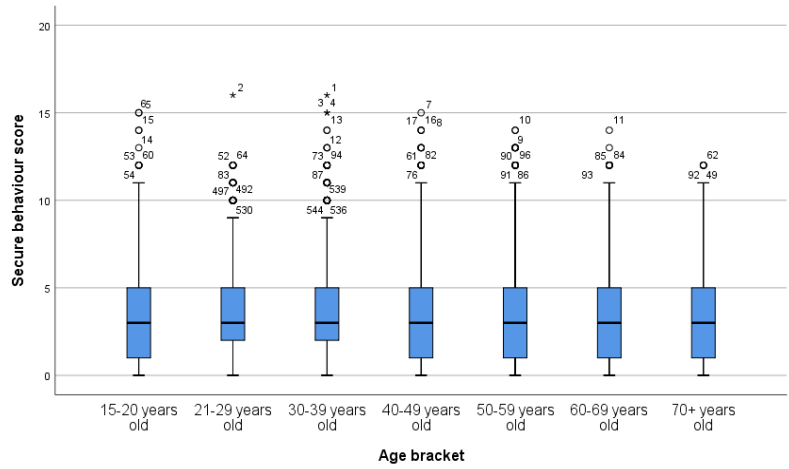

## Distribution histograms

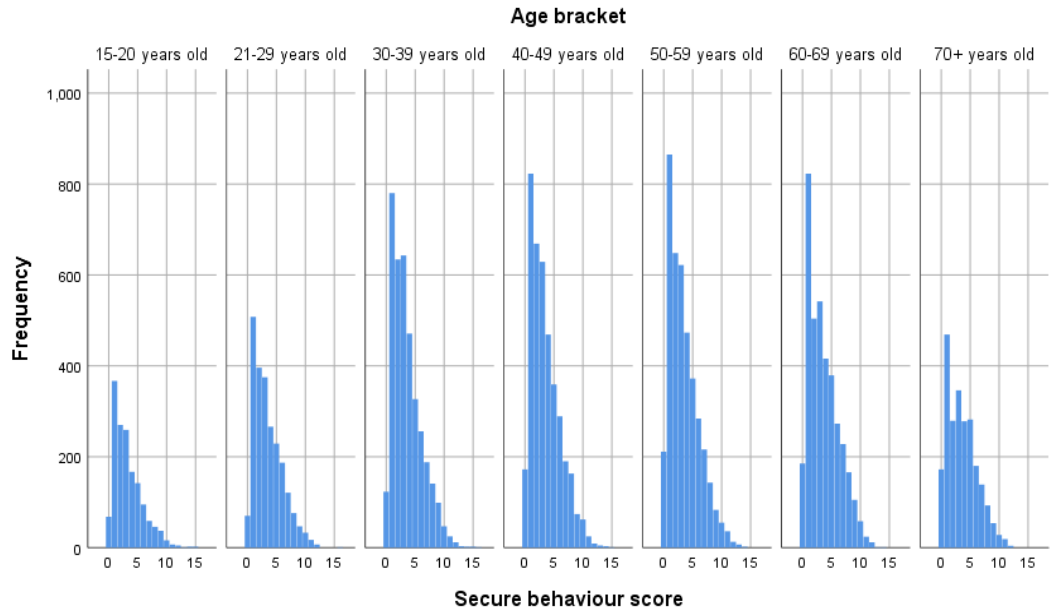

## Normal Q-Q Plots

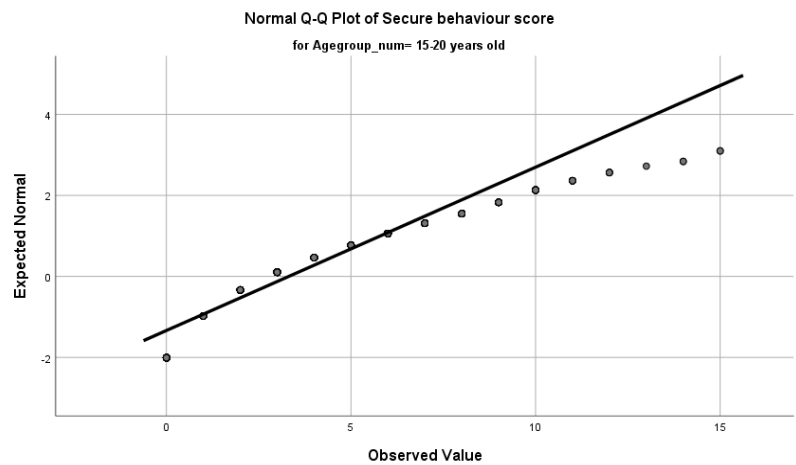

**Normal Q-Q Plot of Secure behaviour score**
for Agegroup_num= 15-20 years old

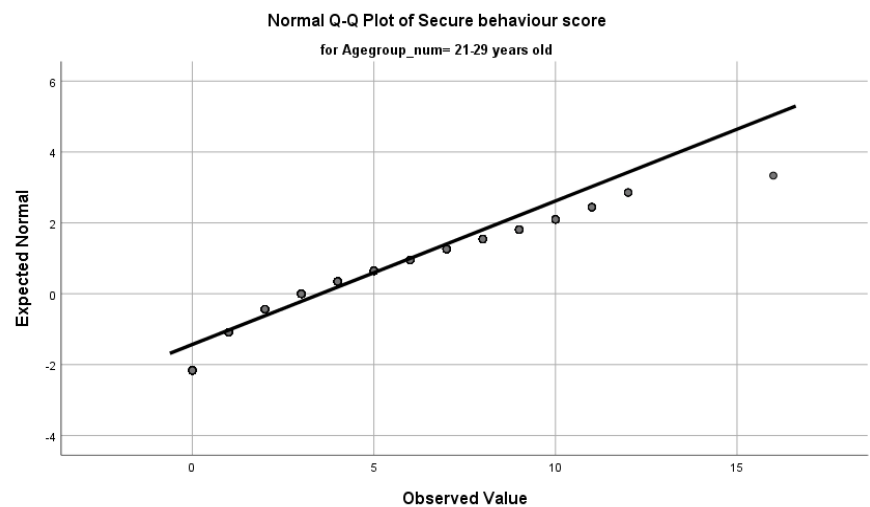

**Normal Q-Q Plot of Secure behaviour score**
for Agegroup_num= 21-29 years old

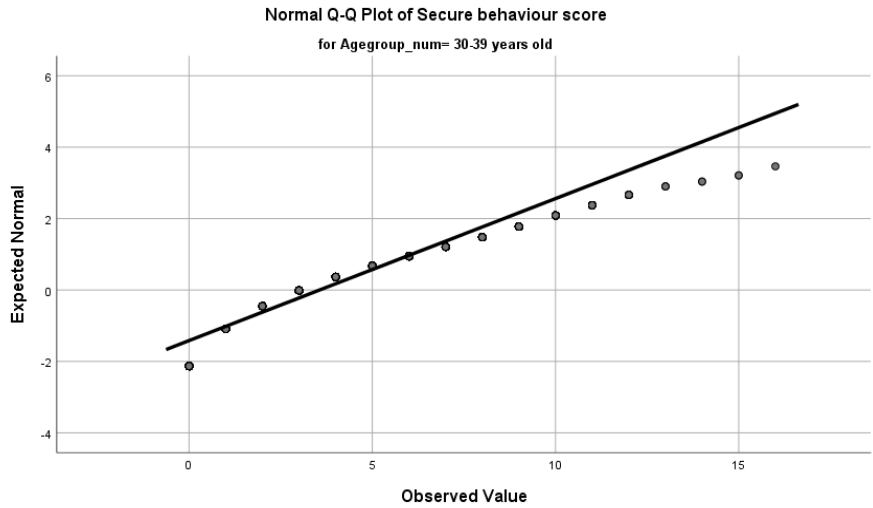

Normal Q-Q Plot of Secure behaviour score
for Agegroup_num= 30-39 years old

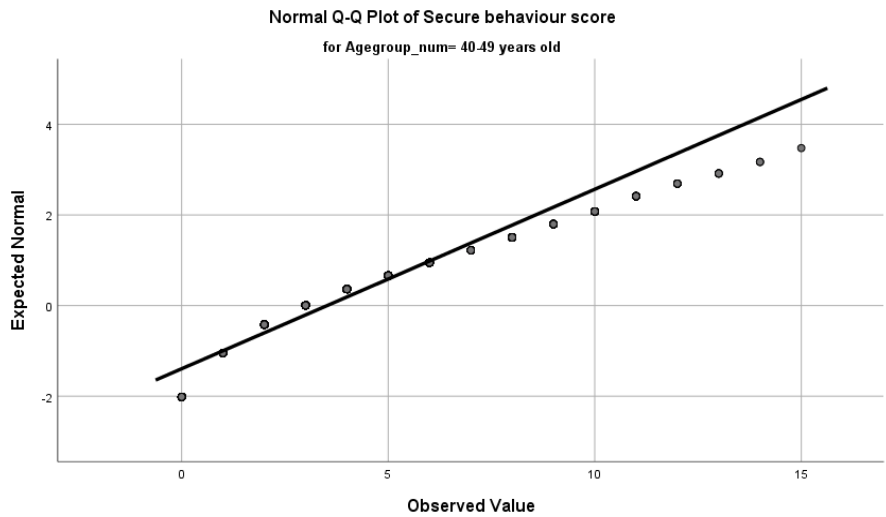

Normal Q-Q Plot of Secure behaviour score
for Agegroup_num= 40-49 years old

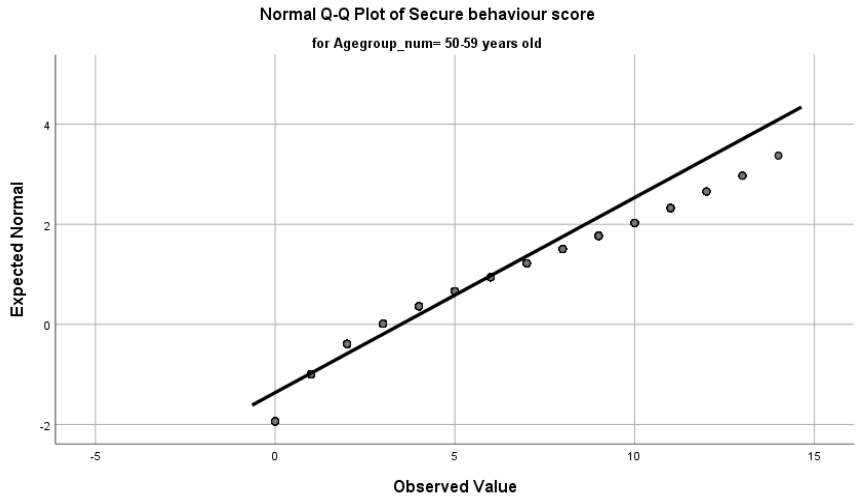

**Normal Q-Q Plot of Secure behaviour score**

for Agegroup_num= 50-59 years old

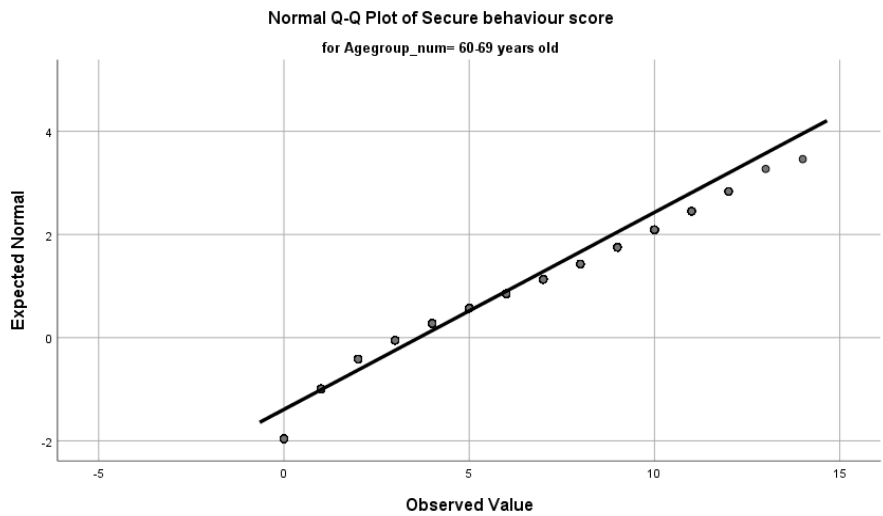

**Normal Q-Q Plot of Secure behaviour score**

for Agegroup_num= 60-69 years old

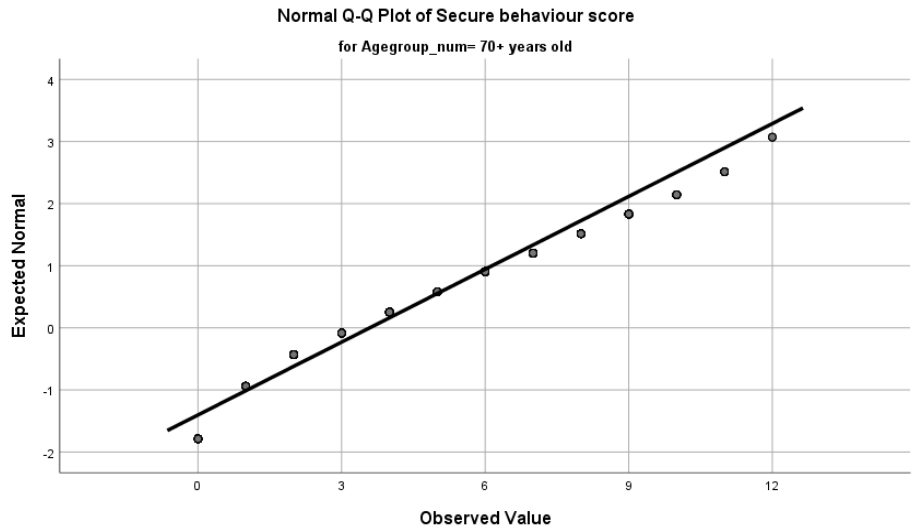

Normal Q-Q Plot of Secure behaviour score
for Agegroup_num= 70+ years old

## 12.1.1.4  Simple regression Age

**Model Summary**

| Model | R | R Square | Adjusted R Square | Std. Error of the Estimate |
|---|---|---|---|---|
| 1 | ,018[a] | ,000 | ,000 | 2,542 |

a. Predictors: (Constant), Age

**ANOVA[a]**

| Model | | Sum of Squares | df | Mean Square | F | Sig. |
|---|---|---|---|---|---|---|
| 1 | Regression | 47,160 | 1 | 47,160 | 7,300 | ,007[b] |
| | Residual | 139935,222 | 21660 | 6,461 | | |
| | Total | 139982,382 | 21661 | | | |

a. Dependent Variable: Secure behaviour score

b. Predictors: (Constant), Age

**Coefficients**[a]

| Model | | Unstandardized Coefficients | | Standardized Coefficients | t | Sig. |
|---|---|---|---|---|---|---|
| | | B | Std. Error | Beta | | |
| 1 | (Constant) | 3,409 | ,051 | | 66,567 | ,000 |
| | Age | ,003 | ,001 | ,018 | 2,702 | ,007 |

a. Dependent Variable: Secure behaviour score

## 12.1.1.5 One-way ANOVA Direct prior experience with security incidents categories

**Descriptives**

Secure behaviour score

| | N | Mean | Std. Deviation | Std. Error | 95% Confidence Interval for Mean | | Minimum | Maximum |
|---|---|---|---|---|---|---|---|---|
| | | | | | Lower Bound | Upper Bound | | |
| Large experience | 169 | 2,36 | 1,533 | ,118 | 2,13 | 2,59 | 0 | 9 |
| Little experience | 19552 | 3,46 | 2,508 | ,018 | 3,42 | 3,49 | 0 | 16 |
| Medium experience | 420 | 3,13 | 2,157 | ,105 | 2,92 | 3,34 | 0 | 12 |
| Some experience | 1521 | 4,82 | 2,783 | ,071 | 4,68 | 4,96 | 0 | 16 |
| Total | 21662 | 3,54 | 2,542 | ,017 | 3,51 | 3,57 | 0 | 16 |

**ANOVA**

Secure behaviour score

| | Sum of Squares | df | Mean Square | F | Sig. |
|---|---|---|---|---|---|
| Between Groups | 2910,455 | 3 | 970,152 | 153,288 | <,001 |
| Within Groups | 137071,926 | 21658 | 6,329 | | |

**Click to navigate**

| | | | | | |
|---|---|---|---|---|---|
| Total | 139982,382 | 21661 | | | |

### ANOVA Effect Sizes[a]

| | | Point Estimate | 95% Confidence Interval | |
|---|---|---|---|---|
| | | | Lower | Upper |
| Secure behaviour score | Eta-squared | ,021 | ,017 | ,025 |
| | Epsilon-squared | ,021 | ,017 | ,024 |
| | Omega-squared Fixed-effect | ,021 | ,017 | ,024 |
| | Omega-squared Random-effect | ,007 | ,006 | ,008 |

a. Eta-squared and Epsilon-squared are estimated based on the fixed-effect model.

**Post Hoc Tests**

### Multiple Comparisons

Dependent Variable:   Secure behaviour score

Tukey HSD

| (I) Direct experience bracket | (J) Direct experience bracket | Mean Difference (I-J) | Std. Error | Sig. | 95% Confidence Interval | |
|---|---|---|---|---|---|---|
| | | | | | Lower Bound | Upper Bound |
| Large experience | Little experience | -1,098[*] | ,194 | <,001 | -1,60 | -,60 |
| | Medium experience | -,770[*] | ,229 | ,004 | -1,36 | -,18 |
| | Some experience | -2,455[*] | ,204 | ,000 | -2,98 | -1,93 |
| Little experience | Large experience | 1,098[*] | ,194 | <,001 | ,60 | 1,60 |
| | Medium experience | ,328[*] | ,124 | ,041 | ,01 | ,65 |
| | Some experience | -1,357[*] | ,067 | ,000 | -1,53 | -1,19 |
| Medium experience | Large experience | ,770[*] | ,229 | ,004 | ,18 | 1,36 |
| | Little experience | -,328[*] | ,124 | ,041 | -,65 | -,01 |
| | Some experience | -1,685[*] | ,139 | ,000 | -2,04 | -1,33 |
| Some experience | Large experience | 2,455[*] | ,204 | ,000 | 1,93 | 2,98 |
| | Little experience | 1,357[*] | ,067 | ,000 | 1,19 | 1,53 |
| | Medium experience | 1,685[*] | ,139 | ,000 | 1,33 | 2,04 |

*. The mean difference is significant at the 0.05 level.

**Homogeneous Subsets**

**Secure behaviour score**

Tukey HSD[a,b]

| Direct experience bracket | N | \multicolumn{3}{c}{Subset for alpha = 0.05} | | |
|---|---|---|---|---|
| | | 1 | 2 | 3 |
| Large experience | 169 | 2,36 | | |
| Medium experience | 420 | | 3,13 | |
| Little experience | 19552 | | 3,46 | |
| Some experience | 1521 | | | 4,82 |
| Sig. | | 1,000 | ,211 | 1,000 |

Means for groups in homogeneous subsets are displayed.

a. Uses Harmonic Mean Sample Size = 444,113.

b. The group sizes are unequal. The harmonic mean of the group sizes is used. Type I error levels are not guaranteed.

Click to navigate

## Boxplot

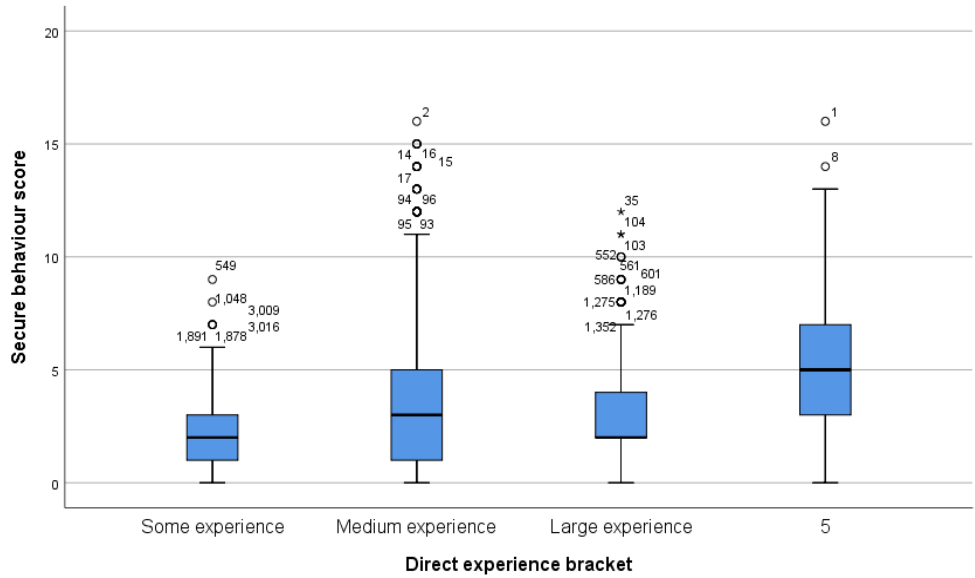

## Distribution histograms

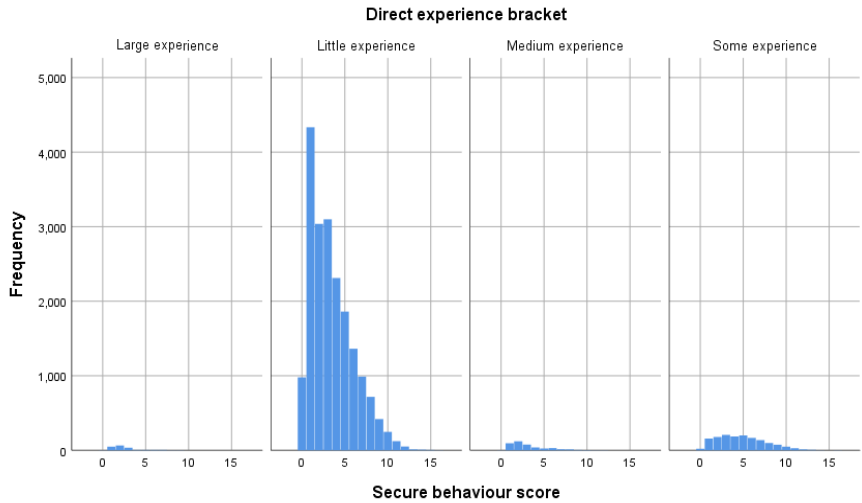

## Normal Q-Q Plots

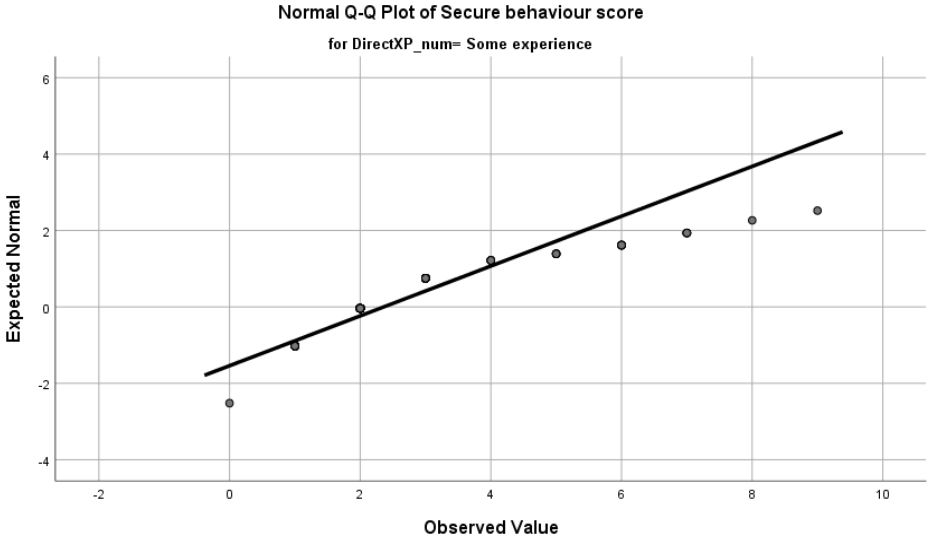

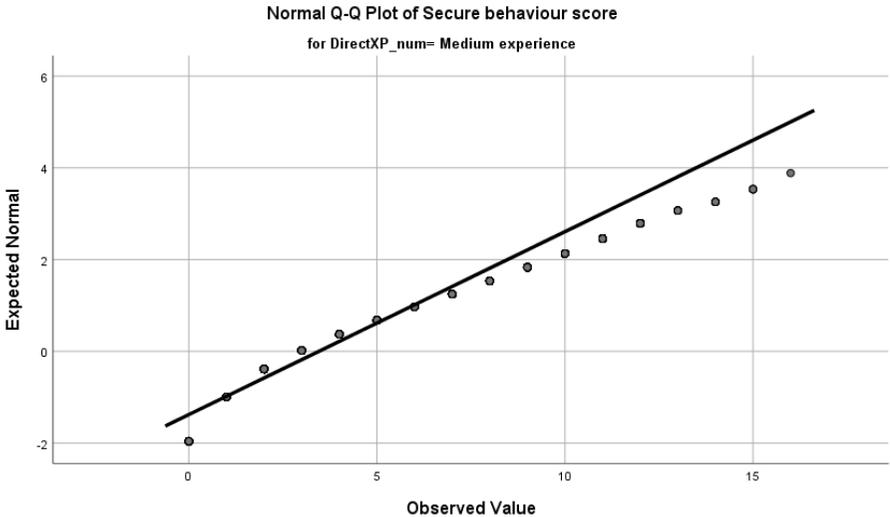

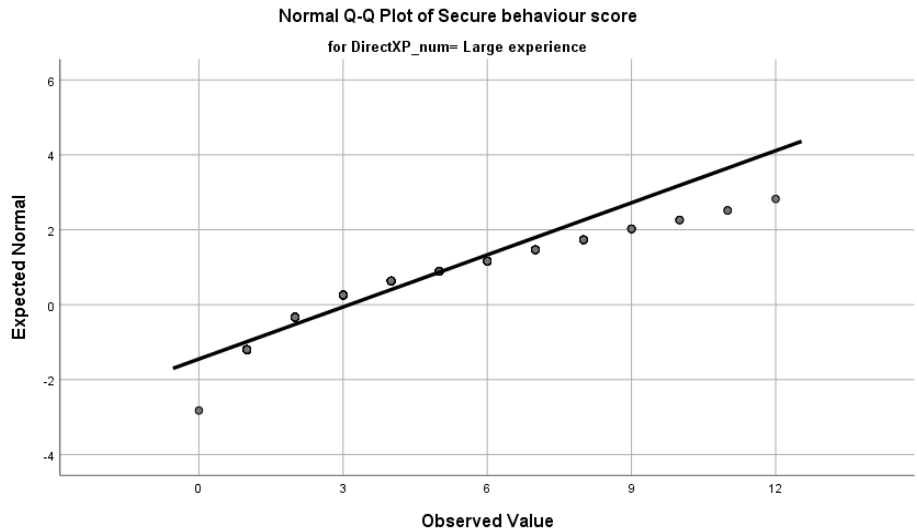

Normal Q-Q Plot of Secure behaviour score
for DirectXP_num= Large experience

## 12.1.1.6 One-way ANOVA Indirect prior experience with security incidents categories

**Descriptives**

Secure behaviour score

| | N | Mean | Std. Deviation | Std. Error | 95% Confidence Interval for Mean | | Minimum | Maximum |
|---|---|---|---|---|---|---|---|---|
| | | | | | Lower Bound | Upper Bound | | |
| Large experience | 21 | 4,57 | 3,544 | ,773 | 2,96 | 6,18 | 0 | 13 |
| Little experience | 18677 | 3,31 | 2,431 | ,018 | 3,27 | 3,34 | 0 | 16 |
| Medium experience | 206 | 5,92 | 3,033 | ,211 | 5,50 | 6,33 | 0 | 14 |
| Some experience | 2758 | 4,93 | 2,689 | ,051 | 4,83 | 5,03 | 0 | 15 |
| Total | 21662 | 3,54 | 2,542 | ,017 | 3,51 | 3,57 | 0 | 16 |

**ANOVA**

Secure behaviour score

Click to navigate

Introduction  Literature review  Research questions  Antecedents of behaviour  Research model  Methodology  Results  Discussion  Conclusion  Limitations  References  Appendix

|  | Sum of Squares | df | Mean Square | F | Sig. |
|---|---|---|---|---|---|
| Between Groups | 7542,058 | 3 | 2514,019 | 411,118 | <,001 |
| Within Groups | 132440,323 | 21658 | 6,115 |  |  |
| Total | 139982,382 | 21661 |  |  |  |

**ANOVA Effect Sizes[a]**

|  |  | Point Estimate | 95% Confidence Interval | |
|---|---|---|---|---|
|  |  |  | Lower | Upper |
| Secure behaviour score | Eta-squared | ,054 | ,048 | ,060 |
|  | Epsilon-squared | ,054 | ,048 | ,060 |
|  | Omega-squared Fixed-effect | ,054 | ,048 | ,060 |
|  | Omega-squared Random-effect | ,019 | ,017 | ,021 |

a. Eta-squared and Epsilon-squared are estimated based on the fixed-effect model.

**Post Hoc Tests**

**Multiple Comparisons**

Dependent Variable:   Secure behaviour score

Tukey HSD

| (I) Indirect experience bracket | (J) Indirect experience bracket | Mean Difference (I-J) | Std. Error | Sig. | 95% Confidence Interval | |
|---|---|---|---|---|---|---|
|  |  |  |  |  | Lower Bound | Upper Bound |
| Large experience | Little experience | 1,265 | ,540 | ,088 | -,12 | 2,65 |
|  | Medium experience | -1,346 | ,566 | ,082 | -2,80 | ,11 |
|  | Some experience | -,359 | ,542 | ,911 | -1,75 | 1,03 |

Click to navigate

Introduction  Literature review  Research questions  Antecedents of behaviour  Research model  Methodology  Results  Discussion  Conclusion  Limitations  References  Appendix

| | | | | | | |
|---|---|---|---|---|---|---|
| Little experience | Large experience | -1,265 | ,540 | ,088 | -2,65 | ,12 |
| | Medium experience | -2,611[*] | ,173 | ,000 | -3,06 | -2,17 |
| | Some experience | -1,625[*] | ,050 | ,000 | -1,75 | -1,49 |
| Medium experience | Large experience | 1,346 | ,566 | ,082 | -,11 | 2,80 |
| | Little experience | 2,611[*] | ,173 | ,000 | 2,17 | 3,06 |
| | Some experience | ,987[*] | ,179 | <,001 | ,53 | 1,45 |
| Some experience | Large experience | ,359 | ,542 | ,911 | -1,03 | 1,75 |
| | Little experience | 1,625[*] | ,050 | ,000 | 1,49 | 1,75 |
| | Medium experience | -,987[*] | ,179 | <,001 | -1,45 | -,53 |

*. The mean difference is significant at the 0.05 level.

**Homogeneous Subsets**

<div align="center">

**Secure behaviour score**

</div>

Tukey HSD[a,b]

| Indirect experience bracket | N | Subset for alpha = 0.05 | | |
|---|---|---|---|---|
| | | 1 | 2 | 3 |
| Little experience | 18677 | 3,31 | | |
| Large experience | 21 | | 4,57 | |
| Some experience | 2758 | | 4,93 | 4,93 |
| Medium experience | 206 | | | 5,92 |
| Sig. | | 1,000 | ,808 | ,067 |

Means for groups in homogeneous subsets are displayed.

a. Uses Harmonic Mean Sample Size = 75,629.

b. The group sizes are unequal. The harmonic mean of the group sizes is used. Type I error levels are not guaranteed.

## Boxplot

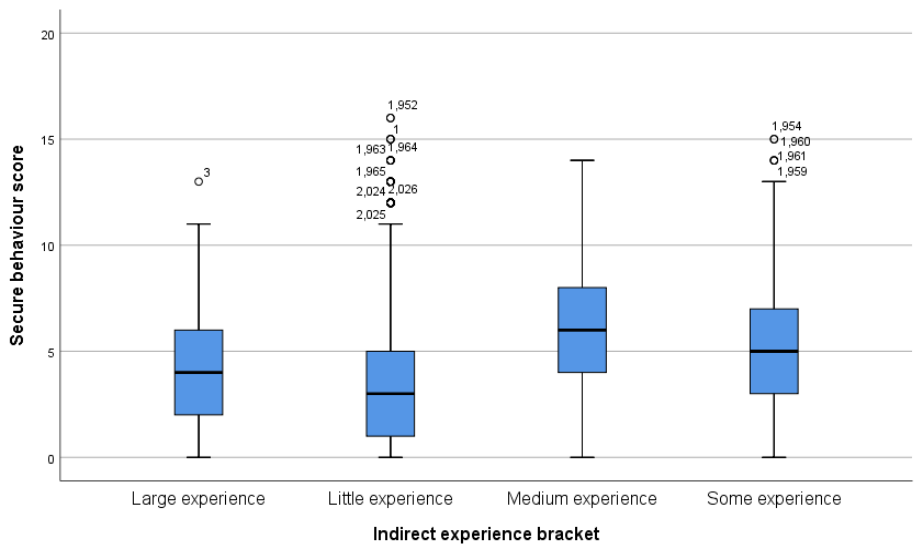

## Distribution histograms

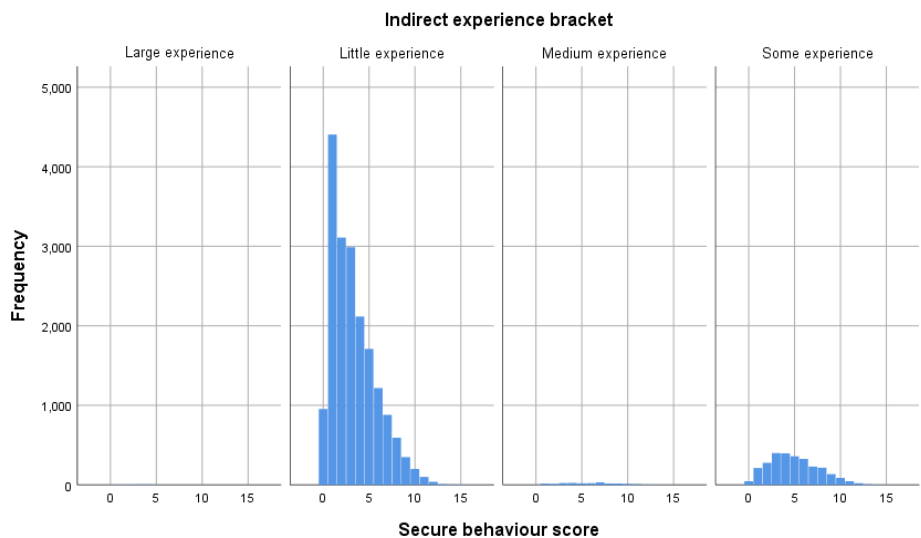

## Normal Q-Q Plots

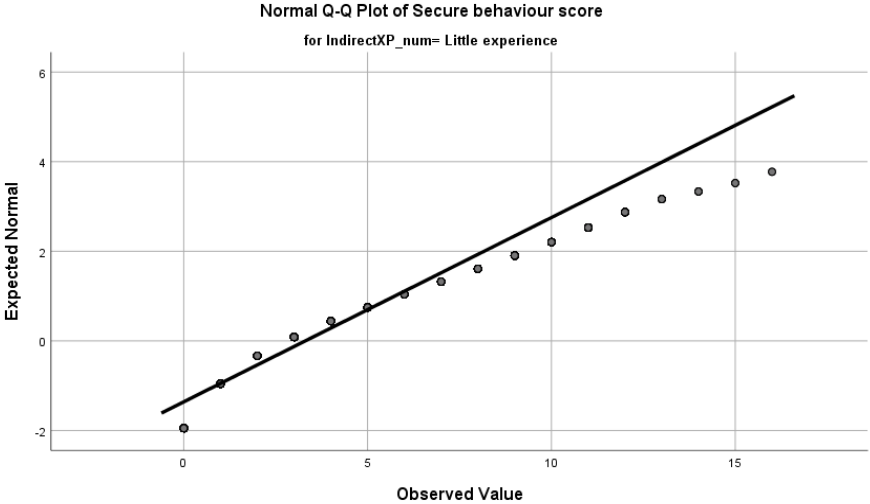

Normal Q-Q Plot of Secure behaviour score
for IndirectXP_num= Little experience

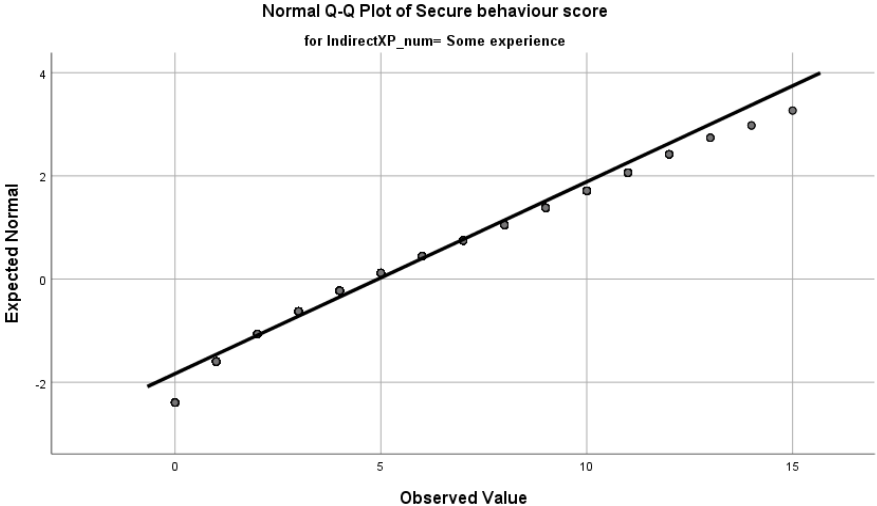

Normal Q-Q Plot of Secure behaviour score
for IndirectXP_num= Some experience

Introduction  Literature review  Research questions  Antecedents of behaviour  Research model  Methodology  Results  Discussion  Conclusion  Limitations  References  Appendix

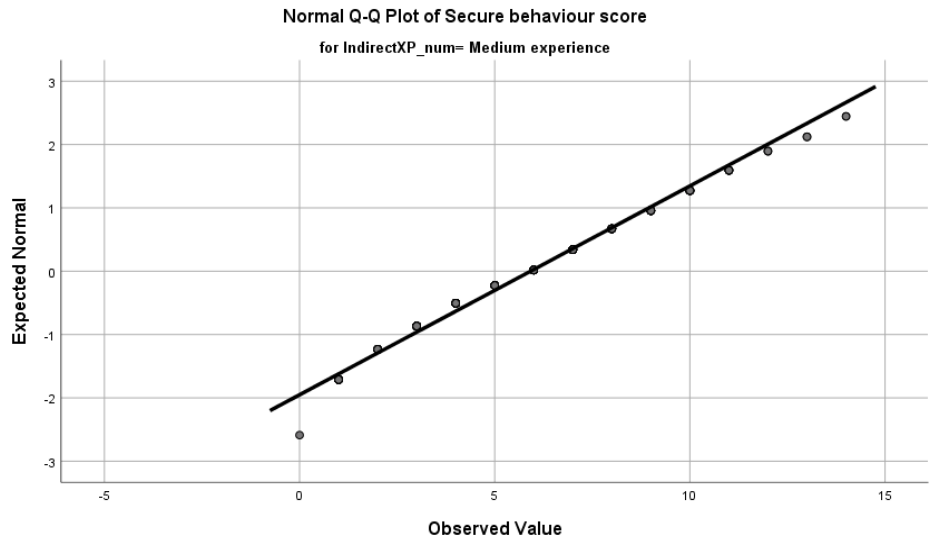

**Normal Q-Q Plot of Secure behaviour score**

for IndirectXP_num= Medium experience

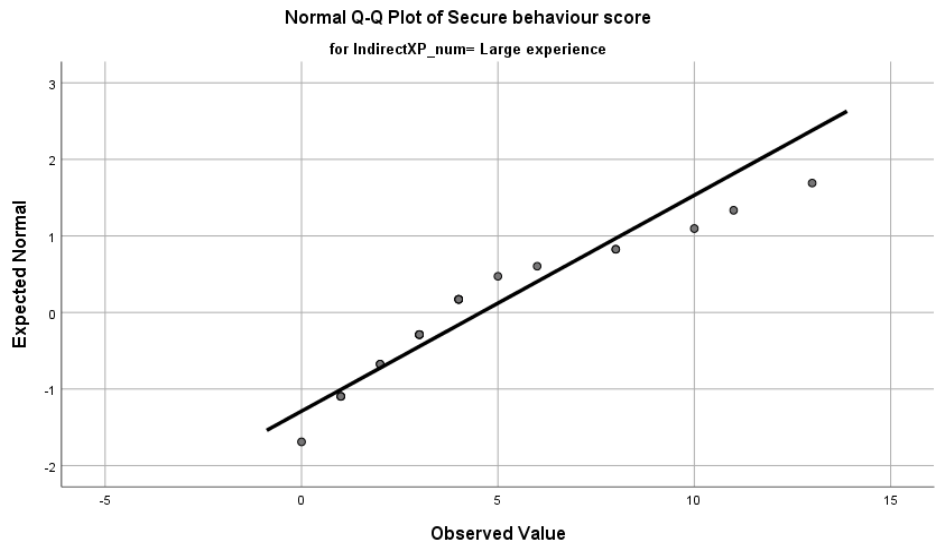

**Normal Q-Q Plot of Secure behaviour score**

for IndirectXP_num= Large experience

### 12.1.1.7 Simple regression Direct prior experience with security

**Model Summary**

| Model | R | R Square | Adjusted R Square | Std. Error of the Estimate |
|---|---|---|---|---|
| 1 | ,164[a] | ,027 | ,027 | 2,508 |

a. Predictors: (Constant), Direct experience score

**ANOVA[a]**

| Model | | Sum of Squares | df | Mean Square | F | Sig. |
|---|---|---|---|---|---|---|
| 1 | Regression | 3767,663 | 1 | 3767,663 | 599,110 | <,001[b] |
| | Residual | 136214,718 | 21660 | 6,289 | | |
| | Total | 139982,382 | 21661 | | | |

a. Dependent Variable: Secure behaviour score

b. Predictors: (Constant), Direct experience score

**Coefficients[a]**

| Model | | Unstandardized Coefficients | | Standardized Coefficients | t | Sig. |
|---|---|---|---|---|---|---|
| | | B | Std. Error | Beta | | |
| 1 | (Constant) | 3,285 | ,020 | | 164,561 | ,000 |
| | Direct experience score | ,138 | ,006 | ,164 | 24,477 | <,001 |

a. Dependent Variable: Secure behaviour score

### 12.1.1.8   Simple regression Indirect prior experience with security

**Model Summary**

| Model | R | R Square | Adjusted R Square | Std. Error of the Estimate |
|---|---|---|---|---|
| 1 | ,249[a] | ,062 | ,062 | 2,462 |

a. Predictors: (Constant), Indirect experience score

**ANOVA[a]**

| Model | | Sum of Squares | df | Mean Square | F | Sig. |
|---|---|---|---|---|---|---|
| 1 | Regression | 8706,631 | 1 | 8706,631 | 1436,561 | <,001[b] |
| | Residual | 131275,751 | 21660 | 6,061 | | |
| | Total | 139982,382 | 21661 | | | |

a. Dependent Variable: Secure behaviour score

b. Predictors: (Constant), Indirect experience score

**Coefficients[a]**

| Model | | Unstandardized Coefficients | | Standardized Coefficients | t | Sig. |
|---|---|---|---|---|---|---|
| | | B | Std. Error | Beta | | |
| 1 | (Constant) | 2,641 | ,029 | | 91,006 | ,000 |
| | Indirect experience score | ,595 | ,016 | ,249 | 37,902 | <,001 |

a. Dependent Variable: Secure behaviour score

Click to navigate

## 12.1.1.9 Simple regression Hofstede dimension Power Distance

**Model Summary**

| Model | R | R Square | Adjusted R Square | Std. Error of the Estimate |
|---|---|---|---|---|
| 1 | ,207[a] | ,043 | ,043 | 2,485 |

a. Predictors: (Constant), Power distance (Hofstede)

**ANOVA[a]**

| Model | | Sum of Squares | df | Mean Square | F | Sig. |
|---|---|---|---|---|---|---|
| 1 | Regression | 5905,949 | 1 | 5905,949 | 956,514 | <,001[b] |
| | Residual | 131633,168 | 21319 | 6,174 | | |
| | Total | 137539,116 | 21320 | | | |

a. Dependent Variable: Secure behaviour score

b. Predictors: (Constant), Power distance (Hofstede)

**Coefficients[a]**

| Model | | Unstandardized Coefficients | | Standardized Coefficients | t | Sig. |
|---|---|---|---|---|---|---|
| | | B | Std. Error | Beta | | |
| 1 | (Constant) | 4,804 | ,044 | | 108,495 | ,000 |
| | Power distance (Hofstede) | -,025 | ,001 | -,207 | -30,928 | <,001 |

a. Dependent Variable: Secure behaviour score

## 12.1.1.10 Simple regression Hofstede dimension Individualism

**Model Summary**

| Model | R | R Square | Adjusted R Square | Std. Error of the Estimate |
|---|---|---|---|---|
| 1 | ,153ª | ,023 | ,023 | 2,510 |

a. Predictors: (Constant), Individualism (Hofstede)

**ANOVAª**

| Model | | Sum of Squares | df | Mean Square | F | Sig. |
|---|---|---|---|---|---|---|
| 1 | Regression | 3207,218 | 1 | 3207,218 | 508,998 | <,001ᵇ |
| | Residual | 134331,899 | 21319 | 6,301 | | |
| | Total | 137539,116 | 21320 | | | |

a. Dependent Variable: Secure behaviour score

b. Predictors: (Constant), Individualism (Hofstede)

**Coefficientsª**

| Model | | Unstandardized Coefficients | | Standardized Coefficients | t | Sig. |
|---|---|---|---|---|---|---|
| | | B | Std. Error | Beta | | |
| 1 | (Constant) | 2,222 | ,061 | | 36,513 | <,001 |
| | Individualism (Hofstede) | ,022 | ,001 | ,153 | 22,561 | <,001 |

a. Dependent Variable: Secure behaviour score

## 12.1.1.11  Simple regression Hofstede dimension Uncertainty Avoidance

**Model Summary**

| Model | R | R Square | Adjusted R Square | Std. Error of the Estimate |
|---|---|---|---|---|
| 1 | ,171[a] | ,029 | ,029 | 2,503 |

a. Predictors: (Constant), Uncertainty avoidance (Hofstede)

**ANOVA[a]**

| Model | | Sum of Squares | df | Mean Square | F | Sig. |
|---|---|---|---|---|---|---|
| 1 | Regression | 4007,391 | 1 | 4007,391 | 639,800 | <,001[b] |
| | Residual | 133531,725 | 21319 | 6,264 | | |
| | Total | 137539,116 | 21320 | | | |

a. Dependent Variable: Secure behaviour score

b. Predictors: (Constant), Uncertainty avoidance (Hofstede)

**Coefficients[a]**

| Model | | Unstandardized Coefficients | | Standardized Coefficients | t | Sig. |
|---|---|---|---|---|---|---|
| | | B | Std. Error | Beta | | |
| 1 | (Constant) | 4,852 | ,055 | | 88,765 | ,000 |
| | Uncertainty avoidance (Hofstede) | -,019 | ,001 | -,171 | -25,294 | <,001 |

a. Dependent Variable: Secure behaviour score

## 12.1.1.12 Simple regression Hofstede dimension Masculinity

**Model Summary**

| Model | R | R Square | Adjusted R Square | Std. Error of the Estimate |
|---|---|---|---|---|
| 1 | ,137[a] | ,019 | ,019 | 2,516 |

a. Predictors: (Constant), Masculinity (Hofstede)

**ANOVA[a]**

| Model | | Sum of Squares | df | Mean Square | F | Sig. |
|---|---|---|---|---|---|---|
| 1 | Regression | 2577,363 | 1 | 2577,363 | 407,129 | <,001[b] |
| | Residual | 134961,753 | 21319 | 6,331 | | |
| | Total | 137539,116 | 21320 | | | |

a. Dependent Variable: Secure behaviour score

b. Predictors: (Constant), Masculinity (Hofstede)

**Coefficients[a]**

| Model | | Unstandardized Coefficients | | Standardized Coefficients | t | Sig. |
|---|---|---|---|---|---|---|
| | | B | Std. Error | Beta | | |
| 1 | (Constant) | 4,173 | ,036 | | 116,476 | ,000 |
| | Masculinity (Hofstede) | -,014 | ,001 | -,137 | -20,177 | <,001 |

a. Dependent Variable: Secure behaviour score

## 12.1.1.13  Simple regression Hofstede dimension Long Term Orientation

**Model Summary**

| Model | R | R Square | Adjusted R Square | Std. Error of the Estimate |
|---|---|---|---|---|
| 1 | ,020[a] | ,000 | ,000 | 2,539 |

a. Predictors: (Constant), Long-term orientation (Hofstede)

**ANOVA[a]**

| Model | | Sum of Squares | df | Mean Square | F | Sig. |
|---|---|---|---|---|---|---|
| 1 | Regression | 55,261 | 1 | 55,261 | 8,569 | ,003[b] |
| | Residual | 137483,855 | 21319 | 6,449 | | |
| | Total | 137539,116 | 21320 | | | |

a. Dependent Variable: Secure behaviour score

b. Predictors: (Constant), Long-term orientation (Hofstede)

**Coefficients[a]**

| Model | | Unstandardized Coefficients | | Standardized Coefficients | t | Sig. |
|---|---|---|---|---|---|---|
| | | B | Std. Error | Beta | | |
| 1 | (Constant) | 3,361 | ,063 | | 52,990 | ,000 |
| | Long-term orientation (Hofstede) | ,003 | ,001 | ,020 | 2,927 | ,003 |

a. Dependent Variable: Secure behaviour score

## 12.1.1.14 Simple regression Hofstede dimension Indulgence

**Model Summary**

| Model | R | R Square | Adjusted R Square | Std. Error of the Estimate |
|---|---|---|---|---|
| 1 | ,238[a] | ,056 | ,056 | 2,469 |

a. Predictors: (Constant), Indulgence (Hofstede)

**ANOVA[a]**

| Model | | Sum of Squares | df | Mean Square | F | Sig. |
|---|---|---|---|---|---|---|
| 1 | Regression | 7907,489 | 1 | 7907,489 | 1296,811 | <,001[b] |
| | Residual | 132074,893 | 21660 | 6,098 | | |
| | Total | 139982,382 | 21661 | | | |

a. Dependent Variable: Secure behaviour score

b. Predictors: (Constant), Indulgence (Hofstede)

**Coefficients[a]**

| Model | | Unstandardized Coefficients | | Standardized Coefficients | t | Sig. |
|---|---|---|---|---|---|---|
| | | B | Std. Error | Beta | | |
| 1 | (Constant) | 2,151 | ,042 | | 51,167 | ,000 |
| | Indulgence (Hofstede) | ,031 | ,001 | ,238 | 36,011 | <,001 |

a. Dependent Variable: Secure behaviour score

## 12.1.1.15  Simple regression Meyer dimension Communicating

**Model Summary**

| Model | R | R Square | Adjusted R Square | Std. Error of the Estimate |
|---|---|---|---|---|
| 1 | ,268[a] | ,072 | ,072 | 2,481 |

a. Predictors: (Constant), Communicating (Meyer)

**ANOVA[a]**

| Model | | Sum of Squares | df | Mean Square | F | Sig. |
|---|---|---|---|---|---|---|
| 1 | Regression | 7701,213 | 1 | 7701,213 | 1251,024 | <,001[b] |
| | Residual | 99455,119 | 16156 | 6,156 | | |
| | Total | 107156,332 | 16157 | | | |

a. Dependent Variable: Secure behaviour score

b. Predictors: (Constant), Communicating (Meyer)

**Coefficients[a]**

| Model | | Unstandardized Coefficients | | Standardized Coefficients | t | Sig. |
|---|---|---|---|---|---|---|
| | | B | Std. Error | Beta | | |
| 1 | (Constant) | 5,807 | ,064 | | 90,485 | ,000 |
| | Communicating (Meyer) | -,429 | ,012 | -,268 | -35,370 | <,001 |

a. Dependent Variable: Secure behaviour score

## 12.1.1.16 Simple regression Meyer dimension Evaluating

### Model Summary

| Model | R | R Square | Adjusted R Square | Std. Error of the Estimate |
|---|---|---|---|---|
| 1 | ,064[a] | ,004 | ,004 | 2,570 |

a. Predictors: (Constant), Evaluating (Meyer)

### ANOVA[a]

| Model | | Sum of Squares | df | Mean Square | F | Sig. |
|---|---|---|---|---|---|---|
| 1 | Regression | 432,681 | 1 | 432,681 | 65,500 | <,001[b] |
| | Residual | 106723,650 | 16156 | 6,606 | | |
| | Total | 107156,332 | 16157 | | | |

a. Dependent Variable: Secure behaviour score

b. Predictors: (Constant), Evaluating (Meyer)

### Coefficients[a]

| Model | | Unstandardized Coefficients | | Standardized Coefficients | t | Sig. |
|---|---|---|---|---|---|---|
| | | B | Std. Error | Beta | | |
| 1 | (Constant) | 3,978 | ,046 | | 86,731 | ,000 |
| | Evaluating (Meyer) | -,094 | ,012 | -,064 | -8,093 | <,001 |

a. Dependent Variable: Secure behaviour score

## 12.1.1.17 Simple regression Meyer dimension Leading

**Model Summary**

| Model | R | R Square | Adjusted R Square | Std. Error of the Estimate |
|---|---|---|---|---|
| 1 | ,271[a] | ,074 | ,074 | 2,479 |

a. Predictors: (Constant), Leading (Meyer)

**ANOVA[a]**

| Model | | Sum of Squares | df | Mean Square | F | Sig. |
|---|---|---|---|---|---|---|
| 1 | Regression | 7892,297 | 1 | 7892,297 | 1284,533 | <,001[b] |
| | Residual | 99264,035 | 16156 | 6,144 | | |
| | Total | 107156,332 | 16157 | | | |

a. Dependent Variable: Secure behaviour score

b. Predictors: (Constant), Leading (Meyer)

**Coefficients[a]**

| Model | | Unstandardized Coefficients B | Unstandardized Coefficients Std. Error | Standardized Coefficients Beta | t | Sig. |
|---|---|---|---|---|---|---|
| 1 | (Constant) | 5,200 | ,048 | | 109,308 | ,000 |
| | Leading (Meyer) | -,282 | ,008 | -,271 | -35,840 | <,001 |

a. Dependent Variable: Secure behaviour score

## 12.1.1.18 Simple regression Meyer dimension Deciding

**Model Summary**

| Model | R | R Square | Adjusted R Square | Std. Error of the Estimate |
|---|---|---|---|---|
| 1 | ,289[a] | ,084 | ,084 | 2,465 |

a. Predictors: (Constant), Deciding (Meyer)

**ANOVA[a]**

| Model | | Sum of Squares | df | Mean Square | F | Sig. |
|---|---|---|---|---|---|---|
| 1 | Regression | 8978,609 | 1 | 8978,609 | 1477,508 | ,000[b] |
| | Residual | 98177,723 | 16156 | 6,077 | | |
| | Total | 107156,332 | 16157 | | | |

a. Dependent Variable: Secure behaviour score

b. Predictors: (Constant), Deciding (Meyer)

**Coefficients[a]**

| Model | | Unstandardized Coefficients | | Standardized Coefficients | t | Sig. |
|---|---|---|---|---|---|---|
| | | B | Std. Error | Beta | | |
| 1 | (Constant) | 5,251 | ,046 | | 113,978 | ,000 |
| | Deciding (Meyer) | -,312 | ,008 | -,289 | -38,438 | ,000 |

a. Dependent Variable: Secure behaviour score

## 12.1.1.19 Simple regression Meyer dimension Trusting

**Model Summary**

| Model | R | R Square | Adjusted R Square | Std. Error of the Estimate |
|---|---|---|---|---|
| 1 | ,289[a] | ,083 | ,083 | 2,466 |

a. Predictors: (Constant), Trusting (Meyer)

**ANOVA[a]**

| Model | | Sum of Squares | df | Mean Square | F | Sig. |
|---|---|---|---|---|---|---|
| 1 | Regression | 8932,590 | 1 | 8932,590 | 1469,247 | <,001[b] |
| | Residual | 98223,741 | 16156 | 6,080 | | |
| | Total | 107156,332 | 16157 | | | |

a. Dependent Variable: Secure behaviour score

b. Predictors: (Constant), Trusting (Meyer)

**Coefficients[a]**

| Model | | Unstandardized Coefficients | | Standardized Coefficients | t | Sig. |
|---|---|---|---|---|---|---|
| | | B | Std. Error | Beta | | |
| 1 | (Constant) | 5,429 | ,050 | | 107,675 | ,000 |
| | Trusting (Meyer) | -,354 | ,009 | -,289 | -38,331 | <,001 |

a. Dependent Variable: Secure behaviour score

## 12.1.1.20 Simple regression Meyer dimension Disagreeing

**Model Summary**

| Model | R | R Square | Adjusted R Square | Std. Error of the Estimate |
|---|---|---|---|---|
| 1 | ,003[a] | ,000 | ,000 | 2,575 |

a. Predictors: (Constant), Disagreeing (Meyer)

**ANOVA[a]**

| Model | | Sum of Squares | df | Mean Square | F | Sig. |
|---|---|---|---|---|---|---|
| 1 | Regression | ,817 | 1 | ,817 | ,123 | ,726[b] |
| | Residual | 107155,515 | 16156 | 6,633 | | |
| | Total | 107156,332 | 16157 | | | |

a. Dependent Variable: Secure behaviour score

b. Predictors: (Constant), Disagreeing (Meyer)

**Coefficients[a]**

| Model | | Unstandardized Coefficients | | Standardized Coefficients | t | Sig. |
|---|---|---|---|---|---|---|
| | | B | Std. Error | Beta | | |
| 1 | (Constant) | 3,661 | ,050 | | 72,547 | ,000 |
| | Disagreeing (Meyer) | -,004 | ,011 | -,003 | -,351 | ,726 |

a. Dependent Variable: Secure behaviour score

## 12.1.1.21 Simple regression Meyer dimension Scheduling

**Model Summary**

| Model | R | R Square | Adjusted R Square | Std. Error of the Estimate |
|---|---|---|---|---|
| 1 | ,225[a] | ,050 | ,050 | 2,510 |

a. Predictors: (Constant), Scheduling (Meyer)

**ANOVA**[a]

| Model | | Sum of Squares | df | Mean Square | F | Sig. |
|---|---|---|---|---|---|---|
| 1 | Regression | 5410,786 | 1 | 5410,786 | 859,169 | <,001[b] |
| | Residual | 101745,546 | 16156 | 6,298 | | |
| | Total | 107156,332 | 16157 | | | |

a. Dependent Variable: Secure behaviour score

b. Predictors: (Constant), Scheduling (Meyer)

**Coefficients**[a]

| Model | | Unstandardized Coefficients | | Standardized Coefficients | t | Sig. |
|---|---|---|---|---|---|---|
| | | B | Std. Error | Beta | | |
| 1 | (Constant) | 4,952 | ,049 | | 101,521 | ,000 |
| | Scheduling (Meyer) | -,296 | ,010 | -,225 | -29,312 | <,001 |

a. Dependent Variable: Secure behaviour score

Click to navigate

## 12.1.1.22 Simple regression Meyer dimension Persuading

**Model Summary**

| Model | R | R Square | Adjusted R Square | Std. Error of the Estimate |
|---|---|---|---|---|
| 1 | ,178[a] | ,032 | ,032 | 2,534 |

a. Predictors: (Constant), Persuading (Meyer)

**ANOVA[a]**

| Model | | Sum of Squares | df | Mean Square | F | Sig. |
|---|---|---|---|---|---|---|
| 1 | Regression | 3400,569 | 1 | 3400,569 | 529,509 | <,001[b] |
| | Residual | 103755,763 | 16156 | 6,422 | | |
| | Total | 107156,332 | 16157 | | | |

a. Dependent Variable: Secure behaviour score

b. Predictors: (Constant), Persuading (Meyer)

**Coefficients[a]**

| Model | | Unstandardized Coefficients | | Standardized Coefficients | t | Sig. |
|---|---|---|---|---|---|---|
| | | B | Std. Error | Beta | | |
| 1 | (Constant) | 2,899 | ,038 | | 76,145 | ,000 |
| | Persuading (Meyer) | ,196 | ,009 | ,178 | 23,011 | <,001 |

a. Dependent Variable: Secure behaviour score

Click to navigate

## 12.1.2 Multiple variable testing: Multiple regression and assumption testing

### 12.1.2.1 Assumption testing Hofstede cultural dimensions

**Model Summary**

| Model | R | R Square | Adjusted R Square | Std. Error of the Estimate |
|---|---|---|---|---|
| 1 | ,303ª | ,092 | ,091 | 2,421 |

a. Predictors: (Constant), Indulgence (Hofstede), Masculinity (Hofstede), Long-term orientation (Hofstede), Uncertainty avoidance (Hofstede), Power distance (Hofstede), Individualism (Hofstede)

**ANOVAª**

| Model | | Sum of Squares | df | Mean Square | F | Sig. |
|---|---|---|---|---|---|---|
| 1 | Regression | 12619,181 | 6 | 2103,197 | 358,850 | ,000ᵇ |
| | Residual | 124919,935 | 21314 | 5,861 | | |
| | Total | 137539,116 | 21320 | | | |

a. Dependent Variable: Secure behaviour score

b. Predictors: (Constant), Indulgence (Hofstede), Masculinity (Hofstede), Long-term orientation (Hofstede), Uncertainty avoidance (Hofstede), Power distance (Hofstede), Individualism (Hofstede)

**Coefficientsª**

| Model | | Unstandardized Coefficients | | Standardized Coefficients | t | Sig. | Collinearity Statistics | |
|---|---|---|---|---|---|---|---|---|
| | | B | Std. Error | Beta | | | Tolerance | VIF |
| 1 | (Constant) | 2,049 | ,153 | | 13,361 | ,000 | | |

| | B | Std. Error | Beta | t | Sig. | Tolerance | VIF |
|---|---|---|---|---|---|---|---|
| Power distance (Hofstede) | -,010 | ,001 | -,080 | -8,650 | ,000 | ,494 | 2,025 |
| Individualism (Hofstede) | -,002 | ,001 | -,017 | -1,782 | ,075 | ,456 | 2,194 |
| Uncertainty avoidance (Hofstede) | -,001 | ,001 | -,011 | -1,269 | ,205 | ,520 | 1,923 |
| Masculinity (Hofstede) | -,010 | ,001 | -,101 | -14,330 | ,000 | ,864 | 1,157 |
| Long-term orientation (Hofstede) | ,022 | ,001 | ,142 | 18,392 | ,000 | ,710 | 1,408 |
| Indulgence (Hofstede) | ,032 | ,001 | ,243 | 27,367 | ,000 | ,540 | 1,852 |

a. Dependent Variable: Secure behaviour score

## Collinearity Diagnostics[a]

| Model | Dimension | Eigenvalue | Condition Index | Variance Proportions | | | | | | |
|---|---|---|---|---|---|---|---|---|---|---|
| | | | | (Constant) | Power distance (Hofstede) | Individualism (Hofstede) | Uncertainty avoidance (Hofstede) | Masculinity (Hofstede) | Long-term orientation (Hofstede) | Indulgence (Hofstede) |
| 1 | 1 | 6,296 | 1,000 | ,00 | ,00 | ,00 | ,00 | ,00 | ,00 | ,00 |
| | 2 | ,329 | 4,373 | ,00 | ,05 | ,02 | ,02 | ,03 | ,00 | ,11 |
| | 3 | ,183 | 5,859 | ,00 | ,03 | ,00 | ,02 | ,87 | ,01 | ,00 |
| | 4 | ,106 | 7,710 | ,00 | ,04 | ,05 | ,03 | ,03 | ,23 | ,22 |
| | 5 | ,048 | 11,448 | ,00 | ,69 | ,01 | ,56 | ,00 | ,00 | ,01 |
| | 6 | ,028 | 15,022 | ,01 | ,02 | ,59 | ,04 | ,06 | ,69 | ,41 |
| | 7 | ,009 | 25,941 | ,99 | ,17 | ,34 | ,33 | ,02 | ,07 | ,24 |

a. Dependent Variable: Secure behaviour score

## Partial regression plots

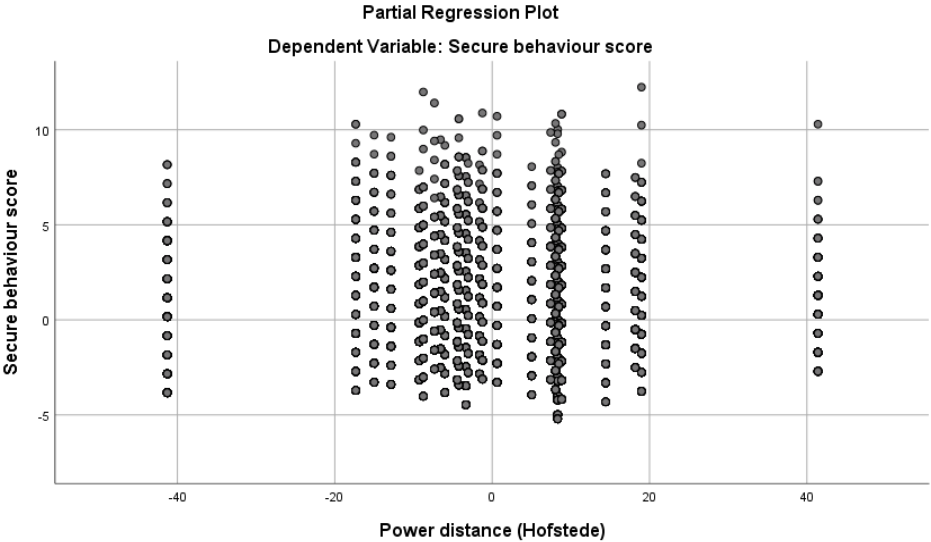

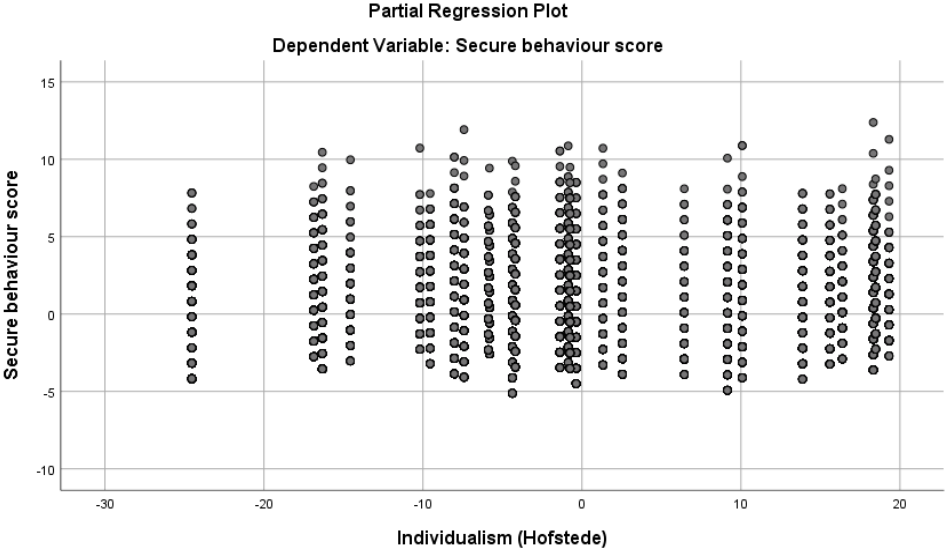

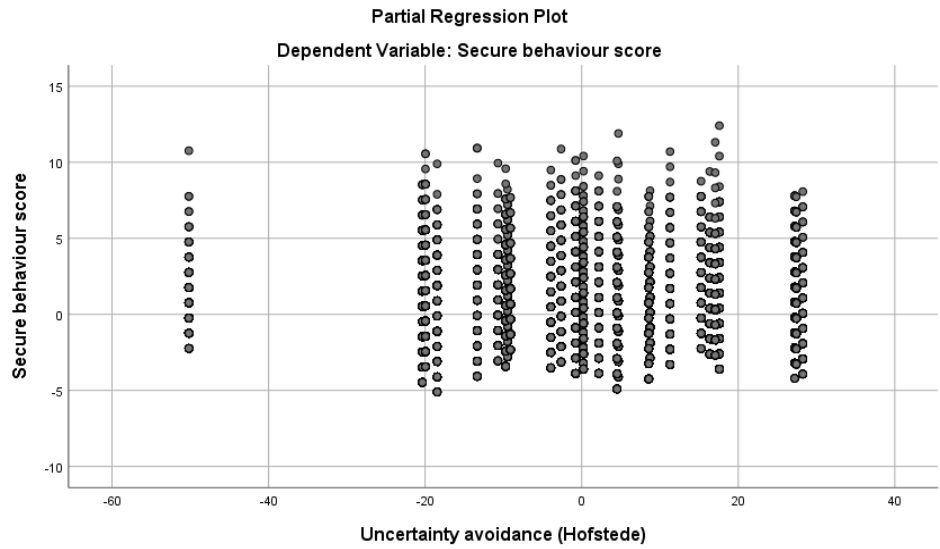

Partial Regression Plot

Dependent Variable: Secure behaviour score

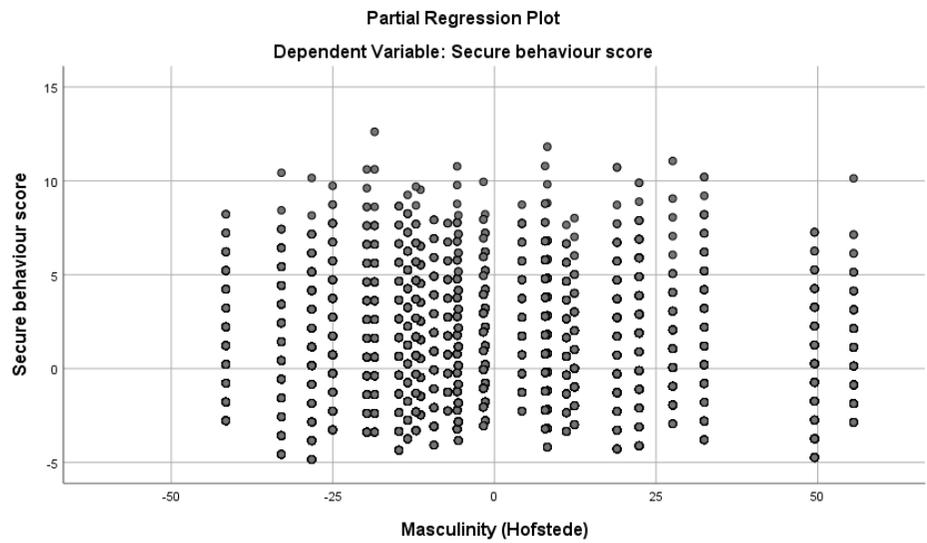

Partial Regression Plot

Dependent Variable: Secure behaviour score

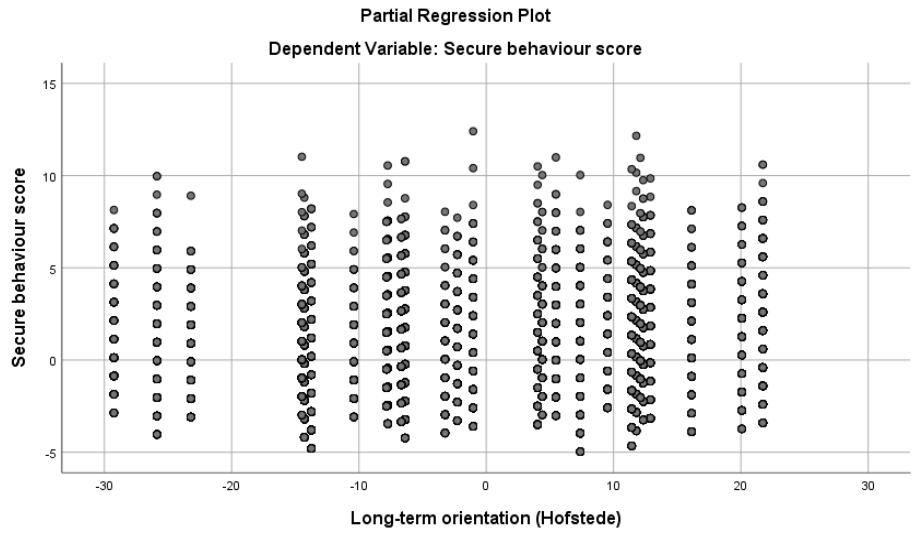

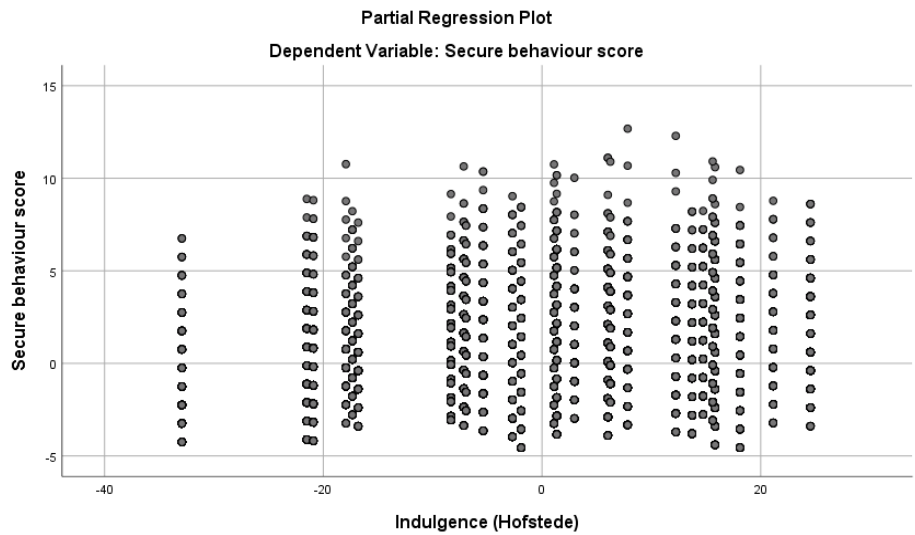

## Regression plot

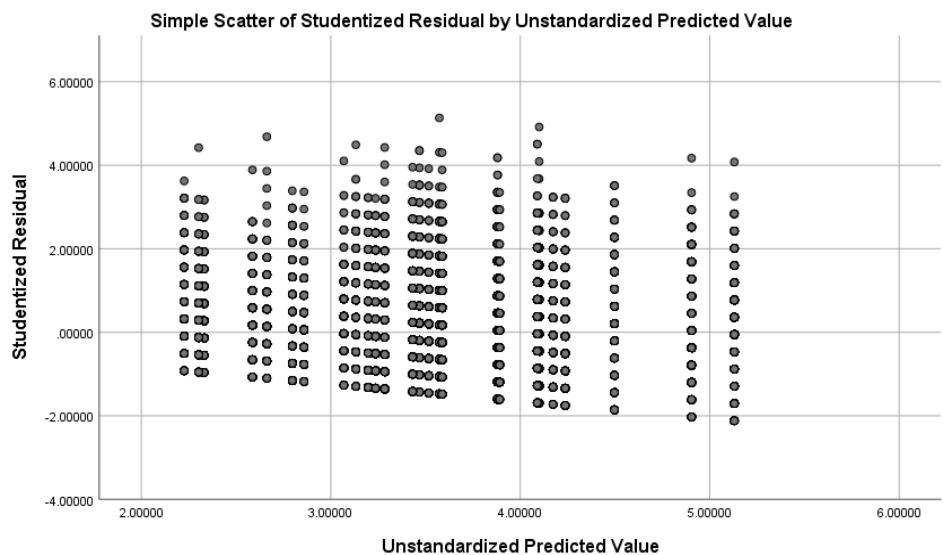

## Distribution histogram

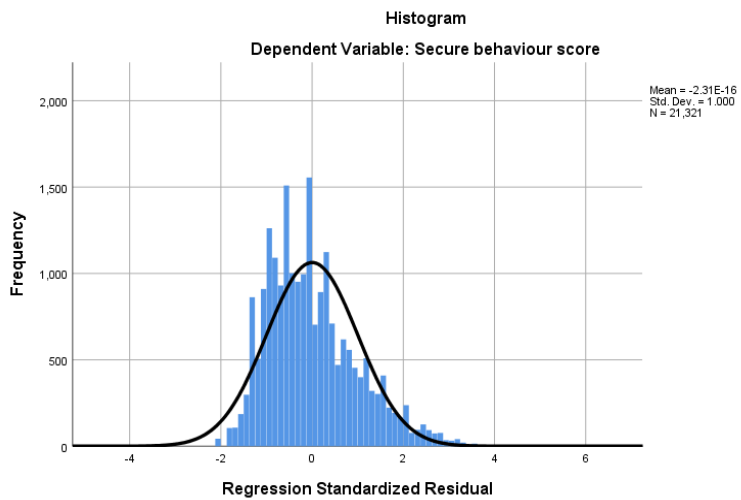

## Normal P-P Plot

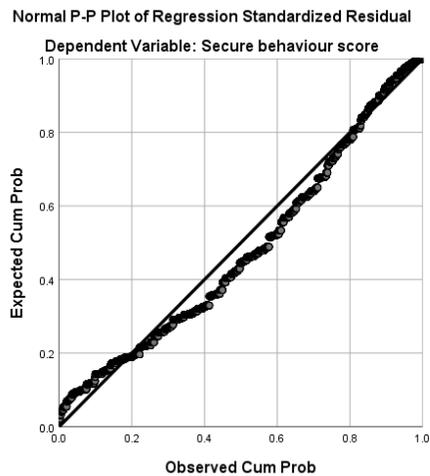

Normal P-P Plot of Regression Standardized Residual
Dependent Variable: Secure behaviour score

## 12.1.2.2  Assumption testing Meyer cultural dimensions

INITIAL MODEL

### Model Summary

| Model | R | R Square | Adjusted R Square | Std. Error of the Estimate |
|---|---|---|---|---|
| 1 | ,317ᵃ | ,100 | ,100 | 2,443 |

a. Predictors: (Constant), Persuading (Meyer), Evaluating (Meyer), Scheduling (Meyer), Deciding (Meyer), Disagreeing (Meyer), Trusting (Meyer), Leading (Meyer), Communicating (Meyer)

### ANOVAᵃ

Click to navigate

Introduction  Literature review  Research questions  Antecedents of behaviour  Research model  Methodology  Results  Discussion  Conclusion  Limitations  References  Appendix

| Model | | Sum of Squares | df | Mean Square | F | Sig. |
|---|---|---|---|---|---|---|
| 1 | Regression | 10768,955 | 8 | 1346,119 | 225,532 | ,000[b] |
| | Residual | 96387,376 | 16149 | 5,969 | | |
| | Total | 107156,332 | 16157 | | | |

a. Dependent Variable: Secure behaviour score

b. Predictors: (Constant), Persuading (Meyer), Evaluating (Meyer), Scheduling (Meyer), Deciding (Meyer), Disagreeing (Meyer), Trusting (Meyer), Leading (Meyer), Communicating (Meyer)

**Coefficients[a]**

| Model | | Unstandardized Coefficients | | Standardized Coefficients | | | Collinearity Statistics | |
|---|---|---|---|---|---|---|---|---|
| | | B | Std. Error | Beta | t | Sig. | Tolerance | VIF |
| 1 | (Constant) | 6,837 | ,207 | | 33,067 | ,000 | | |
| | Communicating (Meyer) | ,065 | ,045 | ,041 | 1,436 | ,151 | ,070 | 14,309 |
| | Evaluating (Meyer) | ,193 | ,024 | ,130 | 7,974 | ,000 | ,208 | 4,798 |
| | Leading (Meyer) | -,197 | ,025 | -,189 | -7,923 | ,000 | ,098 | 10,239 |
| | Deciding (Meyer) | -,068 | ,026 | -,063 | -2,616 | ,009 | ,096 | 10,444 |
| | Trusting (Meyer) | -,160 | ,033 | -,131 | -4,924 | ,000 | ,079 | 12,680 |
| | Disagreeing (Meyer) | -,229 | ,026 | -,158 | -8,871 | ,000 | ,177 | 5,664 |
| | Scheduling (Meyer) | -,156 | ,023 | -,118 | -6,905 | ,000 | ,190 | 5,255 |
| | Persuading (Meyer) | -,091 | ,021 | -,083 | -4,433 | ,000 | ,160 | 6,248 |

a. Dependent Variable: Secure behaviour score

**Collinearity Diagnostics[a]**

Variance Proportions

| Model | Dimension | Eigenvalue | Condition Index | (Constant) | Communicating (Meyer) | Evaluating (Meyer) | Leading (Meyer) | Deciding (Meyer) | Trusting (Meyer) | Disagreeing (Meyer) | Scheduling (Meyer) | Persuading (Meyer) |
|---|---|---|---|---|---|---|---|---|---|---|---|---|
| 1 | 1 | 7,947 | 1,000 | ,00 | ,00 | ,00 | ,00 | ,00 | ,00 | ,00 | ,00 | ,00 |
| | 2 | ,693 | 3,385 | ,00 | ,00 | ,00 | ,00 | ,00 | ,00 | ,01 | ,00 | ,03 |
| | 3 | ,173 | 6,787 | ,01 | ,00 | ,10 | ,00 | ,00 | ,00 | ,03 | ,00 | ,07 |
| | 4 | ,094 | 9,209 | ,00 | ,00 | ,00 | ,06 | ,01 | ,00 | ,01 | ,18 | ,00 |
| | 5 | ,037 | 14,596 | ,07 | ,01 | ,07 | ,00 | ,16 | ,00 | ,05 | ,00 | ,12 |
| | 6 | ,027 | 17,311 | ,02 | ,04 | ,30 | ,15 | ,06 | ,01 | ,21 | ,03 | ,00 |
| | 7 | ,016 | 22,346 | ,02 | ,00 | ,00 | ,00 | ,16 | ,46 | ,08 | ,18 | ,11 |
| | 8 | ,011 | 26,839 | ,06 | ,15 | ,52 | ,17 | ,04 | ,02 | ,58 | ,38 | ,01 |
| | 9 | ,003 | 53,088 | ,82 | ,79 | ,00 | ,62 | ,56 | ,50 | ,02 | ,22 | ,66 |

a. Dependent Variable: Secure behaviour score

## Partial regression plots

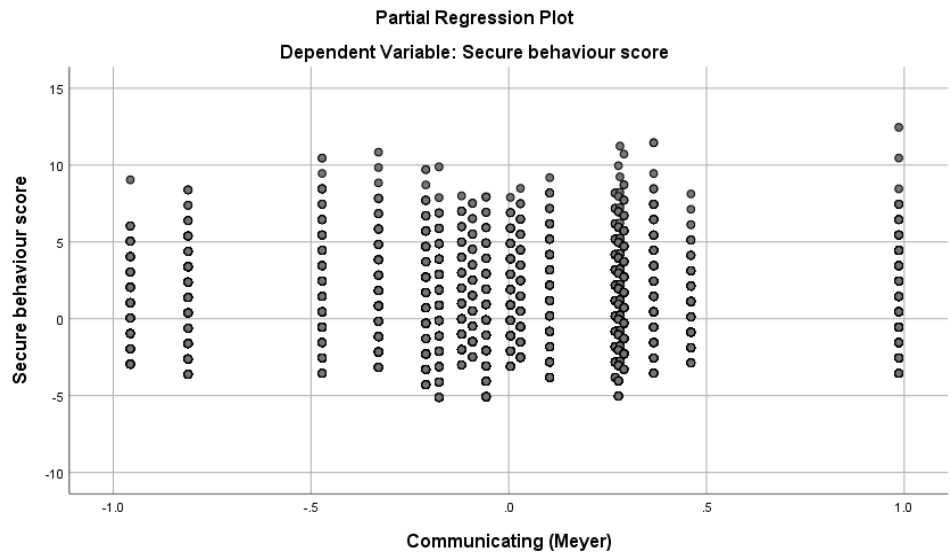

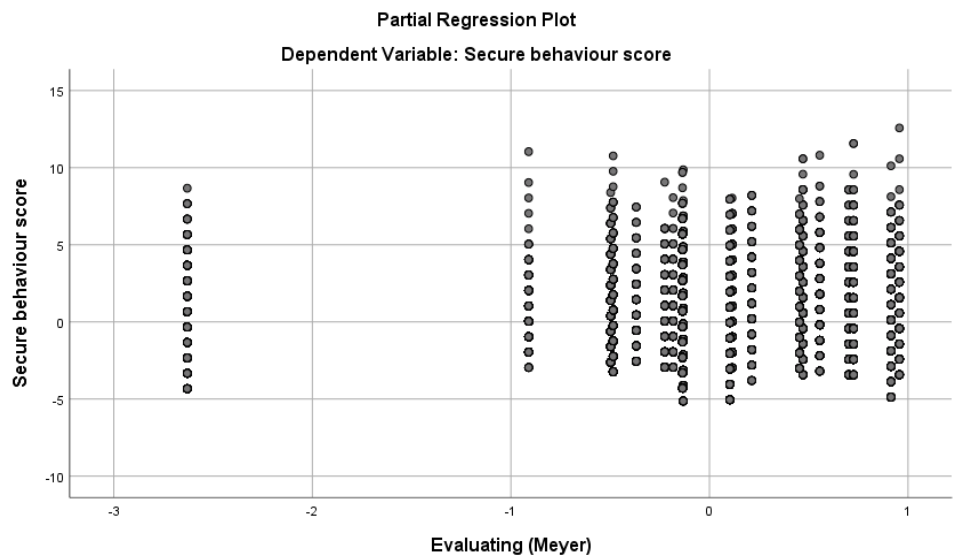

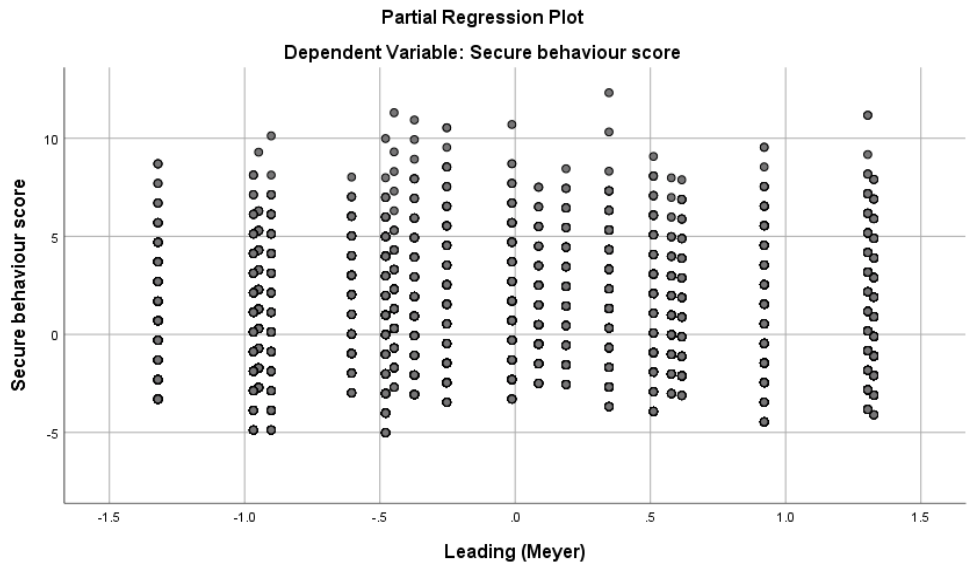

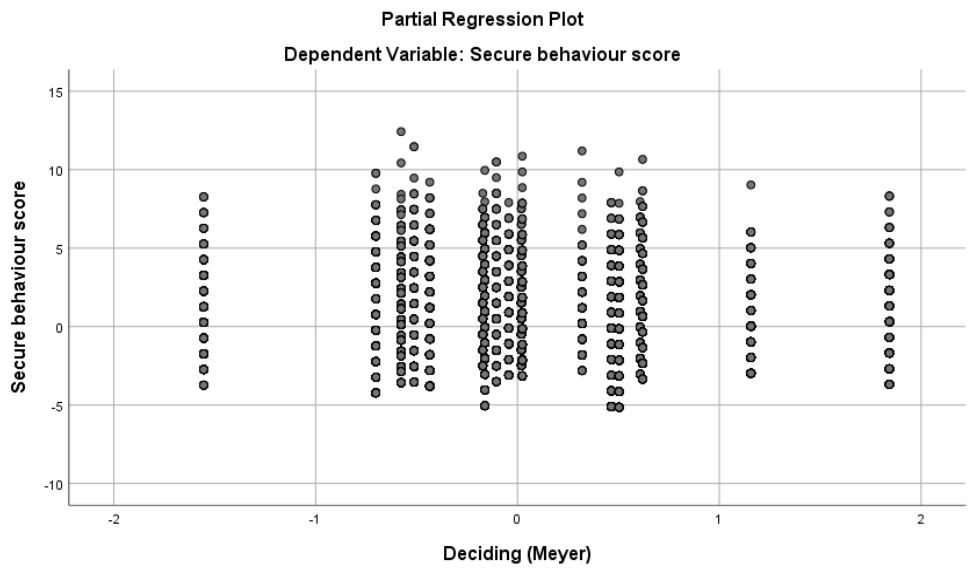

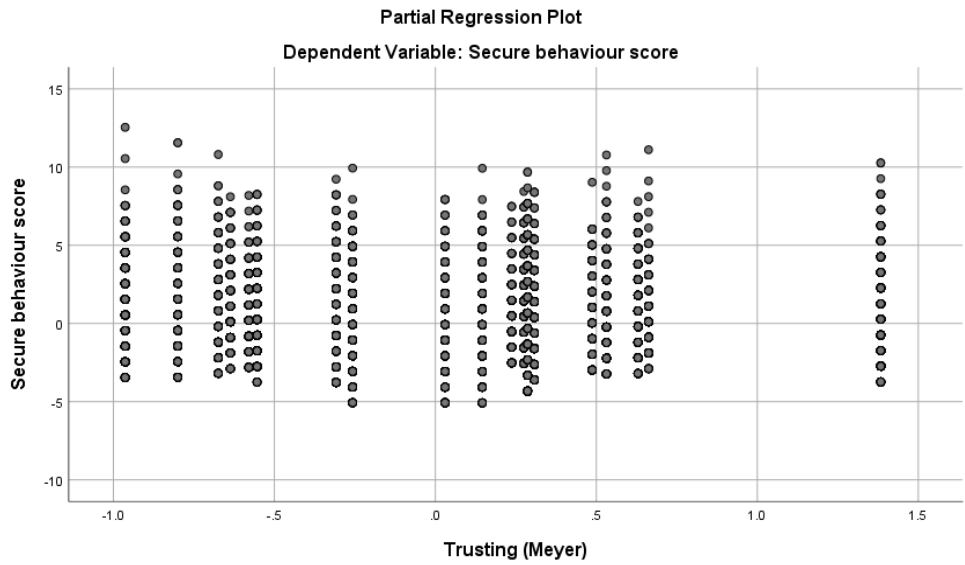

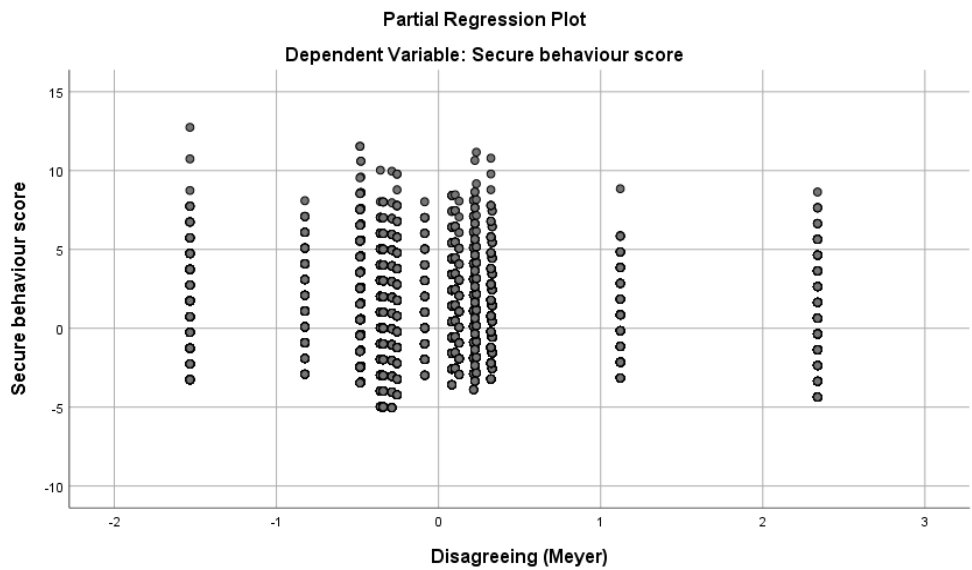

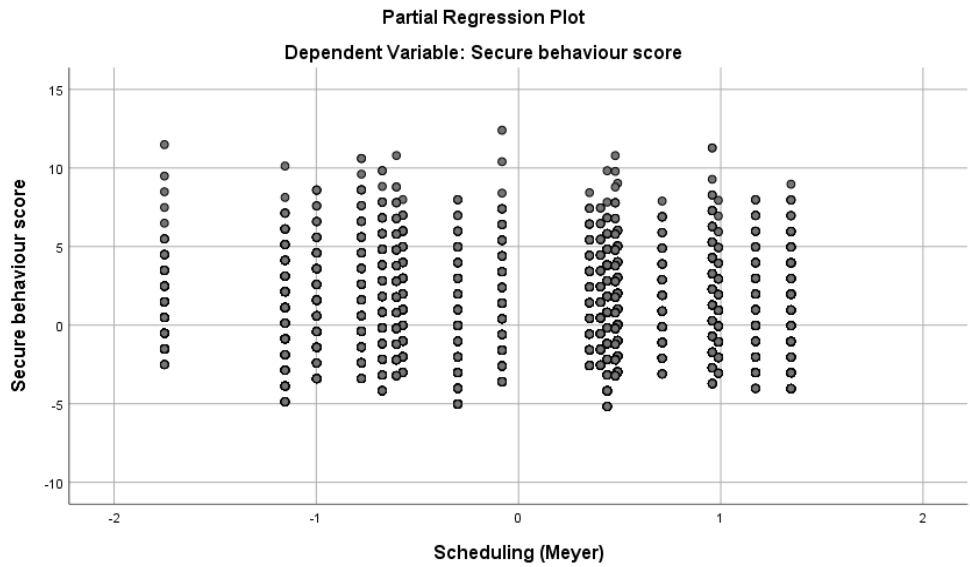

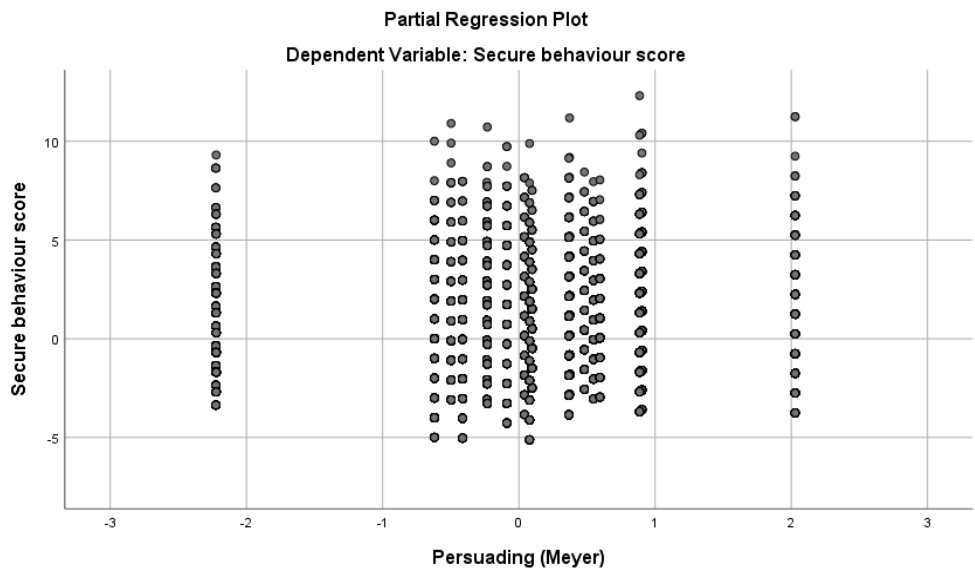

## Regression plot

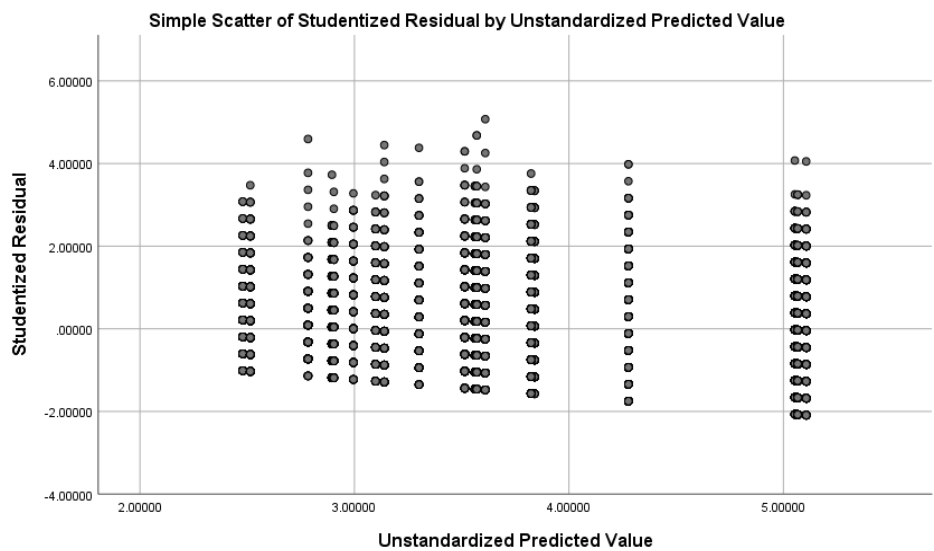

Simple Scatter of Studentized Residual by Unstandardized Predicted Value

## Distribution histogram

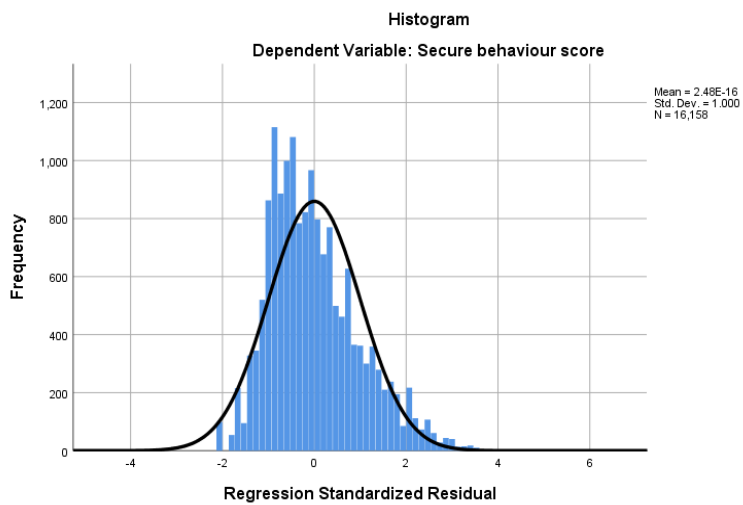

Histogram
Dependent Variable: Secure behaviour score

Mean = 2.48E-16
Std. Dev. = 1.000
N = 16,158

Click to navigate

Introduction   Literature review   Research questions   Antecedents of behaviour   Research model   Methodology   Results   Discussion   Conclusion   Limitations   References   Appendix

## Normal P-P Plot

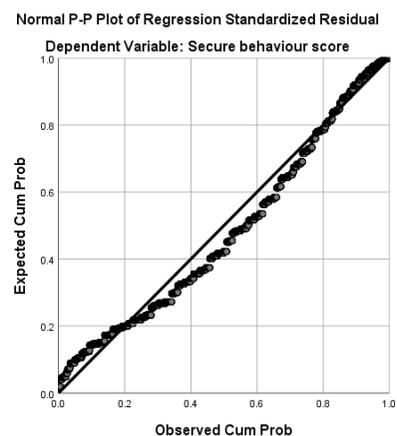

Normal P-P Plot of Regression Standardized Residual
Dependent Variable: Secure behaviour score

## IMPROVED MODEL

### Model Summary

| Model | R | R Square | Adjusted R Square | Std. Error of the Estimate |
|---|---|---|---|---|
| 1 | ,311[a] | ,097 | ,096 | 2,448 |

a. Predictors: (Constant), Scheduling (Meyer), Evaluating (Meyer), Leading (Meyer), Communicating (Meyer), Disagreeing (Meyer), Deciding (Meyer)

### ANOVA[a]

| Model | | Sum of Squares | df | Mean Square | F | Sig. |
|---|---|---|---|---|---|---|
| 1 | Regression | 10371,956 | 6 | 1728,659 | 288,472 | ,000[b] |
| | Residual | 96784,376 | 16151 | 5,992 | | |
| | Total | 107156,332 | 16157 | | | |

a. Dependent Variable: Secure behaviour score

Click to navigate

Introduction | Literature review | Research questions | Antecedents of behaviour | Research model | Methodology | Results | Discussion | Conclusion | Limitations | References | Appendix

b. Predictors: (Constant), Scheduling (Meyer), Evaluating (Meyer), Leading (Meyer), Communicating (Meyer), Disagreeing (Meyer), Deciding (Meyer)

### Coefficientsᵃ

| Model | | Unstandardized Coefficients | | Standardized Coefficients | | | Collinearity Statistics | |
|---|---|---|---|---|---|---|---|---|
| | | B | Std. Error | Beta | t | Sig. | Tolerance | VIF |
| 1 | (Constant) | 6,318 | ,091 | | 69,809 | ,000 | | |
| | Communicating (Meyer) | ,057 | ,028 | ,036 | 2,063 | ,039 | ,186 | 5,374 |
| | Evaluating (Meyer) | ,166 | ,024 | ,112 | 6,899 | ,000 | ,213 | 4,701 |
| | Leading (Meyer) | -,173 | ,017 | -,167 | -10,506 | ,000 | ,222 | 4,510 |
| | Deciding (Meyer) | -,129 | ,020 | -,119 | -6,307 | ,000 | ,156 | 6,410 |
| | Disagreeing (Meyer) | -,265 | ,025 | -,182 | -10,622 | ,000 | ,190 | 5,273 |
| | Scheduling (Meyer) | -,194 | ,019 | -,147 | -10,320 | ,000 | ,274 | 3,652 |

a. Dependent Variable: Secure behaviour score

### Collinearity Diagnosticsᵃ

| Model | Dimension | Eigenvalue | Condition Index | Variance Proportions | | | | | | |
|---|---|---|---|---|---|---|---|---|---|---|
| | | | | (Constant) | Communicating (Meyer) | Evaluating (Meyer) | Leading (Meyer) | Deciding (Meyer) | Disagreeing (Meyer) | Scheduling (Meyer) |
| 1 | 1 | 6,387 | 1,000 | ,00 | ,00 | ,00 | ,00 | ,00 | ,00 | ,00 |
| | 2 | ,394 | 4,028 | ,00 | ,00 | ,03 | ,01 | ,01 | ,03 | ,01 |
| | 3 | ,092 | 8,318 | ,01 | ,00 | ,01 | ,11 | ,01 | ,01 | ,29 |
| | 4 | ,075 | 9,219 | ,37 | ,00 | ,14 | ,02 | ,03 | ,00 | ,00 |
| | 5 | ,026 | 15,597 | ,11 | ,07 | ,24 | ,36 | ,24 | ,21 | ,01 |
| | 6 | ,015 | 20,851 | ,32 | ,48 | ,07 | ,17 | ,69 | ,05 | ,00 |
| | 7 | ,011 | 23,898 | ,19 | ,44 | ,50 | ,33 | ,01 | ,70 | ,68 |

a. Dependent Variable: Secure behaviour score

Click to navigate

Introduction  Literature review  Research questions  Antecedents of behaviour  Research model  Methodology  Results  Discussion  Conclusion  Limitations  References  Appendix

### 12.1.2.3  Assumption testing Prior experience with security incidents

**Model Summary**

| Model | R | R Square | Adjusted R Square | Std. Error of the Estimate |
|---|---|---|---|---|
| 1 | ,262[a] | ,068 | ,068 | 2,454 |

a. Predictors: (Constant), Indirect experience score, Direct experience score

**ANOVA[a]**

| Model | | Sum of Squares | df | Mean Square | F | Sig. |
|---|---|---|---|---|---|---|
| 1 | Regression | 9582,731 | 2 | 4791,366 | 795,832 | ,000[b] |
| | Residual | 130399,651 | 21659 | 6,021 | | |
| | Total | 139982,382 | 21661 | | | |

a. Dependent Variable: Secure behaviour score

b. Predictors: (Constant), Indirect experience score, Direct experience score

**Coefficients[a]**

| Model | | Unstandardized Coefficients | | Standardized Coefficients | t | Sig. | Collinearity Statistics | |
|---|---|---|---|---|---|---|---|---|
| | | B | Std. Error | Beta | | | Tolerance | VIF |
| 1 | (Constant) | 2,620 | ,029 | | 90,424 | ,000 | | |
| | Direct experience score | ,071 | ,006 | ,085 | 12,063 | ,000 | ,869 | 1,151 |
| | Indirect experience score | ,522 | ,017 | ,219 | 31,078 | ,000 | ,869 | 1,151 |

a. Dependent Variable: Secure behaviour score

**Collinearity Diagnostics[a]**

Click to navigate

| Model | Dimension | Eigenvalue | Condition Index | (Constant) | Variance Proportions Direct experience score | Indirect experience score |
|---|---|---|---|---|---|---|
| 1 | 1 | 2,303 | 1,000 | ,05 | ,07 | ,05 |
| | 2 | ,522 | 2,101 | ,14 | ,86 | ,04 |
| | 3 | ,175 | 3,627 | ,81 | ,06 | ,91 |

a. Dependent Variable: Secure behaviour score

## Partial regression plots

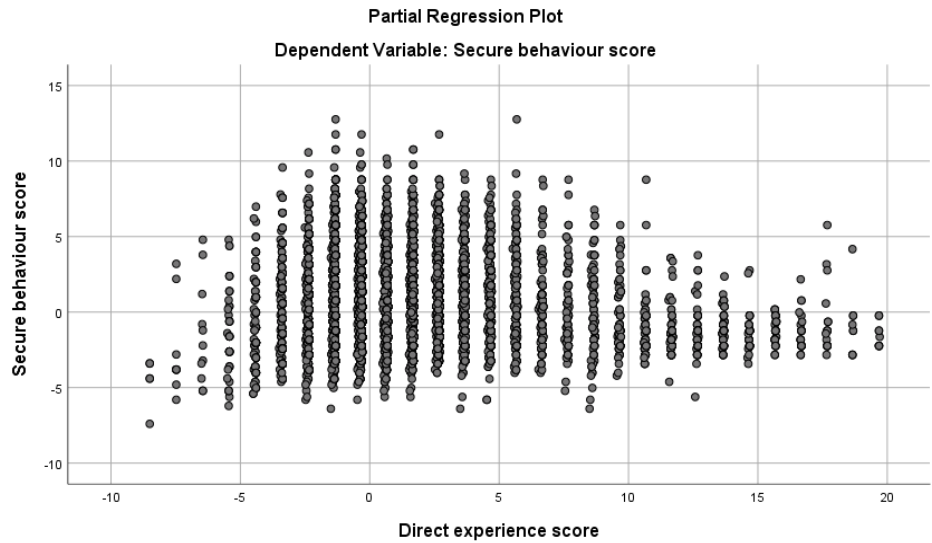

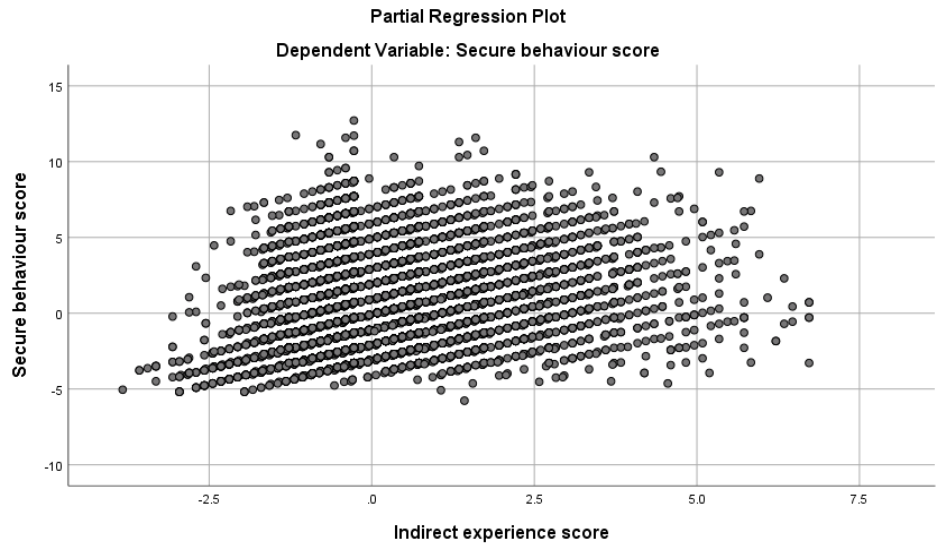

## Regression plot

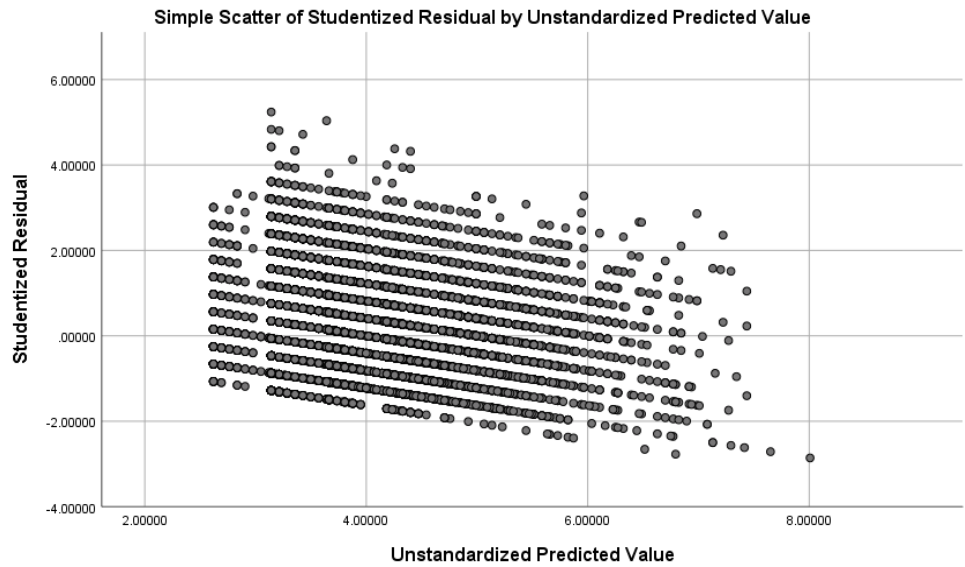

## Distribution histogram

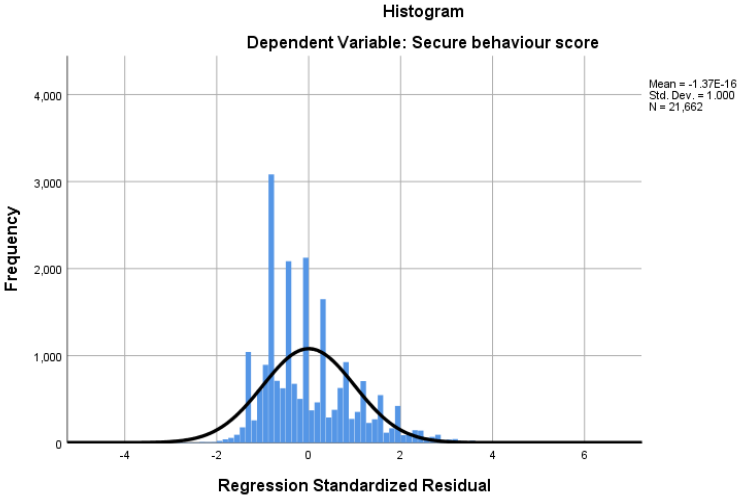

## Normal P-P Plot

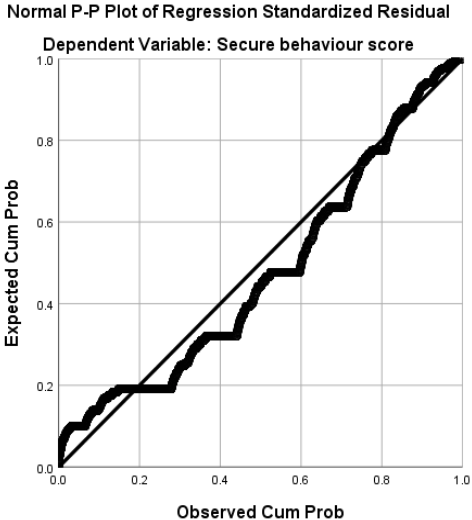

## 12.1.2.4 Assumption testing All metric variables

**Model Summary**

| Model | R | R Square | Adjusted R Square | Std. Error of the Estimate |
|---|---|---|---|---|
| 1 | ,400ᵃ | ,160 | ,159 | 2,361 |

a. Predictors: (Constant), Indirect experience score, Evaluating (Meyer), Age, Power distance (Hofstede), Masculinity (Hofstede), Long-term orientation (Hofstede), Direct experience score, Individualism (Hofstede), Scheduling (Meyer), Communicating (Meyer), Indulgence (Hofstede), Persuading (Meyer), Disagreeing (Meyer), Deciding (Meyer), Uncertainty avoidance (Hofstede), Trusting (Meyer), Leading (Meyer)

**ANOVAᵃ**

| Model | | Sum of Squares | df | Mean Square | F | Sig. |
|---|---|---|---|---|---|---|
| 1 | Regression | 17166,508 | 17 | 1009,795 | 181,110 | ,000ᵇ |
| | Residual | 89989,824 | 16140 | 5,576 | | |
| | Total | 107156,332 | 16157 | | | |

a. Dependent Variable: Secure behaviour score

b. Predictors: (Constant), Indirect experience score, Evaluating (Meyer), Age, Power distance (Hofstede), Masculinity (Hofstede), Long-term orientation (Hofstede), Direct experience score, Individualism (Hofstede), Scheduling (Meyer), Communicating (Meyer), Indulgence (Hofstede), Persuading (Meyer), Disagreeing (Meyer), Deciding (Meyer), Uncertainty avoidance (Hofstede), Trusting (Meyer), Leading (Meyer)

**Coefficients<sup>a</sup>**

| Model | | Unstandardized Coefficients | | Standardized Coefficients | | | Collinearity Statistics | |
|---|---|---|---|---|---|---|---|---|
| | | B | Std. Error | Beta | t | Sig. | Tolerance | VIF |
| 1 | (Constant) | ,510 | ,447 | | 1,139 | ,255 | | |
| | Power distance (Hofstede) | -,039 | ,005 | -,297 | -8,027 | ,000 | ,038 | 26,343 |
| | Individualism (Hofstede) | -,008 | ,004 | -,051 | -2,024 | ,043 | ,081 | 12,376 |
| | Uncertainty avoidance (Hofstede) | ,034 | ,003 | ,337 | 10,067 | ,000 | ,047 | 21,487 |
| | Masculinity (Hofstede) | -,033 | ,005 | -,290 | -6,596 | ,000 | ,027 | 37,228 |
| | Long-term orientation (Hofstede) | ,011 | ,002 | ,074 | 5,907 | ,000 | ,331 | 3,023 |
| | Indulgence (Hofstede) | ,016 | ,004 | ,114 | 4,556 | ,000 | ,084 | 11,954 |
| | Communicating (Meyer) | ,478 | ,091 | ,299 | 5,278 | ,000 | ,016 | 61,801 |
| | Evaluating (Meyer) | ,224 | ,040 | ,151 | 5,610 | ,000 | ,072 | 13,959 |
| | Leading (Meyer) | ,311 | ,060 | ,300 | 5,163 | ,000 | ,015 | 64,739 |
| | Deciding (Meyer) | -,254 | ,047 | -,236 | -5,459 | ,000 | ,028 | 35,973 |
| | Trusting (Meyer) | -,159 | ,063 | -,130 | -2,513 | ,012 | ,019 | 51,385 |
| | Disagreeing (Meyer) | -,349 | ,042 | -,240 | -8,238 | ,000 | ,061 | 16,373 |
| | Scheduling (Meyer) | -,130 | ,028 | -,099 | -4,633 | ,000 | ,114 | 8,773 |
| | Persuading (Meyer) | ,444 | ,047 | ,404 | 9,478 | ,000 | ,029 | 34,869 |
| | Age | -,001 | ,001 | -,007 | -,897 | ,370 | ,898 | 1,114 |
| | Direct experience score | ,063 | ,007 | ,075 | 9,547 | ,000 | ,842 | 1,188 |
| | Indirect experience score | ,415 | ,019 | ,176 | 22,406 | ,000 | ,841 | 1,189 |

a. Dependent Variable: Secure behaviour score

**Collinearity Diagnostics<sup>a</sup>**

Model: 1

| | | Dimension | | | | | | | | | | | | | | | | | |
|---|---|---|---|---|---|---|---|---|---|---|---|---|---|---|---|---|---|---|---|
| | | 1 | 2 | 3 | 4 | 5 | 6 | 7 | 8 | 9 | 10 | 11 | 12 | 13 | 14 | 15 | 16 | 17 | 18 |
| Eigenvalue | | 14,979 | 1,118 | ,692 | ,298 | ,253 | ,225 | ,124 | ,093 | ,084 | ,035 | ,032 | ,023 | ,019 | ,014 | ,006 | ,003 | ,001 | ,001 |
| Condition Index | | 1,000 | 3,661 | 4,651 | 7,091 | 7,687 | 8,166 | 11,001 | 12,722 | 13,360 | 20,551 | 21,589 | 25,267 | 27,946 | 32,152 | 50,065 | 75,233 | 133,658 | 145,532 |
| Variance Proportions | (Constant) | ,00 | ,00 | ,00 | ,00 | ,00 | ,00 | ,00 | ,00 | ,00 | ,00 | ,00 | ,00 | ,00 | ,00 | ,05 | ,18 | ,60 | ,16 |
| | Power distance (Hofstede) | ,00 | ,00 | ,00 | ,00 | ,00 | ,00 | ,00 | ,01 | ,00 | ,01 | ,01 | ,02 | ,00 | ,04 | ,00 | ,19 | ,24 | ,47 |
| | Individualism (Hofstede) | ,00 | ,00 | ,00 | ,00 | ,00 | ,00 | ,00 | ,00 | ,00 | ,00 | ,02 | ,06 | ,05 | ,00 | ,02 | ,14 | ,66 | ,04 |
| | Uncertainty avoidance (Hofstede) | ,00 | ,00 | ,00 | ,00 | ,00 | ,00 | ,00 | ,00 | ,00 | ,02 | ,05 | ,00 | ,01 | ,01 | ,14 | ,22 | ,17 | ,37 |
| | Masculinity (Hofstede) | ,00 | ,00 | ,00 | ,00 | ,00 | ,01 | ,00 | ,00 | ,00 | ,01 | ,01 | ,01 | ,00 | ,00 | ,03 | ,06 | ,80 | ,06 |
| | Long-term orientation (Hofstede) | ,00 | ,00 | ,00 | ,01 | ,00 | ,00 | ,06 | ,02 | ,00 | ,01 | ,07 | ,20 | ,10 | ,08 | ,23 | ,07 | ,10 | ,04 |
| | Indulgence (Hofstede) | ,00 | ,00 | ,00 | ,00 | ,00 | ,00 | ,00 | ,00 | ,04 | ,01 | ,01 | ,01 | ,00 | ,09 | ,00 | ,63 | ,05 | ,14 |
| | Communicating (Meyer) | ,00 | ,00 | ,00 | ,00 | ,00 | ,00 | ,00 | ,00 | ,00 | ,00 | ,01 | ,00 | ,00 | ,01 | ,00 | ,01 | ,21 | ,76 |

| | | | | | | | | | | | | | | | | | | |
|---|---|---|---|---|---|---|---|---|---|---|---|---|---|---|---|---|---|---|
| Evaluating (Meyer) | ,00 | ,00 | ,00 | ,01 | ,00 | ,00 | ,01 | ,00 | ,00 | ,07 | ,00 | ,03 | ,03 | ,04 | ,02 | ,49 | ,01 | ,28 |
| Leading (Meyer) | ,00 | ,00 | ,00 | ,00 | ,00 | ,00 | ,00 | ,00 | ,00 | ,00 | ,00 | ,00 | ,00 | ,00 | ,19 | ,01 | ,36 | ,43 |
| Deciding (Meyer) | ,00 | ,00 | ,00 | ,00 | ,00 | ,00 | ,00 | ,00 | ,00 | ,01 | ,01 | ,05 | ,05 | ,00 | ,03 | ,00 | ,06 | ,79 |
| Trusting (Meyer) | ,00 | ,00 | ,00 | ,00 | ,00 | ,00 | ,00 | ,00 | ,00 | ,00 | ,00 | ,00 | ,06 | ,01 | ,01 | ,04 | ,54 | ,34 |
| Disagreeing (Meyer) | ,00 | ,00 | ,00 | ,01 | ,00 | ,00 | ,01 | ,00 | ,00 | ,06 | ,00 | ,02 | ,00 | ,09 | ,02 | ,30 | ,02 | ,47 |
| Scheduling (Meyer) | ,00 | ,00 | ,00 | ,00 | ,00 | ,01 | ,02 | ,00 | ,02 | ,05 | ,00 | ,00 | ,00 | ,50 | ,07 | ,16 | ,06 | ,12 |
| Persuading (Meyer) | ,00 | ,00 | ,00 | ,00 | ,00 | ,00 | ,01 | ,01 | ,01 | ,00 | ,00 | ,01 | ,02 | ,00 | ,03 | ,00 | ,06 | ,84 |
| Age | ,00 | ,00 | ,00 | ,02 | ,01 | ,04 | ,01 | ,57 | ,22 | ,10 | ,01 | ,00 | ,00 | ,00 | ,01 | ,01 | ,00 | ,00 |
| Direct experience score | ,00 | ,09 | ,58 | ,12 | ,17 | ,02 | ,00 | ,00 | ,00 | ,00 | ,00 | ,00 | ,00 | ,00 | ,00 | ,00 | ,00 | ,00 |
| Indirect experience score | ,00 | ,01 | ,02 | ,10 | ,76 | ,05 | ,00 | ,01 | ,04 | ,00 | ,00 | ,00 | ,00 | ,00 | ,00 | ,00 | ,00 | ,00 |

a. Dependent Variable: Secure behaviour score

## 12.1.2.5  Multiple regression Hofstede cultural dimensions

### Model Summary

| Model | R | R Square | Adjusted R Square | Std. Error of the Estimate |
|---|---|---|---|---|
| 1 | ,243[a] | ,059 | ,059 | 2,464 |
| 2 | ,273[b] | ,074 | ,074 | 2,444 |
| 3 | ,296[c] | ,087 | ,087 | 2,426 |
| 4 | ,303[d] | ,092 | ,091 | 2,421 |

a. Predictors: (Constant), Indulgence (Hofstede)

b. Predictors: (Constant), Indulgence (Hofstede), Long-term orientation (Hofstede)

c. Predictors: (Constant), Indulgence (Hofstede), Long-term orientation (Hofstede), Masculinity (Hofstede)

d. Predictors: (Constant), Indulgence (Hofstede), Long-term orientation (Hofstede), Masculinity (Hofstede), Power distance (Hofstede)

### ANOVA[a]

| Model | | Sum of Squares | df | Mean Square | F | Sig. |
|---|---|---|---|---|---|---|
| 1 | Regression | 8138,218 | 1 | 8138,218 | 1340,784 | ,000[b] |
| | Residual | 129400,899 | 21319 | 6,070 | | |
| | Total | 137539,116 | 21320 | | | |
| 2 | Regression | 10236,665 | 2 | 5118,332 | 857,113 | ,000[c] |
| | Residual | 127302,452 | 21318 | 5,972 | | |
| | Total | 137539,116 | 21320 | | | |
| 3 | Regression | 12028,782 | 3 | 4009,594 | 681,000 | ,000[d] |
| | Residual | 125510,335 | 21317 | 5,888 | | |

Click to navigate

|   |   | B/SS | df | Mean Square | F | Sig. |
|---|---|---|---|---|---|---|
|   | Total | 137539,116 | 21320 |   |   |   |
| 4 | Regression | 12598,005 | 4 | 3149,501 | 537,331 | ,000ᵉ |
|   | Residual | 124941,111 | 21316 | 5,861 |   |   |
|   | Total | 137539,116 | 21320 |   |   |   |

a. Dependent Variable: Secure behaviour score

b. Predictors: (Constant), Indulgence (Hofstede)

c. Predictors: (Constant), Indulgence (Hofstede), Long-term orientation (Hofstede)

d. Predictors: (Constant), Indulgence (Hofstede), Long-term orientation (Hofstede), Masculinity (Hofstede)

e. Predictors: (Constant), Indulgence (Hofstede), Long-term orientation (Hofstede), Masculinity (Hofstede), Power distance (Hofstede)

## Coefficientsᵃ

| Model |   | Unstandardized Coefficients B | Unstandardized Coefficients Std. Error | Standardized Coefficients Beta | t | Sig. | 95,0% Confidence Interval for B Lower Bound | 95,0% Confidence Interval for B Upper Bound |
|---|---|---|---|---|---|---|---|---|
| 1 | (Constant) | 2,125 | ,042 |   | 50,399 | ,000 | 2,042 | 2,207 |
|   | Indulgence (Hofstede) | ,032 | ,001 | ,243 | 36,617 | ,000 | ,030 | ,034 |
| 2 | (Constant) | ,631 | ,090 |   | 7,018 | ,000 | ,455 | ,808 |
|   | Indulgence (Hofstede) | ,039 | ,001 | ,295 | 41,291 | ,000 | ,037 | ,041 |
|   | Long-term orientation (Hofstede) | ,021 | ,001 | ,134 | 18,746 | ,000 | ,018 | ,023 |
| 3 | (Constant) | 1,152 | ,094 |   | 12,234 | ,000 | ,968 | 1,337 |
|   | Indulgence (Hofstede) | ,037 | ,001 | ,283 | 39,657 | ,000 | ,035 | ,039 |
|   | Long-term orientation (Hofstede) | ,022 | ,001 | ,144 | 20,176 | ,000 | ,020 | ,024 |
|   | Masculinity (Hofstede) | -,012 | ,001 | -,116 | -17,446 | ,000 | -,013 | -,010 |

Click to navigate

| 4 | | | | | | | | |
|---|---|---|---|---|---|---|---|---|
| | (Constant) | 1,872 | ,119 | | 15,729 | ,000 | 1,639 | 2,105 |
| | Indulgence (Hofstede) | ,032 | ,001 | ,240 | 28,935 | ,000 | ,029 | ,034 |
| | Long-term orientation (Hofstede) | ,021 | ,001 | ,138 | 19,328 | ,000 | ,019 | ,023 |
| | Masculinity (Hofstede) | -,011 | ,001 | -,105 | -15,590 | ,000 | -,012 | -,009 |
| | Power distance (Hofstede) | -,009 | ,001 | -,078 | -9,855 | ,000 | -,011 | -,008 |

a. Dependent Variable: Secure behaviour score

## Excluded Variables[a]

| Model | | Beta In | t | Sig. | Partial Correlation | Collinearity Statistics Tolerance |
|---|---|---|---|---|---|---|
| 1 | Power distance (Hofstede) | -,108[b] | -13,749 | ,000 | -,094 | ,712 |
| | Individualism (Hofstede) | ,060[b] | 8,135 | ,000 | ,056 | ,817 |
| | Uncertainty avoidance (Hofstede) | -,071[b] | -9,470 | ,000 | -,065 | ,775 |
| | Masculinity (Hofstede) | -,105[b] | -15,779 | ,000 | -,107 | ,981 |
| | Long-term orientation (Hofstede) | ,134[b] | 18,746 | ,000 | ,127 | ,851 |
| 2 | Power distance (Hofstede) | -,098[c] | -12,571 | ,000 | -,086 | ,709 |
| | Individualism (Hofstede) | ,008[c] | ,983 | ,326 | ,007 | ,694 |
| | Uncertainty avoidance (Hofstede) | -,056[c] | -7,462 | ,000 | -,051 | ,765 |
| | Masculinity (Hofstede) | -,116[c] | -17,446 | ,000 | -,119 | ,975 |
| 3 | Power distance (Hofstede) | -,078[d] | -9,855 | ,000 | -,067 | ,689 |
| | Individualism (Hofstede) | ,027[d] | 3,441 | ,001 | ,024 | ,681 |
| | Uncertainty avoidance (Hofstede) | -,037[d] | -4,959 | ,000 | -,034 | ,748 |
| 4 | Individualism (Hofstede) | -,013[e] | -1,415 | ,157 | -,010 | ,526 |
| | Uncertainty avoidance (Hofstede) | -,006[e] | -,662 | ,508 | -,005 | ,601 |

a. Dependent Variable: Secure behaviour score

b. Predictors in the Model: (Constant), Indulgence (Hofstede)

c. Predictors in the Model: (Constant), Indulgence (Hofstede), Long-term orientation (Hofstede)

d. Predictors in the Model: (Constant), Indulgence (Hofstede), Long-term orientation (Hofstede), Masculinity (Hofstede)

e. Predictors in the Model: (Constant), Indulgence (Hofstede), Long-term orientation (Hofstede), Masculinity (Hofstede), Power distance (Hofstede)

## 12.1.2.6    Multiple regression Meyer cultural dimensions

### Model Summary

| Model | R | R Square | Adjusted R Square | Std. Error of the Estimate |
|---|---|---|---|---|
| 1 | ,289[a] | ,084 | ,084 | 2,465 |
| 2 | ,295[b] | ,087 | ,087 | 2,461 |
| 3 | ,300[c] | ,090 | ,090 | 2,457 |
| 4 | ,307[d] | ,094 | ,094 | 2,451 |
| 5 | ,311[e] | ,097 | ,096 | 2,448 |
| 6 | ,311[f] | ,097 | ,096 | 2,448 |

a. Predictors: (Constant), Deciding (Meyer)

b. Predictors: (Constant), Deciding (Meyer), Disagreeing (Meyer)

c. Predictors: (Constant), Deciding (Meyer), Disagreeing (Meyer), Leading (Meyer)

d. Predictors: (Constant), Deciding (Meyer), Disagreeing (Meyer), Leading (Meyer), Scheduling (Meyer)

e. Predictors: (Constant), Deciding (Meyer), Disagreeing (Meyer), Leading (Meyer), Scheduling (Meyer), Evaluating (Meyer)

f. Predictors: (Constant), Deciding (Meyer), Disagreeing (Meyer), Leading (Meyer), Scheduling (Meyer), Evaluating (Meyer), Communicating (Meyer)

### ANOVA[a]

| Model | | Sum of Squares | df | Mean Square | F | Sig. |
|---|---|---|---|---|---|---|
| 1 | Regression | 8978,609 | 1 | 8978,609 | 1477,508 | ,000[b] |
| | Residual | 98177,723 | 16156 | 6,077 | | |
| | Total | 107156,332 | 16157 | | | |
| 2 | Regression | 9296,364 | 2 | 4648,182 | 767,335 | ,000[c] |
| | Residual | 97859,967 | 16155 | 6,058 | | |
| | Total | 107156,332 | 16157 | | | |
| 3 | Regression | 9619,602 | 3 | 3206,534 | 531,065 | ,000[d] |
| | Residual | 97536,730 | 16154 | 6,038 | | |
| | Total | 107156,332 | 16157 | | | |
| 4 | Regression | 10079,440 | 4 | 2519,860 | 419,289 | ,000[e] |
| | Residual | 97076,891 | 16153 | 6,010 | | |
| | Total | 107156,332 | 16157 | | | |
| 5 | Regression | 10346,447 | 5 | 2069,289 | 345,245 | ,000[f] |
| | Residual | 96809,885 | 16152 | 5,994 | | |
| | Total | 107156,332 | 16157 | | | |
| 6 | Regression | 10371,956 | 6 | 1728,659 | 288,472 | ,000[g] |

Click to navigate

| | | | | | |
|---|---|---|---|---|---|
| Residual | 96784,376 | 16151 | 5,992 | | |
| Total | 107156,332 | 16157 | | | |

a. Dependent Variable: Secure behaviour score

b. Predictors: (Constant), Deciding (Meyer)

c. Predictors: (Constant), Deciding (Meyer), Disagreeing (Meyer)

d. Predictors: (Constant), Deciding (Meyer), Disagreeing (Meyer), Leading (Meyer)

e. Predictors: (Constant), Deciding (Meyer), Disagreeing (Meyer), Leading (Meyer), Scheduling (Meyer)

f. Predictors: (Constant), Deciding (Meyer), Disagreeing (Meyer), Leading (Meyer), Scheduling (Meyer), Evaluating (Meyer)

g. Predictors: (Constant), Deciding (Meyer), Disagreeing (Meyer), Leading (Meyer), Scheduling (Meyer), Evaluating (Meyer), Communicating (Meyer)

## Coefficients[a]

| Model | | Unstandardized Coefficients | | Standardized Coefficients | t | Sig. | 95,0% Confidence Interval for B | |
|---|---|---|---|---|---|---|---|---|
| | | B | Std. Error | Beta | | | Lower Bound | Upper Bound |
| 1 | (Constant) | 5,251 | ,046 | | 113,978 | ,000 | 5,161 | 5,342 |
| | Deciding (Meyer) | -,312 | ,008 | -,289 | -38,438 | ,000 | -,328 | -,296 |
| 2 | (Constant) | 5,630 | ,070 | | 80,822 | ,000 | 5,494 | 5,767 |
| | Deciding (Meyer) | -,322 | ,008 | -,299 | -39,173 | ,000 | -,338 | -,306 |
| | Disagreeing (Meyer) | -,080 | ,011 | -,055 | -7,243 | ,000 | -,102 | -,059 |
| 3 | (Constant) | 5,794 | ,073 | | 79,300 | ,000 | 5,651 | 5,937 |
| | Deciding (Meyer) | -,225 | ,016 | -,209 | -14,438 | ,000 | -,256 | -,195 |
| | Disagreeing (Meyer) | -,093 | ,011 | -,064 | -8,274 | ,000 | -,115 | -,071 |
| | Leading (Meyer) | -,111 | ,015 | -,107 | -7,317 | ,000 | -,141 | -,082 |
| 4 | (Constant) | 6,098 | ,081 | | 75,504 | ,000 | 5,940 | 6,257 |
| | Deciding (Meyer) | -,134 | ,019 | -,124 | -7,147 | ,000 | -,171 | -,097 |

| Model | | B | Std. Error | Beta | t | Sig. | Lower Bound | Upper Bound |
|---|---|---|---|---|---|---|---|---|
| | Disagreeing (Meyer) | -,110 | ,011 | -,076 | -9,671 | ,000 | -,132 | -,087 |
| | Leading (Meyer) | -,143 | ,016 | -,138 | -9,175 | ,000 | -,174 | -,113 |
| | Scheduling (Meyer) | -,120 | ,014 | -,091 | -8,747 | ,000 | -,147 | -,093 |
| 5 | (Constant) | 6,351 | ,089 | | 71,270 | ,000 | 6,177 | 6,526 |
| | Deciding (Meyer) | -,113 | ,019 | -,105 | -5,965 | ,000 | -,150 | -,076 |
| | Disagreeing (Meyer) | -,250 | ,024 | -,172 | -10,462 | ,000 | -,297 | -,203 |
| | Leading (Meyer) | -,164 | ,016 | -,158 | -10,329 | ,000 | -,196 | -,133 |
| | Scheduling (Meyer) | -,173 | ,016 | -,132 | -10,930 | ,000 | -,204 | -,142 |
| | Evaluating (Meyer) | ,159 | ,024 | ,107 | 6,674 | ,000 | ,112 | ,206 |
| 6 | (Constant) | 6,318 | ,091 | | 69,809 | ,000 | 6,141 | 6,496 |
| | Deciding (Meyer) | -,129 | ,020 | -,119 | -6,307 | ,000 | -,169 | -,089 |
| | Disagreeing (Meyer) | -,265 | ,025 | -,182 | -10,622 | ,000 | -,314 | -,216 |
| | Leading (Meyer) | -,173 | ,017 | -,167 | -10,506 | ,000 | -,206 | -,141 |
| | Scheduling (Meyer) | -,194 | ,019 | -,147 | -10,320 | ,000 | -,231 | -,157 |
| | Evaluating (Meyer) | ,166 | ,024 | ,112 | 6,899 | ,000 | ,119 | ,213 |
| | Communicating (Meyer) | ,057 | ,028 | ,036 | 2,063 | ,039 | ,003 | ,111 |

a. Dependent Variable: Secure behaviour score

## Excluded Variablesᵃ

| Model | | Beta In | t | Sig. | Partial Correlation | Collinearity Statistics Tolerance |
|---|---|---|---|---|---|---|
| 1 | Communicating (Meyer) | -,081[b] | -5,709 | ,000 | -,045 | ,282 |
| | Evaluating (Meyer) | -,042[b] | -5,591 | ,000 | -,044 | ,994 |
| | Leading (Meyer) | -,089[b] | -6,127 | ,000 | -,048 | ,269 |
| | Disagreeing (Meyer) | -,055[b] | -7,243 | ,000 | -,057 | ,969 |
| | Scheduling (Meyer) | -,058[b] | -5,726 | ,000 | -,045 | ,558 |
| 2 | Communicating (Meyer) | -,064[c] | -4,447 | ,000 | -,035 | ,272 |
| | Evaluating (Meyer) | ,013[c] | ,908 | ,364 | ,007 | ,292 |
| | Leading (Meyer) | -,107[c] | -7,317 | ,000 | -,057 | ,263 |
| | Scheduling (Meyer) | -,069[c] | -6,774 | ,000 | -,053 | ,547 |
| 3 | Communicating (Meyer) | -,055[d] | -3,843 | ,000 | -,030 | ,270 |
| | Evaluating (Meyer) | ,019[d] | 1,340 | ,180 | ,011 | ,291 |
| | Scheduling (Meyer) | -,091[d] | -8,747 | ,000 | -,069 | ,518 |
| 4 | Communicating (Meyer) | ,019[e] | 1,100 | ,271 | ,009 | ,190 |
| | Evaluating (Meyer) | ,107[e] | 6,674 | ,000 | ,052 | ,217 |
| 5 | Communicating (Meyer) | ,036[f] | 2,063 | ,039 | ,016 | ,186 |

a. Dependent Variable: Secure behaviour score

b. Predictors in the Model: (Constant), Deciding (Meyer)

c. Predictors in the Model: (Constant), Deciding (Meyer), Disagreeing (Meyer)

d. Predictors in the Model: (Constant), Deciding (Meyer), Disagreeing (Meyer), Leading (Meyer)

e. Predictors in the Model: (Constant), Deciding (Meyer), Disagreeing (Meyer), Leading (Meyer), Scheduling (Meyer)

f. Predictors in the Model: (Constant), Deciding (Meyer), Disagreeing (Meyer), Leading (Meyer), Scheduling (Meyer), Evaluating (Meyer)

## 12.1.2.7 Additional analysis on results of the multiple regression Meyer cultural dimensions

**Variables Entered/Removed[a]**

| Model | Variables Entered | Variables Removed | Method |
|---|---|---|---|
| 1 | Deciding (Meyer), Evaluating (Meyer), Leading (Meyer), Communicating (Meyer)[b] | . | Enter |

a. Dependent Variable: Secure behaviour score

b. All requested variables entered.

**Model Summary**

| Model | R | R Square | Adjusted R Square | Std. Error of the Estimate |
|---|---|---|---|---|
| 1 | .298[a] | .089 | .089 | 2.458 |

a. Predictors: (Constant), Deciding (Meyer), Evaluating (Meyer), Leading (Meyer), Communicating (Meyer)

## ANOVAᵃ

| Model | | Sum of Squares | df | Mean Square | F | Sig. |
|---|---|---|---|---|---|---|
| 1 | Regression | 9532.642 | 4 | 2383.160 | 394.322 | .000ᵇ |
| | Residual | 97623.690 | 16153 | 6.044 | | |
| | Total | 107156.332 | 16157 | | | |

a. Dependent Variable: Secure behaviour score

b. Predictors: (Constant), Deciding (Meyer), Evaluating (Meyer), Leading (Meyer), Communicating (Meyer)

## Coefficientsᵃ

| Model | | Unstandardized Coefficients | | Standardized Coefficients | t | Sig. | 95.0% Confidence Interval for B | |
|---|---|---|---|---|---|---|---|---|
| | | B | Std. Error | Beta | | | Lower Bound | Upper Bound |
| 1 | (Constant) | 5.725 | .072 | | 79.193 | .000 | 5.583 | 5.867 |
| | Communicating (Meyer) | -.092 | .023 | -.058 | -3.924 | .000 | -.138 | -.046 |
| | Evaluating (Meyer) | -.058 | .012 | -.039 | -4.995 | .000 | -.081 | -.035 |
| | Leading (Meyer) | -.096 | .015 | -.093 | -6.358 | .000 | -.126 | -.067 |
| | Deciding (Meyer) | -.171 | .020 | -.158 | -8.605 | .000 | -.210 | -.132 |

a. Dependent Variable: Secure behaviour score

Click to navigate

Introduction | Literature review | Research questions | Antecedents of behaviour | Research model | Methodology | Results | Discussion | Conclusion | Limitations | References | Appendix

## 12.1.2.8 Partial correlation analysis Communicating

### Descriptive Statistics

|  | Mean | Std. Deviation | N |
|---|---|---|---|
| Secure behaviour score | 3.64 | 2.575 | 16158 |
| Communicating (Meyer) | 5.05 | 1.611 | 16158 |
| Evaluating (Meyer) | 3.53 | 1.735 | 16158 |
| Leading (Meyer) | 5.51 | 2.478 | 16158 |
| Deciding (Meyer) | 5.15 | 2.392 | 16158 |
| Disagreeing (Meyer) | 4.05 | 1.774 | 16158 |
| Scheduling (Meyer) | 4.42 | 1.956 | 16158 |

### Correlations

| Control Variables | | | Secure behaviour score | Communicating (Meyer) | Evaluating (Meyer) | Leading (Meyer) | Deciding (Meyer) | Disagreeing (Meyer) | Scheduling (Meyer) |
|---|---|---|---|---|---|---|---|---|---|
| -none-[a] | Secure behaviour score | Correlation | 1.000 | -.268 | -.064 | -.271 | -.289 | -.003 | -.225 |
|  |  | Significance (2-tailed) | . | .000 | .000 | .000 | .000 | .726 | .000 |
|  |  | df | 0 | 16156 | 16156 | 16156 | 16156 | 16156 | 16156 |
|  | Communicating (Meyer) | Correlation | -.268 | 1.000 | .196 | .740 | .847 | -.052 | .750 |
|  |  | Significance (2-tailed) | .000 | . | .000 | .000 | .000 | .000 | .000 |
|  |  | df | 16156 | 0 | 16156 | 16156 | 16156 | 16156 | 16156 |
|  | Evaluating (Meyer) | Correlation | -.064 | .196 | 1.000 | .015 | .075 | .812 | .156 |
|  |  | Significance (2-tailed) | .000 | .000 | . | .065 | .000 | .000 | .000 |

| | | | | | | | | |
|---|---|---|---|---|---|---|---|---|
| | | df | 16156 | 16156 | 0 | 16156 | 16156 | 16156 | 16156 |
| | Leading (Meyer) | Correlation | -.271 | .740 | .015 | 1.000 | .855 | -.227 | .488 |
| | | Significance (2-tailed) | .000 | .000 | .065 | . | .000 | .000 | .000 |
| | | df | 16156 | 16156 | 16156 | 0 | 16156 | 16156 | 16156 |
| | Deciding (Meyer) | Correlation | -.289 | .847 | .075 | .855 | 1.000 | -.176 | .665 |
| | | Significance (2-tailed) | .000 | .000 | .000 | .000 | . | .000 | .000 |
| | | df | 16156 | 16156 | 16156 | 16156 | 0 | 16156 | 16156 |
| | Disagreeing (Meyer) | Correlation | -.003 | -.052 | .812 | -.227 | -.176 | 1.000 | -.216 |
| | | Significance (2-tailed) | .726 | .000 | .000 | .000 | .000 | . | .000 |
| | | df | 16156 | 16156 | 16156 | 16156 | 16156 | 0 | 16156 |
| | Scheduling (Meyer) | Correlation | -.225 | .750 | .156 | .488 | .665 | -.216 | 1.000 |
| | | Significance (2-tailed) | .000 | .000 | .000 | .000 | .000 | .000 | . |
| | | df | 16156 | 16156 | 16156 | 16156 | 16156 | 16156 | 0 |
| Evaluating (Meyer) & Leading (Meyer) & Deciding (Meyer) & Disagreeing (Meyer) & Scheduling (Meyer) | Secure behaviour score | Correlation | 1.000 | .016 | | | | | |
| | | Significance (2-tailed) | . | .039 | | | | | |
| | | df | 0 | 16151 | | | | | |
| | Communicating (Meyer) | Correlation | .016 | 1.000 | | | | | |
| | | Significance (2-tailed) | .039 | . | | | | | |
| | | df | 16151 | 0 | | | | | |

a. Cells contain zero-order (Pearson) correlations.

## 12.1.2.9 Partial correlation analysis Evaluating

### Descriptive Statistics

|  | Mean | Std. Deviation | N |
|---|---|---|---|
| Secure behaviour score | 3.64 | 2.575 | 16158 |
| Evaluating (Meyer) | 3.53 | 1.735 | 16158 |
| Disagreeing (Meyer) | 4.05 | 1.774 | 16158 |
| Scheduling (Meyer) | 4.42 | 1.956 | 16158 |

### Correlations

| Control Variables | | | | Secure behaviour score | Evaluating (Meyer) | Disagreeing (Meyer) | Scheduling (Meyer) |
|---|---|---|---|---|---|---|---|
| -none-[a] | Secure behaviour score | Correlation | | 1.000 | -.064 | -.003 | -.225 |
| | | Significance (2-tailed) | | . | .000 | .726 | .000 |
| | | df | | 0 | 16156 | 16156 | 16156 |
| | Evaluating (Meyer) | Correlation | | -.064 | 1.000 | .812 | .156 |
| | | Significance (2-tailed) | | .000 | . | .000 | .000 |
| | | df | | 16156 | 0 | 16156 | 16156 |
| | Disagreeing (Meyer) | Correlation | | -.003 | .812 | 1.000 | -.216 |
| | | Significance (2-tailed) | | .726 | .000 | . | .000 |
| | | df | | 16156 | 16156 | 0 | 16156 |
| | Scheduling (Meyer) | Correlation | | -.225 | .156 | -.216 | 1.000 |
| | | Significance (2-tailed) | | .000 | .000 | .000 | . |
| | | df | | 16156 | 16156 | 16156 | 0 |
| Disagreeing (Meyer) & Scheduling (Meyer) | Secure behaviour score | Correlation | | 1.000 | .037 | | |
| | | Significance (2-tailed) | | . | .000 | | |

| | | | 0 | 16154 | | |
|---|---|---|---|---|---|---|
| Evaluating (Meyer) | df | | | | | |
| | Correlation | .037 | 1.000 | | |
| | Significance (2-tailed) | .000 | . | | |
| | df | 16154 | 0 | | |

a. Cells contain zero-order (Pearson) correlations.

## 12.1.2.10 Multiple regression Prior experience with security incidents

### Model Summary

| Model | R | R Square | Adjusted R Square | Std. Error of the Estimate |
|---|---|---|---|---|
| 1 | ,262ª | ,068 | ,068 | 2,454 |

a. Predictors: (Constant), Indirect experience score, Direct experience score

### ANOVAª

| Model | | Sum of Squares | df | Mean Square | F | Sig. |
|---|---|---|---|---|---|---|
| 1 | Regression | 9582,731 | 2 | 4791,366 | 795,832 | ,000ᵇ |
| | Residual | 130399,651 | 21659 | 6,021 | | |
| | Total | 139982,382 | 21661 | | | |

a. Dependent Variable: Secure behaviour score

b. Predictors: (Constant), Indirect experience score, Direct experience score

### Coefficientsª

| Model | | Unstandardized Coefficients | | Standardized Coefficients | t | Sig. | Collinearity Statistics | |
|---|---|---|---|---|---|---|---|---|
| | | B | Std. Error | Beta | | | Tolerance | VIF |
| 1 | (Constant) | 2,620 | ,029 | | 90,424 | ,000 | | |
| | Direct experience score | ,071 | ,006 | ,085 | 12,063 | ,000 | ,869 | 1,151 |
| | Indirect experience score | ,522 | ,017 | ,219 | 31,078 | ,000 | ,869 | 1,151 |

a. Dependent Variable: Secure behaviour score

### Collinearity Diagnostics[a]

| Model | Dimension | Eigenvalue | Condition Index | (Constant) | Variance Proportions Direct experience score | Indirect experience score |
|---|---|---|---|---|---|---|
| 1 | 1 | 2,303 | 1,000 | ,05 | ,07 | ,05 |
|  | 2 | ,522 | 2,101 | ,14 | ,86 | ,04 |
|  | 3 | ,175 | 3,627 | ,81 | ,06 | ,91 |

a. Dependent Variable: Secure behaviour score

# 12.2 Statistical output data analyses KnowBe4 Dataset

## 12.2.1 Single variable testing

### 12.2.1.1 Assumption testing ANOVA Industry type

Box plot

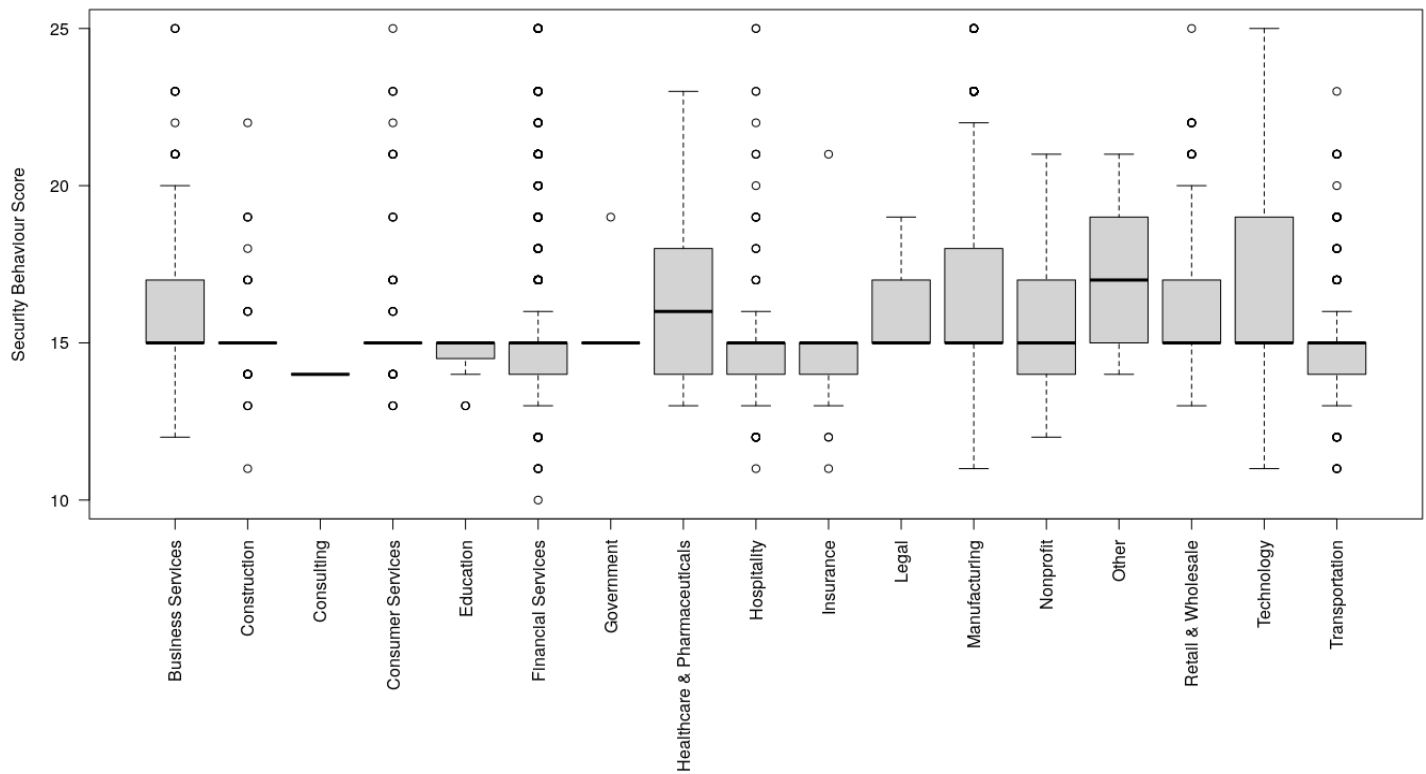

## Distribution histogram

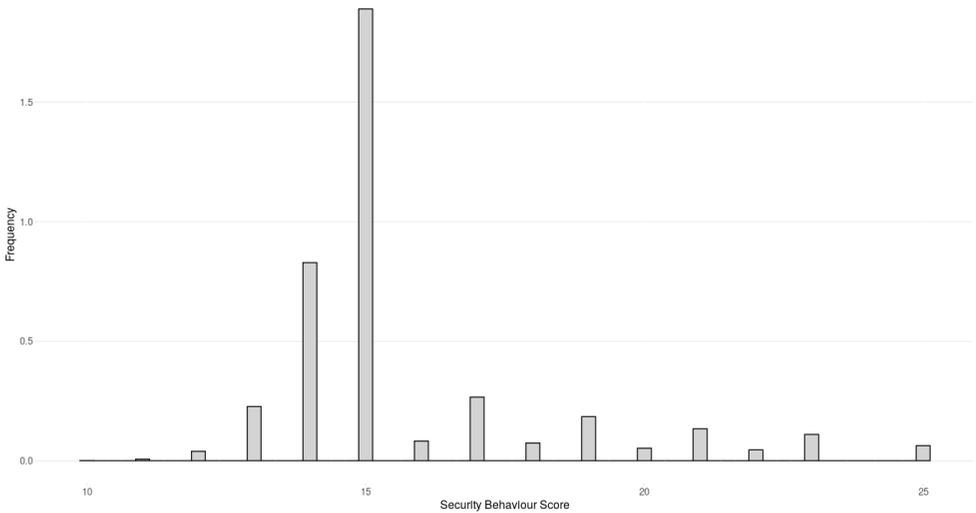

## Normal Q-Q plot

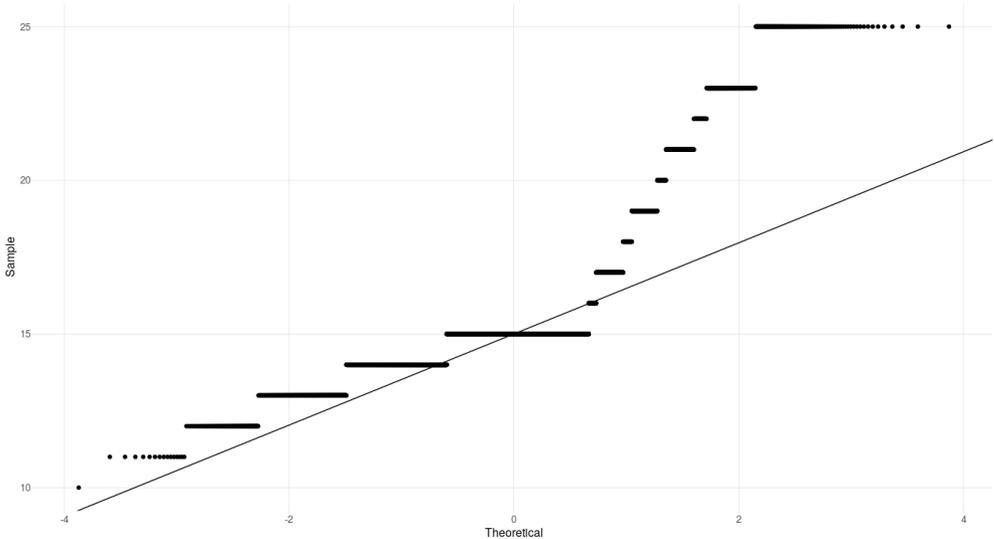

### Levene's test of homogeneity

```
Levene's Test for Homogeneity of Variance (center = median)
        Df F value    Pr(>F)
group   16  43.206 < 2.2e-16 ***
      9191
---
Signif. codes:  0 '***' 0.001 '**' 0.01 '*' 0.05 '.' 0.1 ' ' 1
```

## 12.2.1.2   One-way ANOVA Industry type

```
           Df Sum Sq Mean Sq F value Pr(>F)
Industry   16   6519   407.5   70.38 <2e-16 ***
Residuals 9451  54719     5.8
---
Signif. codes:  0 '***' 0.001 '**' 0.01 '*' 0.05 '.' 0.1 ' ' 1
```

| Industry | Mean Security Behaviour |
| --- | --- |
| Business Services | 16.00000 |
| Construction | 15.21250 |
| Consulting | 14.50000 |
| Consumer Services | 15.66912 |
| Education | 14.71739 |
| Financial Services | 14.95101 |
| Government | 15.66667 |
| Healthcare & Pharmaceuticals | 16.31250 |
| Hospitality | 14.69088 |
| Insurance | 14.40385 |
| Legal | 16.14286 |
| Manufacturing | 16.42518 |
| Nonprofit | 15.50000 |
| Other | 16.66667 |
| Retail & Wholesale | 16.24891 |
| Technology | 16.66656 |
| Transportation | 14.93314 |

Click to navigate

### 12.2.1.3 One-way ANOVA Security Education and Training

```
              Df Sum Sq Mean Sq F value   Pr(>F)
train_group    2    190   95.01   18.38 1.23e-08 ***
Residuals   2091  10812    5.17
---
Signif. codes:  0 '***' 0.001 '**' 0.01 '*' 0.05 '.' 0.1 ' ' 1
```

| | train_group | SecurityBehavior |
|---|---|---|
| 1 | group_12 | 15.33 |
| 2 | group_34 | 15.37 |
| 3 | group_5+ | 14.74 |

### 12.2.1.4 Simple regression Security Culture dimension Attitude

```
Residuals:
    Min      1Q  Median      3Q     Max
-5.5186 -1.5936 -0.7304 -0.3820  9.4752

Coefficients:
             Estimate Std. Error t value Pr(>|t|)
(Intercept) 18.579984   0.526253  35.306  < 2e-16 ***
Attitudes   -0.037486   0.006932  -5.408 6.53e-08 ***
---
Signif. codes:  0 '***' 0.001 '**' 0.01 '*' 0.05 '.' 0.1 ' ' 1

Residual standard error: 2.54 on 9466 degrees of freedom
Multiple R-squared:  0.00308,   Adjusted R-squared:  0.002975
F-statistic: 29.25 on 1 and 9466 DF,  p-value: 6.53e-08
```

Click to navigate

Introduction  Literature review  Research questions  Antecedents of behaviour  Research model  Methodology  Results  Discussion  Conclusion  Limitations  References  Appendix

## 12.2.1.5   Simple regression Security Culture dimension Cognition

```
Residuals:
    Min      1Q  Median      3Q     Max
-5.7174 -1.4502 -0.6274  0.1323  9.5521

Coefficients:
             Estimate Std. Error t value Pr(>|t|)
(Intercept) 22.846790   0.427003   53.51   <2e-16 ***
Cognition   -0.100827   0.006045  -16.68   <2e-16 ***
---
Signif. codes:  0 '***' 0.001 '**' 0.01 '*' 0.05 '.' 0.1 ' ' 1

Residual standard error: 2.507 on 9466 degrees of freedom
Multiple R-squared:  0.02855,   Adjusted R-squared:  0.02845
F-statistic: 278.2 on 1 and 9466 DF,  p-value: < 2.2e-16
```

## 12.2.1.6   Simple regression Security Culture dimension Communication

```
Residuals:
    Min      1Q  Median      3Q     Max
-5.7799 -1.5757 -0.7722 -0.5514  9.4486

Coefficients:
               Estimate Std. Error t value Pr(>|t|)
(Intercept)   14.019296   0.387668  36.163   <2e-16 ***
Communication  0.023157   0.005213   4.442    9e-06 ***
---
Signif. codes:  0 '***' 0.001 '**' 0.01 '*' 0.05 '.' 0.1 ' ' 1

Residual standard error: 2.541 on 9466 degrees of freedom
Multiple R-squared:  0.00208,   Adjusted R-squared:  0.001975
F-statistic: 19.73 on 1 and 9466 DF,  p-value: 9e-06
```

**Click to navigate**

Introduction   Literature review   Research questions   Antecedents of behaviour   Research model   Methodology   Results   Discussion   Conclusion   Limitations   References   Appendix

### 12.2.1.7 Simple regression Security Culture dimension Compliance

```
Coefficients:
            Estimate Std. Error t value Pr(>|t|)
(Intercept) 20.244686   0.267205   75.77   <2e-16 ***
Compliance  -0.062182   0.003669  -16.95   <2e-16 ***
---
Signif. codes:  0 '***' 0.001 '**' 0.01 '*' 0.05 '.' 0.1 ' ' 1

Residual standard error: 2.506 on 9466 degrees of freedom
Multiple R-squared:  0.02945,   Adjusted R-squared:  0.02934
F-statistic: 287.2 on 1 and 9466 DF,  p-value: < 2.2e-16
```

### 12.2.1.8 Simple regression Security Culture dimension Norms

```
Coefficients:
            Estimate Std. Error t value Pr(>|t|)
(Intercept) 17.661903   0.498213   35.450  < 2e-16 ***
Norms       -0.026661   0.006893   -3.868 0.000111 ***
---
Signif. codes:  0 '***' 0.001 '**' 0.01 '*' 0.05 '.' 0.1 ' ' 1

Residual standard error: 2.541 on 9466 degrees of freedom
Multiple R-squared:  0.001578, Adjusted R-squared:  0.001472
F-statistic: 14.96 on 1 and 9466 DF,  p-value: 0.0001105
```

**Click to navigate**

Introduction | Literature review | Research questions | Antecedents of behaviour | Research model | Methodology | Results | Discussion | Conclusion | Limitations | References | Appendix

## 12.2.1.9   Simple regression Security Culture dimension Responsibility

```
Residuals:
    Min     1Q  Median      3Q     Max
-5.7120 -1.7122 -0.7393 -0.6582  9.2992

Coefficients:
                Estimate Std. Error t value Pr(>|t|)
(Intercept)    16.275402   0.604658   26.92   <2e-16 ***
Responsibility -0.007578   0.008511   -0.89    0.373
---
Signif. codes:  0 '***' 0.001 '**' 0.01 '*' 0.05 '.' 0.1 ' ' 1

Residual standard error: 2.543 on 9466 degrees of freedom
Multiple R-squared:  8.374e-05,  Adjusted R-squared:  -2.189e-05
F-statistic: 0.7928 on 1 and 9466 DF,  p-value: 0.3733
```

## 12.2.1.10  Simple regression Security Awareness and Knowledge dimension Passwords & Authentication

```
Call:
lm(formula = total ~ `Passwords & Authentication`, data = score_sapa_scs_phish_1scs)

Residuals:
    Min     1Q  Median      3Q     Max
-5.7866 -1.5643 -0.7866 -0.5643  9.4357

Coefficients:
                            Estimate Std. Error t value Pr(>|t|)
(Intercept)                1.556e+01  5.360e-02 290.370  < 2e-16 ***
`Passwords & Authentication` 3.335e-03 8.983e-04   3.713 0.000206 ***
---
Signif. codes:  0 '***' 0.001 '**' 0.01 '*' 0.05 '.' 0.1 ' ' 1

Residual standard error: 2.541 on 9390 degrees of freedom
  (76 observations deleted due to missingness)
Multiple R-squared:  0.001466,  Adjusted R-squared:  0.001359
F-statistic: 13.78 on 1 and 9390 DF,  p-value: 0.0002063
```

## 12.2.1.11 Simple regression Security Awareness and Knowledge dimension Email Security

```
Call:
lm(formula = total ~ `Email Security`, data = score_sapa_scs_phish_1scs)

Residuals:
    Min      1Q  Median      3Q     Max
-5.4292 -1.4292 -0.6732 -0.1851  9.5708

Coefficients:
                 Estimate Std. Error t value Pr(>|t|)
(Intercept)     15.185079   0.081151 187.121  < 2e-16 ***
`Email Security`  0.007323   0.001015   7.215 5.79e-13 ***
---
Signif. codes:  0 '***' 0.001 '**' 0.01 '*' 0.05 '.' 0.1 ' ' 1

Residual standard error: 2.54 on 9390 degrees of freedom
  (76 observations deleted due to missingness)
Multiple R-squared:  0.005514,  Adjusted R-squared:  0.005408
F-statistic: 52.06 on 1 and 9390 DF,  p-value: 5.795e-13
```

## 12.2.1.12 Simple regression Security Awareness and Knowledge dimension Internet Use

```
Call:
lm(formula = total ~ `Internet Use`, data = score_sapa_scs_phish_1scs)

Residuals:
    Min      1Q  Median      3Q     Max
-5.5133 -1.5133 -0.7071 -0.3195  9.6805

Coefficients:
                Estimate Std. Error t value Pr(>|t|)
(Intercept)    15.319517   0.080392 190.560  < 2e-16 ***
`Internet Use`  0.005814   0.001050   5.536 3.18e-08 ***
---
Signif. codes:  0 '***' 0.001 '**' 0.01 '*' 0.05 '.' 0.1 ' ' 1

Residual standard error: 2.542 on 9390 degrees of freedom
  (76 observations deleted due to missingness)
Multiple R-squared:  0.003253,  Adjusted R-squared:  0.003147
F-statistic: 30.65 on 1 and 9390 DF,  p-value: 3.179e-08
```

Click to navigate

## 12.2.1.13  Simple regression Security Awareness and Knowledge dimension Social Media

```
Call:
lm(formula = total ~ `Social Media`, data = score_sapa_scs_phish_1scs)

Residuals:
    Min      1Q  Median      3Q     Max
-5.8268 -1.6189 -0.7575 -0.5496  9.4504

Coefficients:
                Estimate Std. Error t value Pr(>|t|)
(Intercept)    15.549618   0.077806 199.850   <2e-16 ***
`Social Media`  0.002772   0.001078   2.573   0.0101 *
---
Signif. codes:  0 '***' 0.001 '**' 0.01 '*' 0.05 '.' 0.1 ' ' 1

Residual standard error: 2.545 on 9405 degrees of freedom
  (61 observations deleted due to missingness)
Multiple R-squared:  0.0007033,  Adjusted R-squared:  0.000597
F-statistic: 6.619 on 1 and 9405 DF,  p-value: 0.01011
```

## 12.2.1.14  Simple regression Security Awareness and Knowledge dimension Mobile Devices

```
Call:
lm(formula = total ~ `Mobile Devices`, data = score_sapa_scs_phish_1scs)

Residuals:
    Min      1Q  Median      3Q     Max
-6.0993 -1.4121 -0.8702 -0.1831  9.8169

Coefficients:
                  Estimate Std. Error t value Pr(>|t|)
(Intercept)      1.518e+01  6.591e-02 230.362   <2e-16 ***
`Mobile Devices` 9.162e-03  9.945e-04   9.213   <2e-16 ***
---
Signif. codes:  0 '***' 0.001 '**' 0.01 '*' 0.05 '.' 0.1 ' ' 1

Residual standard error: 2.535 on 9396 degrees of freedom
  (70 observations deleted due to missingness)
Multiple R-squared:  0.008953,  Adjusted R-squared:  0.008848
F-statistic: 84.89 on 1 and 9396 DF,  p-value: < 2.2e-16
```

Click to navigate

## 12.2.1.15 Simple regression Security Awareness and Knowledge dimension Incident Reporting

```
Call:
lm(formula = total ~ `Incident Reporting`, data = score_sapa_scs_phish_1scs)

Residuals:
    Min      1Q  Median      3Q     Max
-5.8292 -1.5860 -0.7076 -0.5860  9.4140

Coefficients:
                      Estimate Std. Error t value Pr(>|t|)
(Intercept)          1.559e+01  5.511e-02 282.796  < 2e-16 ***
`Incident Reporting` 2.432e-03  7.614e-04   3.194  0.00141 **
---
Signif. codes:  0 '***' 0.001 '**' 0.01 '*' 0.05 '.' 0.1 ' ' 1

Residual standard error: 2.544 on 9341 degrees of freedom
  (125 observations deleted due to missingness)
Multiple R-squared:  0.001091,  Adjusted R-squared:  0.0009838
F-statistic:  10.2 on 1 and 9341 DF,  p-value: 0.001409
```

## 12.2.1.16 Simple regression Security Awareness and Knowledge dimension Security Awareness

```
Call:
lm(formula = total ~ `Security Awareness`, data = score_sapa_scs_phish_1scs)

Residuals:
    Min      1Q  Median      3Q     Max
-5.7954 -1.3036 -0.7954 -0.3036  9.6964

Coefficients:
                     Estimate Std. Error t value Pr(>|t|)
(Intercept)         15.303571   0.082525 185.442  < 2e-16 ***
`Security Awareness`  0.004918   0.000871   5.647 1.68e-08 ***
---
Signif. codes:  0 '***' 0.001 '**' 0.01 '*' 0.05 '.' 0.1 ' ' 1

Residual standard error: 2.546 on 9307 degrees of freedom
  (159 observations deleted due to missingness)
Multiple R-squared:  0.003414,  Adjusted R-squared:  0.003307
F-statistic: 31.88 on 1 and 9307 DF,  p-value: 1.685e-08
```

**Click to navigate**

## 12.2.1.17  Assumption testing Security consciousness and confidence

Number of records: 2094

Regression scatter plot

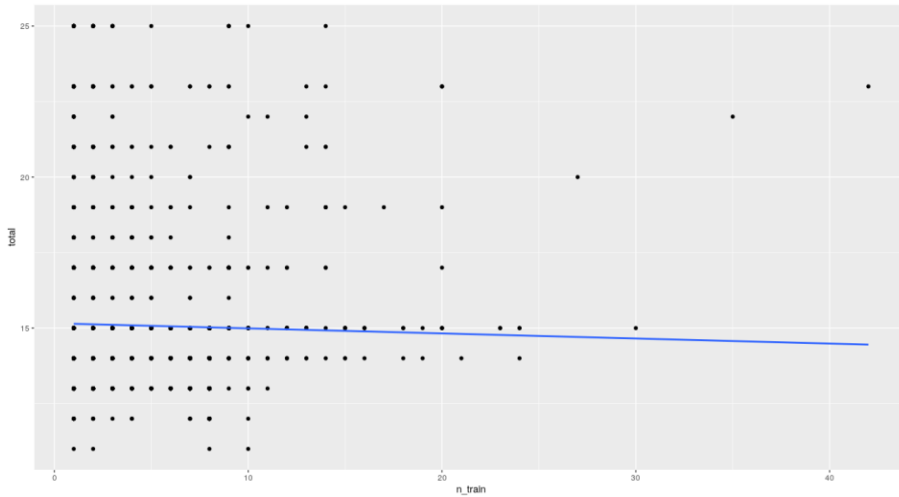

Description of the unusual points and significant outliers

The Studentized deleted residuals (SDR) > 3 are 73

The items that exceed the leverage point value (LEV) of 0.20 are 0

The items that exceed Cook's Distance Value (COO) of 1 are 0

## Histogram of standardized residuals

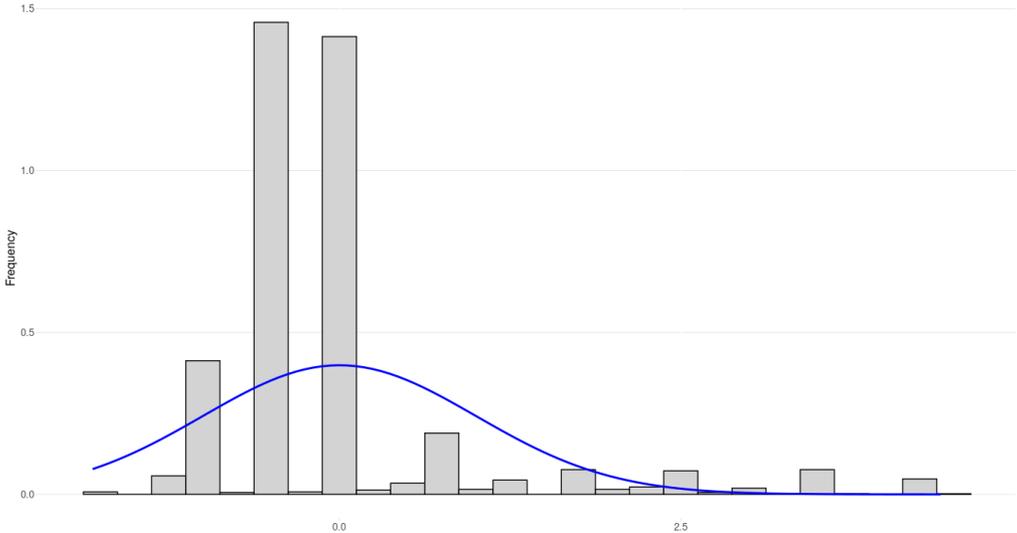

## Normal Q-Q plot

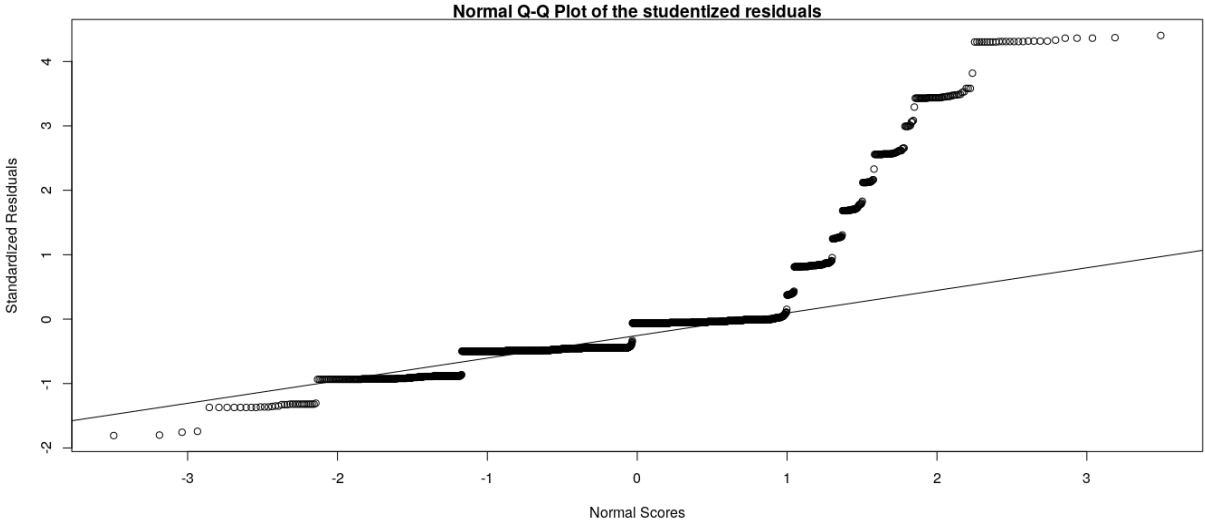

## 12.2.1.18  Simple regression Security consciousness and confidence

```
Call:
lm(formula = total ~ `Level of Confidence`, data = score_sapa_scs_phish)

Residuals:
    Min      1Q  Median      3Q     Max
-5.7696 -1.6925 -0.7696  0.2304  9.3846

Coefficients:
                      Estimate Std. Error t value Pr(>|t|)
(Intercept)          15.461273   0.282473  54.735   <2e-16 ***
`Level of Confidence`  0.003083   0.002914   1.058     0.29
---
Signif. codes:  0 '***' 0.001 '**' 0.01 '*' 0.05 '.' 0.1 ' ' 1

Residual standard error: 2.565 on 9205 degrees of freedom
  (1 observation deleted due to missingness)
Multiple R-squared:  0.0001216,  Adjusted R-squared:  1.295e-05
F-statistic: 1.119 on 1 and 9205 DF,  p-value: 0.2901
```

Number of records: 9208

### Regression scatter plot

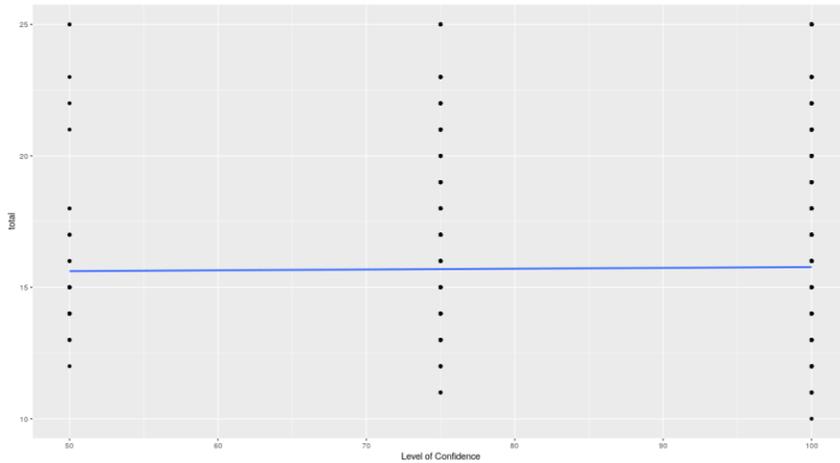

### Description of the unusual points and significant outliers

The Studentized deleted residuals (SDR) > 3 are 145

The items that exceed the leverage point value (LEV) of 0.20 are 0

The items that exceed Cook's Distance Value (COO) of 1 are 0

## Histogram of standardized residuals

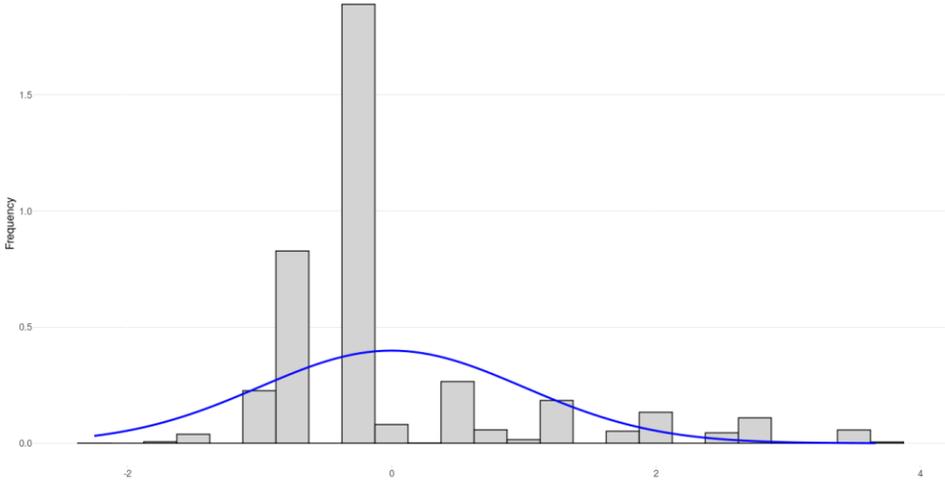

## Normal Q-Q plot

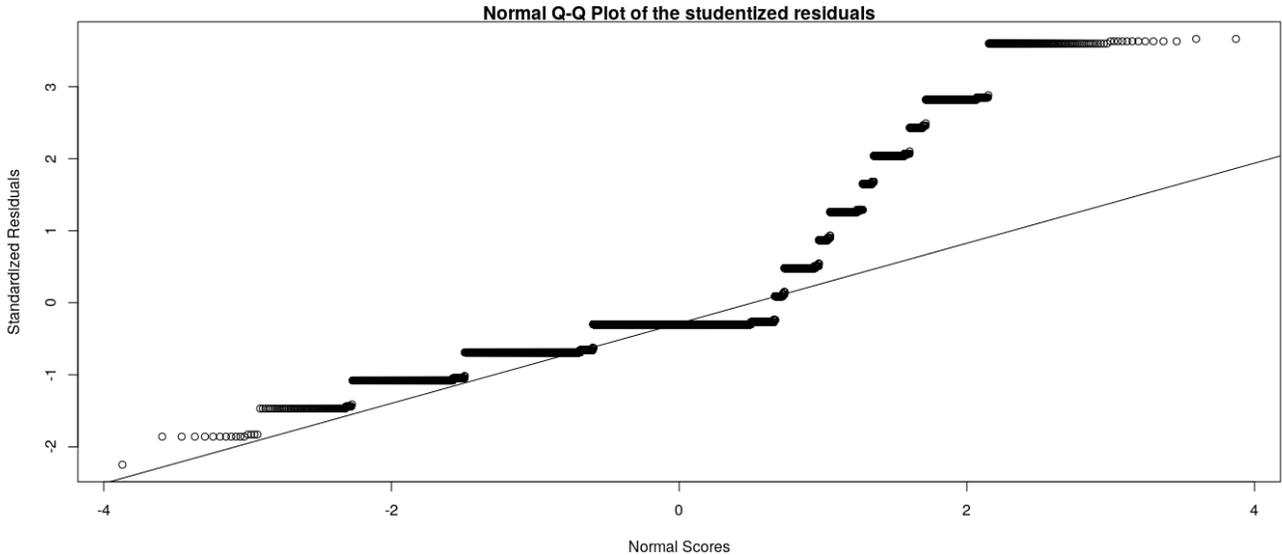

## 12.2.1.20  Simple regression Security education and training

```
Call:
lm(formula = total ~ n_train, data = score_sapa_scs_phish_train_1scs)

Residuals:
    Min      1Q  Median      3Q     Max
-4.1414 -1.1246 -0.1414 -0.0407 10.0768

Coefficients:
            Estimate Std. Error t value Pr(>|t|)
(Intercept) 15.15814    0.07930 191.143   <2e-16 ***
n_train     -0.01678    0.01316  -1.275    0.202
---
Signif. codes:  0 '***' 0.001 '**' 0.01 '*' 0.05 '.' 0.1 ' ' 1

Residual standard error: 2.292 on 2092 degrees of freedom
Multiple R-squared:  0.0007766,  Adjusted R-squared:  0.000299
F-statistic: 1.626 on 1 and 2092 DF,  p-value: 0.2024
```

## 12.2.2 Multiple variable testing

### 12.2.2.1 Assumption testing Security Culture

Number of records: 9208

Variance Inflation Factor (VIF)

```
    Variables Tolerance      VIF
    Attitudes 0.2548567 3.923773
    Cognition 0.4130543 2.420989
Communication 0.5127455 1.950285
   Compliance 0.3957396 2.526914
        Norms 0.6521674 1.533349
Responsibility 0.3244441 3.082195
```

Condition Index

```
Eigenvalue and Condition Index
-------------------------------
  Eigenvalue Condition Index   intercept    Attitudes    Cognition Communication   Compliance        Norms Responsibility
1 6.9879881624        1.00000 3.032561e-05 1.282813e-05 3.075790e-05 4.754461e-05 7.496764e-05 3.664786e-05   1.240722e-05
2 0.0055038156       35.63232 5.160037e-02 3.651621e-04 2.993642e-05 4.821542e-03 4.595782e-01 2.357150e-02   3.670021e-03
3 0.0024311095       53.61346 4.636925e-05 2.385439e-02 1.097433e-03 7.388454e-01 8.113986e-02 5.881428e-02   2.277021e-05
4 0.0015158039       67.89762 2.189183e-02 8.709456e-03 5.046515e-01 3.600669e-02 9.471944e-02 3.977278e-01   1.959824e-02
5 0.0013505441       71.93193 3.655156e-01 1.141783e-02 2.118731e-01 1.792652e-02 9.512823e-02 4.773689e-01   2.124169e-02
6 0.0008332059       91.57984 5.462345e-01 1.929853e-01 2.814928e-01 3.561001e-02 1.128634e-01 2.698525e-05   1.763695e-01
7 0.0003773586      136.08145 1.468095e-02 7.626551e-01 8.244341e-04 1.667422e-01 1.564959e-01 4.245388e-02   7.790854e-01
```

## Partial regression plots

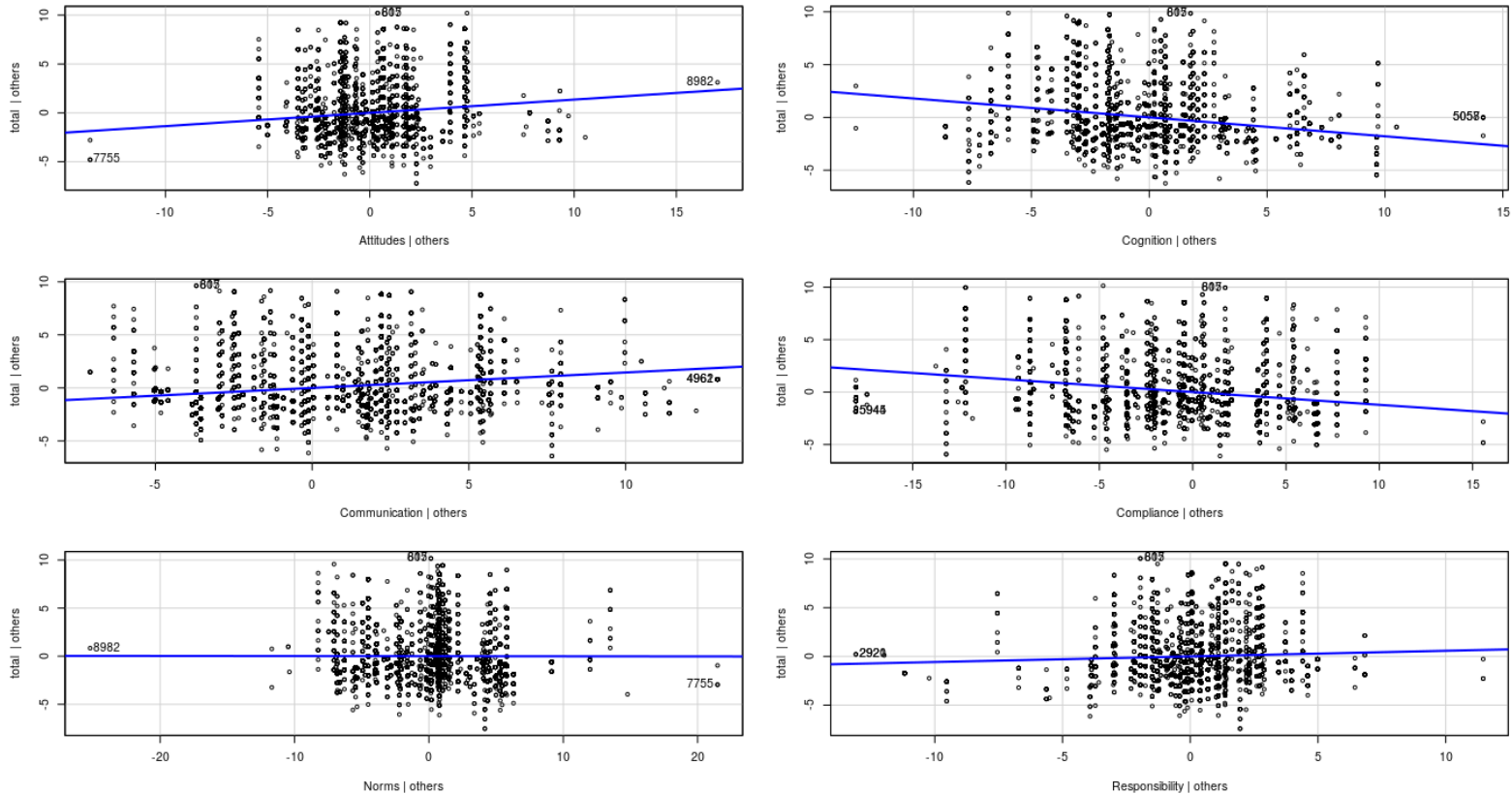

## Description of the unusual points and significant outliers

The Studentized deleted residuals (SDR) > 3 are 184

The items that exceed the leverage point value (LEV) of 0.20 are 0

The items that exceed Cook's Distance Value (COO) of 1 are 0

Click to navigate

Introduction  Literature review  Research questions  Antecedents of behaviour  Research model  Methodology  Results  Discussion  Conclusion  Limitations  References  Appendix

## Histogram of standardized residuals

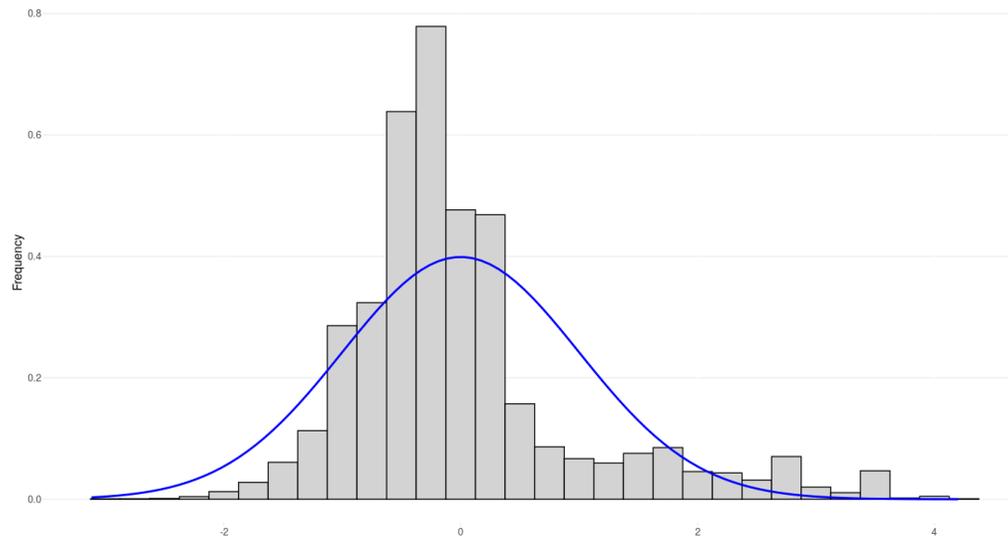

## Normal Q-Q plot

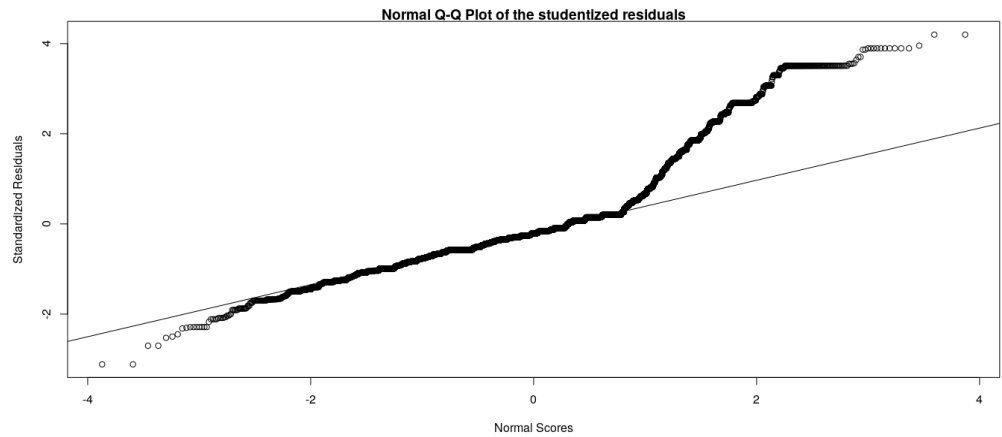

## 12.2.2.2  Multiple regression Security Culture

```
Forward Selection Method
--------------------------

Candidate Terms:

1. Attitudes
2. Cognition
3. Communication
4. Compliance
5. Norms
6. Responsibility
```

Click to navigate

## Step 1: Compliance

```
Forward Selection: Step 1

+ Compliance
```

### Model Summary

```
-------------------------------------------------------------
R               0.172      RMSE              2.506
R-Squared       0.029      Coef. Var        15.922
Adj. R-Squared  0.029      MSE               6.279
Pred R-Squared  0.029      MAE               1.752
-------------------------------------------------------------
 RMSE: Root Mean Square Error
 MSE: Mean Square Error
 MAE: Mean Absolute Error
```

### ANOVA

| | Sum of Squares | DF | Mean Square | F | Sig. |
|---|---|---|---|---|---|
| Regression | 1803.211 | 1 | 1803.211 | 287.188 | 0.0000 |
| Residual | 59435.568 | 9466 | 6.279 | | |
| Total | 61238.779 | 9467 | | | |

### Parameter Estimates

| model | Beta | Std. Error | Std. Beta | t | Sig | lower | upper |
|---|---|---|---|---|---|---|---|
| (Intercept) | 20.245 | 0.267 | | 75.765 | 0.000 | 19.721 | 20.768 |
| Compliance | -0.062 | 0.004 | -0.172 | -16.947 | 0.000 | -0.069 | -0.055 |

## Step 2: Compliance + Communication

Forward Selection: Step 2

+ Communication

```
                    Model Summary
---------------------------------------------------------------
R                0.252    RMSE          2.462
R-Squared        0.063    Coef. Var    15.641
Adj. R-Squared   0.063    MSE           6.059
Pred R-Squared   0.063    MAE           1.669
---------------------------------------------------------------
 RMSE: Root Mean Square Error
 MSE: Mean Square Error
 MAE: Mean Absolute Error
```

```
                         ANOVA
---------------------------------------------------------------
             Sum of
             Squares     DF    Mean Square    F        Sig.
---------------------------------------------------------------
Regression   3886.851     2      1943.425   320.731   0.0000
Residual    57351.929  9465         6.059
Total       61238.779  9467
---------------------------------------------------------------
```

```
                     Parameter Estimates
-----------------------------------------------------------------------
     model       Beta    Std. Error  Std. Beta    t      Sig    lower    upper
-----------------------------------------------------------------------
  (Intercept)   15.188     0.378                40.128   0.000  14.446   15.930
   Compliance   -0.112     0.004      -0.309   -24.909   0.000  -0.121   -0.103
Communication    0.117     0.006       0.230    18.544   0.000   0.104    0.129
-----------------------------------------------------------------------
```

Click to navigate

Step 3: Compliance + Communication + Cognition

Forward Selection: Step 3

+ Cognition

```
                    Model Summary
       --------------------------------------------------
       R              0.288    RMSE              2.436
       R-Squared      0.083    Coef. Var        15.479
       Adj. R-Squared 0.083    MSE               5.934
       Pred R-Squared 0.082    MAE               1.687
       --------------------------------------------------
        RMSE: Root Mean Square Error
        MSE: Mean Square Error
        MAE: Mean Absolute Error
```

```
                          ANOVA
       ---------------------------------------------------------
                   Sum of
                   Squares    DF    Mean Square    F       Sig.
       ---------------------------------------------------------
       Regression  5075.316    3      1691.772  285.077   0.0000
       Residual   56163.464  9464        5.934
       Total      61238.779  9467
       ---------------------------------------------------------
```

```
                        Parameter Estimates
       ----------------------------------------------------------------------
          model      Beta  Std. Error  Std. Beta    t     Sig   lower   upper
       ----------------------------------------------------------------------
       (Intercept)  18.799    0.453               41.479  0.000  17.910  19.687
        Compliance  -0.081    0.005     -0.223   -16.272  0.000  -0.090  -0.071
     Communication   0.144    0.007      0.284    22.087  0.000   0.131   0.157
         Cognition  -0.112    0.008     -0.188   -14.152  0.000  -0.127  -0.096
       ----------------------------------------------------------------------
```

Step 4: Compliance + Communication + Cognition + Attitudes

Forward Selection: Step 4

+ Attitudes

```
                    Model Summary
-----------------------------------------------------------
R                   0.327    RMSE              2.404
R-Squared           0.107    Coef. Var        15.273
Adj. R-Squared      0.107    MSE               5.777
Pred R-Squared      0.106    MAE               1.670
-----------------------------------------------------------
 RMSE: Root Mean Square Error
 MSE: Mean Square Error
 MAE: Mean Absolute Error
```

```
                            ANOVA
-----------------------------------------------------------
                 Sum of
                Squares     DF   Mean Square     F      Sig.
-----------------------------------------------------------
Regression     6568.063      4     1642.016   284.218  0.0000
Residual      54670.716   9463        5.777
Total         61238.779   9467
-----------------------------------------------------------
```

```
                        Parameter Estimates
-----------------------------------------------------------------------
      model     Beta   Std. Error   Std. Beta     t      Sig    lower   upper
-----------------------------------------------------------------------
 (Intercept)  13.099      0.571                22.953   0.000  11.980  14.218
  Compliance  -0.121      0.006      -0.335   -22.033   0.000  -0.132  -0.111
Communication  0.153      0.006       0.301    23.657   0.000   0.140   0.165
   Cognition  -0.175      0.009      -0.293   -20.022   0.000  -0.192  -0.158
   Attitudes   0.164      0.010       0.243    16.074   0.000   0.144   0.184
-----------------------------------------------------------------------
```

Click to navigate

Step 5: Compliance + Communication + Cognition + Attitudes + Responsibility

```
Forward Selection: Step 5

+ Responsibility

                    Model Summary
-----------------------------------------------------------
R                   0.329   RMSE            2.402
R-Squared           0.108   Coef. Var      15.264
Adj. R-Squared      0.108   MSE             5.770
Pred R-Squared      0.107   MAE             1.674
-----------------------------------------------------------
 RMSE: Root Mean Square Error
 MSE: Mean Square Error
 MAE: Mean Absolute Error

                       ANOVA
-----------------------------------------------------------
            Sum of
            Squares      DF   Mean Square      F      Sig.
-----------------------------------------------------------
Regression  6638.570      5     1327.714   230.088   0.0000
Residual   54600.209   9462        5.770
Total      61238.779   9467
-----------------------------------------------------------

                    Parameter Estimates
-------------------------------------------------------------------
      model    Beta  Std. Error  Std. Beta      t     Sig   lower   upper
-------------------------------------------------------------------
  (Intercept) 12.347    0.610               20.254  0.000  11.152  13.542
   Compliance -0.118    0.006    -0.326    -21.128  0.000  -0.129  -0.107
Communication  0.145    0.007     0.285     21.110  0.000   0.131   0.158
    Cognition -0.182    0.009    -0.304    -20.314  0.000  -0.199  -0.164
    Attitudes  0.139    0.012     0.205     11.110  0.000   0.114   0.163
Responsibility  0.049    0.014     0.060      3.496  0.000   0.022   0.077
-------------------------------------------------------------------
```

Click to navigate

## Model summary

```
Final Model Output
------------------

                    Model Summary
    -------------------------------------------------
    R                0.329     RMSE           2.402
    R-Squared        0.108     Coef. Var     15.264
    Adj. R-Squared   0.108     MSE            5.770
    Pred R-Squared   0.107     MAE            1.674
    -------------------------------------------------
     RMSE: Root Mean Square Error
     MSE: Mean Square Error
     MAE: Mean Absolute Error
```

```
                            ANOVA
    -----------------------------------------------------------
                 Sum of
                 Squares      DF    Mean Square     F      Sig.
    -----------------------------------------------------------
    Regression   6638.570      5     1327.714   230.088  0.0000
    Residual    54600.209   9462        5.770
    Total       61238.779   9467
    -----------------------------------------------------------
```

```
                       Parameter Estimates
    ------------------------------------------------------------------
        model      Beta   Std. Error  Std. Beta    t     Sig   lower   upper
    ------------------------------------------------------------------
     (Intercept)  12.347    0.610               20.254  0.000  11.152  13.542
      Compliance  -0.118    0.006      -0.326   -21.128  0.000  -0.129  -0.107
   Communication   0.145    0.007       0.285    21.110  0.000   0.131   0.158
       Cognition  -0.182    0.009      -0.304   -20.314  0.000  -0.199  -0.164
       Attitudes   0.139    0.012       0.205    11.110  0.000   0.114   0.163
  Responsibility   0.049    0.014       0.060     3.496  0.000   0.022   0.077
    ------------------------------------------------------------------
```

Click to navigate

Introduction
Literature review
Research questions
Antecedents of behaviour
Research model
Methodology
Results
Discussion
Conclusion
Limitations
References
Appendix

```
                        Selection Summary
-----------------------------------------------------------------
            Variable              Adj.
Step        Entered      R-Square  R-Square    C(p)       AIC        RMSE
-----------------------------------------------------------------
   1     Compliance       0.0294    0.0293   834.8606  44267.5008   2.5058
   2     Communication    0.0635    0.0633   475.8123  43931.6223   2.4616
   3     Cognition        0.0829    0.0826   271.8777  43735.3619   2.4361
   4     Attitudes        0.1073    0.1069    15.2179  43482.3109   2.4036
   5     Responsibility   0.1084    0.1079     5.0005  43472.0924   2.4022
```

## 12.2.2.3   Assumption testing Security Awareness and Knowledge

**Number of records:** 9208

### Variance Inflation Factor (VIF)

```
`Passwords & Authentication`      `Email Security`        `Internet Use`       `Social Media`
            1.035892                   1.067484              1.036430             1.021362
   `Mobile Devices`      `Incident Reporting`      `Security Awareness`
        1.085669               1.047093                  1.032841
```

### Condition index

| | Eigenvalue | Condition Index | intercept | `Passwords & Authentication` | `Email Security` | `Internet Use` | `Social Media` | `Mobile Devices` | `Incident Reporting` | `Security Awareness` |
|---|---|---|---|---|---|---|---|---|---|---|---|
| 1 | 7.16309793 | 1.000000 | 6.007626e-04 | 3.613511e-03 | 0.001768248 | 0.001742024 | 0.001922924 | 0.002443150 | 0.0034993348 | 0.0017761662 |
| 2 | 0.20986375 | 5.842271 | 1.210397e-04 | 7.505069e-01 | 0.001337250 | 0.001631605 | 0.002050773 | 0.001131817 | 0.3005874615 | 0.0004485171 |
| 3 | 0.18860469 | 6.162745 | 3.231660e-03 | 2.232667e-01 | 0.010539214 | 0.028468517 | 0.032746139 | 0.031393587 | 0.6778911374 | 0.0168733641 |
| 4 | 0.12503086 | 7.569058 | 3.680183e-03 | 5.469516e-07 | 0.002221154 | 0.005118433 | 0.104325551 | 0.879816622 | 0.0001236328 | 0.0605770277 |
| 5 | 0.10365844 | 8.312815 | 2.789904e-04 | 1.248351e-03 | 0.022917271 | 0.214975691 | 0.657081723 | 0.032135749 | 0.0008623541 | 0.0963157445 |
| 6 | 0.09421034 | 8.719692 | 4.669171e-05 | 2.900430e-03 | 0.001642613 | 0.437542211 | 0.016504805 | 0.022522676 | 0.0006387981 | 0.5840636263 |
| 7 | 0.08905308 | 8.968627 | 3.044708e-06 | 1.241978e-03 | 0.837178043 | 0.136591244 | 0.016044093 | 0.016617454 | 0.0059705613 | 0.0750403819 |
| 8 | 0.02648092 | 16.446896 | 9.920376e-01 | 1.722165e-02 | 0.122456111 | 0.173904051 | 0.169323992 | 0.013938944 | 0.0104267201 | 0.1649051721 |
| > | | | | | | | | | | |

## Partial regression plots

Added-Variable Plots

## Description of the unusual points and significant outliers

The Studentized deleted residuals (SDR) > 3 are 158

The items that exceed the leverage point value (LEV) of 0.20 are 0

The items that exceed Cook's Distance Value (COO) of 1 are 0

## Histogram of standardized residuals

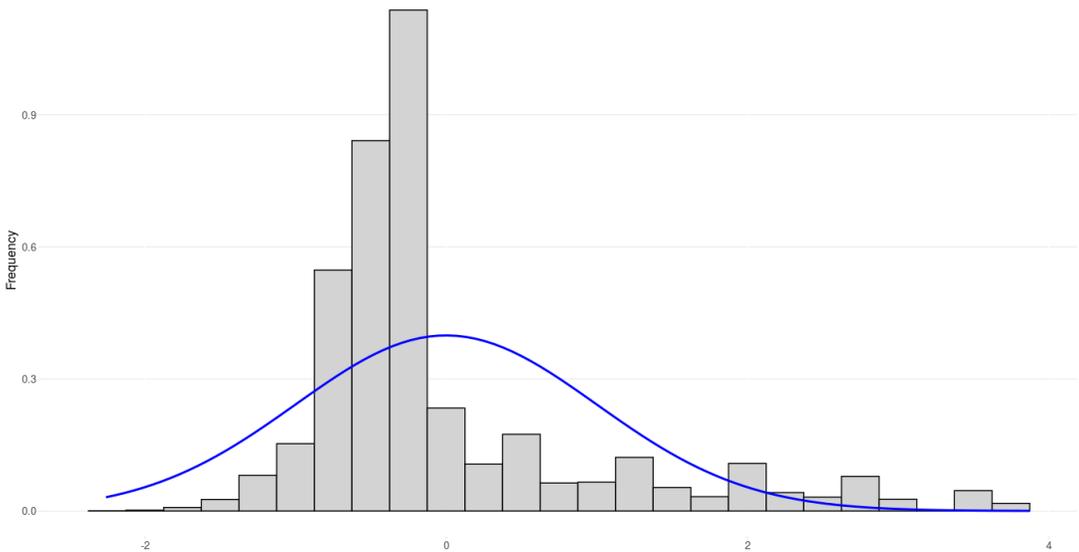

## Normal Q-Q plot

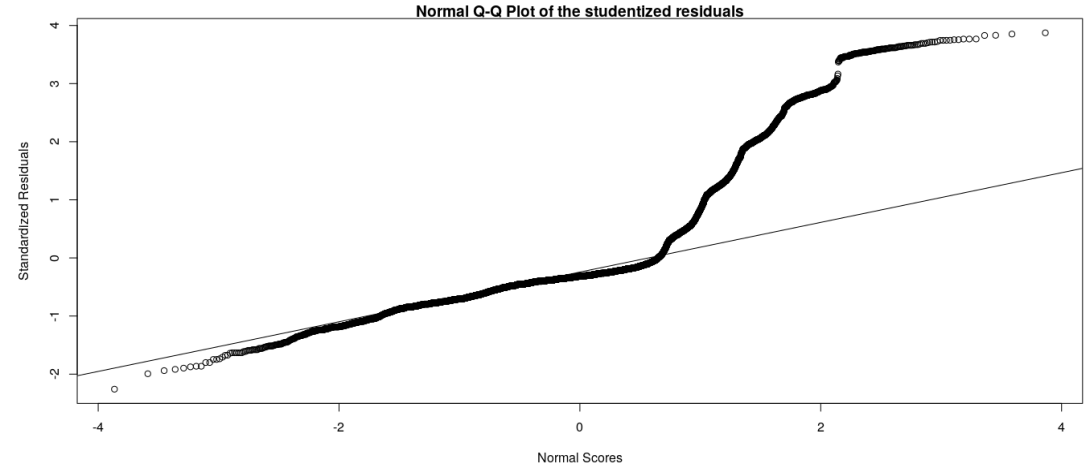

## 12.2.2.4 Multiple regression Security Awareness and Knowledge

Step 1: Mobile Devices

```
Forward Selection Method
---------------------------

Candidate Terms:

1. `Passwords & Authentication`
2. `Email Security`
3. `Internet Use`
4. `Social Media`
5. `Mobile Devices`
6. `Incident Reporting`
7. `Security Awareness`

We are selecting variables based on p value...

Forward Selection: Step 1

+ `Mobile Devices`

                    Model Summary
-----------------------------------------------------
R                  0.098    RMSE           2.557
R-Squared          0.010    Coef. Var     16.220
Adj. R-Squared     0.009    MSE            6.536
Pred R-Squared     0.009    MAE            1.832
-----------------------------------------------------
 RMSE: Root Mean Square Error
 MSE: Mean Square Error
 MAE: Mean Absolute Error

                       ANOVA
---------------------------------------------------------------
            Sum of
            Squares     DF    Mean Square     F       Sig.
---------------------------------------------------------------
Regression   576.810     1       576.810    88.25    0.0000
Residual   59713.551  9136         6.536
Total      60290.360  9137
---------------------------------------------------------------

                   Parameter Estimates
-------------------------------------------------------------------------
         model     Beta   Std. Error  Std. Beta     t       Sig   lower   upper
-------------------------------------------------------------------------
  (Intercept)    15.181     0.067                225.366   0.000  15.049  15.313
`Mobile Devices`  0.010     0.001      0.098      9.394    0.000   0.008   0.012
-------------------------------------------------------------------------
```

**Click to navigate**

Introduction
Literature review
Research questions
Antecedents of behaviour
Research model
Methodology
Results
Discussion
Conclusion
Limitations
References
Appendix

## Step 2: Mobile Devices + Email Security

```
Forward Selection: Step 2

+ `Email Security`

                    Model Summary
-----------------------------------------------------------
R                   0.116     RMSE              2.553
R-Squared           0.013     Coef. Var        16.196
Adj. R-Squared      0.013     MSE               6.516
Pred R-Squared      0.013     MAE               1.826
-----------------------------------------------------------
 RMSE: Root Mean Square Error
 MSE: Mean Square Error
 MAE: Mean Absolute Error

                          ANOVA
-----------------------------------------------------------
              Sum of
              Squares     DF    Mean Square    F       Sig.
-----------------------------------------------------------
Regression    808.248      2       404.124   62.019   0.0000
Residual    59251.540   9093         6.516
Total       60059.788   9095
-----------------------------------------------------------

                      Parameter Estimates
------------------------------------------------------------------------
       model      Beta   Std. Error  Std. Beta      t      Sig    lower    upper
------------------------------------------------------------------------
   (Intercept)   14.773     0.097               153.019   0.000   14.584   14.963
`Mobile Devices`  0.009     0.001      0.089      8.397   0.000    0.007    0.011
`Email Security`  0.006     0.001      0.061      5.740   0.000    0.004    0.008
------------------------------------------------------------------------
```

## Step 3: Mobile Devices + Email Security + Security Awareness

```
Forward Selection: Step 3

+ `Security Awareness`

                    Model Summary
-----------------------------------------------------------
R                 0.122    RMSE              2.554
R-Squared         0.015    Coef. Var        16.199
Adj. R-Squared    0.015    MSE               6.523
Pred R-Squared    0.014    MAE               1.830
-----------------------------------------------------------
 RMSE: Root Mean Square Error
 MSE: Mean Square Error
 MAE: Mean Absolute Error

                        ANOVA
-----------------------------------------------------------
                Sum of
               Squares      DF    Mean Square    F       Sig.
-----------------------------------------------------------
Regression     896.050       3      298.683    45.788   0.0000
Residual     58832.935    9019        6.523
Total        59728.985    9022
-----------------------------------------------------------

                   Parameter Estimates
-----------------------------------------------------------------------
           model     Beta   Std. Error  Std. Beta     t      Sig    lower    upper
-----------------------------------------------------------------------
     (Intercept)   14.502     0.119               122.335   0.000   14.270   14.735
 `Mobile Devices`   0.008     0.001      0.084      7.884   0.000    0.006    0.010
 `Email Security`   0.005     0.001      0.054      5.095   0.000    0.003    0.008
`Security Awareness` 0.004    0.001      0.045      4.301   0.000    0.002    0.006
-----------------------------------------------------------------------
```

## Step 4: Mobile Devices + Email Security + Security Awareness + Internet Use

```
Forward Selection: Step 4

+ `Internet Use`

                    Model Summary
-----------------------------------------------------------
R                 0.128      RMSE              2.553
R-Squared         0.016      Coef. Var        16.191
Adj. R-Squared    0.016      MSE               6.516
Pred R-Squared    0.015      MAE               1.827
-----------------------------------------------------------
 RMSE: Root Mean Square Error
 MSE: Mean Square Error
 MAE: Mean Absolute Error

                       ANOVA
-----------------------------------------------------------
               Sum of
              Squares     DF    Mean Square     F       Sig.
-----------------------------------------------------------
Regression    980.782      4       245.196    37.627   0.0000
Residual    58706.244   9009         6.516
Total       59687.026   9013
-----------------------------------------------------------

                   Parameter Estimates
----------------------------------------------------------------------------------
          model      Beta   Std. Error   Std. Beta      t      Sig    lower    upper
----------------------------------------------------------------------------------
    (Intercept)    14.277      0.134                 106.774   0.000  14.015   14.539
`Mobile Devices`    0.008      0.001       0.079      7.346    0.000   0.006    0.010
 `Email Security`   0.005      0.001       0.051      4.798    0.000   0.003    0.007
`Security Awareness` 0.004     0.001       0.043      4.068    0.000   0.002    0.005
    `Internet Use`  0.004      0.001       0.039      3.664    0.000   0.002    0.006
----------------------------------------------------------------------------------
```

# Step 5: Mobile Devices + Email Security + Security Awareness + Internet Use + Passwords & Authentication

```
Forward Selection: Step 5

+ `Passwords & Authentication`

                    Model Summary
-------------------------------------------------------
R                    0.129    RMSE            2.550
R-Squared            0.017    Coef. Var      16.176
Adj. R-Squared       0.016    MSE             6.502
Pred R-Squared       0.015    MAE             1.825
-------------------------------------------------------
 RMSE: Root Mean Square Error
 MSE: Mean Square Error
 MAE: Mean Absolute Error

                    ANOVA
---------------------------------------------------------------
            Sum of
            Squares      DF    Mean Square      F        Sig.
---------------------------------------------------------------
Regression  992.718       5      198.544     30.534    0.0000
Residual   58489.546   8995        6.502
Total      59482.264   9000
---------------------------------------------------------------

                         Parameter Estimates
------------------------------------------------------------------------------
           model       Beta   Std. Error   Std. Beta     t      Sig    lower    upper
------------------------------------------------------------------------------
       (Intercept)    14.230     0.137                 104.013  0.000  13.962   14.498
    `Mobile Devices`   0.008     0.001      0.077       7.162   0.000   0.006    0.010
    `Email Security`   0.005     0.001      0.049       4.537   0.000   0.003    0.007
 `Security Awareness`  0.004     0.001      0.042       3.923   0.000   0.002    0.005
      `Internet Use`   0.004     0.001      0.039       3.654   0.000   0.002    0.006
`Passwords & Authentication` 0.002 0.001    0.018       1.721   0.085   0.000    0.003
------------------------------------------------------------------------------
```

## Model summary

No more variables to be added.

Variables Entered:

+ `Mobile Devices`
+ `Email Security`
+ `Security Awareness`
+ `Internet Use`
+ `Passwords & Authentication`

Final Model Output
------------------

```
                    Model Summary
-----------------------------------------------------
R                0.129     RMSE         2.550
R-Squared        0.017     Coef. Var   16.176
Adj. R-Squared   0.016     MSE          6.502
Pred R-Squared   0.015     MAE          1.825
-----------------------------------------------------
 RMSE: Root Mean Square Error
 MSE: Mean Square Error
 MAE: Mean Absolute Error
```

```
                    ANOVA
------------------------------------------------------------------
             Sum of
             Squares     DF    Mean Square     F        Sig.
------------------------------------------------------------------
Regression   992.718      5      198.544     30.534    0.0000
Residual    58489.546   8995       6.502
Total       59482.264   9000
------------------------------------------------------------------
```

```
                         Parameter Estimates
----------------------------------------------------------------------------
                 model    Beta   Std. Error  Std. Beta    t      Sig    lower    upper
----------------------------------------------------------------------------
            (Intercept)  14.230    0.137                104.013  0.000  13.962   14.498
          `Mobile Devices`  0.008    0.001      0.077     7.162  0.000   0.006    0.010
          `Email Security`  0.005    0.001      0.049     4.537  0.000   0.003    0.007
       `Security Awareness`  0.004    0.001      0.042     3.923  0.000   0.002    0.005
            `Internet Use`  0.004    0.001      0.039     3.654  0.000   0.002    0.006
  `Passwords & Authentication`  0.002  0.001    0.018     1.721  0.085   0.000    0.003
----------------------------------------------------------------------------
```

```
                              Selection Summary
--------------------------------------------------------------------------------
             Variable                       Adj.
Step         Entered           R-Square   R-Square    C(p)       AIC       RMSE
--------------------------------------------------------------------------------
  1    `Mobile Devices`          0.0096     0.0095    68.1448   43091.6197  2.5566
  2    `Email Security`          0.0135     0.0132    40.9466   42866.8428  2.5527
  3    `Security Awareness`      0.0150     0.0147    51.4377   42533.6126  2.5541
  4    `Internet Use`            0.0164     0.0160    42.9140   42482.7612  2.5527
  5    `Passwords & Authentication`  0.0167  0.0161   24.5197   42403.2142  2.5500
--------------------------------------------------------------------------------
```